\documentclass[aps,prb,twocolumn,superscriptaddress,longbibliography]{revtex4-1}
\usepackage{CJK}
\usepackage{amsfonts}
\usepackage{amsmath}
\usepackage{ dsfont }
\usepackage[colorlinks=true,citecolor=blue,linkcolor=blue,breaklinks=true]{hyperref}
\usepackage{amssymb}
\usepackage{xcolor}
\usepackage{xspace}
\usepackage{ragged2e}
\usepackage{relsize}
\usepackage[bottom]{footmisc}
\newcommand{\sgn}{\text{sgn}}

\usepackage{tikz}
\usetikzlibrary{calc}
\newcommand{\nocontentsline}[3]{}
\newcommand{\tocless}[2]{\bgroup\let\addcontentsline=\nocontentsline#1{#2}\egroup}

\newcommand{\be}{\begin{equation}}
\newcommand{\ee}{\end{equation}}
\newcommand{\bea}{\begin{equation} \begin{aligned}}
\newcommand{\eea}{\end{aligned} \end{equation} }
\newcommand{\bi}{\begin{itemize}}
\newcommand{\ei}{\end{itemize}}

\newcommand{\la}{\lambda}
\renewcommand{\be}{\beta}
\newcommand{\al}{\alpha}
\newcommand{\bpm}{\begin{pmatrix}}
\newcommand{\epm}{\end{pmatrix}}
\newcommand{\eps}{\epsilon}

\renewcommand{\th}{\theta}
\newcommand{\lp}{\left(}
\newcommand{\rp}{\right)}
\newcommand{\del}{\partial}

\newcommand{\Tr}{\text{Tr} \ }

\newcommand{\mbf}[1]{\mathbf{#1}}

\newcommand{\aaa}{{\bf {a}}}
\newcommand{\bbb}{{\bf{b}}}
\newcommand{\kk}{{\bf{k}}}
\newcommand{\KK}{{\bf{K}}}
\newcommand{\rr}{{\bf{r}}}
\newcommand{\RR}{{\bf{R}}}
\newcommand{\qq}{{\bf{q}}}
\newcommand{\QQ}{{\bf{Q}}}
\newcommand{\GG}{{\bf{G}}}

\usepackage{color,graphicx}
\usepackage{mathrsfs}
\usepackage{verbatim}
\usepackage{amsmath}
\usepackage{amssymb}
\usepackage{wasysym}
\usepackage[caption=false]{subfig}
\usepackage{url}
\usepackage{bbold}
\usepackage{slashed}
\usepackage{epstopdf}
\usepackage{braket}
\usepackage{float}
\usepackage[percent]{overpic}

\DeclareRobustCommand{\Sec}[1]{Sec.~\ref{#1}}
\DeclareRobustCommand{\Secs}[2]{Secs.~\ref{#1} and \ref{#2}}

\DeclareRobustCommand{\App}[1]{App.~\ref{#1}}

\DeclareRobustCommand{\Fig}[1]{Fig.~\ref{#1}}

\DeclareRobustCommand{\Eq}[1]{Eq.~(\ref{#1})}
\DeclareRobustCommand{\Eqs}[2]{Eqs.~(\ref{#1}) and (\ref{#2})}
\DeclareRobustCommand{\Ref}[1]{Ref.~\cite{#1}}
\DeclareRobustCommand{\Refs}[2]{Refs.~\cite{#1, #2}}
\DeclareMathAlphabet\mathbfcal{OMS}{cmsy}{b}{n}

\RequirePackage[normalem]{ulem} 
\RequirePackage{color}\definecolor{RED}{rgb}{1,0,0}\definecolor{BLUE}{rgb}{0,0,1} 

\begin{document}

\title{Magnetic Bloch Theorem and Reentrant Flat Bands in Twisted Bilayer Graphene  at $2\pi$ Flux }

\author{Jonah Herzog-Arbeitman}
\affiliation{Department of Physics, Princeton University, Princeton, NJ 08544}
\author{Aaron Chew}
\affiliation{Department of Physics, Princeton University, Princeton, NJ 08544}
\author{B. Andrei Bernevig}
\affiliation{Department of Physics, Princeton University, Princeton, NJ 08544}
\affiliation{Donostia International Physics Center, P. Manuel de Lardizabal 4, 20018
Donostia-San Sebastian, Spain}
\affiliation{IKERBASQUE, Basque Foundation for Science, Bilbao, Spain}

\date{\today}

\begin{abstract}
Bloch's theorem is the centerpiece of topological band theory, which itself has defined an era of quantum materials research. However, Bloch's theorem is broken by a perpendicular magnetic field, making it difficult to study topological systems in strong flux. For the first time, moir\'e materials have made this problem experimentally relevant, and its solution is the focus of this work. We construct gauge-invariant irreps of the magnetic translation group at $2\pi$ flux on infinite boundary conditions,  allowing us to give analytical expressions in terms of the Siegel theta function for the magnetic Bloch Hamiltonian, non-Abelian Wilson loop, and many-body form factors. We illustrate our formalism using a simple square lattice model and the Bistritzer-MacDonald Hamiltonian of twisted bilayer graphene, obtaining reentrant ground states at $2\pi$ flux under the Coulomb interaction. 
\end{abstract}
\maketitle

\section{Introduction}
\label{sec:intro}

Motivated by developments in the fabrication of moir\'e materials with greatly enlarged unit cells \cite{andrei2021marvels,2018Natur.556...80C, Cao2018UnconventionalSI, Kim3364, balents2020superconductivity,liu2021orbital, Chu_2020,2021arXiv210511883Z}, this work revisits the solution of continuum Hamiltonians in strong flux from the modern perspective of topological band theory. The essential difficulty of the problem was identified by Zak who demonstrated that translations do not commute in generic magnetic flux and instead form a projective representation of the translation group  \cite{PhysRev.134.A1602}. As such, Bloch's theorem does not apply. The result is a fractal energy spectrum as a function of magnetic flux known as the Hofstadter butterfly \cite{PhysRevB.14.2239,PhysRevB.14.2239,PhysRevLett.86.147,Hunt1427,moirehofexp}. In this work, we present a new formalism to obtain the exact band structure and topology of a continuum Hamiltonian when the flux through a single unit cell is $2\pi$. At $2\pi$ flux, corresponding to $\sim 25$T in magic angle twisted bilayer graphene (TBG) \cite{2011PNAS..10812233B}, the magnetic translation group commutes due to the Aharonov-Bohm effect, allowing reentrant Hofstadter phases \cite{PhysRev.134.A1602,PhysRevB.14.2239}. Although methods already exist to study the spectrum in arbitrary magnetic fields \cite{PhysRev.134.A1607,BROWN1969313,Streda_1982,1978PSSBR..88..757W,PhysRevB.76.115419,RevModPhys.82.1959,PhysRevB.52.14755,2011PhRvB..84c5440B,toke2005fractional,2011PhRvB..83k5129G,PhysRevB.84.035440,Li1997ThermalPO,Crosse_2020,2021PhRvB.103L1405L,PhysRevB.46.12606,2022PhyE..14215311A}, they are unsuitable for determining the topology and dominant many-body effects essential to moir\'e physics. Our formalism is manifestly gauge-invariant, leading to analytical expressions for the magnetic Bloch Hamiltonian, non-Abelian Berry connection, and many-body form factors. Importantly, numerical implementation is also straightforward, and we are able to study reentrant phases, which have recently become of interest \cite{2021arXiv210501243C,2021arXiv210312083C}, without using simplified models. The methods detailed here were used to study reentrant correlated insulators \cite{2021arXiv211111434H} in twisted bilayer graphene, which have been observed in experiment \cite{2021arXiv211111341D}.

We begin with a general discussion of the symmetry operators in \Sec{sec:setup} which are used to construct gauge-invariant magnetic translation group irreps on infinite boundary conditions in \Sec{sec:MTGirr}. A discussion of the Siegel theta function  \cite{NIST:DLMF,gunning2012riemann,2020JHEP...10..187M}, a multi-dimensional generalization of the Jacobi theta function which appears in our states, may be found in \App{app:siegel}. We provide a general expression for the magnetic Bloch Hamiltonian in \Sec{sec:matel} and compute the band structure for a square lattice model. Then in \Sec{sec:berry}, we define the Berry connection which receives two new magnetic contributions (Abelian and non-Abelian), and we discuss the topological transition between the strong flux or Landau level regime where the kinetic energy dominates and the crystalline regime where the potential dominates. In \Sec{sec:formfactor}, we give convenient expressions for the form factors of generic density-density interactions. Finally in \Secs{sec:BMmodel}{sec:TBGcoulomb}, we study the Bistritzer-MacDonald (BM) Hamiltonian \cite{2011PNAS..10812233B} of twisted bilayer graphene which reaches $2\pi$ flux at $\sim 25$T.  We discuss the symmetries of TBG at $2\pi$ and find that the degree of particle-hole breaking strongly determines the topology of the flat bands, which realize a decomposable elementary band representation \cite{2018PhRvL.120z6401C}. 

We note that the Hofstadter spectrum of tight-binding models under the Peierls substitution \cite{1933ZPhy...80..763P} is periodic in flux with the period equal to an integer multiple of $2\pi$ depending on the orbitals \cite{PhysRevLett.125.236804}. This is because gauge fields on the lattice are compact. Such systems differ from the continuum models considered here where there is no exact periodicity  in $\phi$ (though see \Ref{2022arXiv220408087K} for a discussion of approximate periodicity) and we are not reliant on the validity of the Peierls approximation. Notably, the spectrum and topology of the BM model we obtain at $2\pi$ flux compares well to tight-binding calculations of twisted bilayer graphene at a small commensurate angle \cite{2022arXiv220113062G}.

\section{Symmetry Algebra}
\label{sec:setup}

We consider a two-dimensional Hamiltonian minimally coupled to a background gauge field $\mbf{A}(\mbf{r})$ in the form
\bea
\label{eq:Hkinpot}
H &= h(- i \pmb{\nabla} - e \mbf{A}) + U(\mbf{r}), \qquad \pmb{\nabla} \times \mbf{A} = B > 0 \\
\eea
where we study $h(p) = p^2/2m$ and $h(p) = v_F (p_x \sigma_x + p_y \sigma_y)$ and set $\hbar = 1$. Here $e>0$ is the electron charge, the magnetic field $B$ is perpendicular to the plane, and the cross product is a scalar in two dimensions. We neglect the Zeeman coupling, but it is trivial to add. The potential $U(\mbf{r})$ is periodic: $U(\mbf{r}) = U(\mbf{r}+ \mbf{R})$ where $\mbf{R}$ is on the Bravais lattice with basis vectors $\mbf{a}_1, \mbf{a}_2$ oriented so $\mbf{a}_1 \times \mbf{a}_2 = \Omega > 0$.\footnote{In applications to TBG, $\mbf{a}_i$ will be moir\'e lattice vectors.} The reciprocal lattice is spanned by the vectors $2\pi \mbf{b}_i$ satisfying $\mbf{a}_i \cdot \mbf{b}_j = \delta_{ij}$. The magnetic flux is $\phi = e B \Omega$ which is dimensionless (setting $\hbar = 1$).

In absence of a periodic potential, the Hamiltonian $h(p)$ in flux can be solved in terms of Landau levels by introducing an oscillator algebra. The algebra is formed from the canonical momentum $\pmb{\pi} = - i \pmb{\nabla} - e \mbf{A}$ obeying
\bea
\null [\pi_\mu, \pi_\nu] = i e (\del_\mu A_\nu - \del_\nu A_\mu) = i e B\eps_{\mu \nu}
\eea
where throughout this section, greek letters correspond to cartesian indices, e.g. $\mu,\nu \in \{x,y\}$, and we sum over repeated indices. We define the ladder operators $[a,a^\dag] = 1$ by
\bea
\label{eq:aop}
a &= \frac{\pi_x + i \pi_y}{\sqrt{2eB}}  , a^\dag = \frac{\pi_x - i \pi_y}{\sqrt{2eB}} \ . \\
\eea
In the simplest case of $h(p) = p^2/2m = eB(a^\dag a+ 1/2)$ in magnetic field, the eigenstates are Landau levels given by powers of $a^\dag$. The macroscopic degeneracy of the Landau levels is accounted for by the guiding center momenta $Q_\mu$. The gauge-invariant definition is
\bea
\label{eq:Qoperatorm}
Q_\mu &= \pi_\mu - e B \eps_{\mu\nu} x_\nu = - i \del_\mu - e (A_\mu + B \eps_{\mu \nu} x_\nu) \ . \\
\eea
The guiding center operators commute with the canonical momenta and obey
\bea
\label{eq:abcom}
\null [Q_\mu,\pi_\nu] &= [\pi_\mu - eB \eps_{\mu \rho} x_\rho,\pi_\nu] = ieB\eps_{\mu \nu}  - i e B \eps_{\mu \nu} = 0  \\
[Q_\mu,Q_\nu] &= [\pi_\mu - eB \eps_{\mu \rho} x_\rho ,\pi_\nu - eB\eps_{\nu \sigma} x_\sigma] = - i eB \eps_{\mu \nu} \ . \\
\eea
The guiding centers form a separate oscillator system with $[b,b^\dag] = 1$ defined by (see \App{eq:appopdef})
\bea
\label{eq:b}
b &= \frac{(\mbf{a}_1 - i \mbf{a}_2) \cdot  \mbf{Q}}{\sqrt{2\phi}}, \quad b^\dag = \frac{(\mbf{a}_1   + i \mbf{a}_2) \cdot  \mbf{Q}}{\sqrt{2\phi}},
\eea
Note that the $b$-oscillators commute with the $a$-oscillators by \Eq{eq:abcom}. Comparing \Eq{eq:b} and \Eq{eq:aop}, we see that the $a, a^\dag$ operators are defined using cartesian variables while the $b, b^\dag$ operators are defined using the lattice vectors. This is because the $a,a^\dag$ operators are used to build the continuum kinetic term which has $SO(2)$ rotation symmetry, while the $b,b^\dag$ operators will be used to construct states that respect the lattice periodicity.

The kinetic term $h(\pmb{\pi})$, which is built out of $a$ and $a^\dag$ operators, commutes with $b,b^\dag$. Hence without a potential, every Landau level eigenstate has an infinite degeneracy (on infinite boundary conditions) from acting repeatedly with $b^\dag$ because $[h(\pmb{\pi}),\mbf{Q}] = 0$. A periodic potential breaks this degeneracy. However, we observe that the magnetic translation operators
\bea
\label{eq:Tai}
T_{\mbf{a}_i} &= \exp \lp i \mbf{a}_i \cdot \mbf{Q} \rp
\eea
formed from the $Q_i$ algebra commute with a periodic potential. Using the Baker-Campbell-Hausdorff  (BCH) formula, we check
\bea
e^{ i \mbf{a}_i \cdot \mbf{Q} } U(\mbf{r}) e^{ -i \mbf{a}_i \cdot \mbf{Q} } &= \sum_{n=0}^{\infty} \frac{1}{n!} \Big([i \mbf{a}_i \cdot \mbf{Q}, \Big)^n U(\mbf{r}) ] \\
&= \sum_{n=0}^{\infty} \frac{1}{n!} \big([  i \mbf{a}_i \cdot (-i\pmb{\nabla}), \big)^n U(\mbf{r}) ] \\
&=e^{\mbf{a}_i \cdot \pmb{\nabla}} U(\mbf{r}) e^{- \mbf{a}_i \cdot \pmb{\nabla} } \\
&= U(\mbf{r} + \mbf{a}_i) = U(\mbf{r}) \ .
\eea
where the nested commutator $([X,)^n Y] = [X,[X,\dots,Y]]$ has $n$ factors of $X$ and in the last line we used the lattice periodicity. This is sufficient to prove that $T_{\mbf{a}_i}$ commutes with the whole Hamiltonian $H$ (kinetic plus potential) because $[\mbf{Q}, \pmb{\pi}] = 0$ and the kinetic term only contains $\pmb{\pi}$ operators. Note that $[H,\mbf{Q}] \neq0$ but $[H, e^{i \mbf{a}_i \cdot \mbf{Q}}] = 0$ for a periodic potential. The algebra of the $T_{\mbf{a}_i}$ operators is derived from the BCH formula by
\bea
\label{eq:projrep}
T_{\mbf{a}_1}  T_{\mbf{a}_2}   &= \exp \Big( [i \mbf{a}_1 \cdot \mbf{Q} , i \mbf{a}_2 \cdot \mbf{Q} ]\Big) T_{\mbf{a}_2}  T_{\mbf{a}_1}   = e^{i \phi} T_{\mbf{a}_2}  T_{\mbf{a}_1} \ . \\
\eea
 \Eq{eq:projrep} shows that the magnetic translation operators define a projective representation of the translation group. For generic $\phi \in \mathbb{R}$, $T_{\mbf{a}_1}$ and $T_{\mbf{a}_2}$ do not commute and there is no band structure. The cascade of band splitting that occurs as the flux is increased leads to the fractal Hofstadter energy spectrum \cite{PhysRevB.14.2239}. The $a^\dag$ and $b^\dag$ operators form a basis of the Hilbert space which is used to solve continuum Hamiltonians in terms of degenerate Landau levels. In \Sec{sec:MTGirr}, we will produce basis states which are magnetic translation operator irreps by recombining the $b^\dag$ basis.

So far, the flux $\phi = eB \Omega$ has been unrestricted. In the following sections, we fix $\phi = 2\pi$ where \Eq{eq:projrep} shows that the magnetic translation operators commute. This is an intrinsically quantum mechanical effect because $2\pi$ flux corresponds to one flux quantum $h/e$ piercing each unit cell where $h$ is Planck's constant. In a conventional crystal where the unit cell area is on the order of 10\AA$^2$, $\phi = 2\pi$ corresponds to extreme fields between $10^4$T and $10^5$T. However, moir\'e materials have an effective unit cell which is larger by a factor of $\th^{-2}$ where $\th$ is the twist angle. For angles near $1^\circ$, the moir\'e unit cell is enlarged by a factor of $3000$ allowing $\sim25$T fields to probe the Hofstadter regime.

\section{Magnetic Translation Group Irreps}
\label{sec:MTGirr}

In this section, we construct wavefunctions which are irreps of the magnetic translation group at $\phi = 2\pi$ on infinite boundary conditions in a gauge-invariant manner. These states are the building blocks of all subsequent calculations. To motivate them, we first revisit Bloch's theorem in zero flux.

\subsection{Bloch's Theorem}

Let us briefly recall the traditional Bloch theorem. The translation group in zero flux on infinite boundary conditions is isomorphic to the infinite group $\mathds{Z}^2$ which is Abelian. Hence its irreducible representations (irreps) are all one-dimensional. They are eigenstates of the translation operators labeled by a crystal momentum $\mbf{k} = k_1 \mbf{b}_1 + k_2 \mbf{b}_2$ where $k_i \in (-\pi,\pi)$ defines the Brillouin zone (BZ). It is trivial to construct the first-quantized eigenstates of the zero-flux translation operators $T_\mbf{R} = e^{\mbf{R} \cdot \pmb{\nabla}}$ with eigenvalue $e^{i \mbf{k} \cdot \mbf{R}}$ where $\mbf{R} = R_1 \mbf{a}_1 + R_2 \mbf{a}_2, \ R_i \in \mathds{Z}$: the functions $\psi^{\phi=0}_{\mbf{k},n}(\mbf{r}) = e^{i \mbf{k} \cdot \mbf{r}}u_{\mbf{k},n}(\mbf{r})$ are momentum eigenstates for any periodic function $u_{\mbf{k},n}(\mbf{r}) = u_{\mbf{k},n}(\mbf{r}+ \mbf{a}_i)$ which we normalize to
\bea
\int_\Omega d^2x \, u^*_{\mbf{k},m}(\mbf{x})  u_{\mbf{k},n}(\mbf{x}) = \delta_{mn}
\eea
by integrating over the unit cell $\Omega$. Hence the functions $u^*_{\mbf{k},m}(\mbf{x})$ form a complete basis of periodic functions on the unit cell at each $\mbf{k}$. In this case, the Bloch waves $\psi^{\phi=0}_{\mbf{k},n}(\mbf{r})$ normalized on infinite boundary conditions as
\bea
& \int d^2r \, \psi^{\phi=0}_{\mbf{k},m}(\mbf{r})^* \psi^{\phi=0}_{\mbf{k}',n}(\mbf{r}) \\
& = \sum_\mbf{R} e^{i (\mbf{k}'-\mbf{k}) \cdot \mbf{R} }\int_\Omega d^2x \, e^{i (\mbf{k}'-\mbf{k}) \cdot \mbf{x}}u^*_{\mbf{k},m}(\mbf{x})  u_{\mbf{k}',n}(\mbf{x})  \\
& =(2\pi)^2 \delta(\mbf{k}-\mbf{k}') \int_\Omega d^2x \, u^*_{\mbf{k},m}(\mbf{x})  u_{\mbf{k},n}(\mbf{x}) \\
& =(2\pi)^2 \delta_{mn} \delta(\mbf{k}-\mbf{k}')   \\
\eea
using the identity $(2\pi)^2 \delta(\mbf{k}-\mbf{k}') = \sum_{\mbf{R}} e^{i \mbf{R} \cdot (\mbf{k}-\mbf{k}')}$ with $\mbf{k} - \mbf{k}' \in BZ$. The periodic functions $u_{\mbf{k},n}(\mbf{r})$ form an orthonormal basis of states within a single unit cell, and can be chosen as the eigenstates of the effective Bloch Hamiltonian $e^{-i \mbf{k} \cdot \mbf{r}} H e^{i \mbf{k} \cdot \mbf{r}}$ which is a function of $\mbf{k}$. Note that there are an infinite number of eigenstates $u_{\mbf{k},n}(\mbf{r})$ because the Hilbert space is infinite dimensional. At each $\mbf{k} \in BZ$, $n = 1,2,\dots$ indexes Bloch waves of increasingly high energy. This contrasts the tight-binding approximation where only a finite number of Bloch waves are kept and the local Hilbert space is finite dimensional.

To parallel our construction at $\phi = 2\pi$ in \Sec{eq:MBT2pi}, we now give an alternative representation for the Bloch waves. We introduce the Wannier functions
\bea
\label{eq:wannierdef}
w^{\phi=0}_{\mbf{R},n}(\mbf{r}) \equiv T_\mbf{R} w^{\phi=0}_{n}(\mbf{r}) &= \int \frac{d^2k}{(2\pi)^2} e^{i \mbf{k} \cdot (\mbf{r}+\mbf{R})} u_{\mbf{k},n}(\mbf{r})
\eea
which, being formed from states at different $\mbf{k}$, are generally not energy or momentum eigenstates. Instead the Wannier functions $w^{\phi=0}_{\mbf{R}n}(\mbf{r})$ form a local basis of the Hilbert space which is complementary to the entirely delocalized Bloch wave basis (see \Ref{2012RvMP...84.1419M} for a thorough discussion). A Bloch state can be built from the Wannier functions according to
\bea
\label{eq:blochwan}
\psi^{\phi=0}_{\mbf{k},n}(\mbf{r}) &= \sum_{\mbf{R}} e^{- i \mbf{k} \cdot \mbf{R}} T_\mbf{R} w^{\phi=0}_{n}(\mbf{r}) \\
\eea
which can be proven directly from \Eq{eq:wannierdef}:
\bea
\sum_{\mbf{R}} e^{- i \mbf{k} \cdot \mbf{R}} T_\mbf{R} w^{\phi=0}_{n}(\mbf{r})&= \int \frac{d^2k'}{(2\pi)^2}  \sum_{\mbf{R}} e^{i(\mbf{k} -\mbf{k}') \cdot \mbf{R}}  e^{i \mbf{k}' \cdot \mbf{r}} u_{\mbf{k}',n}(\mbf{r}) \\
&= \psi^{\phi=0}_{\mbf{k},n}(\mbf{r})  \\
\eea
Note that the construction in \Eq{eq:blochwan} is guaranteed to be a momentum eigenstate (if not an energy eigenstate) for any $w^{\phi=0}_n(\mbf{r})$, not necessarily a Wannier function. We now make use of this observation to produce magnetic translation group eigenstates at $\phi = 2\pi$.

\subsection{Magnetic Bloch Theorem at $\phi = 2\pi$}
\label{eq:MBT2pi}

At $2\pi$ flux, the magnetic translation group commutes (see \Eq{eq:projrep}) and is isomorphic to $\mathds{Z}^2$. Hence its irreps are again labeled by $\mbf{k} = k_1 \bbb_1 + k_2 \bbb_2 \in BZ$ which we refer to as the momentum. This quantum number is essential to determining the topology of the Hamiltonian. This differentiates our approach from the open momentum space diagonalization technique developed in \Ref{2021PhRvB.103L1405L} which does not make use of the momentum, but achieves a sparse matrix representation of the Hamiltonian at all fluxes.

To derive a magnetic Bloch Hamiltonian in each $\mbf{k}$ sector, we must construct eigenstates $\psi_{\mbf{k},n}(\mbf{r})$ of the magnetic translation operators. We will do so on infinite boundary conditions so that $\mbf{k}$ is continuous. Using the explicit operators in \Eq{eq:Tai}, there is a natural construction by summing over the infinite Bravais lattice $\mbf{R}$.\footnote{One can also construct states on a finite lattice in the same way. However, in this case one cannot perform the normalization sum in \Eq{eq:Ncalsieg} analytically. Hence we only focus on the infinite case in this work.} Noting that $\mbf{R} \cdot \mbf{b}_i  \in \mathds{Z}$, we define the states
\bea
\label{eq:basis}
\psi_{\mbf{k},n}(\mbf{r}) = \frac{1}{\sqrt{\mathcal{N}(\mbf{k})}} \sum_{\mbf{R}} e^{-i \mbf{k} \cdot \mbf{R}} T_{\mbf{a}_1}^{\mbf{R} \cdot \mbf{b}_1} T_{\mbf{a}_2}^{\mbf{R} \cdot \mbf{b}_2} w_n(\mbf{r})
\eea
where $w_n(\mbf{r})$ is a function to be chosen momentarily. Importantly, the states \Eq{eq:basis} take the same form in any gauge. It is direct to check that $T_{\mbf{a}_i} \psi_{\mbf{k},n}(\mbf{r})  = e^{i \mbf{k} \cdot \mbf{a}_i} \psi_{\mbf{k},n}(\mbf{r})$ because $[T_{\mbf{a}_i},T_{\mbf{a}_j}]=0$ at $\phi = 2\pi$. Hence the states $\psi_{\mbf{k},n}$ are orthogonal in $\mbf{k} \in BZ$.
Similar states have been constructed for tight-binding models in \Ref{PhysRevLett.125.236804}. To achieve orthogonality in $n$, we use the $a,a^\dag$ operators which commute with $T_{\mbf{a}_i}$ to define
\bea
\label{eq:LLstates}
w_n(\mbf{r}) = \frac{a^{\dag n}}{\sqrt{n!}} \psi_{0}(\mbf{r}), \qquad a \psi_{0}(\mbf{r}) = b \psi_{0}(\mbf{r}) = 0 \ . \\
\eea
It follows that the states $\psi_{\mbf{k},n}(\mbf{r})$ are orthogonal because they are eigenstates of the Hermitian Landau level operator $a^\dag a$ with eigenvalue $n$. We will not need an explicit expression for the Landau level groundstate $\psi_{0}(\mbf{r})$, but one can be obtained because $a$ and $b$ are commuting linear differential operators, so the first order differential equations in \Eq{eq:LLstates} can be directly integrated.\footnote{In the symmetric gauge, it is well known \cite{1999tald.conf...53G,fradkin_2013} that $\psi_0(\mbf{r}) \sim \exp (-\phi \frac{r^2}{4 \Omega}) = \exp(-|z|^2/4 \ell_B^2)$ where $z = x + i y$ is the holomorphic coordinate and $\ell_B = 1/\sqrt{eB}$ is the magnetic length.}

Lastly, the normalization $\mathcal{N}(\mbf{k})$ in \Eq{eq:basis} is defined by requiring
\bea
\label{eq:normwave}
\int d^2r \, \psi^\dag_{\mbf{k},m}(\mbf{r}) \psi_{\mbf{k}',n}(\mbf{r}) = (2\pi)^2 \delta(\mbf{k}-\mbf{k}') \delta_{mn}
\eea
which, after a detailed calculation contained in \App{app:siegelnorm}, yields
\bea
\label{eq:Ncalsieg}
\mathcal{N}(\mbf{k}) = \vartheta \lp \left. \frac{(k_1,k_2)}{2\pi} \right| \Phi \rp , \quad \Phi =  \frac{i}{2} \bpm 1 & i \\  i & 1 \epm \ .
\eea
The function $\vartheta \lp\!\left.\mbf{z} \right| \!\Phi \rp$ is called the Siegel theta function.\footnote{The Siegel theta function, also known as the Riemann theta function, is implemented in Mathematica. } It is a multi-dimensional generalization of the Jacobi theta function defined for $\mbf{z} \in \mathds{C}^2$ by
\bea
\label{eq:siegeldefm}
\vartheta \lp \mbf{z} \left| \Phi \right. \rp = \sum_{\mbf{n}\in\mathds{Z}^2} e^{2\pi i \lp \frac{1}{2} \mbf{n} \cdot \Phi  \cdot \mbf{n}- \mbf{z} \cdot \mbf{n} \rp} \ . \\
\eea
The matrix $\Phi$ which defines the Siegel theta function is sometimes called the Riemann matrix. For the sum in \Eq{eq:siegeldefm} to converge, $\text{Im } \Phi$ must be a positive definite matrix.  In \App{app:greens}, we show that $\Phi$ is a special ``self-dual" Riemann matrix which permits the Siegel theta function to be written in terms of Jacobi theta functions at $\phi = 2\pi$. It is apparent from \Eq{eq:siegeldefm} that $\mathcal{N}(\mbf{k} + 2\pi \mbf{b}_i) = \mathcal{N}(\mbf{k})$, which matches the periodicity of the BZ. The Siegel theta function is quasi-periodic for complex $\mbf{z}$. A self-contained derivation of the quasi-periodicity may be found in \App{app:siegelprop}. We show in \App{app:greens} that $\mathcal{N}(\mbf{k}) \geq 0$ for $\mbf{k} \in BZ$ but at $\pi \mbf{b}_1 + \pi \mbf{b}_2$, $\mathcal{N}(\mbf{k}) $ has a quadratic zero. Thus the states $\psi_{\mbf{k},n}$ do not exist exactly at $\mbf{k}^* = \pi \mbf{b}_1 + \pi \mbf{b}_2$. We show in \App{eq:appopdef} that the wavefunction can be defined in patches by shifting the operator $\mbf{Q} \to \mbf{Q} + \mbf{p}$ which shifts the undefined states to $\mbf{k}^* + \mbf{p}$. In fact, the existence of a zero is topologically protected because the states $\psi_{\mbf{k},n}$ carry nonzero Chern number (see \Sec{sec:berry}) and hence cannot be well-defined and periodic everywhere in the BZ.  We will show in \Sec{sec:matel} that the magnetic Bloch Hamiltonian used to compute the spectrum is an analytic function of $\mbf{k}$, so the zero in $\mathcal{N}(\mbf{k})$ only introduces a removable singularity in the Hamiltonian. Lastly, we give a gauge-invariant proof in \App{app:completeness} that the $\psi_{\mbf{k},n}$ basis is complete when acting on suitable test functions.

For brevity, we now define braket notation for the magnetic translation operator eigenstates \Eq{eq:basis}:
\bea
\label{eq:ketabsis}
\ket{\mbf{k},n} \equiv \frac{1}{\sqrt{\mathcal{N}(\mbf{k})}} \sum_{\mbf{R}} e^{-i \mbf{k} \cdot \mbf{R}} T_{\mbf{a}_1}^{\mbf{R} \cdot \mbf{b}_1} T_{\mbf{a}_2}^{\mbf{R} \cdot \mbf{b}_2} \ket{n} , \ \ket{n} = \frac{a^{\dag n}}{\sqrt{n!}} \ket{0},
\eea
and $a \ket{0} = b\ket{0} = 0$. For Hamiltonians with additional degrees of freedom indexed by $\al$, such as spin, sublattice, valley, or layer (see \Sec{sec:BMmodel}), the basis states of the Hamiltonian can be defined $\ket{\mbf{k},n, \al} = \ket{\al} \otimes \ket{\mbf{k},n}$. In braket notation, \Eq{eq:normwave} reads
\bea
\braket{\mbf{k},m|\mbf{k}',n} = (2\pi)^2  \delta_{mn} \delta(\mbf{k}-\mbf{k}') \
\eea
and it should be implicitly understood that $\mbf{k} = \pi \mbf{b}_1 + \pi \mbf{b}_2$ is excluded from the basis. While discussing single-particle physics in \Sec{sec:matel} and \Sec{sec:berry}, the braket notation is useful for shortening expressions. Lastly, the structure of the states in \Eq{eq:ketabsis}  generalizes to the $q$-dimensional irreps of the magnetic translation group at rational flux $\phi = \frac{2\pi p}{q}$. We leave this construction to future work.

Before concluding this section, we will emphasize the difference between our gauge invariant construction and the commonly used Landau gauge states (see e.g. \Ref{PhysRevB.46.12606,Crosse_2020,BROWN1969313}). In the Landau gauge $\mbf{A} = B(0,x)$ which preserves translation along the $y$ direction for instance, a basis of ``Landau level states" can be labeled by $k_y$ and a Landau level index $n$. These states are fully delocalized along $y$ and localized on the scale of the magnetic length in harmonic oscillator wavefunctions along $x$ \cite{PhysRevB.46.12606}. To form eigenstates of the magnetic translation group, these states are resummed to obtain magnetic translation invariance along $x$. This process is somewhat involved and obscures the physical symmetry of the system since it treats $x$ and $y$ differently due to the asymmetry of the Landau gauge. In contrast, our gauge-invariant construction in \Eq{eq:ketabsis} is manifestly symmetric under the magnetic translation group and is immediately valid for arbitrary lattices. It has many practical advantages: all calculations can be performed using the oscillator algebra \Eq{eq:abcom}, and the singularity due to the Chern number of the states is made explicit. This latter feature in particular has not been discussed in earlier treatments, and makes it possible for us to apply the tools of topological band theory in direct analogy to the Bloch wave formalism at zero flux. 

\section{Matrix Elements}
\label{sec:matel}

Because the Hamiltonian $H^{\phi = 2\pi}$ commutes with the magnetic translation group, it must be diagonal in $\mbf{k}$ because of the selection rule
\bea
\label{eq:selectionrule}
 \braket{\mbf{k}',m| H^{\phi = 2\pi}| \mbf{k},n} &= e^{i (\mbf{k} - \mbf{k}')\cdot \mbf{a}_i} \braket{\mbf{k}',m| H^{\phi = 2\pi}| \mbf{k},n}
\eea
which shows that if $k_i - k_i' \neq 0 \mod 2\pi$, then $ \braket{\mbf{k}',m| H^{\phi = 2\pi}| \mbf{k},n} = 0$. \Eq{eq:selectionrule} follows from inserting $1 = T_{\mbf{a}_i}^\dag T_{\mbf{a}_i}$ and commuting $T_{\mbf{a}_i}$ through $H^{\phi = 2\pi}$. Having constructed a basis of states which is diagonal in $\mbf{k}$, we define an effective ``Bloch" Hamiltonian $H^{\phi = 2\pi}_{mn}(\mbf{k})$ according to
\bea
\label{eq:blochham}
(2\pi)^2 \delta(\mbf{k}-\mbf{k}') H^{\phi = 2\pi}_{mn}(\mbf{k}) = \braket{\mbf{k}',m| H^{\phi = 2\pi}| \mbf{k},n} \\
\eea
which can be diagonalized after imposing a Landau level cutoff. To compute the effective Hamiltonian, we need formulas for the matrix elements of  \Eq{eq:Hkinpot}. The kinetic term is simple because $h(\pmb{\pi})$ is composed of $a, a^\dag$ operators, so it only acts on the $m,n$ indices and its matrix elements will not depend on $\mbf{k}$ (see \Sec{sec:square} for an example). Hence we focus on the potential term $U(\mbf{r})$ which causes scattering between different Landau levels. Recall that $U(\mbf{r})$ is periodic so can be expanded as a Fourier series. Hence we need to compute the general scattering amplitude
\bea
\braket{\mbf{k},m| e^{-2\pi i \mbf{G} \cdot \mbf{r} } | \mbf{k},n}, \quad  \mbf{G} = G_1 \bbb_1 + G_2 \bbb_2, ~G_1,G_2 \in \mathds{Z} \ .
\eea
It is possible to perform the calculation exactly without choosing a gauge for $\mbf{A}(\mbf{r})$ because $\mbf{G} \cdot \mbf{r}$ can be expressed simply in terms of $\pmb{\pi}$ and $\mbf{Q}$ using
\bea
(eB)^{-1}\eps_{\mu \nu}(Q_\nu - \pi_\nu) = -\eps_{\mu \nu}\eps_{\nu \rho} x_\rho = x_\mu \\
\eea
which allows the us to perform the calculation using BCH. The details may be found in \App{app:siegelcomplete}. The result is
\bea
\label{eq:Hmatm}
\braket{\mbf{k}',m| e^{-2\pi i \mbf{G} \cdot \mbf{r} } | \mbf{k},n}  &= \\
 (2\pi)^2 \delta(\mbf{k}-\mbf{k}')  & e^{- i \pi G_1 G_2 - i (G_1k_2-G_2k_1)} \mathcal{H}^{2\pi \mbf{G}}_{mn} \\
 \eea
where we have defined the Landau level scattering matrix for a general momentum $\mbf{q}$ with $q_i= \mbf{q} \cdot \mbf{a}_i$ and $z_j = (\hat{x} + i \hat{y}) \cdot \mbf{a}_j /\sqrt{\Omega}$:
\bea
\label{eq:Z}
\mathcal{H}^{\mbf{q}}_{mn} =  \bra{m} \exp \lp i \eps_{ij} q_i Z_j \rp \ket{n},
\ Z_j &= \frac{\bar{z}_j a + z_j a^\dag }{\sqrt{2 \phi}} \ .
\eea
Here $i,j \in \{1,2\}$ are the crystalline indices which are summed over. A closed-form expression for the unitary matrix $\mathcal{H}^{\mbf{q}}$ in terms of Laguerre polynomials is provided in Eq. 140 of \App{app:siegelcomplete}. With \Eq{eq:Hmatm}, the action of any periodic potential on the magnetic translation group eigenbasis is easily obtained. The kinetic term in \Eq{eq:blochham} does not depend on $\mbf{k}$ because it only contains $a,a^\dag$ operators and creates flat Landau levels. We observe that all the $\mbf{k}$-dependence of \Eq{eq:blochham} is contained in the potential term matrix elements \Eq{eq:Hmatm} in the form $\exp (i \Omega \mbf{k} \times \mbf{G} ) = \exp (-i (G_1 k_2 - G_2 k_1) )$ and hence $H^{\phi = 2\pi}(\mbf{k})$ is analytic in $\mbf{k}$. From the $\mbf{k}$-dependence of \Eq{eq:ketabsis}, we deduce that $\ket{\mbf{k}+2\pi \mbf{G},n} = \ket{\mbf{k},n}$. Thus
$H^{\phi = 2\pi}_{mn}(\mbf{k}+2\pi \mbf{G}) = H^{\phi = 2\pi}_{mn}(\mbf{k})$ is explicitly periodic in $\mbf{k}$, so no embedding matrices \cite{PhysRevLett.125.236804} are required.

\section{Berry Connection}
\label{sec:berry}

\begin{figure}
\includegraphics[width=0.45\textwidth,trim = 0 0 0 0]{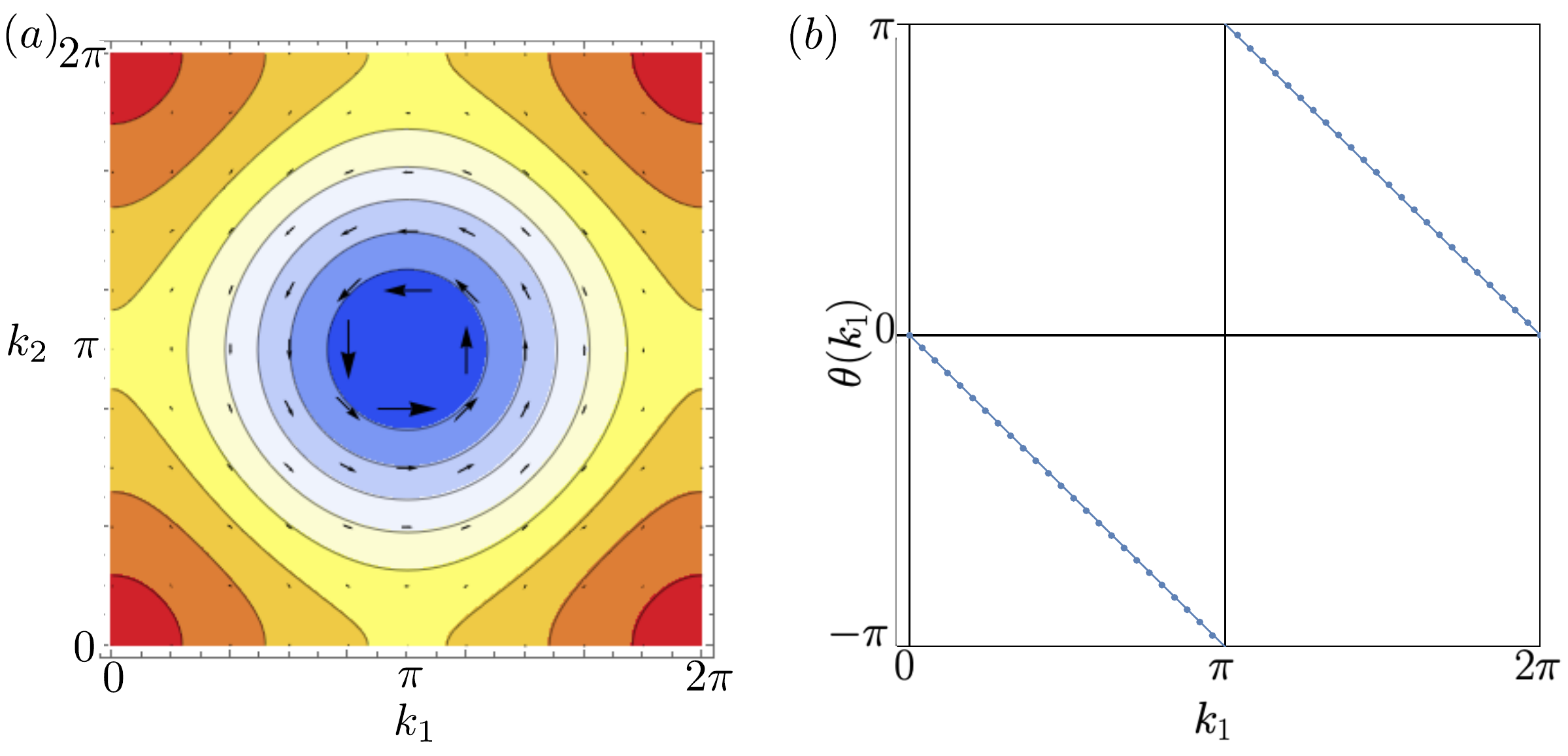}
\caption{(a) The Siegel theta function $\mathcal{N}(\mbf{k})$ (see \Eq{eq:Ncalsieg}) is plotted with arrows denoting the vector field ${\cal A}_{nn}(\kk)$.  The winding in ${\cal A}$ around the zero located at $k_1 = \pi, k_2 = \pi$ leads to a Chern number in the \emph{basis} states. (b) The Wilson loop $W(k_1) = e^{i \th(k_1)}$ of a single Landau level integrated along $k_2$ is plotted as a function of $k_1$. The Wilson loop is computed analytically in \App{app:berry} to be $W(k_1) = e^{-i k_1}$ (shown in solid blue) which winds once crossing the vortex at $(\pi, \pi)$. The numerical approximation of $W(k_1)$ is dotted.}
\label{fig:siegelwinding}
\end{figure}

Our basis of magnetic translation eigenstates (\Eq{eq:basis}) is built from continuum Landau levels. These states are known to carry a Chern number \cite{PhysRevLett.49.405}, and it will be important to see how this arises in our formalism. To study the topology, we need to compute the continuum Berry connection:
\bea
(2\pi)^2 \delta(\mbf{k}-\mbf{k}') \mathcal{A}^{mn}(\mbf{k}) &= \braket{\mbf{k}',m| \mbf{r}| \mbf{k},n} \ . \\
\eea
In zero flux where the basis states are plane waves or Fourier transforms of localized orbitals, $\mathcal{A}^{mn}(\mbf{k})$ would be trivial. However, the basis states at $2\pi$ flux are built out of Landau levels,  which by themselves are topologically nontrivial. We can see this directly by computing $\braket{\mbf{k}',n| \mbf{r}| \mbf{k},n}$ (here the Landau level index $n$ is unsummed), the Abelian Berry connection of the $n$th Landau level, using the oscillator algebra. The result from \App{app:berry} is
\bea
\label{eq:Annabe}
\null \mathcal{A}^{nn}_i(\mbf{k}) &= - \frac{1}{2} \eps_{ij}  \del_{j} \log \vartheta \lp \left. \frac{(k_1,k_2)}{2\pi} \right| \Phi \rp
 \eea
where $\del_i = \frac{\del}{\del k_i}$ here for brevity, $\mathcal{A}_i = \mbf{b}_i \cdot \mathcal{A}$, and we emphasize that $\mathcal{A}_{nn}(\mbf{k})$ is independent of $n$. Interestingly, a similar formula has appeared recently in flat band Chern states in \Ref{2021arXiv210507491W}. We now show that the connection \Eq{eq:Annabe} has Chern number $-1$.\footnote{In many texts, the sign of the Chern number is made positive by orienting the $B$ field in the $-\hat{z}$ direction. This is just a matter of convention.} In \App{app:berry}, we show with a direct computation that the Berry curvature is given by
\bea
\label{eq:Berrycurvature}
\eps_{ij} \del_i \mathcal{A}_j^{nn} =  \frac{1}{2} \del^2 \log \vartheta = -\frac{1}{2\pi} + 2\pi \delta(\mbf{k} - \pi \mbf{b}_1 - \pi \mbf{b}_2) \\
\eea
and has two contributions. The $-1/2\pi$ term in \Eq{eq:Berrycurvature} is the constant and nonzero Berry curvature of a Landau level \cite{2021arXiv210507491W,PhysRevB.46.12606}. The delta function appearing at $\mbf{k}^* = \pi \mbf{b}_1 + \pi \mbf{b}_2$ is an artifact of the undefined basis states at $\mbf{k}^*$  where $\mathcal{N}(\mbf{k}^*) = 0$ and is discussed fully in \App{app:berry}. In fact, the $2\pi$ delta function is unobservable in the Wilson loop winding because the Berry phase is only defined mod $2\pi$. To see this, we explicitly calculate the Abelian Wilson loop (or Berry phase) in \App{app:berry} and show the result in \Fig{fig:siegelwinding}(b) where we see that the Wilson loop eigenvalues are indeed continuous mod $2\pi$. Hence we can think of the the basis states in \Eq{eq:basis} as lattice-regularized Landau levels. We also see that the zero in the normalization factor $\mathcal{N}(\mbf{k})$ (see \Sec{sec:MTGirr}) is an \emph{essential} feature of the basis rather than a pathological one: it is a manifestation of the topology of the basis states. If there were no zero, then we would have written down wavefunctions which were periodic and differentiable on the entire BZ, hence precluding a Chern number \cite{Brouder_2007}.

Finally, we obtain an explicit expression for the non-Abelian Berry connection $\mathcal{A}^{MN}(\mbf{k})$ in the occupied bands indexed by $M,N$:
\bea
(2\pi)^2 \delta(\mbf{k}-\mbf{k}') \mathcal{A}^{MN}(\mbf{k}) \!=\! \sum_{mn} [U^\dag(\mbf{k}') ]^{M}_{m}\!\braket{\mbf{k}',m| \mbf{r}| \mbf{k},n} U_{n}^N(\mbf{k})
\eea
 where $U(\mbf{k})$ is the $N_{LL} \times N_{occ}$ matrix of eigenvectors. $N_{occ}$ is the number of occupied bands and $N_{LL}$ is the dimension of the matrix Hamiltonian, which is truncated at $N_{LL}$ Landau levels. Leaving the details of the calculation to \App{app:berry}, we give the general formula
\bea
\label{eq:nonabA}
\mathcal{A}_i^{MN}(\mbf{k}) &= [U^\dag (i \del_i - \eps_{ij} \tilde{Z}_j) U]^{MN}  \\
&\qquad - \frac{\delta^{MN}}{2} \eps_{ij}   \del_{j} \log \vartheta \lp \left. \frac{(k_1,k_2)}{2\pi} \right| \Phi \rp \ .
\eea

The Abelian term in the second line of \Eq{eq:nonabA} describes the Chern numbers of the basis states as in \Eq{eq:Annabe}. Note that it is proportional to the identity $\delta^{MN}$ and so can be factored out of the Wilson loop to give an overall winding factor per Landau level as shown in \Fig{fig:siegelwinding}(b). The new non-Abelian term $U^\dag \tilde{Z}_j U$ of \Eq{eq:nonabA} describes coupling between Landau levels where the Hermitian matrix $[\tilde{Z}_j]_{mn} = \braket{m|Z_j|n}$ is given in \Eq{eq:Z}. Returning to \Eq{eq:nonabA}, we write the non-Abelian Wilson loop as the path-ordered matrix exponential
\bea
\label{eq:WC}
\null [W_{\mathcal{C}}]^{MN} &= \left[ \exp \lp i \oint_{\mathcal{C}} d\mbf{k} \cdot \mathcal{A}(\mbf{k})  \rp \right]^{MN} \\
&= e^{-i \oint_{\mathcal{C}} d\mbf{k} \times \frac{1}{2}\pmb{\nabla} \log \vartheta \lp \left. \frac{(k_1,k_2)}{2\pi} \right| \Phi \rp }   \\
&\qquad \times \left[ \exp \lp i \oint_{\mathcal{C}} dk_i \, U^\dag (i \del_i - \eps_{ij}  \tilde{Z}_j) U  \rp \right]^{MN}
\eea
with a sum over $i,j$ implied. For numerical computations, \Eq{eq:WC} should be expanded into an ordered product form using the projectors $P_{\mbf{k}} = U(\mbf{k}) U^\dag(\mbf{k})$. This procedure can be carried through exactly (the details may be found in \App{app:berry}) and the result is
\bea
W_{\mathcal{C}} &= \exp \left[-i \oint_{\mathcal{C}} d\mbf{k} \times \frac{1}{2}\pmb{\nabla} \log \vartheta \lp \left. \frac{(k_1,k_2)}{2\pi} \right| \Phi \rp \right]   \\
&\qquad \times  U^\dag(\mbf{k}_{L})  \mathcal{H}^{-d\mbf{k}_{L}} \lp \prod_{n}^{(L-1) \leftarrow 1}  P(\mbf{k}_{n})  \mathcal{H}^{-d\mbf{k}_n} \rp U(\mbf{k}_{0})
\eea
where $\mathcal{C}$ is a closed path with starting at $\mbf{k}_1$ which is broken into $L$ segments labeled by $\mbf{k}_n$, and $d\mbf{k}_n = \mbf{k}_{n} - \mbf{k}_{n-1}$. The insertions of non-Abelian terms $\mathcal{H}^{d\mbf{k}} = e^{i \eps_{ij} dk_i \tilde{Z}_j}$ act off-diagonally on the Landau level index (see \Eq{eq:Z}). The appearance of these non-Abelian terms reflects the fact that the Landau level states in \Eq{eq:basis} are not localized below the magnetic length, which is $1/\sqrt{\phi}$ in dimensionless units.  In \Sec{sec:square}, we use the results of this section to calculate the Wilson loop in a square lattice model tuned through a topological phase transition at $2\pi$ flux by increasing the strength of the crystalline potential.

\section{Square lattice Example}

\label{sec:square}
\begin{figure}
\includegraphics[width=0.22\textwidth]{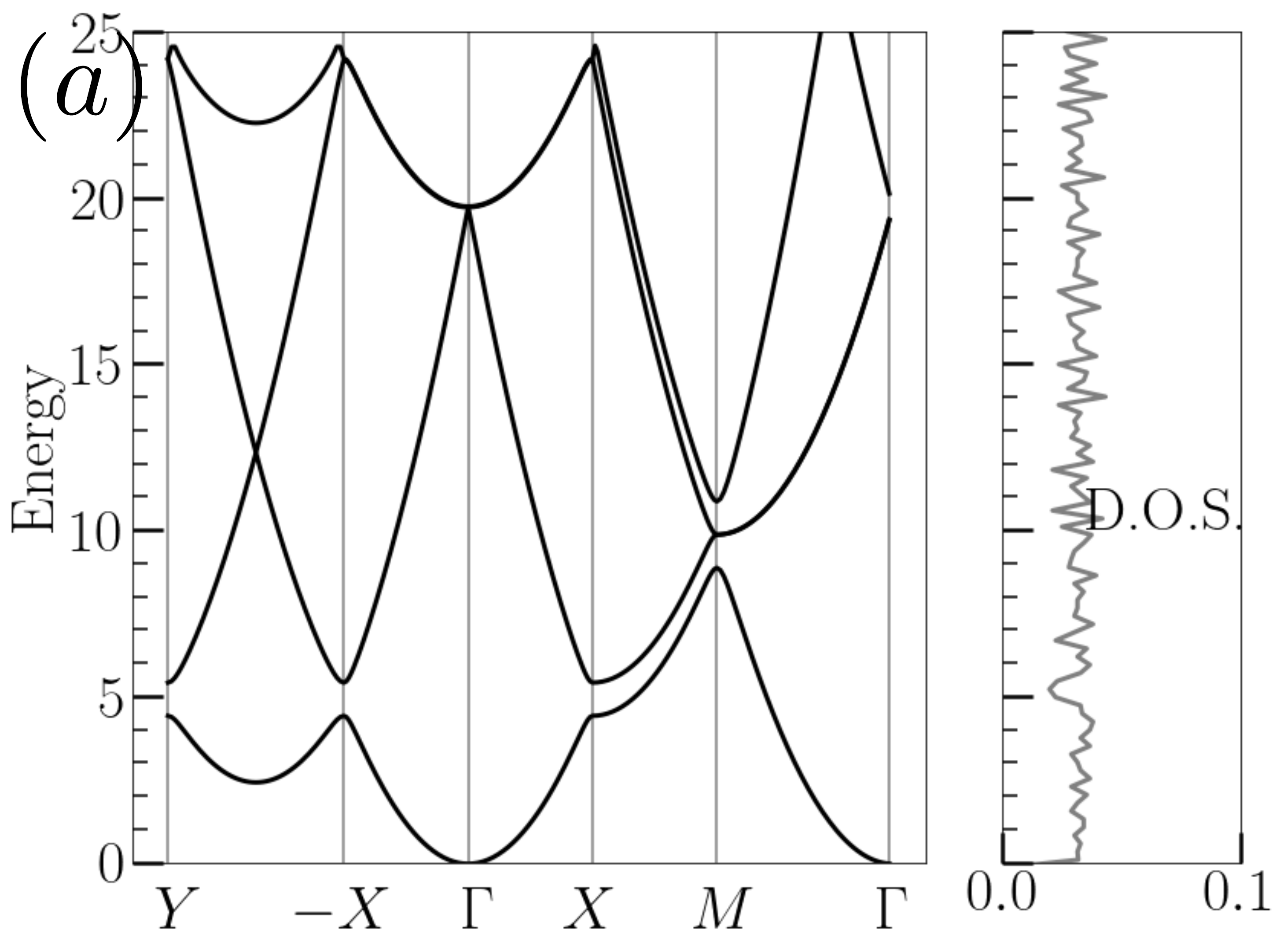}
\includegraphics[width=0.22\textwidth]{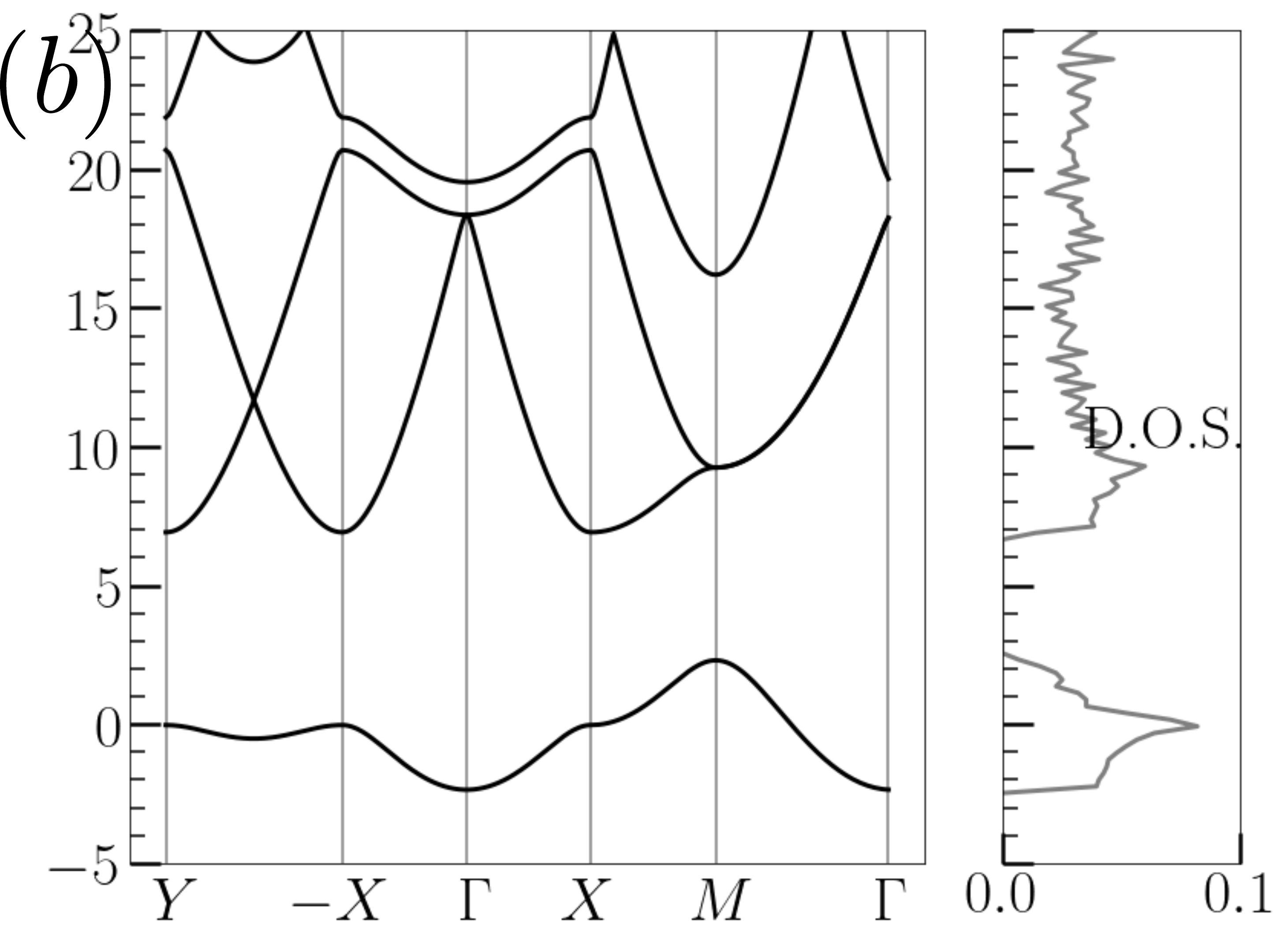}
\includegraphics[width=0.22\textwidth]{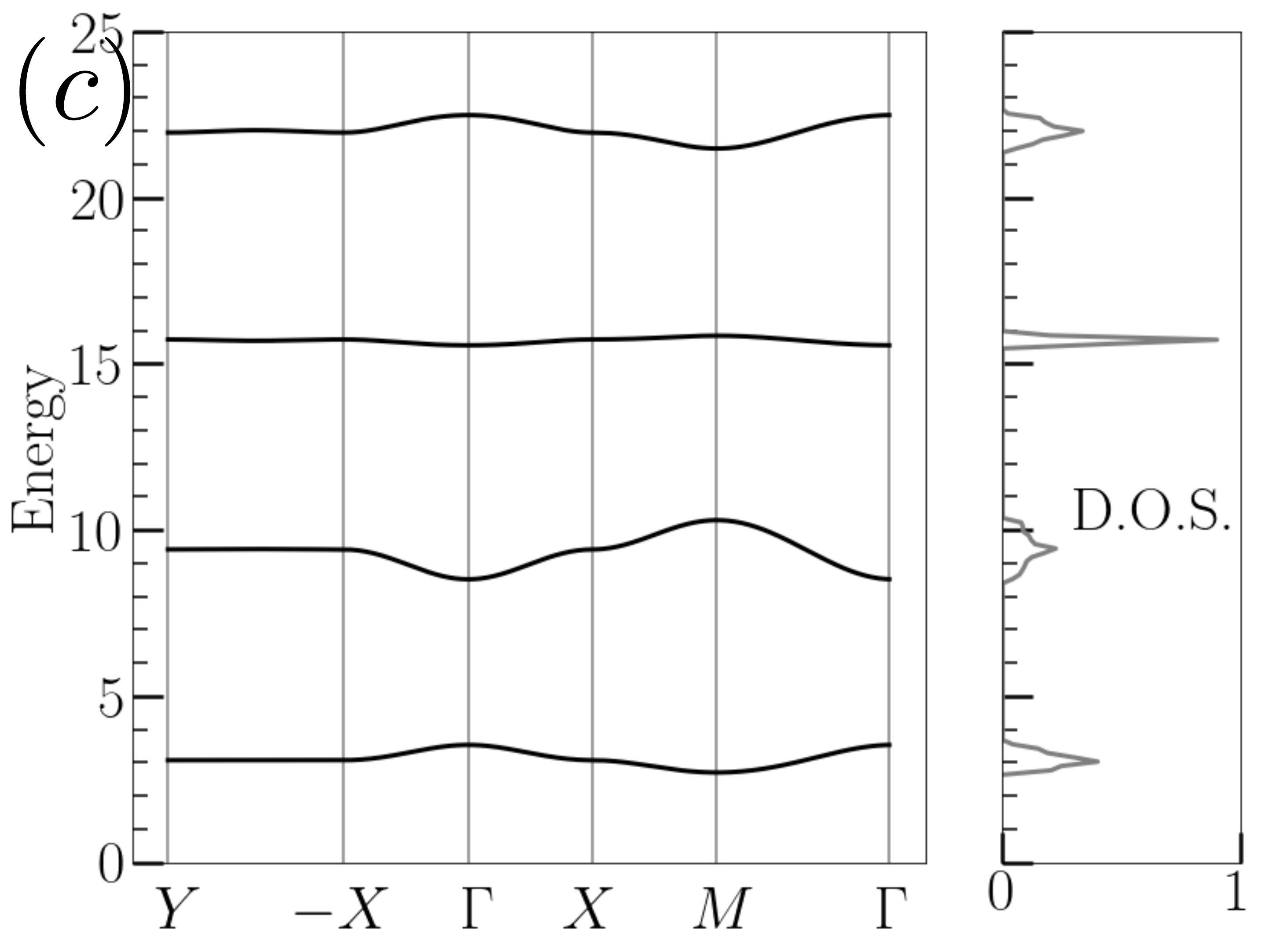}~
\includegraphics[width=0.22\textwidth]{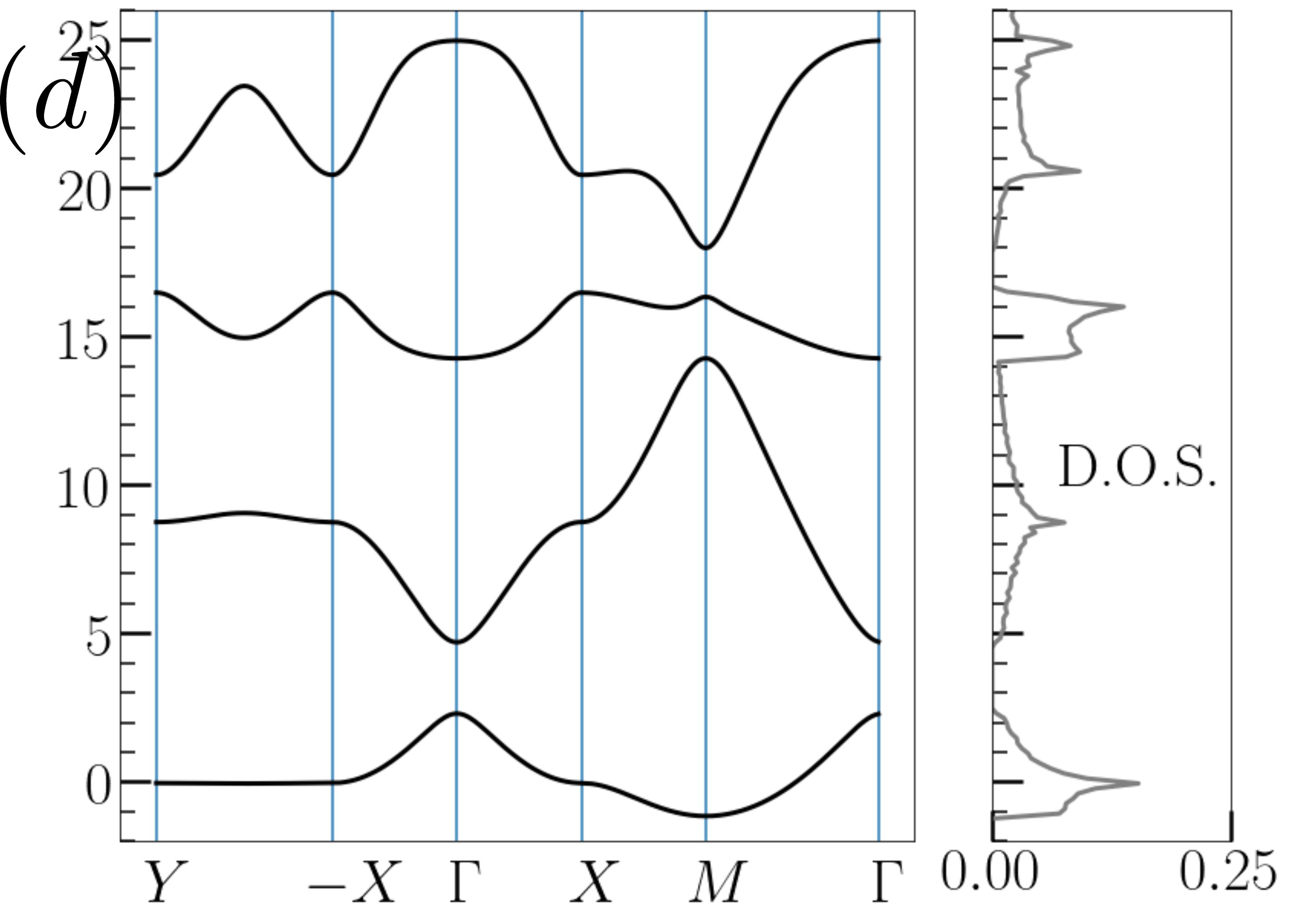}
\includegraphics[width=0.22\textwidth]{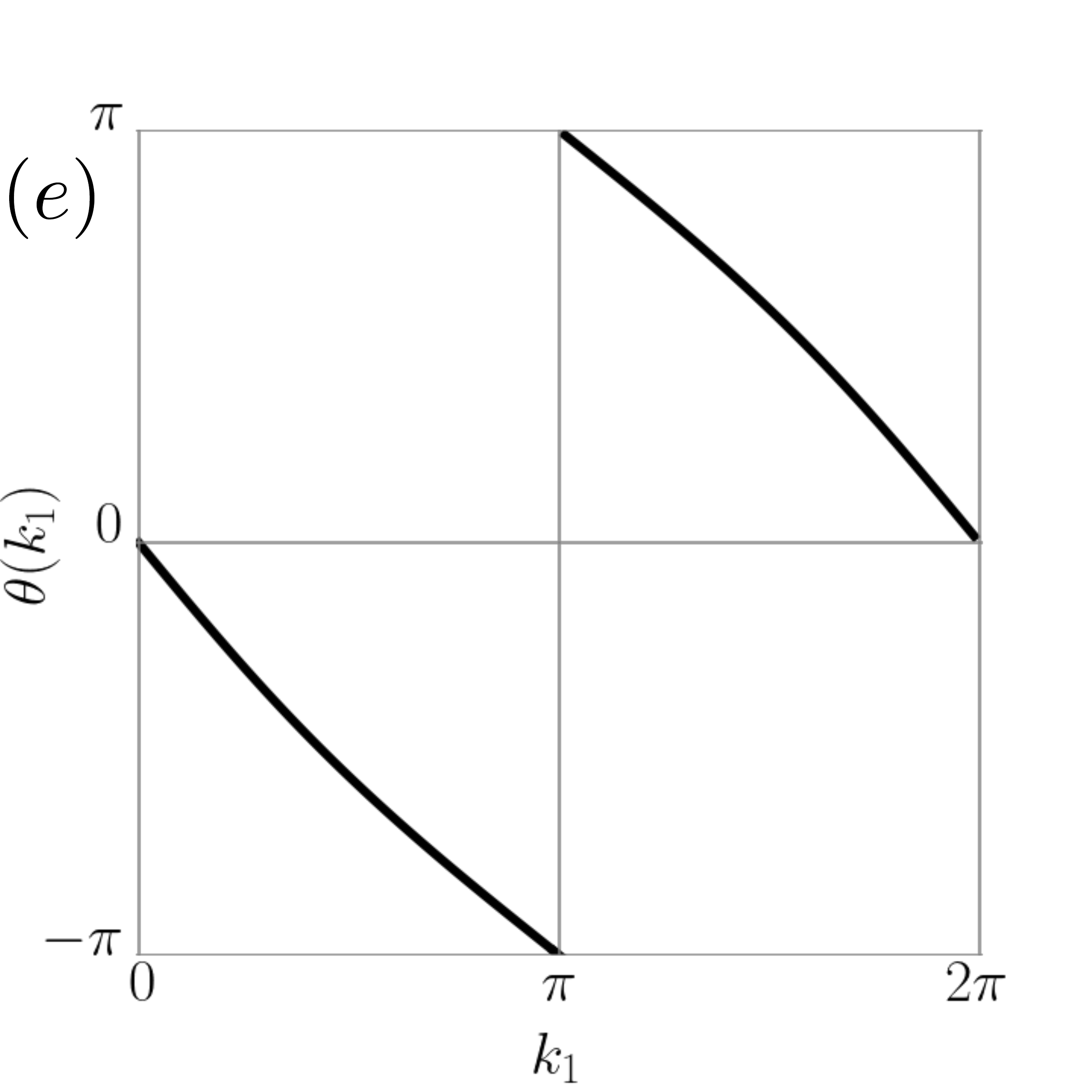}
\includegraphics[width=0.22\textwidth]{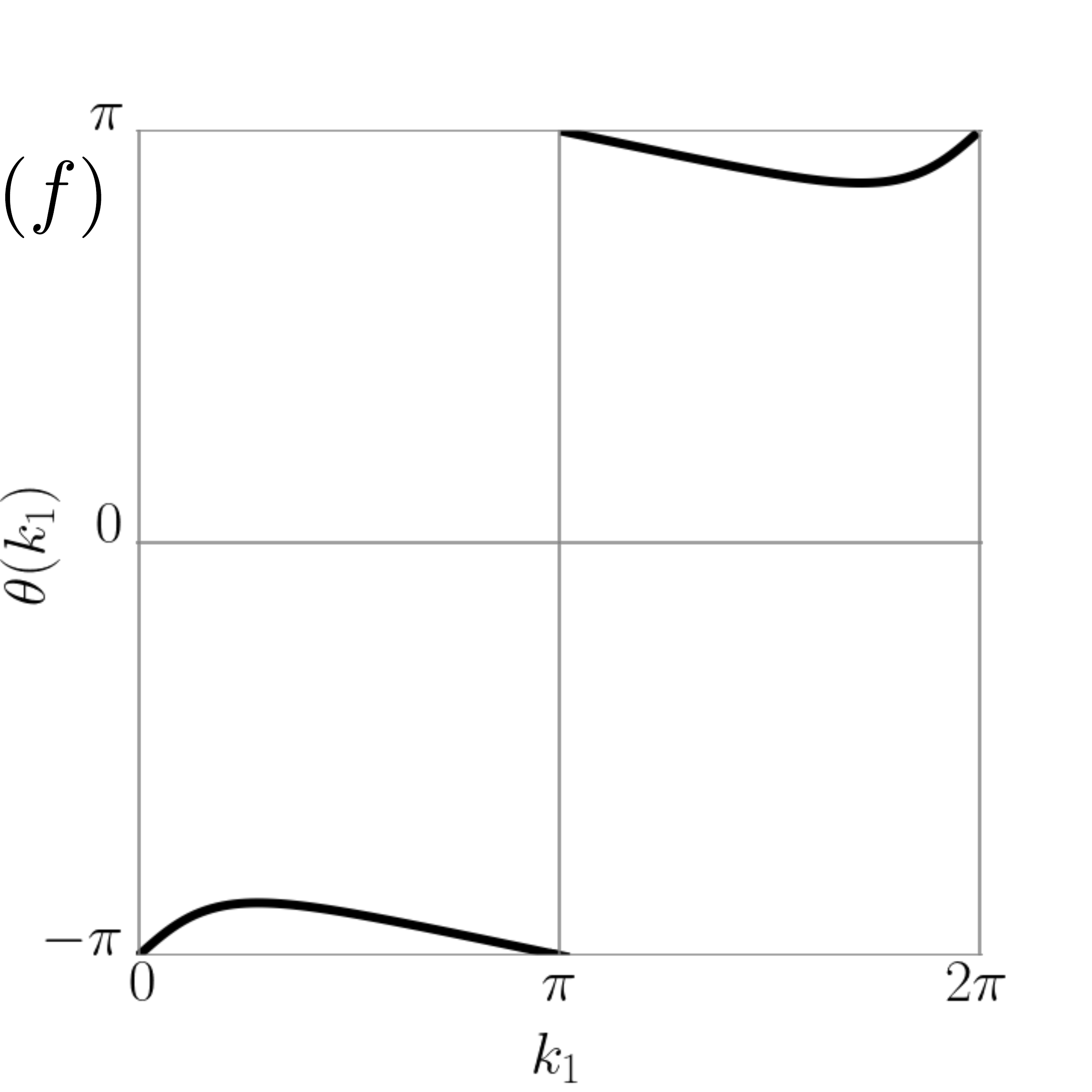}
\caption{$\mathbf(a,b)$ Square lattice in zero flux, at low potential $w = 1$ and high potential $w = 7$ respectively.  $\mathbf(c,d)$ Square lattice in $2\pi$ flux, at low potential and high potential respectively.
$\mathbf(e,f)$ Wilson loops of the square lattice in flux.  At low hoppings, the Hamiltonian resembles a Landau level system, resulting in nearly flat bands and a winding in the Wilson loop for the lowest band.  At large hoppings, a gap closing occurs and allows the lowest band to have Chern number zero.}
\label{fig:squareLatticeData}
\end{figure}

The simplicity of implementing our formalism is illustrated with a model of a scalar particle mass $m=1$ which feels a square lattice cosine potential. While it may be possible to simulate this type of model on an optical lattice \cite{2012Natur.483..302T,2015PhRvA..91f3628Y,PhysRevLett.111.185301}, we intend this example to be pedagogical rather than physically motivated. We take the lattice vectors and reciprocal vectors to be $\aaa_{1} = \bbb_{1} = (1, 0), \aaa_{2} = \bbb_{2} = (0, 1)$ so $\Omega = 1$ and define the zero-flux Hamiltonian as
\begin{align}
H^{\phi=0} = -\frac{1}{2}{\pmb{\nabla}}^2 + \frac{w}{2}(e^{-2\pi i\bbb_1 \cdot \rr} + e^{-2\pi i\bbb_2 \cdot \rr} + H.c.),
\label{eq:squarelatH}
\end{align} where we have taken $\hbar = 1$. When $w = 0$, the Hamiltonian  $H^{\phi=0}$ has continuous translation symmetry and solutions can be labeled by momentum $\kk$.  When $w$ is nonzero, the continuous translation symmetry is broken to a discrete symmetry which weakly couples the plane wave states and opens gaps at the corners of the BZ. By Bloch's theorem, the states are labeled by momentum $\kk$ in the BZ and the effective Hamiltonian reads
\begin{align}
H^{\phi=0}_{\mbf{G},\mbf{G}'}(\kk) &= \frac{1}{2} (\kk - \mbf{G})^2 \delta_{\mbf{G}\mbf{G}'} \\
&\qquad + \frac{w}{2}(\delta_{\mbf{G},\mbf{G}'-2\pi\bbb_1} + \delta_{\mbf{G},\mbf{G}'-2\pi\bbb_2} + H.c.) \nonumber
\label{}
\end{align} and $\mbf{G} = G_1 \mbf{b}_1 + G_2 \mbf{b}_2, G_i \in \mathds{Z}$ (see \App{app:square_lattice} for details). We show the Bloch band structure in \Fig{fig:squareLatticeData} in the weak and strong potential regimes. In flux, the Hamiltonian \Eq{eq:squarelatH} is written in terms of the canonical momentum
\begin{align}
H^\phi = \frac{1}{2}{\pmb{\pi}}^2 + \frac{w}{2}(e^{-2\pi i\bbb_1 \cdot \rr} + e^{-2\pi i\bbb_2 \cdot \rr} + H.c.),
\label{}
\end{align} that is, Landau levels in a lattice potential. In $2\pi$ flux using the matrix elements in Eq. 139 of \Ref{SM}, the magnetic Bloch Hamiltonian is
\bea
H^{\phi=2\pi}_{mn}(\mbf{k}) &= \phi(m+\frac{1}{2}) \delta_{mn} \\
&\qquad + \frac{w}{2} (e^{- i k_2}  \mathcal{H}^{2\pi \mbf{b}_1}_{mn} + e^{i k_1} \mathcal{H}^{2\pi \mbf{b}_2}_{mn} + H.c.)
\label{}
\eea
and recalling that the kinetic term acts on the $\ket{\kk,m}$ basis as $\frac{1}{2}\pmb{\pi}^2 = \phi (a^\dagger a + \frac{1}{2})$. The potential term $\mathcal{H}^{2\pi \mbf{G}}_{mn}$ couples the Landau levels, giving nontrivial dispersion. We numerically calculate the band structure in the weak coupling ($w = 1$) and strong coupling ($w = 7$) regimes.  The Landau level regime in weak coupling exhibits nearly flat bands (\Fig{fig:squareLatticeData}(c)), and its lowest band carries a Chern number, as exemplified by the winding of the Wilson loop shown in \Fig{fig:squareLatticeData}(e).  Increasing $w$ pushes the model through a phase transition with a band touching at the $\Gamma$ point. At strong coupling ($w = 7$), the $2\pi$ flux spectrum is gapped (\Fig{fig:squareLatticeData}(d)) and its lowest band has zero Chern number (\Fig{fig:squareLatticeData}(f)). Hence the lowest band cannot be interpreted as a Landau level, despite the strong flux.

\section{Many-body Form Factors}
\label{sec:formfactor}

Thus far, we have discussed the single-particle spectrum and Wilson loop topology of continuum Hamiltonians at $2\pi$ flux. In this section, we extend our formalism to many-body physics and derive a convenient expression for the Coulomb Hamiltonian
\bea
H_{int} &= \frac{1}{2} \int d^2r d^2r' \, n(\mbf{r}) V(\mbf{r}-\mbf{r}') n(\mbf{r}') \\
\eea
 in terms of the magnetic translation operator eigenbasis \Eq{eq:basis}. Here $n(\mbf{r}) = c^\dag (\mbf{r}) c(\mbf{r})$ is the local density operator at $\mbf{r}$ and $c(\mbf{r}),c^\dag(\mbf{r})$ are the continuum fermion operators satisfying $\{c^\dag(\mbf{r}), c(\mbf{r}') \} = \delta(\mbf{r}-\mbf{r}')$. In \Sec{sec:TBGcoulomb}, we will project the Coulomb interaction on the flat bands of TBG in order to study its many-body insulating groundstates, as done in zero flux in \Refs{2020arXiv200913530L}{2020arXiv200914200B}. The calculation for TBG is more involved because there are additional indices corresponding to valley and spin (see \App{eq:appcoulomb} for details). For simplicity, we focus on models with only a single orbital per unit cell in this section and study the projected Coulomb Hamiltonian at $2\pi$ flux.

\begin{figure}
\includegraphics[width=0.5\textwidth]{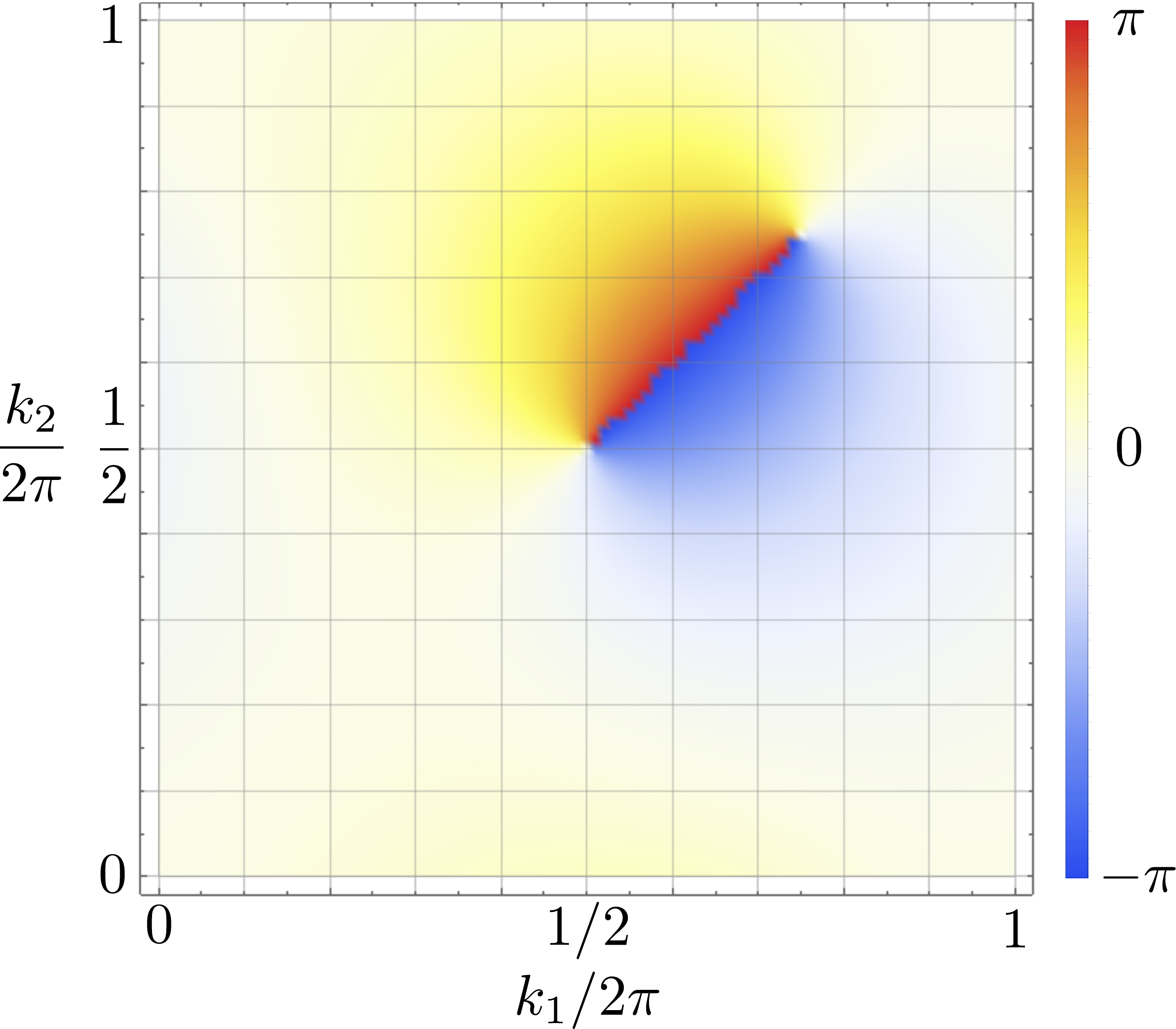}
\caption{Phase $\xi_\qq(\kk)$ in Eq.~\ref{eq:siegelPhaseFactor} for $\qq = \frac{\pi}{2} \mbf{b}_1 + \frac{\pi}{2} \mbf{b}_2$, plotted as a density.  Note the branch cut linking $(1/2,1/2)$ to $(3/4,3/4)$.}
\label{fig:siegelPhase}
\end{figure}

To avoid confusion with the Fock space braket notation in many-body calculations, we return to a wavefunction notation for the magnetic translation group eigenstates:
\begin{align}
\psi_{\kk,n}(\mbf{r}) = \frac{1}{\sqrt{{\cal N}(\kk)}} \sum_\RR e^{-i\kk \cdot \RR} T_{\mbf{a}_1}^{\mbf{b}_1 \cdot \mbf{R}}T_{\mbf{a}_2}^{\mbf{b}_2 \cdot \mbf{R}} \frac{a^{\dagger n}}{\sqrt{n!}} \psi_0(\mbf{r}),
\label{}
\end{align} where $\psi_0$ is the zeroth Landau level $a \psi_0 = b \psi_0 = 0$.  Throughout this section, $\ket{0} $ is the Fock vacuum satisfying $c(\mbf{r})\ket{0}=0$ (not the Landau level vacuum) as is clear from context. The second-quantized creation operators $\psi_{\kk,n}^\dag$ are defined by
\bea
\bra{\mbf{r}} \psi_{\kk,n}^\dag \ket{0} = \braket{0| c_{\mbf{r}} \psi_{\kk,n}^\dag|0}  = \psi_{\kk,n}(\mbf{r})
\eea
and $\{\psi_{\kk',m}^\dag,\psi_{\kk,n}\} = (2\pi)^2 \delta_{mn} \delta(\mbf{k}-\mbf{k}')$. We study the a general density-density interaction (essentially the Coulomb interaction with arbitrary screening) which can be put into the form
\bea
\label{eq:Hintcoul}
H_{int} &= \frac{1}{2} \int d^2r d^2r' \, n(\mbf{r}) V(\mbf{r}-\mbf{r}') n(\mbf{r}') \\
&= \frac{1}{2} \int \frac{d^2q}{(2\pi)^2} \,  V(\mbf{q}) \rho_{-\mbf{q}} \rho_{\mbf{q}}, \quad \rho_{\mbf{q}}=  \int d^2r \, e^{-i \mbf{q} \cdot \mbf{r}} n(\mbf{r})
\eea
where $V(\mbf{q})$ is the Fourier transform of the position-space potential. Throughout, we use $\mbf{q} = \mbf{k} + 2\pi \mbf{G}$ to denote a continuum momentum. We assume that $V(\mbf{q})>0$ but is otherwise fully general. Our goal is to express the Fourier modes $\rho_{\mbf{q}}$ in terms of the $\psi^\dag_{\mbf{k},m}$ operators. This is accomplished by calculating the matrix elements $\braket{0|\psi_{\mbf{k},m} \rho_\mbf{q} \psi^\dag_{\mbf{k}',n}|0}$ because $\rho_{\mbf{q}}$ is a one-body operator. The calculation is performed in \App{app:gcgii}, and yields
\begin{align}
\rho_\qq  &= \sum_{mn} \int \dfrac{d^2 k}{(2\pi)^2} e^{i\xi_\qq(\kk)} \psi_{\kk-\qq,m}^\dagger {\cal H}^\qq_{mn} \psi_{\kk,n},
\label{eq:rhoqm}
\end{align}
with the phase factor $\xi_\qq(\mbf{k})$ defined by
\begin{align}
e^{i\xi_\qq(\kk)} = \frac{e^{- \frac{ \bar{q} q}{4\phi}}  \vartheta \lp \left. \frac{(k_1 - q/2 ,k_2+ i q/2)}{2\pi} \right| \Phi \rp  }{\sqrt{\vartheta \lp \left. \frac{(k_1 ,k_2)}{2\pi} \right| \Phi \rp \vartheta \lp \left. \frac{(k_1 - q_1 ,k_2 - q_2)}{2\pi} \right| \Phi \rp}}.
\label{eq:siegelPhaseFactor}
\end{align}
 with $q = (\mbf{a}_1 + i \mbf{a}_2) \cdot \mbf{q}$. The unitary matrix $\mathcal{H}^\mbf{q}$ defined in \Eq{eq:Z}. We prove analytically that $e^{i\xi_\qq(\kk)}$ is a pure phase at the end of \App{app:siegelcomplete}. At $\mbf{k} = \pi \mbf{b}_1 + \pi \mbf{b}_2$ and $\mbf{k} = \pi \mbf{b}_1 + \pi \mbf{b}_2 + \mbf{q}$, the denominator of \Eq{eq:siegelPhaseFactor} has zeroes which are exactly canceled by the zeros of the numerator (they are removable singularities), so $\xi_\mbf{q}(\mbf{k})$ is always real. We plot $\xi_\mbf{q}(\mbf{k})$ in Fig.~\ref{fig:siegelPhase} which shows that a branch cut connects the removable singularities at $(\pi,\pi)$ and $(\pi + q_1, \pi+q_2)$.

So far, we have developed an expression for the density operators (\Eq{eq:rhoqm}) and thus for the many-body Coulomb Hamiltonian in terms of the single-particle magnetic translation group eigenstates. This will make it possible to perform a projection onto a set of low-energy bands. To do so, define the energy eigenstate operator $\gamma_{\kk,N}^\dagger$ that creates state at momentum $\kk$ in band $N$:
\begin{align}
\label{eq:gammavecsum}
\gamma_{\kk,N}^\dagger = \sum_m U^N_m(\kk) \psi^\dagger_{\kk,m},
\end{align} with $U^N(\mbf{k})$ the eigenvector of the Hamiltonian corresponding to band $N$. (In models with more orbitals indexed by $\al$, \Eq{eq:gammavecsum} would also contain a sum over $\al$.) In second quantized notation, we arrive at the general expression
\begin{align}
\rho_\qq = \int \dfrac{d^2 k}{(2\pi)^2} \sum_{MN} \gamma_{\kk-\qq,M}^\dagger M_{MN} (\kk,\qq) \gamma_{\kk,N},
\label{eq:densityFormFactor}
\end{align}
where the form factor matrix $M(\mbf{k},\mbf{q})$ obtained from \Eq{eq:densityFormFactor} is defined as
\begin{align}
M_{MN} (\kk,\qq) = e^{i\xi_\qq(\kk)} [U^\dag(\kk - \qq) {\cal H}^\qq U(\kk)]_{MN} \ .
\label{eq:formFactor}
\end{align}
Note that $M(\mbf{k},\mbf{q})$ is not a gauge-invariant quantity because the eigenvectors in the matrices $U(\mbf{k})$ and $U(\mbf{k} - \mbf{q})$ are only defined up to overall phases (or in general unitary transformations if there are degeneracies in the bands). \App{app:coulham} contains a complete discussion, which we summarize by noting the ``gauge freedom" taking $M(\mbf{k},\mbf{q}) \to W^\dag(\mbf{k} - \mbf{q}) M(\mbf{k},\mbf{q}) V(\mbf{k})$ where $W(\mbf{k} - \mbf{q}),V(\mbf{k})$ are arbitrary unitary matrices.  There are gauge-invariant quantities determined from $M(\mbf{k},\mbf{q}) $ such as its singular values, which are the eigenvalues of $M^\dag(\mbf{k},\mbf{q}) M(\mbf{k},\mbf{q})$. We will use the singular values to study the flat metric condition \cite{2020arXiv200911301B} in \Sec{sec:groundstates}.

Having discussed the form factors, we emphasize that \Eq{eq:densityFormFactor} is an exact expression for the density operator. To define a projected density operator, we restrict the indices $M, N$ to a subset of low-energy bands so that $\rho_{\mbf{q}}$ annihilates all other bands. Our result in Eq.~\ref{eq:densityFormFactor} is structurally similar to the form factor expression obtained in \Ref{2020arXiv200911301B} in zero flux. We discuss the behavior of the form factor in \App{eq:appcoulomb}.

\section{Twisted Bilayer Graphene: Single-particle Physics}
\label{sec:BMmodel}

\begin{figure}
\includegraphics[width=0.4\textwidth]{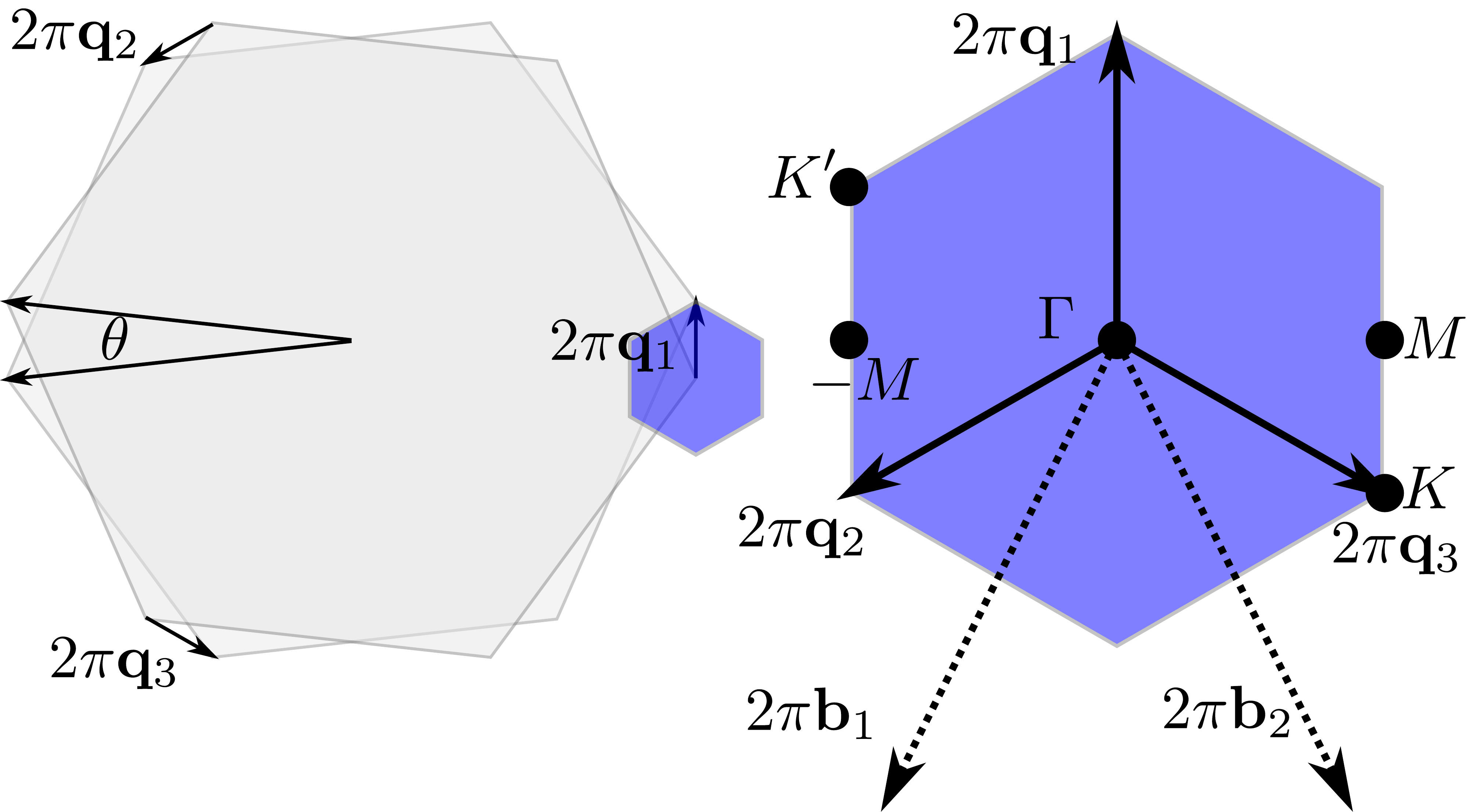}
\caption{Construction and conventions of the moir\'e BZ, blue hexagon, from the graphene layers with relative twist $\theta$.}
\label{fig:MoireSetup}
\end{figure}

Twisted bilayer graphene (TBG) is a metamaterial formed from twisting two graphene sheets by a relative angle $\th$ \cite{2011PNAS..10812233B,PhysRevB.98.085435,Song_2019}. The resulting moir\'e pattern is responsible for the very large unit cell that allows experimental access to $\phi = 2\pi$. Let us set our conventions for the geometry of the moir\'e twist unit cell. First, the graphene unit cell has a lattice vector of length $a_{g} = .246$nm and an area $\Omega_{g}  = a_{g}^2 \frac{\sqrt{3}}{2}$. The monolayer graphene K point is $\mbf{K}_{g} = \frac{2\pi}{a_{g}}(0, 2/3)$. The moir\'e vectors $\mbf{q}_j$ are defined by the difference in momentum space of the rotated layers' $K$ points:
\bea
\label{eq:kth}
2\pi \mbf{q}_1 &= (R_{\theta/2}-R_{-\theta/2})\mbf{K}_{g}, \quad \mbf{q}_j = C_3 \mbf{q}_{j-1}, \\
 2\pi |\mbf{q}_j| &\equiv  k_\th = 2 |\mbf{K}_{g}|  \sin \frac{\th}{2} \, =  \frac{8\pi  \sin \frac{\th}{2} }{3a_{g}}  \\
\eea
where $R_\th$ is a 2D rotation matrix. The moir\'e reciprocal lattice vectors are defined
\bea
\mbf{b}_j = \mbf{q}_j - \mbf{q}_3, \quad \mbf{b}_1 \times \mbf{b}_2 &= \frac{(2 \sin \frac{\th}{2})^2}{\Omega_{g}} \ .
\eea
The moir\'e lattice is defined by $\mbf{a}_i \cdot \mbf{b}_j = \delta_{ij}$ which yields
\bea
\mbf{a}_1 = \frac{a_{g}}{2 \sin \frac{\th}{2}}  \{-\frac{\sqrt{3}}{2}, - \frac{1}{2}\} \quad \mbf{a}_2 = \frac{a_{g}}{2 \sin \frac{\th}{2}} \{\frac{\sqrt{3}}{2},  -\frac{1}{2}\} \ . \\
\eea
Finally, the moir\'e unit cell has area
\bea
\Omega = \mbf{a}_1 \times \mbf{a}_2 = \frac{\Omega_{g}}{(2 \sin \frac{\th}{2})^2} \ .
\eea  The moir\'e Brillouin zone is depicted in Fig.~\ref{fig:MoireSetup}.
At the magic angle where $\theta = 1.05^\circ$, the moir\'e unit cell is $\sim3000$ times larger than the graphene unit cell. The magnetic translation group commutes when $\phi = \frac{e B \Omega }{\hbar} = 2\pi$, which occurs at $B \in  (25,32)$T for $\theta \in (1.03^\circ, 1.15^\circ)$. These fields are experimentally accessible, making it possible to explore the Hofstadter regime of TBG. \Ref{2021arXiv211111434H} focuses on TBG at the magic angle, as well as the evolution of the spectrum in flux.

\begin{figure}
\includegraphics[trim=0 0 0 0,clip,width=0.5\textwidth]{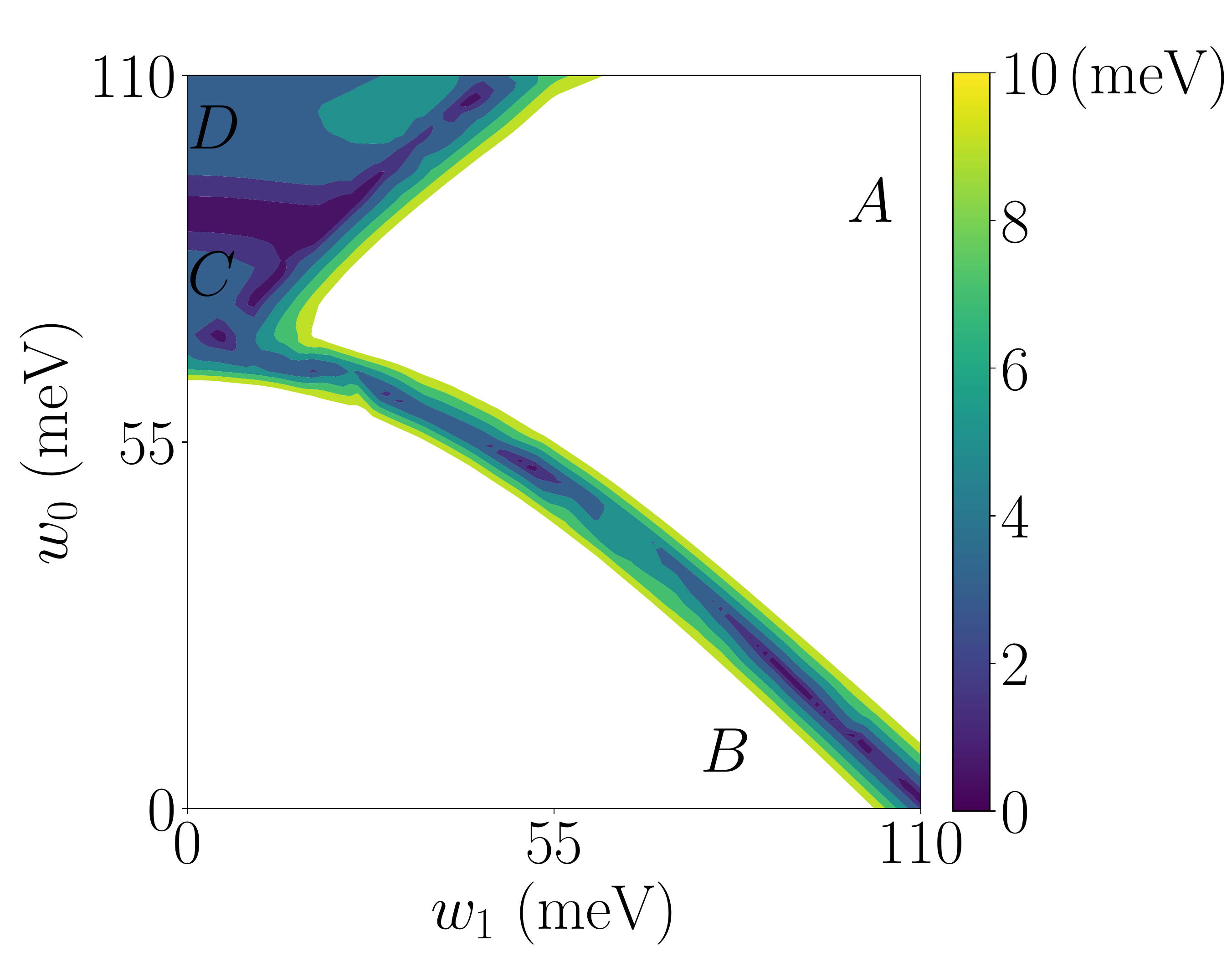}
\caption{Phase diagram of TBG in $2\pi$ flux at magic angle.  We plot the gap between the flat and passive bands as a function of parameters $w_0, w_1$. Phase $A$, containing the physical TBG parameters is in the crystalline regime where the flat bands have zero Chern number, while phase $B$ is connected to the Landau level limit where each flat band has Chern number $-1$. The first chiral limit where $w_0$ is in phase $B$ at $w_1 = 110$meV. C and D are phases connected to the second chiral limit $w_1 = 0$ where the bands have strong dispersion (see \Fig{fig:manyTBGPlots}).}
\label{fig:phaseDiagram}
\end{figure}

The following sections contain a thorough treatment of TBG at $2\pi$ flux. We discuss the Bistritzer-MacDonald (BM) Hamiltonian in \Sec{sec:TBGbs} and show the phase diagram of TBG, identifying a crystalline regime (including the physical TBG parameters) where the flat bands have vanishing Chern number and a Landau level regime (including the first chiral limit) where the flat bands each have Chern number $-1$, denoted by $A$ and $B$ respectively in the phase diagram Fig.~\ref{fig:phaseDiagram}. In \Sec{sec:TBGsymtop}, we discuss the symmetries, topology, and Wannier functions which are different than at zero flux. Importantly, we find that the $C_{2z}\mathcal{T}$ symmetry, which is essential in protecting the nontrivial topology at $\phi=0$, is broken. At $\phi = 2\pi$, we find that the TBG flat band structure can be obtained from atomic limits but still has Wannier functions pinned to the corners of the moir\'e unit cells. In \Sec{sec:TBGchiral}, we focus on the chiral limit of TBG where the chiral anomaly, a well-studied feature of relativistic gauge theory \cite{fradkin2021quantum, peskin2018introduction, PhysRev.177.2426,bell1969pcac,Fujikawa:2004cx,2019PhRvB..99w5144L,Niemi:1984vz,PhysRevB.94.195150,RevModPhys.83.1057}, protects a pair of perfectly flat bands in TBG at all angles at $2\pi$ flux.

\begin{figure}
\includegraphics[width=0.47\textwidth]{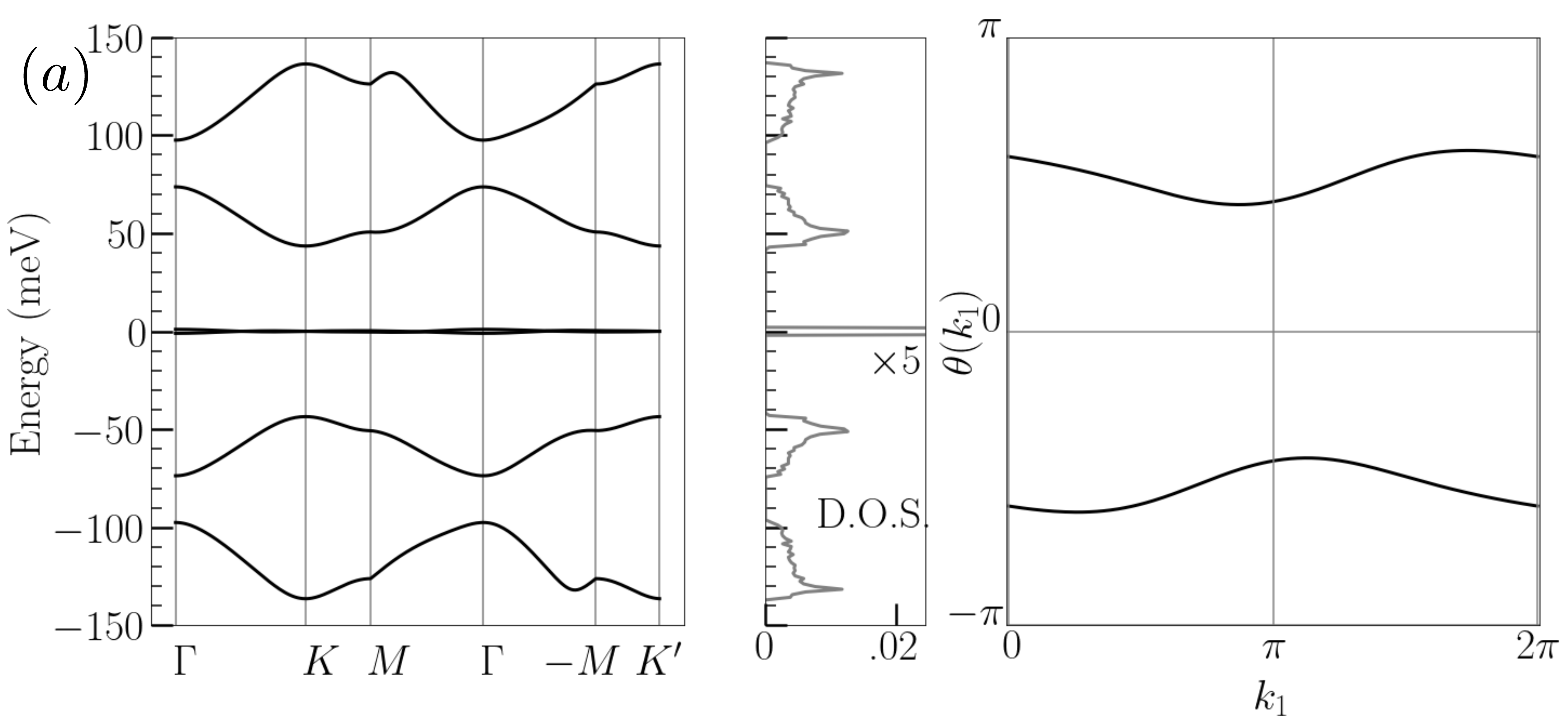}
\includegraphics[width=0.47\textwidth]{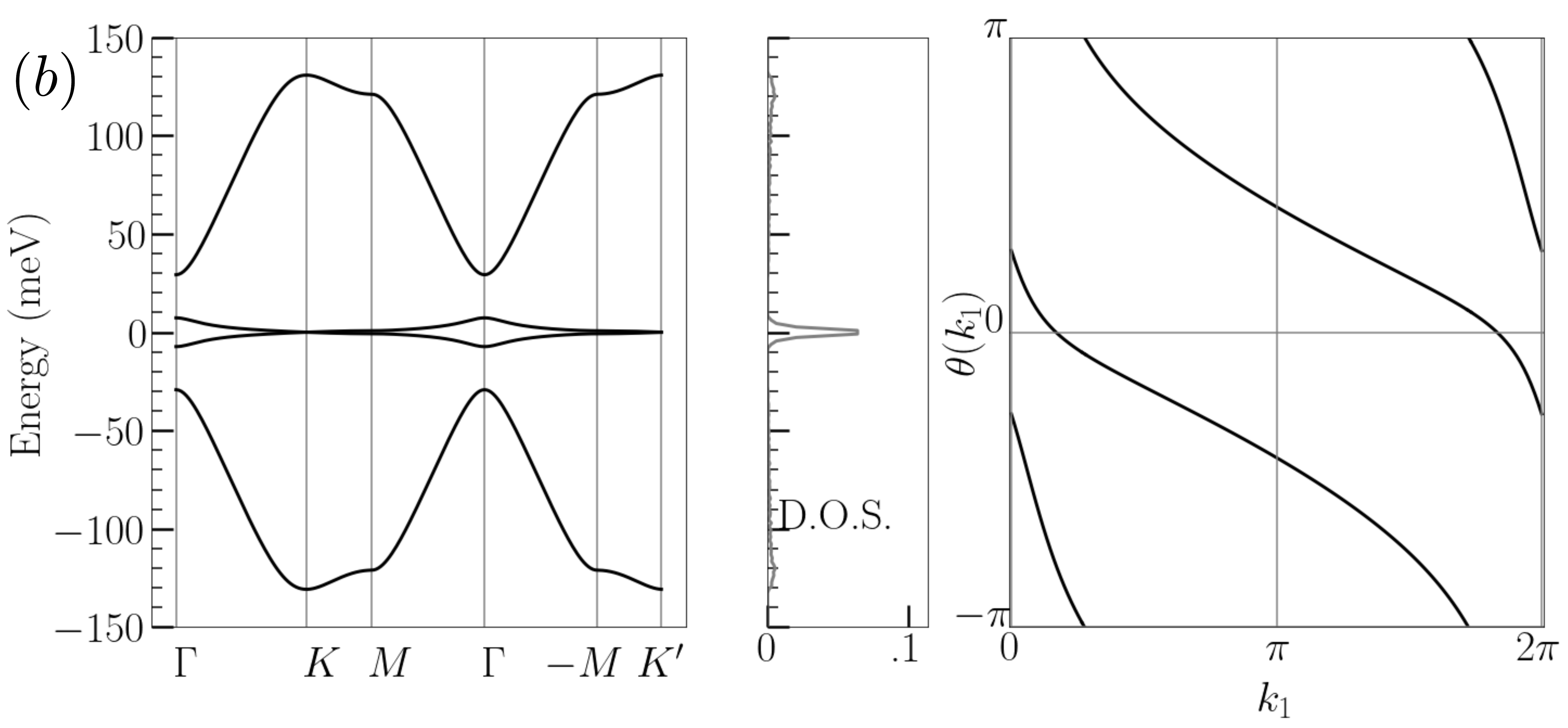}
\includegraphics[width=0.48\textwidth]{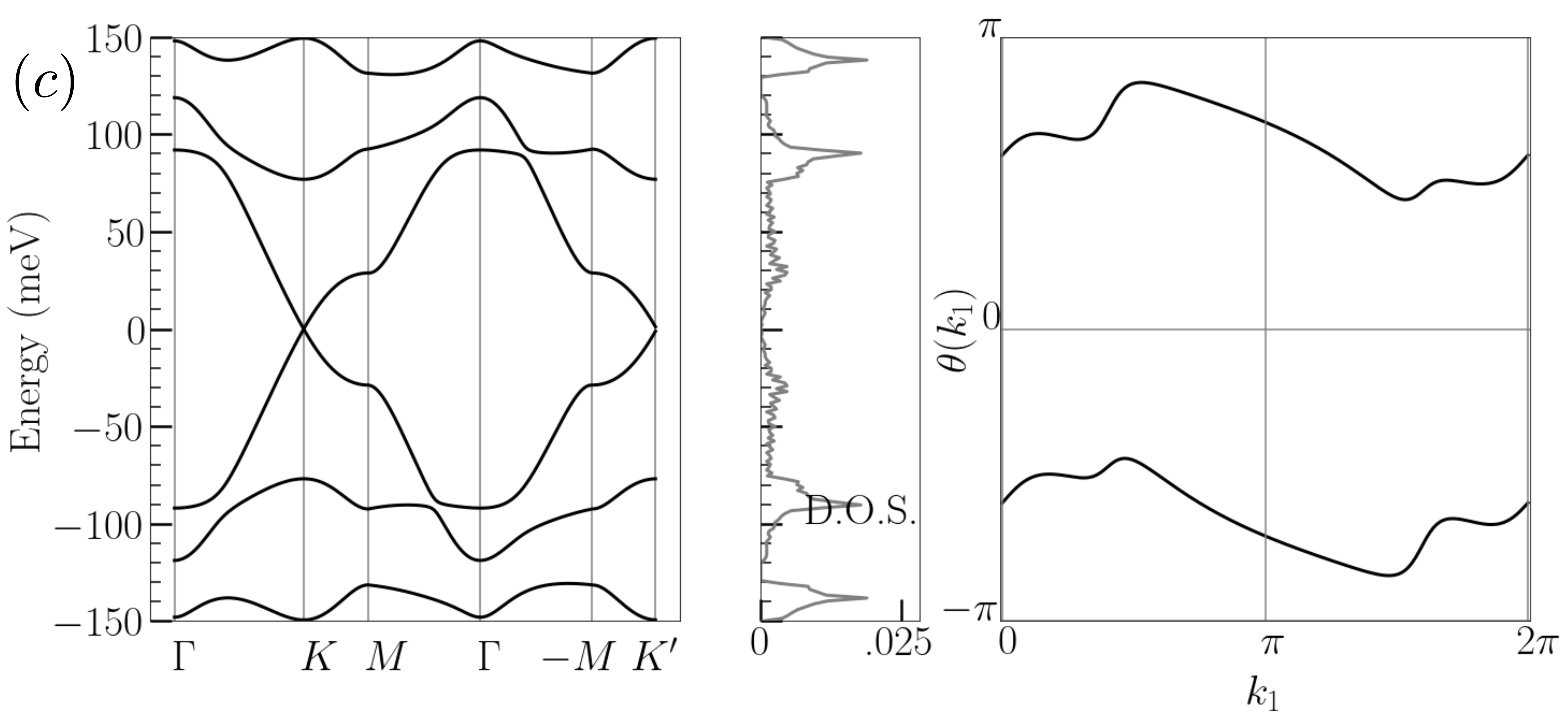}
\includegraphics[width=0.48\textwidth]{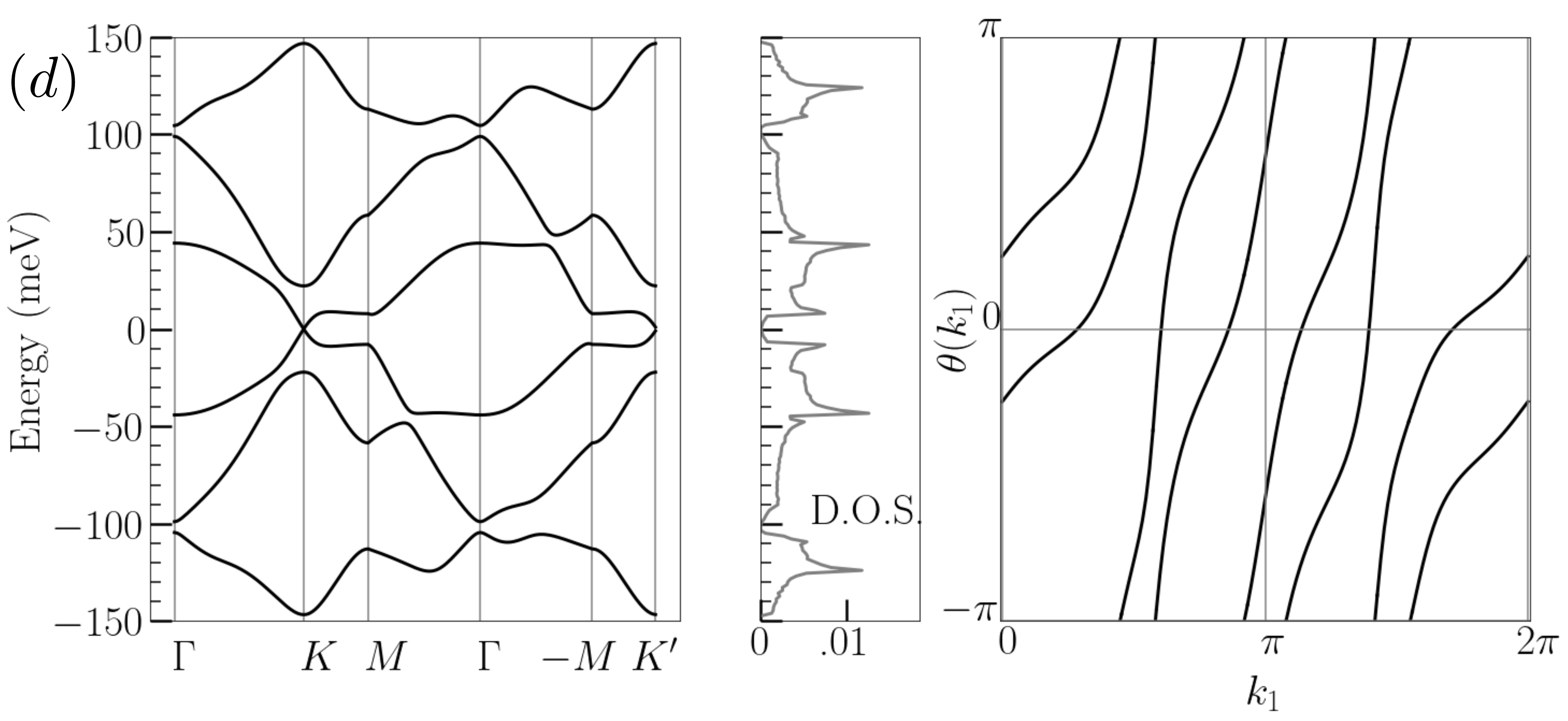}
\includegraphics[width=0.48\textwidth]{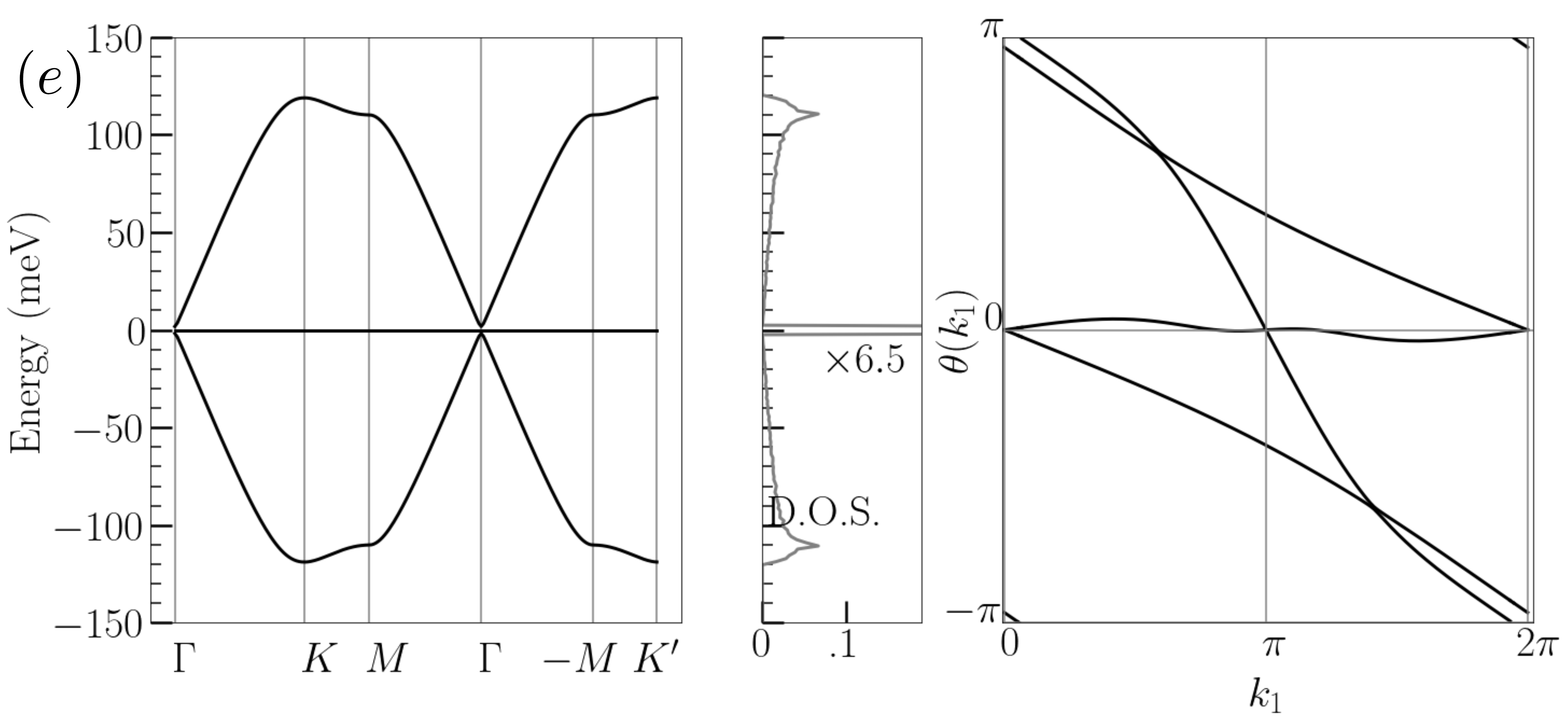}
\caption{Band structures (left), density of states (middle), and Wilson loops (right) of TBG at $2\pi$ flux. The parameters $\big(\sqrt{3}w_0/(v_Fk_\th) ,\sqrt{3}w_1/(v_Fk_\th) \big) $  given by (a): $(0.8,1)$, (b): $(0.05, 0.8)$, (c): $(0.7, 0.15)$, (d): $(0.97, 0.32)$, (e): $(0.0,1.0)$. (a-d) are chosen to be connected to phases $A- D$ (see \Fig{fig:phaseDiagram}), and $(e)$, the chiral limit, is connected to $B$ but has a very small gap ($< 2$meV). The very small gap makes the flat band Wilson loop ill-conditioned, so we compute the Wilson loop of the middle $4$ bands. }
\label{fig:manyTBGPlots}
\end{figure}

\subsection{Band structure}

\label{sec:TBGbs}
We begin with the Bistritzer-MacDonald model of twisted bilayer graphene in the untwisted graphene $K$ valley (and arbitrary spin) at zero flux:
\begin{align}
H_{BM} = \begin{pmatrix}
-i \hbar v_F \pmb{\sigma} \cdot \pmb{\nabla} & T^\dagger(\rr) \\
T(\rr) & -i \hbar v_F \pmb{\sigma} \cdot \pmb{\nabla}
\end{pmatrix},
\label{eq:HBMmodel}
\end{align} with $\sigma$ labeling the sublattice degree of freedom and the $2 \times 2$ matrix notation labeling the layer index. Note that $H_{BM}$ neglects the twist angle dependence in the kinetic term and thus has an exact particle-hole symmetry \cite{Song_2019}. For simplicity, we work in this approximation, but we note that incorporating the twist angle dependence poses no essential difficulty for our formalism. The moir\'e potential is $T(\rr) = \sum_{j=1}^3 e^{2\pi i\qq_j \cdot \rr} T_j$ where
\begin{align}
T_{j+1}= w_0 \sigma_0 &+ w_1 \lp \sigma_1 \cos \frac{2\pi}{3}j  +\sigma_2  \sin \frac{2\pi}{3} j \rp.
\label{}
\end{align}
To add flux into $H_{BM}$, we employ the canonical substitution $-i\hbar\pmb{\nabla} \rightarrow \pmb{\pi}$.  As written, $H_{BM}$ is not translation-invariant: the $\qq_i$ vectors which appear in the moir\'e potential are not reciprocal lattice vectors. However, $H_{BM}$ can be made translation invariant by a unitary transformation:
\bea
V_1 = \begin{pmatrix}
e^{i\pi \qq_1 \cdot \rr} & 0 \\ 0 & e^{-i\pi \qq_1 \cdot \rr}
\end{pmatrix} \\
\eea
which acts only on the layer index.\footnote{See \App{eq:BMsymm} for a discussion of $V_1$ in the $K'$ valley.} Acting on the states, $V_1$ shifts the momentum in the different layers by $2\pi \mbf{q}_1$, reflecting separation of the Dirac points in \Fig{fig:MoireSetup}. We then define the Hamiltonian in flux by
\bea
\label{eq:BM2pi}
& H^{\phi}_{BM}(\rr) = V_1 \begin{pmatrix}
 v_F \pmb{\sigma} \cdot \pmb{\pi} & T^\dagger(\rr) \\
T(\rr) & v_F \pmb{\sigma} \cdot \pmb{\pi}
\end{pmatrix} V_1^\dagger  \\
&= \begin{pmatrix}
 v_F \pmb{\sigma} \cdot \pmb{\pi} - \pi v_F \qq_1\cdot \pmb{\sigma} & {\tilde T}^\dagger(\rr) \\
{\tilde T}(\rr) & v_F \pmb{\sigma} \cdot \pmb{\pi}+ \pi v_F \qq_1\cdot \pmb{\sigma}
\end{pmatrix}
\eea
with $\tilde{T}(\rr) = T_1 + T_2 e^{2\pi i\bbb_1 \cdot \rr} + T_3 e^{2\pi i\bbb_2 \cdot \rr}$. In this form, the matrix elements of $\tilde{T}(\rr)$ in the magnetic translation operator basis can be directly obtained with Eq. 139 in \Ref{SM} in a sublattice/Landau level tensor product basis. An explicit expression is given in Eq. 228 of \Ref{SM}. The kinetic term can be expressed simply with Landau level operators. Expanding the Pauli matrices, we find
\bea
\label{eq:Diracham}
v_F \pmb{\sigma} \cdot \pmb{\pi}   &= v_F \sqrt{2eB} \bpm 0 & a^\dag \\ a & 0 \epm = v_F\sqrt{2\phi/\Omega}   \bpm 0 & a^\dag \\ a & 0 \epm \\
&= v_F k_\th \lp\frac{3\sqrt{3}}{2\pi} \rp^{1/2}  \bpm 0 & a^\dag \\ a & 0 \epm \\
\eea
using $\phi = 2\pi$ and the moir\'e wavevector $k_\th$ in \Eq{eq:kth}. The numerical factor coming from the unit cell geometry is $(3\sqrt{3}/2\pi)^{1/2} \simeq .91$. Lastly, the momentum shift $\pi v_F \qq_1\cdot \pmb{\sigma}$ in \Eq{eq:BM2pi} acts as the identity on the Landau level index, and $\pi v_F \qq_1\cdot \pmb{\sigma} = \frac{v_F k_\th}{2} \sigma_2$ using $2\pi\mbf{q}_1 = k_\theta \hat{y}$. The Dirac Hamiltonian \Eq{eq:Diracham} in flux is well-studied.  At $2\pi$ flux and $\theta = 1.05^\circ$, the low energy spectrum of \Eq{eq:Diracham} consists of a zero mode and states at $\pm E_1 = \pm .91 v_F k_\th = \pm170$meV. This is on the same scale as the potential strength $w_1 = 110$meV.

\begin{figure}
\includegraphics[width=0.45\textwidth]{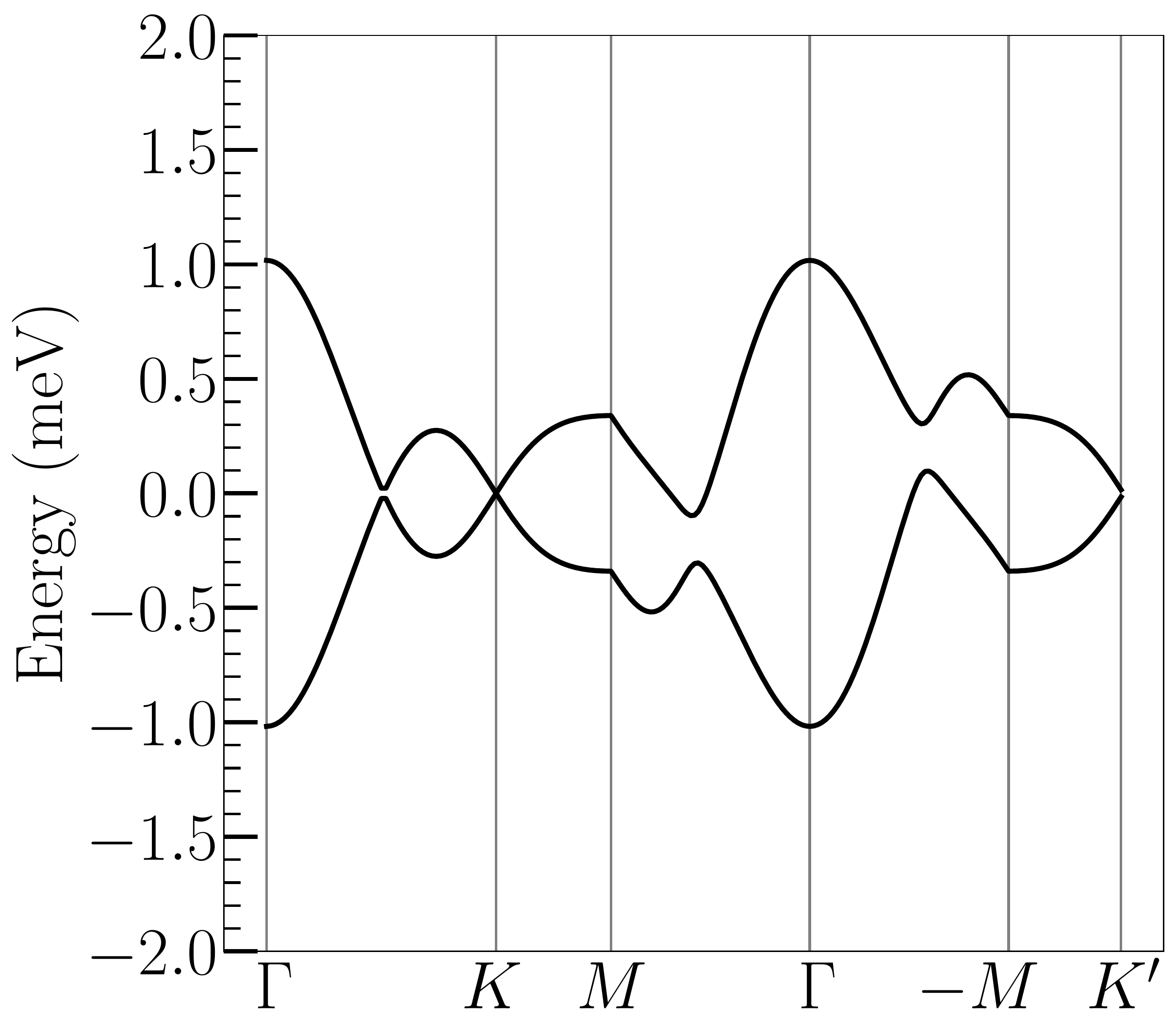}
\caption{Close-up of the flat bands of TBG in flux at magic angle.  Note the protected Dirac points at $K, K'$ due to the different $C_3$ eigenvalues of the flat bands (see \Eq{eq:actsym}) and $MT, P$ symmetries, as well as the maximal gap at $\Gamma$ where the $C_3$ eigenvalues are the same.}
\label{fig:zoomedFlatBands}
\end{figure}

Numerical analysis of the band structure is straightforward and yields two flat bands (per valley and spin, or $8$ total) gapped from the dispersive bands by approximately 40 meV.
See Fig.~\ref{fig:manyTBGPlots}(a) for the band structure, density-of-states, and the Wilson loop of the flat bands for TBG, Fig.~\ref{fig:manyTBGPlots}(b-e) for other choices of parameters $w_0, w_1$.  For a close-up of the flat-band dispersion at the magic angle see Fig.~\ref{fig:zoomedFlatBands}.
\label{sec:TBGsymtop}
\begin{figure}
\includegraphics[width=0.22\textwidth]{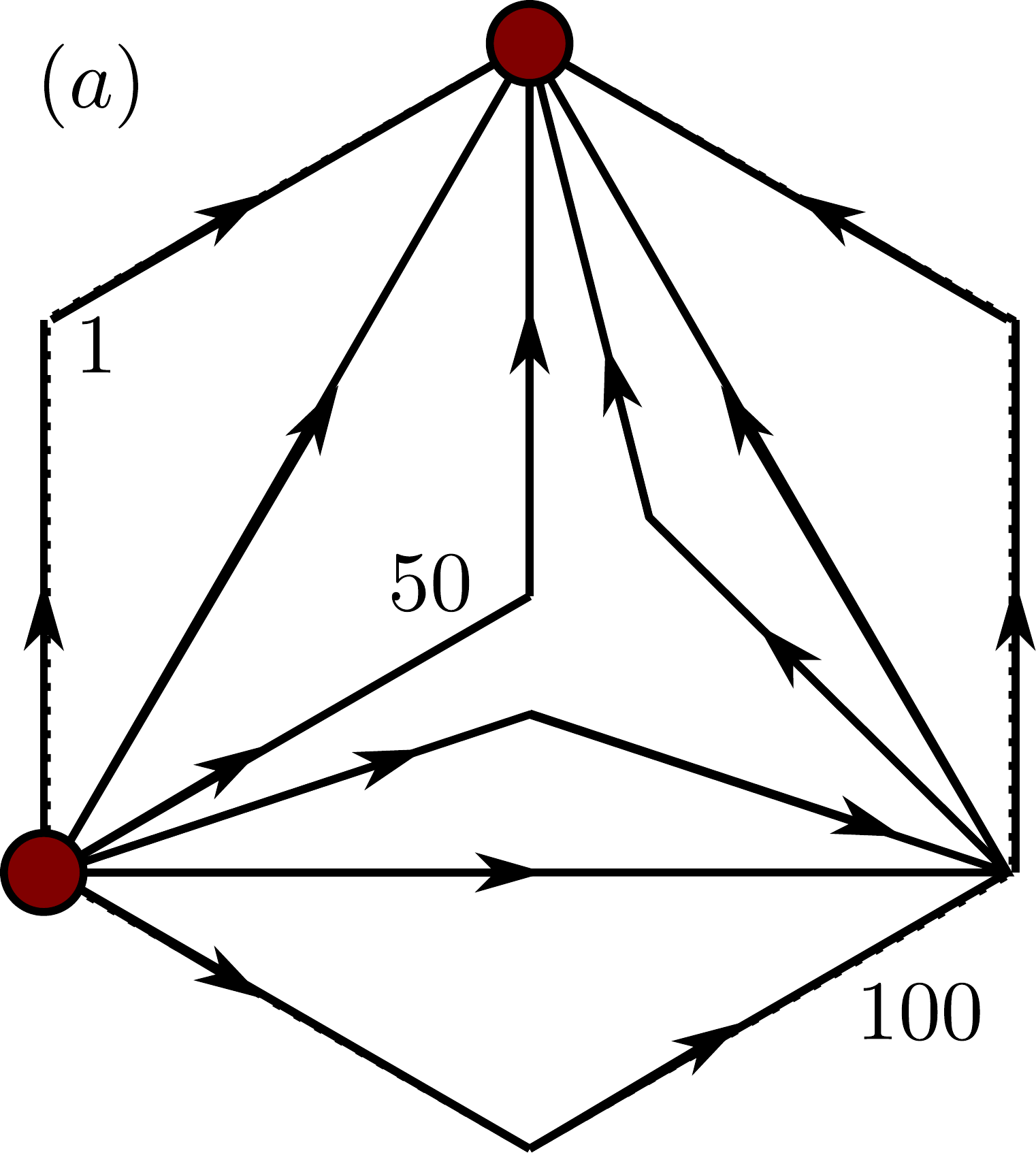}
\includegraphics[width=0.22\textwidth]{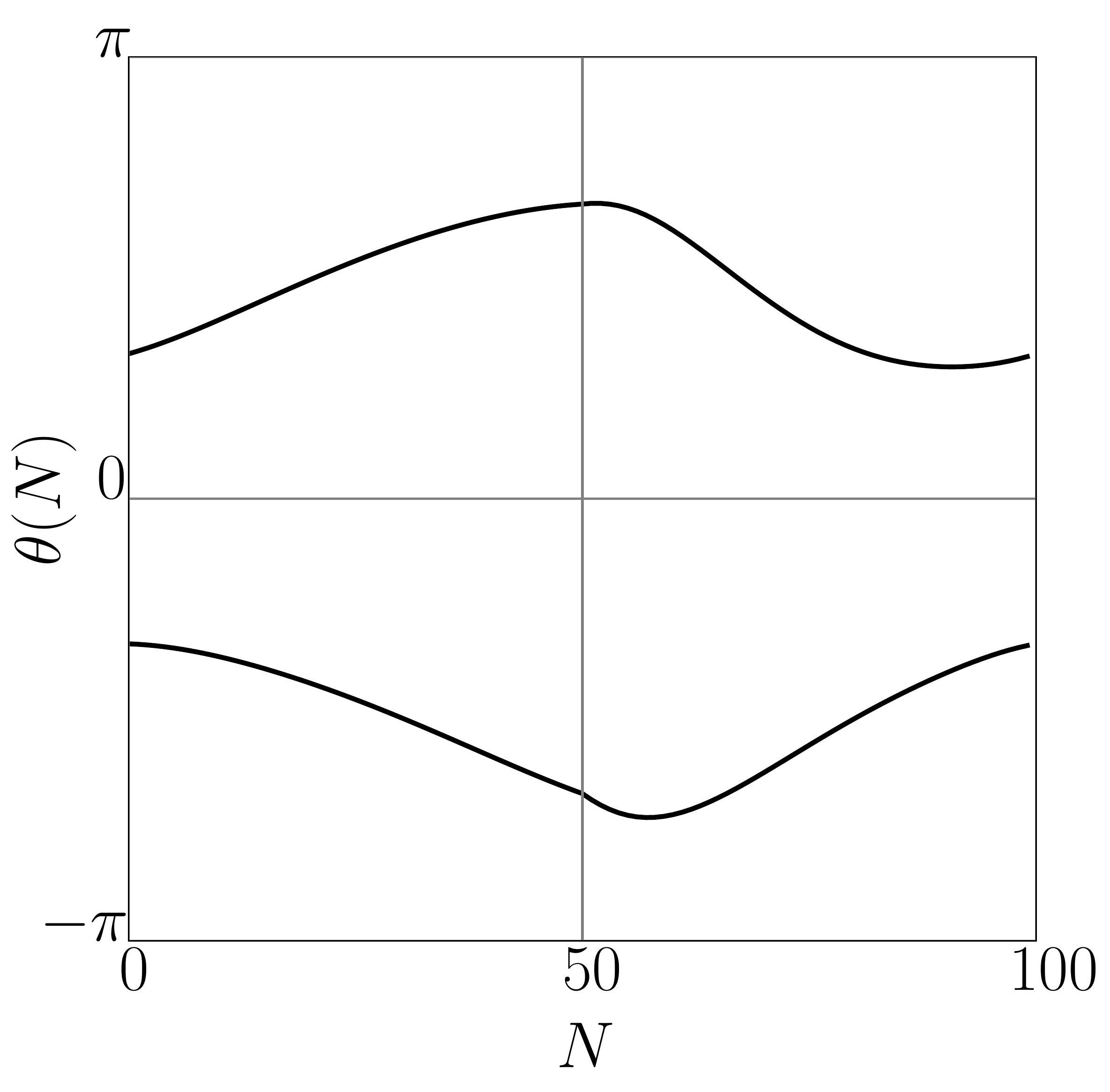}
\includegraphics[width=0.22\textwidth]{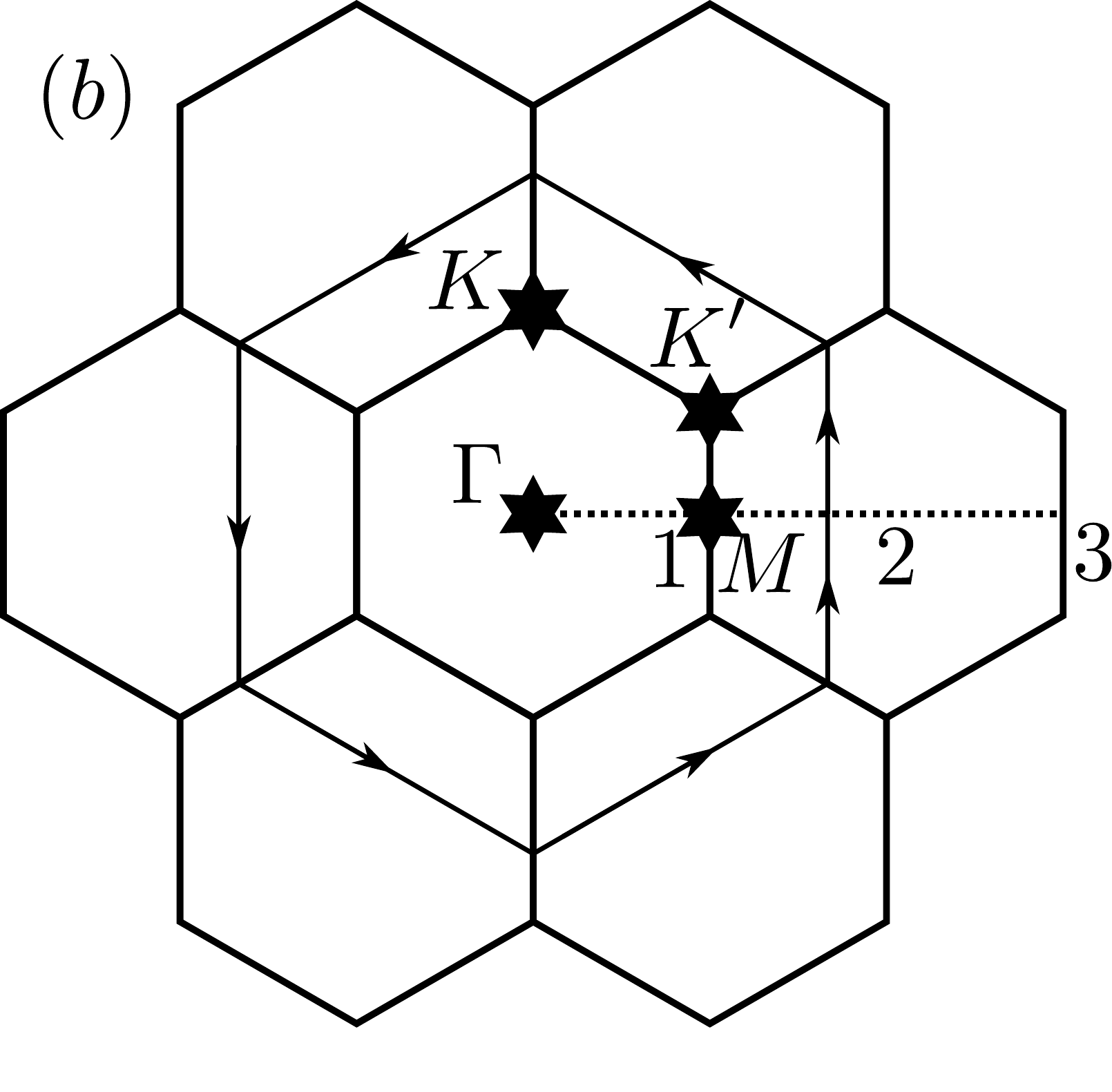}
\includegraphics[width=0.22\textwidth]{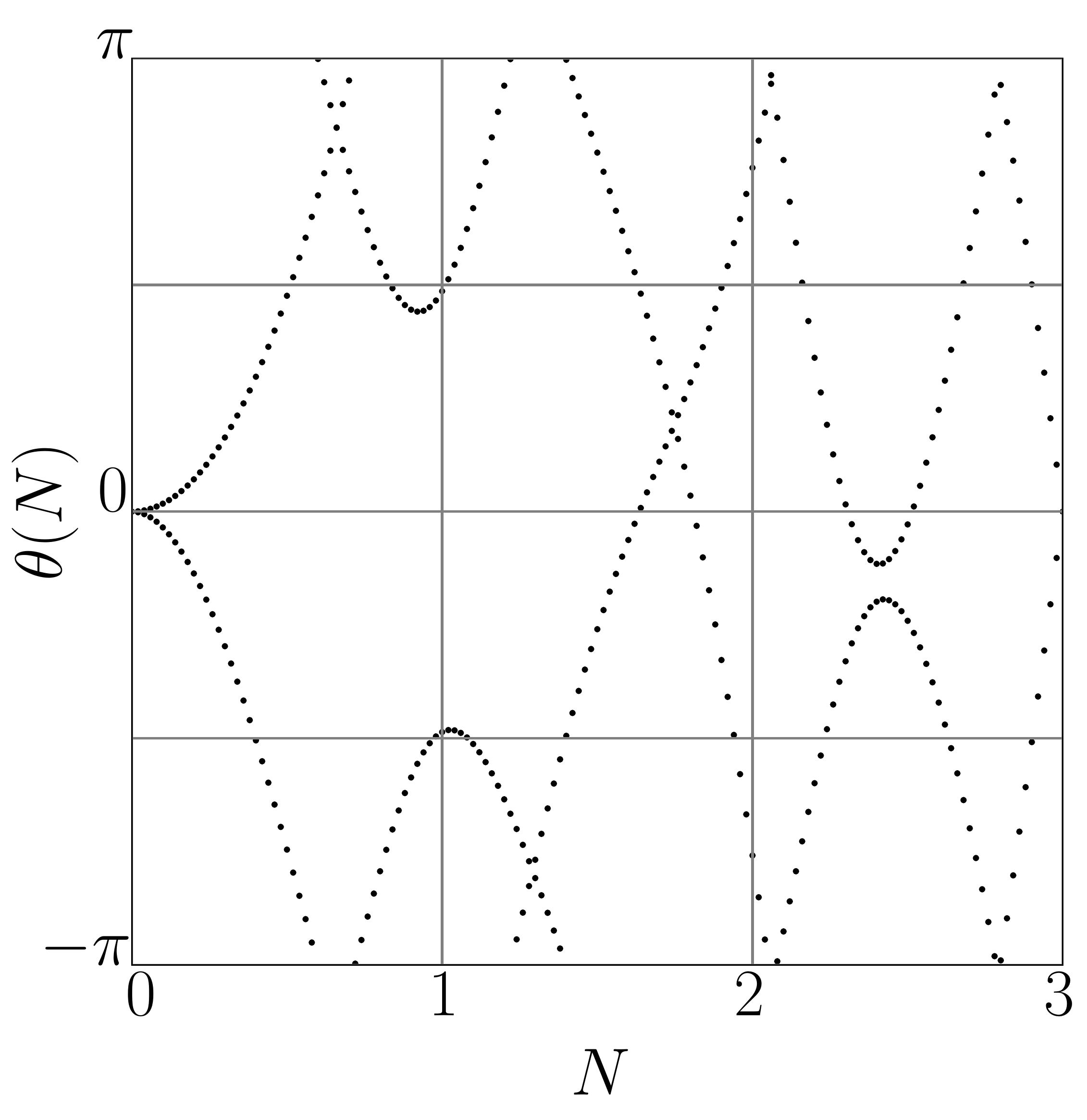} \\
(c) \includegraphics[width=0.32\textwidth]{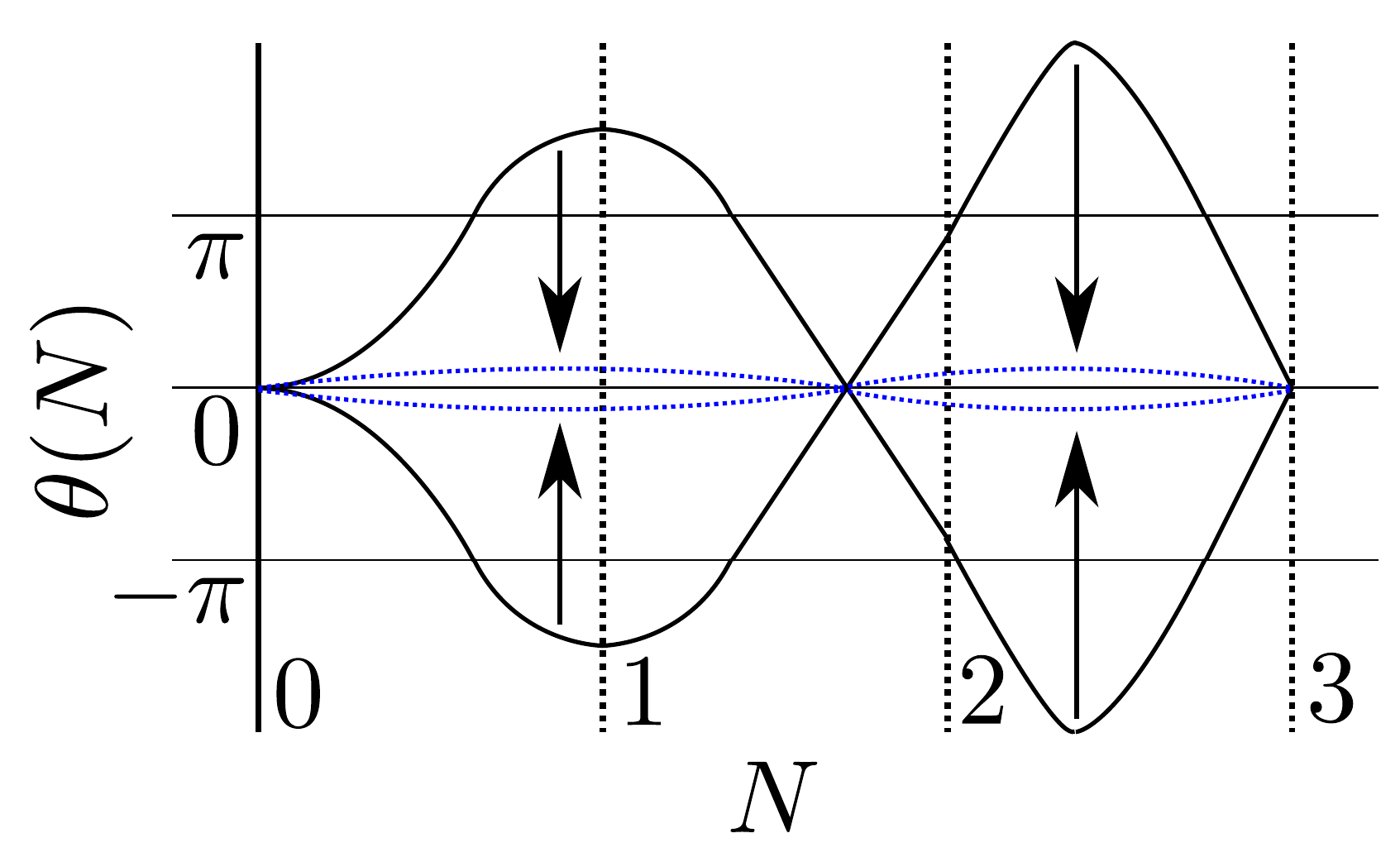}
\caption{$C_3$-symmetric Wilson loops, discussed in Refs.~\cite{Song_2019,Bradlyn_2019}.  In (a), the path 1 begins at $K$, goes to $K'$, then back to $K$.  The midpoint of the path is continuously changed until $\Gamma$ at $50$; further paths then follow a more complicated trajectory linking $K$ back to $K$ and then back again. In $(b)$, Wilson loops are taken in successively larger hexagons surrounding the $\Gamma$ point. Neither loop has nontrivial winding because there are no symmetries that protect crossings at $\pm \pi$, so the Wilson loops can be deformed to flat lines as depicted in $(c)$ which shows a caricature of the deformation process.}
\label{fig:C3WilsonLoops}
\end{figure}

\subsection{Symmetries and Topology}

In zero flux, the topology of the TBG flat bands is protected by $C_{2z}\mathcal{T}$ symmetry \cite{Song_2019,Song_2021,2018arXiv180409719B}. However, $C_{2z}\mathcal{T}$ is broken in nonzero flux because $\mathcal{T}$ reverses the magnetic field and $C_{2z}$ preserves it \cite{PhysRevLett.125.236804}. On the lattice in the Peierls approximation, $C_{2z}\mathcal{T}$ is restored as a (projective) symmetry at certain values of the flux \cite{PhysRevLett.125.236804}, but we do not consider this approximation here. In this section, we show that the band representation of TBG at $\phi = 2\pi$ can be obtained from inducing atomic orbitals at the corners of the moir\'e unit cell, so the fragile topology at $\phi = 0$ is broken by magnetic field. However, we find that band representation is decomposable \cite{2017Natur.547..298B,2018PhRvL.120z6401C,2018PhRvB..97c5139C,Bradlyn_2019}, so the flat bands are topologically nontrivial when gapped from each other via a particle-hole breaking term.

First we review the topology in zero flux which is discussed comprehensively in \Refs{Song_2019}{Song_2021}. The space group of TBG is $p6'2'2$ which is generated by $C_{3}, C_{2z} \mathcal{T}, $ and $C_{2x}$.\footnote{Technically $p6'2'2$ is a 3D space group. We only consider the $k_z= 0$ plane, which is equivalent to the 2D wallpaper group $p6'm'm$ because the action of $M_y$ and $C_{2x}$ is identical in 2D.} The symmetries are: three-fold rotations around the $AA$ moir\'e site $C_3$, spacetime inversion $C_{2z} \mathcal{T}$, and two-fold rotation around the $x$-axis $C_{2x}$.  Note that in 2D, $C_{2x}$ is indistinguishable from $M_y$, a mirror taking $y\to -y$. The band representation of the flat bands is

\bea
\mathcal{B}^{\phi = 0} = \Gamma_1 + \Gamma_2 + K_2K_3 + M_1 + M_2
\eea

and the irreps are defined at the high symmetry momenta $\Gamma = (0,0), K = 2\pi \mbf{q}_1, M = \pi \mbf{b}_1$ by

\bea
\begin{array}{c|ccc}
6'm'm &1& C_3 & M_y \\
\hline
\Gamma_1 &1&1 &1\\
\Gamma_2 &1 & 1 & -1  \\
\end{array}, \ \begin{array}{r|rr}
6' &1& C_3 \\
\hline
K_2 K_3 &2 & -1 \\
\end{array},\ \begin{array}{c|cc}
2'm'm&1& C_{2x} \\
\hline
M_1 &1&1\\
M_2 &1 & -1 \\
\end{array} \ .
\eea

The presence of the anti-unitary $C_{2z}\mathcal{T}$ ($PC_{2z}\mathcal{T}$) symmetry in the space group is required to prove that the band representation $\mathcal{B}^{\phi=0}$ is fragile (stable) topological \cite{Song_2019,Song_2021}.

At $2\pi$ flux, the $C_{2z}\mathcal{T}$ and $C_{2x}$ symmetries are broken because they reverse the magnetic field \cite{PhysRevLett.125.236804}. The resulting band topology is mentioned in \Ref{2021arXiv211111434H}, which we review here for completeness. Without $C_{2z}\mathcal{T}$, the topology of the flat bands is not protected. The most direct way to see this is from the Wilson loop (see \Eq{eq:WC}) integrated along $\mbf{b}_2$ in \Fig{fig:C3WilsonLoops}(a) which shows no relative winding. The same Wilson loop at zero flux has $C_{2z}\mathcal{T}$-protected relative winding \cite{Song_2019}. We also plot the $C_3$-symmetric Wilson loops discussed in Refs.~\onlinecite{Song_2019, Bradlyn_2019} and find no winding, as shown in Fig.~\ref{fig:C3WilsonLoops}(a,b). The lack of winding in any Wilson loop suggests that localized, symmetry-respecting Wannier states may be formed from the two TBG flat bands at $2\pi$ flux (per valley per spin) \cite{2017Natur.547..298B, 2020arXiv201000598E}. Below, we discuss the flat bands in detail from the perspective of topological quantum chemistry.

At $2\pi$ flux, the 2D space group is reduced to $p31m'$ (the $k_z = 0$ plane of the 3D space group 157.55 in the BNS setting) generated by $C_{3}$ and $ M\mathcal{T} \equiv C_{2x}C_{2z}\mathcal{T}$. The full algebra, including the anti-commuting unitary $P$ symmetry,  is

\begin{align}
M\mathcal{T} C_3 &= C^\dag_3 M\mathcal{T}, ~~C_3^3 = 1 \nonumber \\
[P, C_3] &= 0,~~ P^2 = -1 \nonumber \\
\{P, M \mathcal{T}\} &= 0 ~~ (M \mathcal{T})^2 = +1 \nonumber
\label{eq:symmetryRelations}
\end{align}

and their action on the Hamiltonian is

\bea
\label{eq:actsym}
C_3 H^{\phi=2\pi}(\mbf{k}) C_3^\dag &= H^{\phi=2\pi}(C_3\mbf{k}) ,\\
M\mathcal{T} H^{\phi=2\pi}(k_x,k_y) (M\mathcal{T})^{-1} &= H^{\phi=2\pi}(k_x, -k_y) , \\
P H^{\phi=2\pi}(\mbf{k}) P^\dag &= - H^{\phi=2\pi}(-\mbf{k})  . \\
\eea

The operator ${\cal P} = P M\mathcal{T}$ squares to $+1$ and satisfies ${\cal P} C_3 = C_3^2 {\cal P}$. ${\cal P}$ sends $(k_x, k_y) \to (-k_x, k_y)$ and hence is local at the $K$ and $K'$ points.  Because ${\cal P}$ anticommutes with the Hamiltonian at $\Gamma, K$, and $K'$, it switches the two flat bands if they are at nonzero energies $\pm E$.  If ${\cal P} \ket{\Psi_{+E}} = \ket{\Psi_{-E}}$ and $\ket{\Psi_{+E}}$ carries $C_3$ eigenvalue $\omega$, then $\ket{\Psi_{-E}}$ also carries eigenvalue $\omega$.  For the $\Gamma$ point this is indeed what happens -- we find the $\Gamma$ point is gapped in \Fig{fig:zoomedFlatBands} -- but the $K, K'$ points cannot gap, as a Dirac cone carries different $C_3$ eigenvalues in the two flat bands.

\begin{figure}
\includegraphics[width=0.5\textwidth]{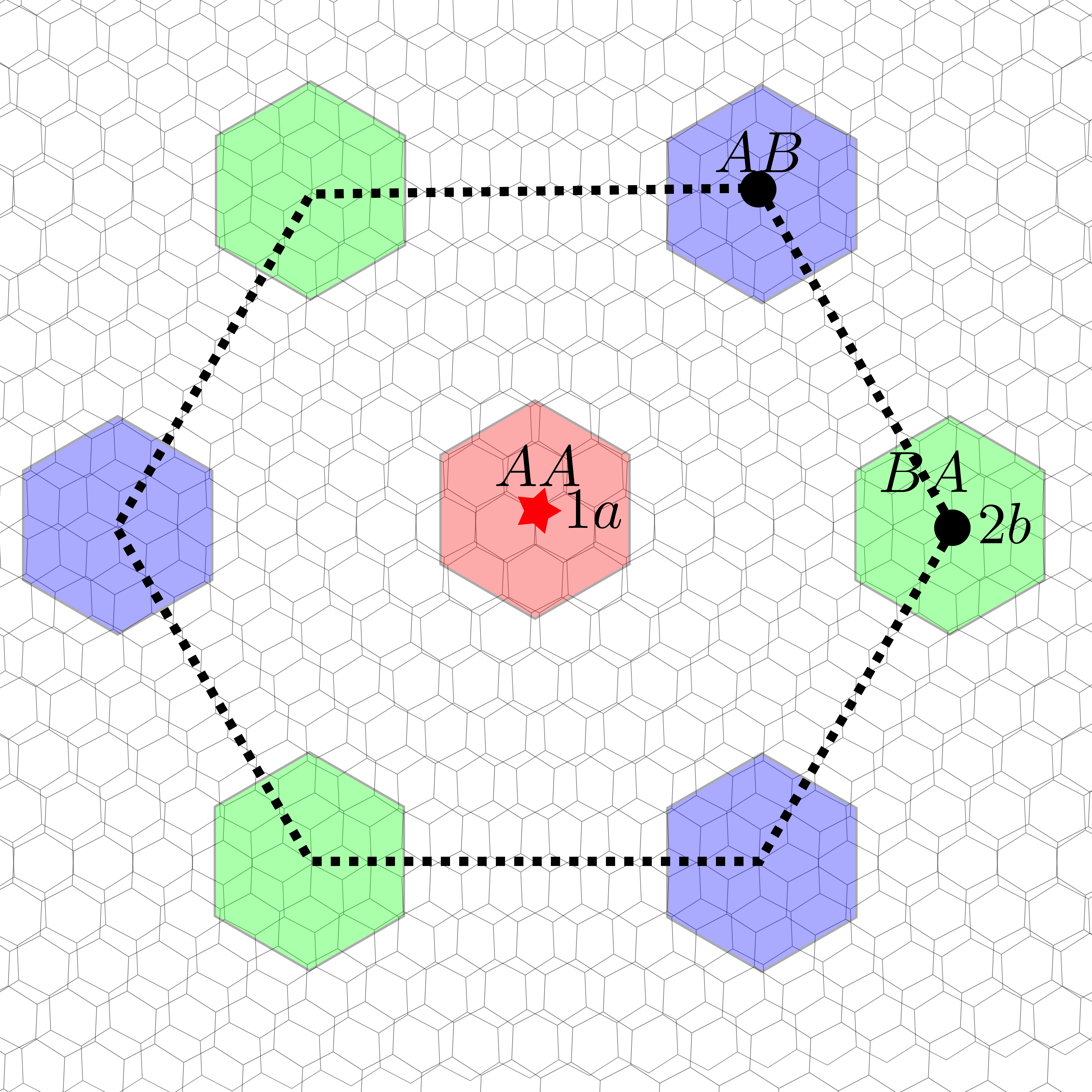}
\caption{Moir\'e lattice in real space, with colored regions denoting the AA and AB, BA stacking regions. The band representation $\mathcal{B}^{\phi=2\pi}$ can be induced from $s$ orbitals at the 2b position, which is composed of the AB and BA moir\'e sites.}
\label{fig:moirePositions}
\end{figure}

\Ref{PhysRevLett.125.236804} demonstrated that no symmetries or topology protect a gap closing between the flat bands and passive bands at nonzero flux, matched by experimental evidence in \Refs{2021arXiv211111341D}{2020arXiv200713390D}. As such, the irreps in nonzero flux are obtained from $\mathcal{B}^{\phi =0}$ by reduction to the $p31m'$ subgroup of $p6'2'2$. We use the Bilbao Crystallographic Server \cite{Aroyo:firstpaper,Aroyo:xo5013} to determine the irreps and elementary band representations of $p31m'$. They may be found at \url{https://www.cryst.ehu.es/cgi-bin/cryst/programs/mbandrep.pl}. The irreps of $p31m'$ are very simple: the high symmetry momenta are $\Gamma, K$, and $K'$ where all irreps are those of the point group $3$, so irreps at $\phi=0$ reduce to their $C_3$ eigenvalues at $\phi \neq 0$. We find
\bea
\label{eq:BR2pimom}
\mathcal{B}^{\phi = 2\pi} = \mathcal{B}^{\phi = 0}  \downarrow p31m' = 2 \Gamma_1 + K_2+K_3 + K'_2+K'_3
\eea
where the irreps in $p31m'$ that appear in \Eq{eq:BR2pimom} are defined
\bea
\begin{array}{r|rr}
3m'&1& C_3 \\
\hline
\Gamma_1 &1&1\\
\end{array}, \quad \begin{array}{r|rr}
3 &1& C_3 \\
\hline
K_2 &1 & e^{\frac{2\pi i}{3}}\\
K_3 &1 & e^{-\frac{2\pi i}{3}}\\
\end{array},\quad \begin{array}{r|rr}
3 &1& C_3 \\
\hline
K'_2 &1 & e^{-\frac{2\pi i}{3}}\\
K'_3 &1 & e^{\frac{2\pi i}{3}}\\
\end{array} \ .
\eea
As discussed, the particle-hole symmetry $\mathcal{P}$ ensures that the irreps at the $K$ and $K'$ points are degenerate, so $K_2+K_3$ and $K_2'+K_3'$ should be thought of as co-irreps. We can induce $\mathcal{B}^{\phi=2\pi}$ from the elementary band representations of $p31m'$:
\bea
\label{eq:BR2pi}
\mathcal{B}^{\phi = 2\pi} = A_{2b} \uparrow p31m'
\eea
where $2b$ is the Wyckoff position consisting of the $M\mathcal{T}$-related corners of the moir\'e unit cell (the AB and BA positions shown in \Fig{fig:moirePositions}) and the two-dimensional $A$ irrep is two $s$ orbitals, i.e. the representation of $C_3$ is $\mathbb{1}_{2\times2}$. From \Eq{eq:BR2pi}, we see that the band representation of TBG at $2\pi$ flux can be obtained from elementary band representations. This fact, coupled with the calculation of trivial Wilson loops, demonstrates the elementary band representation is not topological. Note that the unitary particle-hole symmetry $P$ acts as inversion in real space, and is implemented on the $A_{2b}$ irrep by exchanging the $s$ orbitals at AB and BA sites. Because there is no obstruction to locally realizing all symmetries of TBG at $2\pi$ flux, lattice model approaches \cite{2021arXiv210605670V,PhysRevX.8.031088} can faithfully capture the the topology. However, although $\mathcal{B}^{\phi = 2\pi} $ is an elementary band representation, the Bilbao crystallographic server reveals that it is decomposable into two topological bands with Chern numbers $\pm 1$ if the particle-hole symmetry $P$ is broken and the flat bands gap. This case is discussed in \Ref{2021arXiv211111434H}.

\subsection{Chiral Anomaly in TBG}
\label{sec:TBGchiral}

\Ref{PhysRevLett.122.106405} first identified a special region in the TBG parameter space called the chiral limit where $w_0 = 0$ ($w_1$ is unrestricted). In the chiral limit, an anti-commuting symmetry $C = \tau_0 \otimes \sigma_3 \otimes \mathbb{1} $ ($\tau_0$ is the $2\times2$ identity on the layer indices and $\mathbb{1}$ is the identity on the Landau level indices) appears which obeys
\bea
\{ C, H_{BM}^{\phi} \} = 0
\eea
for all flux $\phi$. We see this from Eq. 228 of \Ref{SM} because only $\sigma_1$ and $\sigma_2$ matrices appear when $w_0 = 0$ (see \App{eq:BMsymm}). In zero flux, \Ref{PhysRevLett.122.106405} identifies a discrete series of $w_1$ values where the two bands become \emph{exactly} flat and have opposite chirality.

We now show that in chiral TBG at $2\pi$ flux, there are two exactly flat bands for \emph{all} values of $w_1$, as we observe in \Fig{fig:manyTBGPlots}(e). We will prove this is protected by the two flat bands having the \emph{same} chirality. This is known as the chiral anomaly, which is a non-crystalline representation of chiral symmetry and cannot be realized in zero flux. First, recall that any state $\ket{E}$ at energy $E\neq0$ yields a distinct state $\ket{-E} = C \ket{E}$ of energy $-E$, and the chiral eigenvalues on the basis $\ket{E}, \ket{-E}$ are $\pm1$ because they are exchanged by $C$. We can determine the chiral eigenvalues of the flat bands in TBG analytically in the small $w_1$ limit where the kinetic term dominates and
\bea
H^{\phi}_{BM}(\rr) \to \begin{pmatrix}
 v_F \pmb{\sigma} \cdot \pmb{\pi} & 0 \\
0& v_F \pmb{\sigma} \cdot \pmb{\pi}
\end{pmatrix} , \quad \text{as } w_1 \to 0 \ .
\eea
The eigenstates are in the form $(\ket{E_n}, \pm \ket{E_n})^T$ where the $\pm$ states are orthogonal (so there are two states of energy $E_n$ to account for the two layers) and the Dirac Hamiltonian eigenstates are defined
\bea
\ket{E_0} = \bpm \ket{0} \\ 0  \epm , \quad \ket{E_n} = \frac{1}{\sqrt{2}} \bpm \ket{|n|} \\ \sgn (n) \ket{|n|-1}\epm , n \neq 0
\eea
with energies $\pmb{\sigma} \cdot \pmb{\pi} \ket{E_n} = \sgn(n) \sqrt{2 |n| \phi/\Omega} \ket{E_n}$ and $\sgn(0) =0$. The chirality operator on the Dirac states obeys
\bea
\sigma_3 \ket{E_0}  = + \ket{E_0}, \quad \sigma_3 \ket{E_n} = \ket{E_{-n}} \ .
\eea
In the $w_1\to 0$ limit, the zero energy flat band eigenstates of $H_{BM}$ in the chiral limit are
\bea
\label{eq:fbzec}
\frac{1}{\sqrt{\mathcal{N}(\mbf{k})}} \ \sum_{\mbf{R}} e^{-i \mbf{k} \cdot \mbf{R}} T_{\mbf{a}_1}^{\mbf{b}_1\cdot\mbf{R}} T_{\mbf{a}_2}^{\mbf{b}_2\cdot\mbf{R}} \bpm \ket{E_0} \\ \pm \ket{E_0} \epm
\eea
at every $\mbf{k} \in BZ$. The bands in \Eq{eq:fbzec} carry chiral eigenvalues $+1,+1$. Note that the chiral eigenvalues protect the perfectly flat bands at all $\mbf{k}$: if the energy of either of the flat bands states was not exactly zero, then $C\ket{E}$ would be a distinct state and the pair would have chiral eigenvalues $\pm1$. Hence the $+1,+1$ eigenvalues pin the states to zero energy. We now show this is true for $w_1 \neq 0$. The proof is by contradiction. First, we increase $w_1$ away from zero so the flat band eigenstates are superpositions of many Landau levels. However, the chiral eigenvalues cannot change from $+1,+1$. All gap closings occur as states $\ket{\pm E}$ touch the zero energy flat bands, but a pair of states $\ket{\pm E}$ necessarily has chiral eigenvalues $\pm1$ so the sum of the chiralities of the occupied bands is always $2$. Thus two states are always pinned to zero energy at every $\mbf{k}$ and all $w_1$, yielding exactly flat bands at \emph{all} angles. We emphasize that this situation is very different than at zero flux where the chiral eigenvalues of the flat bands are $\pm1$ which allows them to gap at generic values of $w_1$.

The $+1,+1$ chiral eigenvalues are called the chiral anomaly because the trace of $C$ over all bands at fixed $\mbf{k}$ formally satisfies
\bea
\label{eq:chiralsum}
\Tr \, C &= \sum_{N=-\infty}^{\infty} U^\dag_N(\mbf{k}) \sigma_3 U_N(\mbf{k}) \\
&=  \sum_{N=\pm1} U^\dag_N(\mbf{k}) \sigma_3 U_N(\mbf{k})  = 2
\eea
which is anomalous because $\Tr \sigma_3 = 0$. As in \Eq{eq:gammavecsum}, $U_N(\mbf{k})$ is the eigenvector of the $N$th band at momentum $\mbf{k}$. In the second line of \Eq{eq:chiralsum}, we used the $\pm1$ chiral eigenvalues of states at $E\neq0$ to cancel them from the sum, leaving only the passive bands. The fact that $\Tr C = 2$ can be understood from the Atiyah-Singer index theorem \cite{atiyah1968index,eguchi1980gravitation} which states that each Dirac Hamiltonian contributes $\phi/(2\pi)$ to the trace of the chirality operator, so $\Tr C = 2$ at $\phi = 2\pi$ because there are two layers \cite{2019PhRvB..99w5144L}. Strictly speaking, we cannot apply the index theorem because we have constructed the spectrum on an infinite plane which is not compact. However, we can effectively compactify the spectrum by taking $\mbf{k}$ to be discrete with $L^2$ values in the BZ corresponding to an $L \mbf{a}_1 \times L \mbf{a}_2$ torus in real space. Then there are a total of $2 L^2$ zero modes of $+1$ chirality from \Eq{eq:chiralsum}, so  $\Tr C = 2$ at each $\mbf{k}$.

We can also consider the second chiral limit of TBG identified in \Ref{2020arXiv200911301B} where $w_0 \neq 0$ and $w_1 = 0$. This limit has the chiral symmetry $C' = \tau_3\sigma_3$ where $\tau_3$ is the Pauli matrix acting on the layer index. Numerically, we do not find zero-energy bands in the second chiral limit. This is because the Dirac zero modes in the top and bottom layers have opposite chiralities due to $\tau_3$, so there is no chiral anomaly to protect the exact flatness.

\section{Twisted Bilayer Graphene: Many-body Physics}
\label{sec:TBGcoulomb}

The rich single-particle physics of TBG at $2\pi$ flux, discussed at length in \Sec{sec:BMmodel}, is characterized by the presence of low-energy flat bands. At the magic angle $\theta = 1.05$, the theoretically predicted small bandwidth $\sim 2$meV means that the Coulomb interaction, which is $\sim 24$meV, is the dominant term in the TBG Hamiltonian \cite{2020arXiv200912376B}. The large gap to the passive bands of $\sim 40$meV makes a strong coupling approximation viable where the Coulomb Hamiltonian is projected into the flat bands and the flat band kinetic energy is neglected.  This strategy has been used to great effect in predicting the groundstate properties of TBG near zero flux \cite{2020arXiv200912376B, 2020arXiv200913530L,2020arXiv200914200B,PhysRevLett.122.246401,2021arXiv210401145K,2021arXiv210508858L}.

Because the kinetic band energy is $< 2$meV and the Zeeman spin splitting is also $\sim 2$meV at $30$T, it is consistent to neglect both terms in the Hamiltonian at $2\pi$ flux. In this case, a $U(4)$ symmetry emerges in the strong coupling approximation just like at $\phi = 0$. Briefly, the spin and valley degeneracies act locally on the momentum $\mbf{k}$ and lead to a $U(2) \times U(2)$ symmetry group, which is expanded in the strong coupling approximation to $U(4)$ by the operator $C_{2z} P$ which also acts locally on $\mbf{k}$ (see \App{eq:BMsymm}). Note that $C_{2z} P$ commutes with the Coulomb term in \Eq{eq:Hintcoul} but anti-commutes with the single-particle Hamiltonian $H_0$ which is why only the enhanced symmetry appears only in the strong coupling approximation where $H_0$ is set to zero in the flat bands. This is briefly reviewed in \App{eq:appcoulomb} and explained in depth in \Ref{2020arXiv200912376B}.

We now apply the results of \Sec{sec:formfactor} to TBG, setting the screened Coulomb interaction to
\bea
V(\mbf{q}) = \pi \xi^2 U_\xi \frac{\tanh \xi |\mbf{q}|/ 2}{\xi |\mbf{q}|/ 2} \\
\eea
where the parameters of the screened Coulomb interaction are $\xi = 10$nm, $U_{\xi} = e^2/(\eps \xi) = 24$meV where $\eps$ is the dielectric constant \cite{2020arXiv200912376B}.

\subsection{Many-body Insulator Eigenstates}
\label{sec:groundstates}

\begin{figure}
\includegraphics[width=0.4\textwidth]{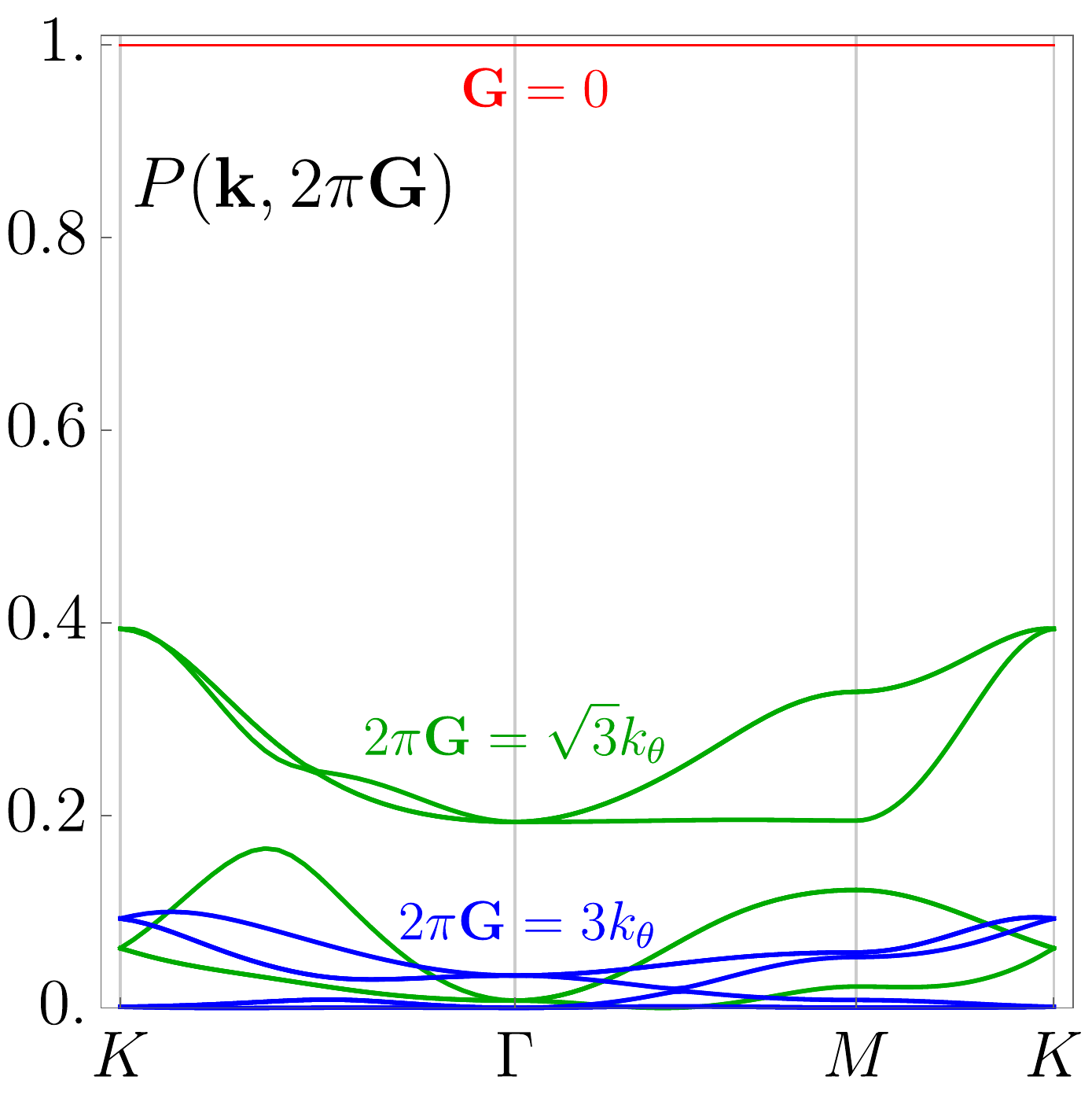}
\caption{The validity of the flat metric condition can be evaluated by examining the eigenvalues of $P(\mbf{k},2\pi\mbf{G}) = M^\dag(\mbf{k},2\pi\mbf{G}) M(\mbf{k},2\pi\mbf{G})$ as a function of $\mbf{k}$. At $\mbf{G} = 0$ (red), $M(\mbf{k}, 0) = U^\dag(\mbf{k}) U(\mbf{k})$ is the identity matrix so the flat metric condition is exactly satisfied. Because the form factor $M(\mbf{k}, 2\pi \mbf{G})$ decays exponentially in $\mbf{G}$, the flat metric condition is very nearly true for $|\mbf{G}| \geq 3$ (blue) because the eigenvalues are quite small. Thus the validity of the flat metric condition is determined to very good approximation by only the first momentum shell composed of $\mbf{G} = \pm \mbf{b}_1, \pm\mbf{b}_2, \pm(\mbf{b}_1 - \mbf{b}_2)$ (green). We see that, while $M(\mbf{k},2\pi \mbf{G})$ is not proportional to the identity, the differences between the eigenvalues of $P(\mbf{k},2\pi\mbf{G})$ are $\lesssim .33$ which is only a small violation of the flat metric condition. We used the parameters $\th = 1.05$ and $w_0 = .8 w_1$, but we checked that the flat metric condition is reliable over a range of parameters.}
\label{fig:FMC}
\end{figure}

Because the flat bands, approximate spin rotation, and valley symmetry survive the addition of $2\pi$ flux, one may add Coulomb interactions in the same manner as TBG in zero flux: by projecting density-density terms into the $8$ flat bands. These 8 bands have the creation operators $\gamma^\dag_{\mbf{k},M, \eta,s}$ where $M = \pm1$ is the band, $\eta$ is the valley, and $s$ is the spin. We note that $\gamma^\dag_{\mbf{k}+2\pi \mbf{G},M, \eta,s} = \gamma^\dag_{\mbf{k},M, \eta,s}$ because the eigenstates are periodic in $\mbf{k}$ (see \Sec{eq:MBT2pi}). Just as in zero-flux, the density-density form of the Coulomb interaction in \Eq{eq:Hintcoul} (that has neither spin nor valley dependence) takes the positive-semidefinite form
\begin{align}
H_{int} = \frac{1}{2 \Omega_{tot}} \sum_{\mbf{q} \in BZ} \sum_{\mbf{G}} \, O_{-\qq, -\mbf{G}} O_{\qq,\mbf{G}},
\label{eq:Hcoulma}
\end{align}
where $\Omega_{tot}$ is the total area of the sample and the operators $O_{\mbf{q},\mbf{G}} = O^\dag_{-\mbf{q},-\mbf{G}}$ are
\begin{align}
O_{\mbf{q},\mbf{G}} &= \sqrt{V(\mbf{q}+2\pi \mbf{G})} \sum_{\mbf{k}\in BZ} \sum_{\eta, s} \sum_{MN} \bar{M}^\eta_{MN}(\mbf{k},\mbf{q}+2\pi\mbf{G})  \nonumber \\
&\qquad \times (\gamma^\dag_{\mbf{k}- \mbf{q},M, \eta,s}  \gamma_{\mbf{k},N, \eta,s}  - \frac{1}{2} \delta_{MN} \delta_{\mbf{q},0}) \ .
\end{align}
An expression for the form factor $\bar{M}^\eta_{MN}(\mbf{k},\mbf{q})$ is given in Eq. 282 of \App{app:BMHam}. The term $\frac{1}{2} \delta_{MN} \delta_{\mbf{q},0}$ is added to make $H_{int}$ symmetric about charge neutrality as in \Ref{2020arXiv200912376B}. To project in the flat bands, we merely restrict $M,N$ to the flat bands which we label $\pm 1$. If all flat band states of a given valley $\eta$ and spin $s$ are filled, $O_{\qq,\mbf{G}}$ annihilates the state for all $\qq \neq 0 \text{ mod } 2\pi \mathbf{G}$.  This allows for the construction of exact eigenstates at filling $\nu = -4, -2, 0, 2, 4$:
\bea
\label{eq:ketnu}
\ket{\Psi_{\nu}} &= \prod_{\mbf{k}}  \prod_{j}^{(\nu+4)/2}  \gamma^\dag_{\mbf{k},+, s_j, \eta_j}  \gamma^\dag_{\mbf{k},-, s_j, \eta_j}  \ket{0},
\eea
where $\gamma^\dagger_{\kk,\pm,s_j,\eta_j}$ operators create flat band eigenstates with spin $s_j$ and valley $\eta_j$ which are arbitrary. Different choices of $j$ are related by $U(4)$ \cite{2020arXiv200913530L}. The states $\ket{\Psi_{\nu}}$ all have zero Chern number because the two flat bands have no total winding (see \Sec{sec:TBGsymtop}). The operators $O_\qq$ act simply on these states as calculated in \App{eq:TBGBMgs}:
\bea
 O_{\mbf{q}, \mbf{G}} \ket{\Psi_\nu}  = \delta_{\mbf{q},0} \la_{\mbf{G}}  \ket{\Psi_\nu} \, \\
 \eea
 where $\qq$ here is restricted to the BZ and
 \bea
 \label{eq:Mtretano}
 \la_{\mbf{G}} =  \nu \sqrt{V(2\pi\mbf{G})} \sum_{\mbf{k}\in BZ}  \frac{1}{2} \Tr \bar M(\mbf{k},2\pi\mbf{G}) \ .
 \eea
 We prove in \App{eq:TBGBMgs} that $\bar M^\eta(\mbf{k},2\pi\mbf{G}) $ and $\bar M^{-\eta}(\mbf{k},2\pi\mbf{G})$ are related by a unitary transform, so we drop the $\eta$ label on quantities which are independent of valley, such as $\Tr \bar M^{\eta}(\mbf{k},2\pi\mbf{G})$. Appealing to \Eq{eq:Hcoulma}, we show in \App{eq:TBGBMgs} that the energy of the eigenstates is
 \bea
 H_{int} \ket{\Psi_\nu} &= \lp  \frac{1}{2 \Omega_{tot}} \sum_{\mbf{G}}|\la_{\mbf{G}}|^2 \rp \ket{\Psi_\nu} \\
 \eea
which vanishes at the charge neutrality point $\nu =0$ because $\la_\mbf{G} \propto \nu$. Because $H_{int}$ is positive semi-definite, $\ket{\Psi_0}$ must be a groundstate because it has zero energy at $\nu = 0$. Additionally, the $\nu = \pm4$ eigenstates are trivially groundstates because they are fully filled/fully empty. Whether the $\ket{\Psi_{\nu}}$ are true groundstates for $\nu = \pm 2$ is still in question. One way to assess the groundstates at $\nu =2$ is with the flat metric condition \cite{2020arXiv200911301B}, which is the approximation
\bea
\label{eq:FMC}
\bar{M}^\eta(\mbf{k}, 2\pi \mbf{G}) = m_{\mbf{G}} \mathbb{1}_{2\times2},
\eea
in other words that $\bar{M}(\mbf{k}, 2\pi \mbf{G})$ is multiple of the identity matrix which does not depend on $\mbf{k}$ at each $\mbf{G}$. In \Ref{2020arXiv200913530L} it was shown that if the flat metric condition is satisfied, then $\ket{\Psi_\nu}$ are necessarily groundstates. \App{eq:TBGBMgs} contains a detailed review of this claim. In \Fig{fig:FMC}, we numerically calculate the singular values of $M(\mbf{k}, 2\pi\mbf{G})$ as in \Ref{2020arXiv200913530L} and argue that \Eq{eq:FMC} holds to a high degree of accuracy for all $2\pi |\mbf{G}| \neq \sqrt{3} k_\th$, as is also the case at $\phi = 0$. For six $\mbf{G}$ momenta $ \pm \mbf{b}_1, \pm\mbf{b}_2, \pm(\mbf{b}_1 - \mbf{b}_2)$ where $2\pi |\mbf{G}| = \sqrt{3} k_\th$, the flat metric condition is still an acceptable approximation to an accuracy in energy of $\Omega^{-1} V(2\pi \sqrt{3} k_\th) \sim 10$meV times a numerical $O(1)$ constant depending on the violation of \Eq{eq:FMC}. From \Eq{fig:FMC}, the difference of the eigenvalues of $M^\dag(\mbf{k},2\pi\mbf{G}) M(\mbf{k},2\pi\mbf{G})$ is $\lesssim .33$, whereas if the flat metric condition held, the difference would be zero. Hence we estimate that the flat metric condition holds within $\Omega^{-1} V(2\pi \sqrt{3} k_\th) \times \sqrt{.33} \sim 5$meV. Unless states other than $\ket{\Psi_{\nu}}$ are very competitive in energy, we can assume that $\ket{\Psi_{\nu}}$ is a groundstate at $\nu = \pm 2$. The excitation spectrum above these ground states at $2\pi$ flux is studied in \Ref{2021arXiv211111434H}. \Ref{2021arXiv211208620W} uses a complimentary technique to study the strong coupling excitations in small magnetic fields. 

\section{Discussion}
The techniques developed in this paper allow for an analysis of general periodic Hamiltonians in $2\pi$ flux --- most notably the continuum models of moir\'e meta-materials --- generalizing Bloch's theorem in a way that allows theoretical access to non-Peierls physics. We derived formulae for matrix elements, Wilson loops and Berry curvature, and projected density-density interactions. These tools expand the reach of modern topological band theory to the strong flux limit, opening Hofstadter topology to analytical and numerical study in the continuum.

Using these techniques, we build a physical picture of twisted bilayer graphene in $2\pi$ flux --- a tantalizing experimental setup as the large moir\'e unit cell allows for laboratory access to the Hofstadter limit for intermediate and large flux \cite{2021arXiv211111341D,2021arXiv210800009Y}. We find that in magic angle twisted bilayer graphene, the flat bands are reenter at $2\pi$ flux after splitting and broadening into Hofstadter bands at intermediate flux. The chiral limit of TBG, although physically inaccessible, showcases the chiral anomaly and exemplifies the non-crystalline properties of Hofstadter phases.

A natural development of this work is the extension of our gauge-invariant method to study the topology of band structures at general rational flux, which we pursue in future work. Such a development would be a powerful tool to study non-Peierls physics in topological magnetic systems, particularly with the ability to perform gauge-invariant Wilson loop calculations within our formalism. Investigations of strongly correlated phases like superconductivity and the fractional quantum hall effect are also made possible due to our expressions for the form factors. 

During the preparation of this work, \Ref{2021arXiv210610650S} independently studied the chiral limit in magnetic field. They find exact eigenstates for the zero-energy flat bands protected by chiral symmetry at \emph{all} flux, but their techniques do not generalize to non-chiral Hamiltonians. We identify the same phase transition in \Fig{fig:manyTBGPlots}(e) as described in their work.

\section{Acknowledgements}

We thank Zhi-Da Song and Dmitri Efetov for their insight. B.A.B. and A.C. were supported by the ONR Grant No. N00014-20-1-2303, DOE Grant No. DESC0016239, the Schmidt Fund for Innovative Research, Simons Investigator Grant No. 404513, the Packard Foundation, the Gordon and Betty Moore Foundation through Grant No. GBMF8685 towards the Princeton theory program, and a Guggenheim Fellowship from the John Simon Guggenheim Memorial Foundation. Further support was provided by the NSF-MRSEC Grant No. DMR-1420541 and DMR-2011750, BSF Israel US foundation Grant No. 2018226, and the Princeton Global Network Funds. JHA is supported by a Marshall Scholarship funded by the Marshall Aid Commemoration Commission.

\bibliography{finalbib}
\bibliographystyle{apsrev4-1}

\onecolumngrid
\appendix

\begin{center}
 \textbf{Supplementary Material for ``Magnetic Bloch Theorem and Reentrant Flat Bands in Twisted Bilayer Graphene  at $2\pi$ Flux"}
 \end{center}

The following Appendices provide self-contained, detailed calculations for readers interested in the theoretical developments described in the Main Text. Alternatively, the important computational results are highlighted for readers wishing to simply make use of the formalism. We provide a brief outline of the Appendices and summarize the final results below. \\

\App{app:siegel} fully develops the gauge-invariant formalism of the magnetic Bloch theorem at $2\pi$ flux. \App{eq:appopdef} defines the Landau level and guiding center operators. \App{app:siegelnorm} defines the magnetic translation group irreps. \App{app:siegelprop} introduces the Siegel theta function and relevant identities. \App{app:greens} proves an important holomorphic/anti-holomorphic factorization formula for the Siegel theta function appearing in the basis states. \App{app:completeness} discusses the completeness of the magnetic translation group irrep basis states. \App{app:siegelcomplete} provides general formulas for the matrix elements of the magnetic Bloch Hamiltonian. \App{app:berry} contains an expression for the non-Abelian Wilson loop. \\

\App{app:square_lattice} exemplifies the magnetic Bloch formalism with a simple Hamiltonian. \App{eq:zerofluxsquare} reviews Bloch's theorem at zero flux, and \App{app:squarelattice2pifluxHam} analogously discusses the magnetic Bloch theorem at $2\pi$ flux. \\

\App{app:gcgii} develops a gauge-invariant formula for the Coulomb interaction and an expression for the many-body form factors. \\

\App{app:BMHam} applies the magnetic Bloch theorem to Twisted Bilayer Graphene. \App{app:spBMham} defines the conventions of the Bistritzer-MacDonald Hamiltonian. \App{eq:BMsymm} discusses the symmetries of the continuum model in magnetic flux. \App{app:coulham} reviews the strong coupling expansion used to study the many-body physics. \App{eq:TBGBMgs} derives exact many-body groundstates at even integer fillings.

\tableofcontents

\newpage

\section{Ladder Operator Calculations and the Siegel Theta function}
\label{app:siegel}

In this Appendix, we define the operators and special functions used to derive the magnetic Bloch theorem at $2\pi $ flux. This section is technical, and the important results are highlighted for the convenience of readers.  \App{eq:appopdef} contains a derivation of the symmetry operators in flux. In \App{app:siegelnorm}, we prove a central formula for the magnetic translation operator eigenstates. The Siegel theta function, a generalization of the Jacobi theta function, arises in this construction and is discussed in \App{app:siegelprop}. In \App{app:greens}, we derive a simple expression (related to the Green's function of the Laplacian on the torus) for the Siegel theta function appearing as a normalization factor. With this result, we then prove the completeness of the magnetic translation group irreps in \App{app:completeness}. Moving on to the spectrum, we then prove a general formula for the scattering amplitude between eigenstates in \App{app:siegelcomplete} which is used to compute the magnetic Bloch Hamiltonian and many-body form factors. Finally, \App{app:berry} derives expressions for the non-Abelian Berry connection and demonstrates that the Landau levels forming the basis carry nonzero Chern numbers.

\subsection{Operator Definitions}
\label{eq:appopdef}

First, we must set our index conventions. We sum over repeated indices and use greek letters $\mu,\nu,\rho,\sigma$ for cartesian indices and roman letters $i,j,k,l$ for the crystalline indices of the lattice vectors $\mbf{a}_i$. For instance, the reciprocal lattice vectors $\mbf{b}_i$ are defined by $\mbf{a}_i \cdot \mbf{b}_j = \delta_{ij}$, where we used the dot product as a shorthand for contracting the cartesian indices of the vectors $\mbf{a}_i,\mbf{b}_i$.

We recall the definitions of the canonical momentum and ladder operators defined in \Sec{sec:setup} of the Main Text.  The canonical momentum with $\hbar =1$ is $\pmb{\pi} = - i \pmb{\nabla} - e \mbf{A}$ obeying
\bea
\null [\pi_\mu, \pi_\nu] = i e (\del_\mu A_\nu - \del_\nu A_\mu) = i e B\eps_{\mu \nu}
\eea
and the Landau level ladder operators $ [a,a^\dag]$ are defined
\bea
\label{eq:aop}
a &= \frac{\pi_x + i \pi_y}{\sqrt{2eB}}  , a^\dag = \frac{\pi_x - i \pi_y}{\sqrt{2eB}} \quad \implies \quad \pi_x = \sqrt{\frac{eB}{2}}(a^\dag+a), \pi_y = i \sqrt{\frac{eB}{2}} (a^\dag-a) \ .
\eea
The guiding center momenta $Q_\mu$ are defined in a gauge-invariant way by
\bea
\label{eq:Qoperator}
Q_\mu &= \pi_\mu - e B \eps_{\mu\nu} x_\nu = - i \del_\mu - e (A_\mu + B \eps_{\mu \nu} x_\nu) \ . \\
\eea
We note that $Q_\mu = A_\mu + B \eps_{\mu \nu} x_\nu$ can be interpreted as the canonical momentum corresponding to a system with opposite magnetic field $-B$. At zero flux, where the time-reversal is not broken, $Q_\mu$ and $\pi_\mu$ are identical (up to a gauge choice). We will make frequent use of the simple identity
\bea
(eB)^{-1}\eps_{\mu \nu}(Q_\nu - \pi_\nu) = -\eps_{\mu \nu }\eps_{\nu \rho} x_\rho = x_\mu  \\
\eea
which gives a gauge-independent formula for $x_\mu$. The guiding center operators commute with the canonical momenta and obey
\bea
\label{eq:abcom}
\null [Q_\mu,\pi_\nu] &= [\pi_\mu - eB \eps_{\mu \rho} x_\rho,\pi_\nu] = ieB\eps_{\mu \nu}  - i e B \eps_{\mu \nu} = 0  \\
[Q_\mu,Q_\nu] &= [\pi_\mu - eB \eps_{ \mu \rho} x_\rho ,\pi_\nu - eB\eps_{\nu \sigma} x_\sigma] = - i eB \eps_{\mu \nu} \ . \\
\eea
The guiding centers form a separate oscillator system with $[b,b^\dag] = 1$ defined by
\bea
\label{eq:b}
b &= \frac{(\mbf{a}_1 - i \mbf{a}_2) \cdot  \mbf{Q}}{\sqrt{2\phi}}, \quad b^\dag = \frac{(\mbf{a}_1   + i \mbf{a}_2) \cdot  \mbf{Q}}{\sqrt{2\phi}} \quad \implies \quad \sqrt{\phi/2} (b + b^\dag) = \mbf{a}_1 \cdot \mbf{Q}, \  i \sqrt{\phi/2} (b - b^\dag) = \mbf{a}_2 \cdot \mbf{Q} \ . \\
\eea
To prove $[b,b^\dag] = 1$, it is useful to note $[\mbf{a}_1 \cdot \mbf{Q}, \mbf{a}_2 \cdot \mbf{Q}] = -ieB \mbf{a}_1 \times \mbf{a}_2 = - i \phi$. From \Eq{eq:abcom}, we have $[a,b]=[a^\dag,b] = [a,b^\dag]=[a^\dag,b^\dag] = 0$. Because $a,b$ commute, we can find simultaneous eigenstates of $a^\dag a$ and $b^\dag b$. In particular, they have a simultaneous groundstate $\ket{0}$ satisfying $a\ket{0} = b\ket{0} = 0$.

We define the magnetic translation operators to be
\bea
T_{\mbf{a}_i} &= \exp ( i \mbf{a}_i \cdot \mbf{Q})
\eea
which obey the projective algebra
\bea
T_{\mbf{a}_1}  T_{\mbf{a}_2}   &= \exp \Big( [i \mbf{a}_1 \cdot \mbf{Q} , i \mbf{a}_2 \cdot \mbf{Q} ]\Big) T_{\mbf{a}_2}  T_{\mbf{a}_1}   = e^{i \phi} T_{\mbf{a}_2}  T_{\mbf{a}_1}  \\
\eea
obtained using the Baker-Campbell-Hausdorff (BCH) identity $\exp X \exp Y = \exp([X,Y]) \exp Y \exp X$ when $[X,Y]$ is a $c$-number: since $[[Q_\mu, Q_\nu], Q_\gamma] = 0$ for all $\mu,\nu,\gamma$, higher-order commutators in the BCH formula disappear. Lastly, it is also useful to define the gauge-invariant \footnote{As the operator is built out of $\pmb{Q}, \pmb{\pi}$ operators and makes no reference to the gauge field $\mbf{A}$, it is manifestly gauge-invariant.} angular momentum operator
\bea
\label{eq:defLz}
L_z = \frac{\mbf{Q}^2-\pmb{\pi}^2}{2eB} \ . \\
\eea
This choice is convenient because it decouples the $a$ and $b$ oscillators. To motivate the form of \Eq{eq:defLz}, we check that
\bea
 \frac{\mbf{Q}^2-\pmb{\pi}^2}{2eB} &= \frac{1}{2eB}(Q_\mu-\pi_\mu)(\pi_\mu + Q_\mu) \\
&= -\frac{1}{2}\eps_{\mu \rho} x_\rho (2\pi_\mu - e B \eps_{\mu \sigma} x_\sigma) \\
&= \eps_{\mu \nu} x_\mu (-i \del_\nu) + e \eps_{\mu \rho} x_\rho \left[ A_\mu + \frac{B}{2} \eps_{\mu \sigma} x_\sigma \right]
\eea
and the term in brackets is pure gauge (curl-free), i.e. $\eps_{\mu \nu} \del_\mu (A_\nu + \frac{B}{2} \eps_{\nu \sigma} x_\sigma) = 0$. Thus in the symmetric gauge $A_\mu = -\frac{B}{2} \eps_{\mu \nu} x_\nu$ where $\pi_\mu = - i \del_\mu + \frac{eB}{2} \eps_{\mu \nu} x_\nu, Q_\mu = - i \del_\mu - \frac{eB}{2} \eps_{\mu \nu} x_\nu $, we find
\bea
 \frac{\mbf{Q}^2-\pmb{\pi}^2}{2eB}  &= \eps_{\mu \nu}x_\mu (-i \del_\nu) \\
\eea
which is the canonical angular momentum operator. We now check the algebra of rotations with the magnetic translation operators. These results will be useful when we derive the real-space form of the symmetry operators in flux. The first identity we need is
\bea
\null [\mbf{Q}^2, Q_\mu] = Q_\nu [Q_\nu, Q_\mu] + [Q_\nu, Q_\mu] Q_\nu = 2 e B i \eps_{\mu \nu} Q_\nu
\eea
which leads to
\bea
\label{eq:rotationQ}
\lp [\frac{\mbf{Q}^2}{2eB}, \rp^n Q_\mu] &=  (i \eps_{\mu \mu_1})(i \eps_{\mu_1 \mu_2}) \dots (i \eps_{\mu_{n-1} \mu_n}) Q_{\mu_n} \\
\lp [\frac{\mbf{Q}^2}{2eB}, \rp^n \mbf{Q}]  &= \sigma_2^n \mbf{Q}
\eea
where $\sigma_2$ is the Pauli matrix acting on the $\mu,\nu$ indices. Defining $([A,)^n B] = [A, [A, \cdots [A,B]]],$ and using an alternate version of BCH:
\bea
e^X e^Y e^{-X}= e^{Y + [X,Y] + \frac{1}{2!}[X,[X,Y]] \dots}
\label{}
\eea
we perform the following calculation using \Eq{eq:rotationQ}:
\bea
e^{i \th L_z} T_{\RR} e^{-i \th L_z}  &= e^{i \th L_z} \exp[i \RR \cdot \mbf{Q}] e^{-i \th L_z} \\
&= \exp \sum_{n=0}^\infty \frac{1}{n!} \big( [i \th \frac{\mbf{Q}^2}{2eB}, \big)^n i \RR \cdot \mbf{Q} ] \\
&= \exp \sum_{n=0}^\infty \frac{1}{n!} (i\theta)^n i \RR \cdot \big( [\frac{\mbf{Q}^2}{2eB}, \big)^n \mbf{Q} ] \\
&= \exp \sum_{n=0}^\infty \frac{1}{n!} (i\theta)^n i \RR \cdot \sigma_2^n \mbf{Q} \\
&= \exp \lp i \RR \cdot  R_{-\th} \mbf{Q}  \rp \\
&=T_{R_\th \RR} \ .
\eea

This result shows that the algebra of rotations with translations is the same as at zero flux. Trivially, $[L_z, \pmb{\pi}^2] = 0$, so $L_z$ commutes with the scalar kinetic term. For a Dirac-type Hamiltonian, we compute directly
\bea
\null [L_z + \frac{1}{2}\sigma_3, \pi_\mu \sigma_\mu] &= [- \frac{\pmb{\pi}^2}{2eB} + \frac{1}{2}\sigma_3, \pi_\mu \sigma_\mu] \\
&= i \eps_{\mu \nu} \pi_\nu \sigma_\mu  + i \eps_{\mu \nu} \pi_\mu \sigma_\nu  = 0 \\
\eea
which shows that $L_z + \frac{1}{2}\sigma_3$ is the conserved angular momentum, appropriate for a particle with Berry phase. Finally, we need the action of functions of $\mbf{r}$ to show that $L_z$ commutes with rotationally symmetric potential terms. We use
\bea
\null [L_z, x_\mu] &= \frac{1}{eB} \eps_{\mu \nu}[L_z, Q_\nu - \pi_\nu] \\
&= \frac{1}{2(eB)^2} \eps_{ \mu \nu} ( [\mbf{Q}^2, Q_\nu]  + [\pmb{\pi}^2, \pi_\nu]) \\
&= \frac{1}{eB} \eps_{\mu \nu} (i\eps_{\nu \rho} Q_\rho -i \eps_{\nu \rho} \pi_\rho ) \\
&= \frac{1}{eB} \eps_{\mu \nu} ( i\eps_{\nu \rho} (Q_\rho-\pi_\rho) ) \\
&=  i\eps_{\mu \nu} x_\nu \\
\eea
which is the same action as the canonical angular momentum $[\eps_{\nu \rho} x_\nu (-i\del_\rho), x_\mu] = -i \eps_{\nu \mu} x_\mu$. Hence a $C_n$-symmetric potential at $B=0$ is also symmetric under rotations by \Eq{eq:defLz} in nonzero magnetic field. This is just the classical statement that a perpendicular magnetic field does not break rotation symmetry.

\subsection{Eigenstate Normalization}
\label{app:siegelnorm}

The calculation of the basis states in \Eq{eq:basis} of the Main Text is straightforward but involved. To streamline the notation, we use $\ket{n} = \frac{1}{\sqrt{n!}} a^{\dag n} \ket{0}$ to denote the Landau level wavefunctions, which obey $a\ket{0} = b\ket{0} = 0$ and $\braket{m|n}$. The magnetic translation eigenstates are
\bea
\label{eq:statedefAppa}
\boxed{
\ket{\mbf{k},n} = \frac{1}{\sqrt{\mathcal{N}(\mbf{k})}} \sum_{\mbf{R}} e^{-i \mbf{k} \cdot \mbf{R}} T_{\mbf{a}_1}^{\mbf{R} \cdot \mbf{b}_1} T_{\mbf{a}_2}^{\mbf{R} \cdot \mbf{b}_2} \ket{n}
}
\eea
and $T_{\mbf{a}_i} = e^{i \mbf{a}_i \cdot \mbf{Q}}$. We have defined $\RR = R_1 \mbf{a}_1 + R_2 \mbf{a}_2$ for $R_i \in \mathds{Z}$.

All calculations can be performed with the BCH identity $\exp X \exp Y = \exp(\frac{1}{2} [X,Y] ) \exp (X+Y)$ when $[X,Y]$ is a $c$-number. We also set $\phi = 2\pi$. The first step is
\bea
T_{\mbf{a}_1}^{\mbf{R} \cdot \mbf{b}_1} T_{\mbf{a}_2}^{\mbf{R} \cdot \mbf{b}_2} &= e^{i R_1 \mbf{a}_1 \cdot \mbf{Q}} e^{i R_2 \mbf{a}_2 \cdot \mbf{Q}} = e^{i\mbf{R} \cdot \mbf{Q} + i \frac{\phi}{2} R_1 R_2}= e^{i\mbf{R} \cdot \mbf{Q} + i \frac{\phi}{2} (\mbf{R} \cdot \mbf{b}_1)(\mbf{R} \cdot \mbf{b}_2)}
\eea
using $[\mbf{a}_1 \cdot \mbf{Q},\mbf{a}_2 \cdot \mbf{Q}] = -i \phi$. Hence we can write our states
\bea
\label{eq:knexplicit}
\ket{\mbf{k}, n} = \frac{1}{\sqrt{\mathcal{N}(\mbf{k})}} \sum_{\mbf{R}} e^{-i \mbf{k} \cdot \mbf{R} +  i \frac{\phi}{2} (\mbf{R} \cdot \mbf{b}_1)(\mbf{R} \cdot \mbf{b}_2)} e^{i\mbf{R} \cdot \mbf{Q} }\ket{n} \ . \\
\eea
Recall that the $\mbf{Q}$ operators are built of $b, b^\dagger$ operators, and $\ket{n}$ is a $b$-vacuum because $a^\dag$ and $b$ commute. We now use the oscillator variables
\bea
\mbf{R} \cdot \mbf{Q} &= \sqrt{\phi/2} (b + b^\dag) R_1 + i \sqrt{\phi/2} (b - b^\dag) R_2 = \sqrt{\phi/2} (R b + \bar{R} b^\dag), \qquad R = R_1 + i R_2, \bar{R} = R_1 - i R_2  \\
\label{eq:rdotq}
\eea
along with the BCH identity to compute
\bea
\bra{m} e^{i \mbf{R} \cdot \mbf{Q}} \ket{n} &= e^{- \phi/4 \bar{R} R} \bra{m} e^{i \sqrt{\phi/2} \bar{R} b^\dag} e^{i \sqrt{\phi/2} R b} \ket{n} = e^{- \frac{\phi}{4} \bar{R}R} \delta_{mn} \ .
\label{eqTranslationExpectation}
\eea
With this expression, a direct calculation yields
\bea
\label{eq:TpT}
\sqrt{\mathcal{N}(\mbf{k})\mathcal{N}(\mbf{k}')} \braket{\mbf{k}', m|\mbf{k}, n} &= \sum_{\mbf{R},\mbf{R}'} e^{-i \mbf{k} \cdot \mbf{R} +i \mbf{k}' \cdot \mbf{R}' +  i \frac{\phi}{2} (\mbf{R} \cdot \mbf{b}_1)(\mbf{R} \cdot \mbf{b}_2)- i \frac{\phi}{2} (\mbf{R}' \cdot \mbf{b}_1)(\mbf{R}' \cdot \mbf{b}_2)} \bra{m} e^{-i\mbf{R}' \cdot \mbf{Q}} e^{i\mbf{R} \cdot \mbf{Q} }\ket{n} \\
&=  \sum_{\mbf{R},\mbf{R}'} e^{-i \mbf{k} \cdot \mbf{R} +i \mbf{k}' \cdot \mbf{R}' +  i \frac{\phi}{2} (\mbf{R} \cdot \mbf{b}_1)(\mbf{R} \cdot \mbf{b}_2)- i \frac{\phi}{2} (\mbf{R}' \cdot \mbf{b}_1)(\mbf{R}' \cdot \mbf{b}_2)} \bra{m}e^{-i\sqrt{\phi/2}(R' b + \bar{R'} b^\dag)}  e^{i\sqrt{\phi/2}(R b + \bar{R} b^\dag)} \ket{n} \\
&= \sum_{\mbf{R},\mbf{R}'} e^{-i \mbf{k} \cdot \mbf{R} +i \mbf{k}' \cdot \mbf{R}' +  i \frac{\phi}{2} (\mbf{R} \cdot \mbf{b}_1)(\mbf{R} \cdot \mbf{b}_2)- i \frac{\phi}{2} (\mbf{R}' \cdot \mbf{b}_1)(\mbf{R}' \cdot \mbf{b}_2) - \frac{\phi}{4} \bar{R}R- \frac{\phi}{4} \bar{R}'R'} \\
& \qquad \times \bra{m} e^{-i\sqrt{\phi/2} (\bar{R'} b^\dag)} e^{-i\sqrt{\phi/2}(R' b)}  e^{i\sqrt{\phi/2}(\bar{R} b^\dag)} e^{i\sqrt{\phi/2}(R b)} \ket{n} \\
&= \sum_{\mbf{R},\mbf{R}'} e^{-i \mbf{k} \cdot \mbf{R} +i \mbf{k}' \cdot \mbf{R}' + i \frac{\phi}{2} (\mbf{R} \cdot \mbf{b}_1)(\mbf{R} \cdot \mbf{b}_2)- i \frac{\phi}{2} (\mbf{R}' \cdot \mbf{b}_1)(\mbf{R}' \cdot \mbf{b}_2) - \frac{\phi}{4} \bar{R}R- \frac{\phi}{4} \bar{R}'R'} \bra{m} e^{-i \sqrt{\phi/2} R' b} e^{i \sqrt{\phi/2} \bar{R} b^\dag} \ket{n} \\
&=\delta_{mn} \sum_{\mbf{R},\mbf{R}'} e^{-i \mbf{k} \cdot \mbf{R} +i \mbf{k}' \cdot \mbf{R}' +   i \frac{\phi}{2} (\mbf{R} \cdot \mbf{b}_1)(\mbf{R} \cdot \mbf{b}_2)- i \frac{\phi}{2} (\mbf{R}' \cdot \mbf{b}_1)(\mbf{R}' \cdot \mbf{b}_2)  - \frac{\phi}{4} \bar{R}R- \frac{\phi}{4} \bar{R}'R'+ \frac{\phi}{2} R' \bar{R} } \ . \\
\eea
We have reduced the calculation to a double infinite sum over the lattice vectors. A sum of this form with quadratic term in the exponential is given by a generalized theta function, called a Siegel theta function (also called a Riemann theta function). The trick to computing the sum is noticing that the sign of the imaginary terms is arbitrary because $\frac{\phi}{2} R_1 R_2'$ is a multiple of $\pi$ so $e^{i \frac{\phi}{2} R_1 R_2'} = \pm 1 = e^{-i \frac{\phi}{2} R_1 R_2'}$. Expanding out the quadratic terms, we find that
\bea
&  i \frac{\phi}{2} (\mbf{R} \cdot \mbf{b}_1)(\mbf{R} \cdot \mbf{b}_2)- i \frac{\phi}{2} (\mbf{R}' \cdot \mbf{b}_1)(\mbf{R}' \cdot \mbf{b}_2)  - \frac{\phi}{4} \bar{R}R- \frac{\phi}{4} \bar{R}'R'+ \frac{\phi}{2} R' \bar{R}  \\
 & \qquad\qquad\qquad\qquad\qquad\qquad\qquad = - \frac{\phi}{4} \bpm R_1 & R_2 & R'_1 & R'_2 \epm \left(
\begin{array}{cccc}
 1 & i & -1 & -i \\
 i & 1 & -i & -1 \\
 -1 & -i & 1 & i \\
 -i & -1 & i & 1 \\
\end{array}
\right) \bpm R_1 \\ R_2 \\ R'_1 \\ R'_2 \epm \mod 2\pi i
\eea
This matrix is a quadratic form has two zero modes and two eigenvalues with positive real part (not counting the $-\phi/4$ prefactor) which ensures the convergence of the sum. The positive eigenvalues introduce Gaussian decay in the terms as $\RR, \RR' \rightarrow \infty$. Note that we have assumed $\phi >0$ in \Eq{eq:aop}. The zero modes disappear from the quadratic form and summing over them enforces momentum conservation as we will show. (This is identical to the Fourier transform of $1$ being a delta function). We start by introducing the center-of-mass variables
\bea
\label{eq:cofv}
s_i = \frac{R_i+R_i'}{2}, \quad d_i = R_i-R_i' \ . \\
\eea
The $\mbf{R},\mbf{R}'$ variables each lie on the lattice $\mathds{Z}^2$, and hence $\mbf{s} \in \frac{1}{2} \mathds{Z}$ and $\mbf{d} \in \mathds{Z}$. These two variables are not independent: if $s_i$ is half-integer, then $d_i$ is odd, and if $s_i$ is integer, then $d_i$ is even. In terms of these variables, the sum in \Eq{eq:TpT} can be simplified because the summand of \Eq{eq:TpT} factors into $\mbf s$-dependent and $\mbf d$-dependent terms:
\bea
\label{eq:TpTsym}
e^{-is_1(k_1 - k_1') -is_2(k_2 - k_2')} \exp \lp -\frac{i}{2} d_1 (k_1+ k'_1) -\frac{i}{2} d_2 (k_2+ k'_2) - \frac{\phi}{4} \bpm d_1 & d_2 \epm \bpm 1 & i \\  i & 1 \epm \bpm d_1 \\ d_2 \epm \rp \ .
\eea
Here $k_i = \aaa_i \cdot \kk.$  In the original sum all four variables $R_1, R_2, R_1',R_2'$ are all interconnected by the Riemann matrix.  However, after switching to center of mass variables only the variables $d_1, d_2$ are connected.  This is made obvious by noting the original $4 \times 4$ matrix can be written as a tensor product
\begin{align}\left(
  \begin{array}{cccc}
   1 & i & -1 & -i \\
   i & 1 & -i & -1 \\
   -1 & -i & 1 & i \\
   -i & -1 & i & 1
  \end{array}\right) =   \left(\begin{array}{cccc}
     1 & -1 \\
     -1 & 1
   \end{array}\right) \otimes   \left(\begin{array}{cccc}
       1 & i \\
       i & 1
      \end{array}\right)
\label{}
\end{align}
and so its eigenstates factor. Using the center of mass variables \Eq{eq:cofv}, we can write the double sum as
\bea
\label{eq:doubsum}
\sum_{\mbf{R},\mbf{R}'} = \sum_{\mbf{s} \in (\mathds{Z}+\frac{1}{2},\mathds{Z}+\frac{1}{2})}  \sum_{\mbf{d}  \in(2\mathds{Z}+1,2\mathds{Z}+1)} + \sum_{\mbf{s} \in (\mathds{Z},\mathds{Z})}  \sum_{\mbf{d} \in(2\mathds{Z},2\mathds{Z})}  +\sum_{\mbf{s} \in (\mathds{Z}+\frac{1}{2},\mathds{Z})}  \sum_{\mbf{d}  \in(2\mathds{Z} + 1,2\mathds{Z})}  +\sum_{\mbf{s} \in (\mathds{Z},\mathds{Z}+\frac{1}{2})}  \sum_{\mbf{d}  \in(2\mathds{Z},2\mathds{Z}+1)}  \\
\eea
which separates into decoupled sums of odd ($2\mathds{Z}+1$) and even $(2\mathds{Z})$ lattices. We consider the two cases. If the $s_i$ sum is over $\mathds{Z}$ and the $d_i$ sum is over $2\mathds{Z}$, so the $k_i$-dependent terms can be simplified to
\bea
\sum_{s_i \in \mathds{Z}} e^{i s_i(k_i-k_i')}\sum_{d_i \in 2\mathds{Z}} e^{-\frac{i}{2} d_{i} (k_i+ k'_i)} &= 2\pi\delta( k_i - k_i' \mod 2\pi)\sum_{d_i \in 2\mathds{Z}} e^{-\frac{i}{2} d_{i} (k_i+ k'_i)} \\
&= 2\pi\delta( k_i - k_i' \mod 2\pi)\sum_{d \in \mathds{Z}} e^{-\frac{i}{2} 2d (k_i+ k_i + 2\pi \mathds{Z})} \\
&= 2\pi\delta( k_i - k_i' \mod 2\pi)\sum_{d \in \mathds{Z}} e^{-i d (2k_i+2\pi \mathds{Z})} \\
&= 2\pi\delta( k_i - k_i' \mod 2\pi)\sum_{d \in \mathds{Z}} e^{-i d (2k_i)} \\
&= 2\pi\delta( k_i - k_i' \mod 2\pi)\sum_{d_i \in 2\mathds{Z}} e^{-i d_i k_i} \\
\eea
If the $s_i$ sum is over $\mathds{Z}+1/2$, then the $d_i$ sum is over $2\mathds{Z}+1$ and
\bea
\sum_{s_i \in \mathds{Z}+1/2} e^{i s_i(k_i-k_i')}\sum_{d_i \in 2\mathds{Z}+1} e^{-\frac{i}{2} d_i (k_i+ k'_i)}  &= 2\pi\delta( k_i - k_i' \mod 2\pi) e^{\frac{i}{2}(k_1 - k_i')} \sum_{d_i \in 2\mathds{Z}+1} e^{-\frac{i}{2} d_i (k_i+ k'_i)} \\
&= 2\pi\delta( k_i - k_i' \mod 2\pi) e^{\frac{i}{2}(2 \pi \mathds{Z})}  \sum_{d \in \mathds{Z}} e^{-\frac{i}{2} (2d+1) (k_i+ k_i + 2\pi \mathds{Z})} \\
&= 2\pi\delta( k_i - k_i' \mod 2\pi)\sum_{d  \in \mathds{Z}} e^{-\frac{i}{2} 2d(k_i+ k_i + 2\pi \mathds{Z}) -\frac{i}{2} (k_i+ k_i + 2\pi \mathds{Z}) + \frac{i}{2}(2 \pi \mathds{Z})} \\
&= 2\pi\delta( k_i - k_i' \mod 2\pi)\sum_{d  \in \mathds{Z}} e^{-\frac{i}{2} 2d(2k_i) -\frac{i}{2} (2k_i) } \\
&= 2\pi\delta( k_i - k_i' \mod 2\pi)\sum_{d_i \in2\mathds{Z}+1} e^{-i d_i k_i } \ . \\
\eea
We have shown that both the even and odd type sums give the $2\pi$ periodic delta function that enforces momentum conservation. Hence we arrive at
\bea
\label{eq:siegel_sum_simplified}
&\sum_{\mbf{R},\mbf{R}'} e^{-i \mbf{k} \cdot \mbf{R} +i \mbf{k}' \cdot \mbf{R}' +   i \frac{\phi}{2} (\mbf{R} \cdot \mbf{b}_1)(\mbf{R} \cdot \mbf{b}_2)- i \frac{\phi}{2} (\mbf{R}' \cdot \mbf{b}_1)(\mbf{R}' \cdot \mbf{b}_2)  - \frac{\phi}{4} \bar{R}R- \frac{\phi}{4} \bar{R}'R'+ \frac{\phi}{2} R' \bar{R} } \\
&= (2\pi)^2 \delta(\mbf{k}-\mbf{k}') \lp \sum_{d_1 \in 2\mathds{Z}} + \sum_{d_1 \in 2\mathds{Z}+1} \rp\lp \sum_{d_2 \in 2\mathds{Z}} + \sum_{d_2 \in 2\mathds{Z}+1} \rp \exp \lp - i d_1 k_1 - i d_2 k_2 - \frac{\phi}{4} \bpm d_1 & d_2 \epm \bpm 1 & i \\  i & 1 \epm \bpm d_1 \\ d_2 \epm \rp \\
&= (2\pi)^2 \delta(\mbf{k}-\mbf{k}') \sum_{d_1,d_2} \exp \lp - i d_1 k_1 - i d_2 k_2 - \frac{\phi}{4} \bpm d_1 & d_2 \epm \bpm 1 & i \\  i & 1 \epm \bpm d_1 \\ d_2 \epm \rp \ . \\
\eea
We will omit further mentions of `mod $2\pi$' in delta functions relating momenta. The remaining $\mbf{d}$ sum is a Seigel theta function (sometimes called the Riemann theta function) defined by the quadratic form $\Phi$:
\bea
\vartheta \lp \mbf{z} \left| \Phi \right. \rp = \sum_{\mbf{n}\in\mathds{Z}^2} e^{2\pi i \lp \frac{1}{2} \mbf{n}\cdot \Phi \cdot \mbf{n}- \mbf{z} \cdot \mbf{n} \rp} \\
\eea
which is a multi-dimensional generalization of the Jacobi theta function (see \App{app:siegelprop}).  We can write our normalization factor explicitly in terms of this special function
\bea
\label{eq:Phi}
\sqrt{\mathcal{N}(\mbf{k})\mathcal{N}(\mbf{k}')} \braket{\mbf{k}', m|\mbf{k}, n}  &= (2\pi)^2 \delta(\mbf{k} - \mbf{k}') \vartheta \lp \left. \frac{(k_1,k_2)}{2\pi} \right| \Phi \rp, \qquad \Phi =  \frac{i \phi}{4\pi} \bpm 1 & i \\  i & 1 \epm
\eea
where we defined the matrix $\Phi$ for later convenience. We now prove that $\vartheta \lp \mbf{z} \left| \Phi \right. \rp$ is real for real $\mbf{z}$ for $\phi = 2\pi$. This is because
\bea
\label{eq:realtheta}
\vartheta \lp \mbf{z} \left| \Phi \right. \rp^* &= \sum_{\mbf{n}\in\mathds{Z}^2} e^{-2\pi i \lp \frac{1}{2} \mbf{n}\cdot \Phi^* \cdot \mbf{n}- \mbf{z}^* \cdot \mbf{n} \rp}  \\
&= \sum_{\mbf{n}\in\mathds{Z}^2} e^{2\pi i \lp \frac{1}{2} \mbf{n}\cdot (-\Phi^*) \cdot \mbf{n}- \mbf{z}^* \cdot \mbf{n} \rp}  \\
&= \vartheta \lp \mbf{z}^* \left| \Phi \right. \rp \\
\eea
where in the second line we took $\mbf{n} \to - \mbf{n}$ in the sum, and in the third line we used $\frac{1}{2} \mbf{n}\cdot \Phi \cdot \mbf{n} = \frac{1}{2} \mbf{n}\cdot (-\Phi^*) \cdot \mbf{n} \mod 1$. If $\mbf{k}$ is real, then $\mbf{z} = (\mbf{a}_1 \cdot \mbf{k},\mbf{a}_2 \cdot \mbf{k})/(2\pi)$ is real.  As will be shown later in \Eq{eq:siegeljac} of \App{app:siegelprop}, $\vartheta \lp \left. \frac{(k_1,k_2)}{2\pi} \right| \Phi \rp$ for $\phi = 2\pi$ is non-negative for $k_i \in \mathbb{R}$ so we can take
\bea
\boxed{
\mathcal{N}(\mbf{k}) = \vartheta \lp \left. \frac{(k_1,k_2)}{2\pi} \right| \Phi \rp \ .
}
\eea
In fact, we find that $\mathcal{N}(\mbf{k})$ has a quadratic zero at $\mbf{k} = \pi \mbf{b}_1+\pi \mbf{b}_2$ (we prove this in \App{app:siegelprop}), so the states at $(k_1, k_2) = (\pi,\pi)$ are not well-defined. However, this zero must exist in the BZ for topological reasons: when a band has nonzero Chern number, it is impossible to define a smooth gauge that makes the wavefunction well-defined everywhere throughout the BZ.  Our basis states $\ket{\kk, n}$ correspond to Landau levels with Chern number $-1$ (see \App{eq:Hmat}), and hence there must be a point in the BZ where $\ket{\kk, n}$ is ill-defined. In our basis this occurs at $\kk = (\pi, \pi)$, but it can be shifted arbitrarily (allowing the wavefunction to be defined in patches) by taking $\mbf{Q} \to \tilde{\mbf{Q}} =\mbf{Q} + \mbf{p}$ for any complex number $\mbf{p}$. We define $\tilde{T}_{\mbf{a}_i} = \exp (i \mbf{a}_i \cdot \tilde{\mbf{Q}}) = e^{i \mbf{a}_i \cdot \mbf{p}}T_{\mbf{a}_i}$ so that from \Eq{eq:statedefAppa}, we see that the state
\bea
\label{eq:shiftp}
\frac{1}{\sqrt{\mathcal{N}(\mbf{k}-\mbf{p})}} \sum_{\mbf{R}} e^{-i \mbf{k} \cdot \mbf{R}} \tilde{T}_{\mbf{a}_1}^{\mbf{R} \cdot \mbf{b}_1} \tilde{T}_{\mbf{a}_2}^{\mbf{R} \cdot \mbf{b}_2} \ket{n} = \frac{1}{\sqrt{\mathcal{N}(\mbf{k}-\mbf{p})}} \sum_{\mbf{R}} e^{-i (\mbf{k}-\mbf{p}) \cdot \mbf{R}} T_{\mbf{a}_1}^{\mbf{R} \cdot \mbf{b}_1} T_{\mbf{a}_2}^{\mbf{R} \cdot \mbf{b}_2} \ket{n}
\eea
is a properly normalized eigenstate of $\tilde{T}_{\mbf{a}_i}$ with an undefined point at $\mbf{k} = \pi \mbf{b}_1 + \pi \mbf{b}_2 + \mbf{p}$. Moreover, we will show in \App{app:siegelcomplete} that the zero does not impact the spectrum because magnetic Bloch Hamiltonian only has a \emph{removable} singularity at $(\pi,\pi)$ and thus is analytic in $\mbf{k}$ with a smooth spectrum.

\subsection{Properties of the Siegel Theta Functions}
\label{app:siegelprop}

We will prove some useful properties of the Siegel theta functions used throughout the rest of the work. Recall that the one-dimensional theta functions $\vartheta(z|\tau)$ enjoy the (quasi-)periodicity relations
\bea
\vartheta(z+1|\tau) =  \vartheta(z|\tau) , \qquad \vartheta(z+\tau |\tau)  =  e^{- 2\pi i (z + \tau/2)}
\vartheta(z|\tau)
\eea
which define the function in the whole complex plane from the principal domain. We now prove the analogous properties for the Siegel theta functions.

First recall the definition of the Siegel theta function for a symmetric matrix $M$ where $\text{Im } M >0$ as the convergent sum:
\bea
\vartheta \lp \mbf{z} \left| M \right. \rp = \sum_{\mbf{n}\in\mathds{Z}^2} e^{2\pi i \lp \frac{1}{2} \mbf{n} \cdot  M  \cdot \mbf{n}- \mbf{z} \cdot \mbf{n} \rp} \ . \\
\eea
By shifting variables in the sum, we find for any integer vector $\bf{n}$
\bea
\label{eq:thetaquasiper}
\vartheta \lp \mbf{z} \left| M \right. \rp &= \sum_{\mbf{m}\in\mathds{Z}^2} e^{2\pi i \lp \frac{1}{2} (\mbf{m} - \mbf{n}) \cdot M \cdot (\mbf{m}-\mbf{n})- \mbf{z} \cdot (\mbf{m} - \mbf{n}) \rp} \\
&= e^{i \pi \mbf{n} \cdot M \cdot \mbf{n} + 2\pi i \mbf{z} \cdot \mbf{n}}\sum_{\mbf{m}\in\mathds{Z}^2} e^{2\pi i \lp \frac{1}{2} \mbf{m} \cdot M \cdot \mbf{m} - \mbf{n} \cdot M \mbf{m}-  \mbf{z} \cdot \mbf{m} \rp} \\
&= e^{i \pi \mbf{n} \cdot M \cdot \mbf{n} + 2\pi i \mbf{z} \cdot \mbf{n}}\sum_{\mbf{m}\in\mathds{Z}^2} e^{2\pi i \lp \frac{1}{2} \mbf{m} \cdot M \cdot \mbf{m} - (\mbf{z}+M \mbf{n}) \cdot \mbf{m} \rp} \\
&= e^{i \pi \mbf{n} \cdot M\cdot  \mbf{n} + 2\pi i \mbf{z} \cdot \mbf{n}} \vartheta \lp \mbf{z}+M \mbf{n} \left| M \right. \rp  \\
\eea
which gives the quasi-periodicity associated with the theta function. There is also the simpler periodicity $\vartheta \lp \mbf{z} + \mbf{n} \left| M \right. \rp =\vartheta \lp \mbf{z} \left| M \right. \rp$ for integer vectors $\mbf{n}$. In our application, we have $M= \Phi$ (\Eq{eq:Phi}) for $\phi = 2\pi$ which gives the elementary relations
\bea
\vartheta \lp \mbf{z} + \frac{1}{2}(i, -1) \left| \Phi \right. \rp  &= e^{\pi/2 - 2\pi i z_1} \vartheta \lp \mbf{z} \left| \Phi \right. \rp  \\
\vartheta \lp \mbf{z} + \frac{1}{2}(-1, i)\left| \Phi \right. \rp  &= e^{\pi/2 - 2\pi i z_2} \vartheta \lp \mbf{z} \left| \Phi \right. \rp \ .  \\
\eea
We emphasize that this identity only holds at $2\pi$ flux.

We now use a property known as the addition formula \cite{NIST:DLMF} to prove that  $\vartheta \big( (1/2,1/2)| \Phi\big) = 0$. This property is crucial for proving the Chern number of the Landau levels. Let the vector $\pmb{\nu}$ take values in $(0,0),(1,0),(0,1),(1,1)$. The addition formula (\url{https://dlmf.nist.gov/21.6}) reads
\bea
\label{eq:addition}
\vartheta(\mbf{x} | M)\vartheta(\mbf{y} | M) = \sum_{\pmb{\nu}} \vartheta_{\pmb{\nu}/2}(\mbf{x}+\mbf{y}|2M) \vartheta_{\pmb{\nu}/2}(\mbf{x}-\mbf{y}|2M) , \qquad  \vartheta_{\pmb{\nu}/2}(\mbf{x}|M) = \sum_{\mbf{n}} e^{2\pi i ( \frac{1}{2} (\mbf{n}+\pmb{\nu}/2)\cdot M \cdot (\mbf{n}+\pmb{\nu}/2) - \mbf{x} \cdot (\mbf{n}+\pmb{\nu}/2))}
\eea
where $\pmb{\nu}/2$ is often called a semi-integer characteristic. This addition formula follows from straightforward algebraic manipulations of the definitions of the theta functions \cite{gunning2012riemann}. Expanding, we find
\bea
\label{eq:thetacharnu}
 \vartheta_{\pmb{\nu}/2}(\mbf{x}|M) &= \sum_{\mbf{n}} e^{2\pi i \big( \frac{1}{2} (\mbf{n}+\pmb{\nu}/2)\cdot M \cdot (\mbf{n}+\pmb{\nu}/2) - \mbf{x} \cdot (\mbf{n}+\pmb{\nu}/2) \big)} \\
 &= e^{-\pi i\mbf{x} \cdot \pmb{\nu} + \frac{\pi i}{4} \pmb{\nu} \cdot M \cdot \pmb{\nu} } \sum_{\mbf{n}} e^{2\pi i ( \frac{1}{2} \mbf{n}\cdot M \cdot \mbf{n} + (M \pmb{\nu}/2- \mbf{x}) \cdot \mbf{n})} \\
  &= e^{-\pi i\mbf{x} \cdot \pmb{\nu} + \frac{\pi i}{4} \pmb{\nu} \cdot M \cdot \pmb{\nu} } \vartheta(\mbf{x} - M \pmb{\nu}/2 |M)
\eea
We now apply the addition formula applied to $M = \Phi$ and $\mbf{x} = \mbf{y} = (1/2,1/2)$ and use the invariance of $\vartheta(\mbf{z}|\Phi)$ under $\mbf{z} \to \mbf{z} + \mathds{Z}^2$ and $\mbf{z} \to - \mbf{z}$ to find
\bea
\label{eq:idused}
\vartheta((1/2,1/2) | \Phi)\vartheta((1/2,1/2) | \Phi) &= \sum_{\pmb{\nu}} e^{-\pi i(1,1) \cdot \pmb{\nu} + \frac{\pi i}{4} \pmb{\nu} \cdot (2\Phi) \cdot \pmb{\nu} } \vartheta((1,1)-\Phi \pmb{\nu} |2\Phi) e^{\pi i(0,0) \cdot \pmb{\nu} + \frac{\pi i}{4} \pmb{\nu} \cdot (2\Phi) \cdot \pmb{\nu} } \vartheta(\Phi \pmb{\nu} |2\Phi) \\
&= \sum_{\pmb{\nu}} e^{-\pi i(1,1) \cdot \pmb{\nu} + i \pi \pmb{\nu} \cdot \Phi \cdot \pmb{\nu} } \vartheta(\Phi \pmb{\nu}|2\Phi) \vartheta(\Phi \pmb{\nu} |2\Phi) \\
&= \vartheta((0,0)|2\Phi)^2 - e^{-\pi/2}\vartheta((i/2,-1/2)|2\Phi)^2- e^{-\pi/2}\vartheta((-1/2,i/2)|2\Phi)^2
\\
&\qquad - e^{-\pi} \vartheta((-1/2 + i /2,-1/2 + i /2)|2\Phi)^2 \ . \\
\eea
The next important fact is that $\vartheta(\mbf{z}|2\Phi) = \theta(z_1|i)\theta(z_2|i)$ where the Jacobi theta function is
\bea
\label{eq:jacobi_def}
\theta(z|\tau) = \sum_{\mbf{n}} e^{2\pi i ( \frac{\tau}{2} n^2 - z n)} , \quad \theta(z|\tau) = \theta(-z|\tau)
\eea
which follows because the off-diagonal term in $2\Phi$ disappears:
\bea
2\pi i \frac{\mbf{n} \cdot (2 \Phi) \cdot \mbf{n}}{2} = -\pi n_1^2 - \pi n_2^2 - 2\pi i n_1 n_2 = -\pi n_1^2 - \pi n_2^2  \mod 2\pi \ .
\eea
Using the Jacobi theta functions in the result of \Eq{eq:idused}, we find
\bea
\vartheta((1/2,1/2) | \Phi)^2 &= \theta(0|i)^4 - 2e^{-\pi/2} \theta\big(\frac{i}{2}|i\big)^2\theta\big(\frac{1}{2}|i\big)^2 \\
\eea
where the term involving $\vartheta((-1/2 + i /2,-1/2 + i /2)|2\Phi)$ has vanished because $-\frac{1}{2} + \frac{i}{2}$ is a zero of the Jacobi theta functions. We now use the sum-of-squares identity \cite{NIST:DLMFjac} (\url{https://dlmf.nist.gov/20.7})
\bea
\label{eq:sumofsq}
 \theta(0|\tau)^2  \theta(z|\tau)^2 &= \theta(1/2|\tau)^2  \theta(z+1/2|\tau)^2 + e^{2\pi i z + i \pi \tau}\theta(i/2|\tau)^2  \theta(z+i/2|\tau)^2 \\
\eea
which implies $ \theta(0|i)^4 = \theta(1/2|i)^4+ e^{-\pi}\theta(i/2|i)^4$ and hence
\bea
\label{eq:Sthetamod}
\vartheta((1/2,1/2) | \Phi)^2 &= \lp \theta(1/2|i)^2 - e^{-\pi/2}\theta(i/2|i)^2 \rp^2 \ .
\eea
Lastly, we use the modular identity $\sqrt{-i \tau} \th(z|\tau) = e^{- i \pi z^2/\tau} \th(z/\tau|-1/\tau)$ at the self-dual point $\tau = i = -1/i$, giving
\bea
\label{eq:Jthetamod}
 \theta(1/2|i) = e^{- \pi/4} \th(-i/2|i) = e^{- \pi/4} \th(i/2|i)
\eea
using the evenness of $\theta(z,\tau)$ in the second equality:$\theta(-z,\tau) = \theta(z,\tau)$. Squaring \Eq{eq:Jthetamod} proves \Eq{eq:Sthetamod} is equal to zero. Numerically, we check that the zero is second order via the winding in \Fig{fig:siegelwinding} of the Main Text. Note that ${\cal N}(\kk)$ is reflection symmetric about $(\pi,\pi)$ in both $k_1, k_2$ directions, indicating a quadratic (or higher) zero. This follows because $\vartheta \lp (1/2,1/2) + \mbf{z} \left| \Phi \right. \rp = \vartheta \lp- (1/2,1/2)  - \mbf{z}  \left| \Phi \right. \rp = \vartheta \lp (1/2,1/2) - \mbf{z} \left| \Phi \right. \rp$ using the periodicity and evenness in $\mbf{z}$.

In fact, we can prove a more general identity relating $\vartheta(\mbf{z}|\Phi)$ to Jacobi theta functions. Returning to \Eq{eq:addition}, we have
\bea
\label{eq:sumofsz}
\vartheta(\mbf{z} | \Phi)^2 &= \sum_{\pmb{\nu}} \vartheta_{\pmb{\nu}/2}(2\mbf{z}|2\Phi) \vartheta_{\pmb{\nu}/2}(0|2\Phi) \ . \\
\eea
To fit with the usual conventions for the Jacobi theta functions (see \url{https://en.wikipedia.org/wiki/Theta_function}), we define
\bea
\label{eq:thetadef}
\th_1(z | \tau) &\equiv -i e^{i \pi z + i \pi \tau/4} \th(z + \frac{1}{2} + \frac{\tau}{2} | \tau) \\
\th_2(z | \tau) &\equiv e^{i \pi z + i \pi \tau/4} \th(z + \frac{\tau}{2} | \tau) \\
\th_3(z | \tau) &\equiv \th(z | \tau) \\
\th_4(z | \tau) &\equiv \th(z + \frac{1}{2} | \tau) \\
\eea
which are convenient parameterizations of the theta functions defined in Eq.~\eqref{eq:jacobi_def} shifted by $0, \frac{1}{2}, \frac{i}{2}, \frac{1}{2}(1+i)$. These shifted theta functions will be useful when calculating the four semi-integer characteristics. It is helpful to remember that $\th_1(z | \tau)$ is odd in $z$ and $\th_2(z | \tau),\th_3(z | \tau),\th_4(z | \tau)$ are even.

 Using \Eq{eq:thetacharnu}, we check that
\bea
\label{eq:Jacthetid}
\vartheta_{(0,0)}(\mbf{z}|2\Phi)  &= \theta_3(z_1|i) \theta_3(z_2|i) \\
\vartheta_{(1/2,0)}(\mbf{z}|2\Phi)  &= e^{-\pi i z_1 - \pi/4 } \vartheta(\mbf{z}- (i/2,-1/2)|2\Phi) =  \th_2(z_1|i) \th_4(z_2|i) \\
\vartheta_{(0,1/2)}(\mbf{z}|2\Phi)  &= e^{-\pi i z_2 - \pi/4 } \vartheta(\mbf{z}- (-1/2,i/2)|2\Phi) =  \th_4(z_1|i) \th_2(z_2|i)  \\
\vartheta_{(1/2,1/2)}(\mbf{z}|2\Phi)  &= e^{-\pi i (z_1 + z_2) - \frac{\pi}{2} - i \frac{\pi}{2}} \vartheta((z_1 + 1/2 - i/2, z_2 + 1/2 - i/2)|2\Phi) = i  \th_1(z_1|i)\th_1(z_2|i) \ .
\eea
and hence the Siegel theta function with our matrix $\Phi$ can be expressed in terms of the Jacobi theta functions by plugging into \Eq{eq:sumofsz}
\bea
\label{eq:siegeljac1}
\vartheta(\mbf{z} | \Phi)^2 &= \theta_3(2z_1|i) \theta_3(2z_2|i)  \theta_3(0|i)^2 + \Big( \theta_2(2z_1|i) \theta_4(2z_2|i) + \theta_4(2z_1|i) \theta_2(2z_2|i) \Big) \theta_2(0|i)\theta_4(0|i) \ .
\eea
No $\theta_1(z|i)$ term appears because $\theta_1(0|i) = 0$ because it is odd. We emphasize that \Eq{eq:siegeljac1} relies on the precise form of $\Phi$ at $\phi = 2\pi$. We now need the brief identity
\bea
\label{eq:shortide}
\frac{ \theta_3(0|i)^2}{ \theta_2(0|i) \theta_4(0|i)} = e^{\pi/4} \frac{ \theta(0|i)^2}{ \theta(1/2|i) \theta(i/2|i)}  =  \frac{ \theta(0|i)^2}{ \theta(1/2|i)^2}  = \sqrt{2}
\eea
which is proved using the sum-of-squares identity in \Eq{eq:sumofsq} and the definitions \Eq{eq:thetadef}. We simplify \Eq{eq:siegeljac1} to find
\bea
\label{eq:siegeljac}
\boxed{
\vartheta(\mbf{z} | \Phi) =  \theta_3(0|i) \sqrt{ \theta_3(2z_1|i) \theta_3(2z_2|i)  + \frac{1}{\sqrt{2}} \big( \theta_2(2z_1|i) \theta_4(2z_2|i) + \theta_4(2z_1|i) \theta_2(2z_2|i) \big) } \ . }
\eea
The sign of the square root above (\Eq{eq:siegeljac}) is positive for real arguments $\mbf{z} \in \mathbb{R}^2$, as we know $\vartheta(\mbf{z} | \Phi), \theta_3(0|i)$ are non-negative.  In \App{app:greens}, we give another expression for $\vartheta(\mbf{z} | \Phi) $ in terms of Jacobi theta functions which is useful for numerical implementation because many common software packages, such as Python, do not implement Siegel theta (while they do implement the Jacobi theta function).

\Eq{eq:siegeljac} is also useful for proving a useful property of the form factors.  In Eq.~\ref{eq:siegelPhaseFactor} of the Main Text, we give an important expression for a phase factor that appears in our calculations:
\begin{align}
e^{i\xi_\qq(\kk)} = \frac{e^{- \frac{ \bar{q} q}{4\phi}}  \vartheta \lp \left. \frac{(k_1 - q/2 ,k_2+ i q/2)}{2\pi} \right| \Phi \rp  }{\sqrt{\vartheta \lp \left. \frac{(k_1 ,k_2)}{2\pi} \right| \Phi \rp \vartheta \lp \left. \frac{(k_1 - q_1 ,k_2 - q_2)}{2\pi} \right| \Phi \rp}}, \qquad  q = q_1 + i q_2.
\label{eq:xiphasevariousproofs}
\end{align} At $k_1, k_2 = \pi$, the denominator tends to zero as $\sqrt{\vartheta \lp \left. \frac{(\pi ,\pi)}{2\pi} \right| \Phi \rp} = 0.$  In order for this phase factor to be well defined, we must show the numerator $\vartheta \lp \left. \frac{(\pi - q/2 ,\pi+ i q/2)}{2\pi} \right| \Phi \rp = 0$.  We perform this calculation below.

Let $z_1 = 1/2 - q/2, z_2 = 1/2 + i q/2$ where $q \in \mathbb{C}$. (Note that Eq.~\ref{eq:siegelPhaseFactor} generally has complex arguments in its theta functions.) Then we find
\bea
\label{eq:siegelcomplexsqrt}
\vartheta((1/2 - q/2, 1/2 + i q/2) | \Phi) &= \theta_3(0|i) \sqrt{ \theta_3(q|i) \theta_3(iq |i) - \frac{1}{\sqrt{2}} \big( \theta_2(q|i) \theta_4(iq|i) + \theta_4(q|i) \theta_2(i q|i) \big) }
\eea
where we used the elementary properties $\theta_3(1-q|\tau) = \theta_3(q|\tau), \theta_4(1-q|\tau) = \theta_4(q|\tau), $ and $\theta_2(1-q|\tau) = -\theta_2(q|\tau)$ \cite{NIST:DLMFjac}. For general complex arguments, there is branch cut in the square root of \Eq{eq:siegelcomplexsqrt}. This will not concern us here because we will prove both sides of \Eq{eq:siegelcomplexsqrt} are zero.

We now use the modular identities at $\tau = i$ which, using evenness, read
\bea
\label{eq:Jacmodid}
\theta_2(q|i) &= e^{- \pi q^2} \theta_4(iq|i) \\
\theta_3(q|i) &= e^{- \pi q^2} \theta_3(iq|i) \\
\theta_4(q|i) &= e^{- \pi q^2} \theta_2(iq|i) \\
\eea
to find
\bea
\label{eq:siegeqgone}
\vartheta\Big((1/2 - q/2, 1/2 + i q/2) \Big| \Phi\Big) &= \theta_3(0|i) e^{\pi q^2/2} \sqrt{  \theta_3(q|i)^2 - \frac{1}{\sqrt{2}} \big( \theta_2(q|i)^2 + \theta_4(q|i)^2 \big) } \ .
\eea
Once again, we use the sum-of-squares identity in \Eq{eq:sumofsq} which can be written
\bea
\theta_3(0|i)^2 \theta_3(q|i)^2 &= \theta_2(0|i)^2 \theta_2(q|i)^2 +\theta_4(0|i)^2 \theta_4(q|i)^2 \\
 \theta_3(q|i)^2 &= \frac{1}{\sqrt{2}} \theta_2(q|i)^2 + \frac{1}{\sqrt{2}} \theta_4(q|i)^2 \\
\eea
where we used \Eq{eq:shortide} and $ \theta_2(0|i) =  \theta_4(0|i)$. Plugging in to \Eq{eq:siegeqgone}, we find the surprising consequence
\bea
\label{eq:suprisingid}
\vartheta\Big((1/2 - q/2, 1/2 + i q/2) \Big| \Phi\Big)  = 0 \quad \forall q \in \mathbb{C} \ . \\
\eea
Thus $\xi_\mbf{q}(\mbf{k})$ (see \Eq{eq:siegelPhaseFactor} of the Main Text) is well-defined. We now prove that the ratio of theta functions is a pure phase, i.e. it has magnitude 1, so $\xi_\mbf{q}(\mbf{k})$ is real. We need to show $e^{i \xi_{\mbf{q}}(\mbf{k})} (e^{i \xi_{\mbf{q}}(\mbf{k})})^* = 1$, which is equivalent to
\bea
\label{eq:NTSxi}
\frac{e^{- \frac{ \bar{q} q}{2\phi}}  \vartheta \lp \left. \frac{(k_1 - q/2 ,k_2+ i q/2)}{2\pi} \right| \Phi \rp\vartheta \lp \left. \frac{(k_1 - \bar{q}/2 ,k_2-  i \bar{q}/2)}{2\pi} \right| \Phi \rp  }{\vartheta \lp \left. \frac{(k_1 ,k_2)}{2\pi} \right| \Phi \rp \vartheta \lp \left. \frac{(k_1 - q_1 ,k_2 - q_2)}{2\pi} \right| \Phi \rp}  &= 1\\
\eea
where we used $\vartheta(\mbf{z}|\Phi)^* = \vartheta(\mbf{z}^*|\Phi)$. Both the numerator and denominator are products of theta functions which are amenable to the addition formula. Using \Eq{eq:addition}, the numerator of \Eq{eq:NTSxi} is
\bea
\label{es:xinum}
& \vartheta \lp \left. \frac{(k_1 - q/2 ,k_2+ i q/2)}{2\pi} \right| \Phi \rp\vartheta \lp \left. \frac{(k_1 - \bar{q}/2 ,k_2-  i \bar{q}/2)}{2\pi} \right| \Phi \rp \\
 & \qquad \qquad = \sum_{\pmb{\nu}} \vartheta_{\pmb{\nu}/2}\lp \left. \frac{(2k_1 - (q+\bar{q})/2,2k_2+ i (q-\bar{q})/2)}{2\pi} \right| 2\Phi \rp \vartheta_{\pmb{\nu}/2}\lp \left. \frac{( (\bar{q}- q)/2,i (q+\bar{q})/2)}{2\pi} \right| 2\Phi \rp \\
 & \qquad \qquad = \sum_{\pmb{\nu}} \vartheta_{\pmb{\nu}/2}\lp \left. \frac{(2k_1 - q_1,2k_2 - q_2)}{2\pi} \right| 2\Phi \rp \vartheta_{\pmb{\nu}/2}\lp \left. \frac{(-iq_2,i q_1)}{2\pi} \right| 2\Phi \rp \\
\eea
and the denominator of \Eq{eq:NTSxi} is written
\bea
\label{es:xidom}
\vartheta \lp \left. \frac{(k_1 ,k_2)}{2\pi} \right| \Phi \rp\vartheta \lp \left. \frac{(k_1 - q_1,k_2- q_2)}{2\pi} \right| \Phi \rp = \sum_{\pmb{\nu}} \vartheta_{\pmb{\nu}/2}\lp \left. \frac{(2k_1 - q_1,2k_2 - q_2)}{2\pi} \right| 2\Phi \rp \vartheta_{\pmb{\nu}/2}\lp \left. \frac{(q_1, q_2)}{2\pi} \right| 2\Phi \rp \ . \\
\eea
The first $\mbf{k}$-dependent factors in \Eqs{es:xinum}{es:xidom} are identical, so we only need to prove
\bea
\label{eq:siegelmodNTS}
e^{- \frac{ \bar{q} q}{2\phi}} \vartheta_{\pmb{\nu}/2}\lp \left. \frac{(-iq_2,i q_1)}{2\pi} \right| 2\Phi \rp &= \vartheta_{\pmb{\nu}/2}\lp \left. \frac{(q_1, q_2)}{2\pi} \right| 2\Phi \rp
\eea
for each of the four $\pmb{\nu}$. To do so, we only need to use \Eq{eq:Jacthetid} to write the Siegel theta functions in terms of Jacobi theta functions and use the modular identities in \Eq{eq:Jacmodid}. It is direct to show that
\bea
\label{eq:javsigelist}
\vartheta_{\pmb{\nu}/2}\lp \left. \frac{(-iq_2,i q_1)}{2\pi} \right| 2\Phi \rp &= \begin{cases}
\theta_3(-i q_2/(2\pi)| i) \theta_3(i q_1/(2\pi)| i), & \pmb{\nu} = (0,0) \\
\theta_2(-i q_2/(2\pi)| i) \theta_4(i q_1/(2\pi)| i), & \pmb{\nu} = (1,0) \\
\theta_4(-i q_2/(2\pi)| i) \theta_2(i q_1/(2\pi)| i), & \pmb{\nu} = (0,1) \\
\end{cases} \\
&= e^{\pi (q_1^2 + q_2^2)/(2\pi)^2} \begin{cases}
 \theta_3(q_2/(2\pi)| i) \theta_3(q_1/(2\pi)| i), & \pmb{\nu} = (0,0) \\
\theta_4(q_2/(2\pi)| i) \theta_2(q_1/(2\pi)| i), & \pmb{\nu} = (1,0) \\
\theta_2(q_2/(2\pi)| i) \theta_4(q_1/(2\pi)| i), & \pmb{\nu} = (0,1) \\
\end{cases} \\
&= e^{\frac{ \bar{q} q}{2\phi}} \vartheta_{\pmb{\nu}/2}\lp \left. \frac{(q_1,q_2)}{2\pi} \right| 2\Phi \rp
\eea
which directly proves three of the four cases in \Eq{eq:siegelmodNTS}. For the last case where $\theta_1(z|i)$ is an odd function of $z$, we use \Eq{eq:Jacthetid} and the modular identity $\th_1(z|i) = - i e^{- \pi z^2} \theta_1(i z|i)$ to show
\bea
\vartheta_{(1/2,1/2)} \lp \left. \frac{(-iq_2,i q_1)}{2\pi} \right| 2\Phi \rp &= i  \th_1(-iq_2/(2\pi)|i)\th_1(i q_1/(2\pi)|i) \\
&= - i i^2 e^{\pi (q_1^2+q_2^2)/(2\pi)^2} \th_1(q_2/(2\pi)|i)\th_1(q_1/(2\pi)|i) \\
&= e^{\frac{ \bar{q} q}{2\phi}} \vartheta_{(1/2,1/2)} \lp \left. \frac{(q_1,q_2)}{2\pi} \right| 2\Phi \rp  \\
\eea
which takes the same form as \Eq{eq:javsigelist} and completes the proof. Hence we have shown that
\bea
\label{eq:xiexpl}
e^{i \xi_\mbf{q}(\mbf{k})}  = \frac{e^{- \frac{ \bar{q} q}{4\phi}}  \vartheta \lp \left. \frac{(k_1 - q/2 ,k_2+ i q/2)}{2\pi} \right| \Phi \rp  }{\sqrt{\vartheta \lp \left. \frac{(k_1 ,k_2)}{2\pi} \right| \Phi \rp \vartheta \lp \left. \frac{(k_1 - q_1 ,k_2 - q_2)}{2\pi} \right| \Phi \rp}}
\eea
is a pure phase. This is numerically verified in \Fig{fig:siegelPhase} of the Main Text, which also shows there is a branch cut connecting $(k_1,k_2)$ and $(k_1+q_1,k_2+q_2)$.

\subsection{Siegel Theta as the Green's Function on the Torus}
\label{app:greens}

In \App{app:siegelprop}, we found an expression (\Eq{eq:siegeljac}) for the Siegel theta function $\vartheta(\mbf{z}|\Phi)$ in terms of Jacobi theta functions, which we used to prove some helpful identities. We now prove a different expression in terms of holomorphic variables $k = k_1 + i k_2, \bar{k} = k_1 - i k_2$ which we need to study the completeness of the magnetic translation group basis states in \Eq{eq:knexplicit}. We will also show that $\log \vartheta((k_1,k_2)/2\pi|\Phi)$ is a Green's function on the torus \cite{2006math......8358L,Mamode2014FundamentalSO,Polchinski:1998rq,francesco2012conformal}, which is essential for proving that the basis states have constant nonzero Berry curvature in \App{app:berry}.

The major technical result of this section is a proof of the following claim:
\bea
\label{eq:GFstari}
\vartheta \lp \left. \frac{(k_1-\pi,k_2-\pi)}{2\pi} \right| \Phi \rp = \sqrt{2} \left| \th_1\lp \left. \frac{k}{2\pi} \right| i \rp \th_1 \left. \lp \frac{i \bar{k}}{2\pi} \right|  i \rp \right| \exp \lp - \frac{k \bar{k}}{4\pi} \rp \ .
\eea
We start by considering the theta function product on the righthand side. First let us state the Jacobi theta addition formula (see \url{https://dlmf.nist.gov/20.7})
\bea
\th_4(0|i)^2 \th_1(w +z|i)  \th_1(w - z|i) &= \th_3^2(w|i)\th_2^2(z|i) - \th_2^2(w|i)\th_3^2(z|i)
\eea
which in concert with the modular identity $\th_1(z|i) = - i e^{- \pi z^2} \theta_1(i z|i)$ yields
\bea
- i \th_4(0|i)^2 \th_1(w +z|i)  e^{- \pi (w-z)^2} \theta_1(i (w-z)|i)&= \th_3^2(w|i)\th_2^2(z|i) - \th_2^2(w|i)\th_3^2(z|i) \ .
\eea
Upon identifying $w = k_1/(2\pi)$ and $z = i k_2/(2\pi)$, we find
\bea
\th_1\lp \left. \frac{k}{2\pi} \right| i \rp \th_1 \left. \lp \frac{i \bar{k}}{2\pi} \right|  i \rp &= \frac{i}{\th_4(0|i)^2} e^{\bar{k}^2/(4\pi)} \lp \th_3^2(\frac{k_1}{2\pi}|i)\th_2^2(\frac{i k_2}{2\pi}|i) - \th_2^2(\frac{k_1}{2\pi}|i)\th_3^2(\frac{i k_2}{2\pi}|i) \rp \\
&= \frac{i}{\th_4(0|i)^2} e^{\bar{k}^2/(4\pi)} e^{2\pi k_2^2/(2\pi)^2}\lp \th_3^2(\frac{k_1}{2\pi}|i)\th_4^2(\frac{k_2}{2\pi}|i) - \th_2^2(\frac{k_1}{2\pi}|i)\th_3^2(\frac{ k_2}{2\pi}|i) \rp \\
&= \frac{i}{\th_4(0|i)^2} e^{(k_1^2+k_2^2)/4\pi - i k_1 k_2/2\pi}\lp \th_3^2(\frac{k_1}{2\pi}|i)\th_4^2(\frac{k_2}{2\pi}|i) - \th_2^2(\frac{k_1}{2\pi}|i)\th_3^2(\frac{ k_2}{2\pi}|i) \rp \\
\eea
where in the second line we used the modular identities for the $\th_2$ and $\th_3$ functions: $
\th_3(i z|i) = e^{\pi z^2} \th_3(z|i)$ and $\th_2(i z|i) = e^{\pi z^2} \th_4(z|i)$. The Jacobi theta functions $ \th_i(z|i)$ are real for real arguments, so we conclude
\bea
\left| \th_1\lp \left. \frac{k}{2\pi} \right| i \rp \th_1 \left. \lp \frac{i \bar{k}}{2\pi} \right|  i \rp \right|^2 &= \frac{1}{\th_4(0|i)^4} e^{(k_1^2+k_2^2)/2\pi} \lp \th_3^2(\frac{k_1}{2\pi}|i)\th_4^2(\frac{k_2}{2\pi}|i) - \th_2^2(\frac{k_1}{2\pi}|i)\th_3^2(\frac{ k_2}{2\pi}|i) \rp^2
\eea
assuming that $k_1, k_2 \in \mathbb{R}$. Returning to \Eq{eq:GFstari} with this identity and using \Eq{eq:siegeljac1}, the claim in \Eq{eq:GFstari} is equivalent to
\bea
\label{eq:thetasquaredverify}
\vartheta \lp \left. \frac{(k_1-\pi,k_2-\pi)}{2\pi} \right| \Phi \rp^2 &= \frac{2}{\th_4(0|i)^4} \lp \th_3^2(\frac{k_1}{2\pi}|i)\th_4^2(\frac{k_2}{2\pi}|i) - \th_2^2(\frac{k_1}{2\pi}|i)\th_3^2(\frac{ k_2}{2\pi}|i) \rp^2  \ .
\eea
Using the quasi-periodicities $\th_2(z -1|i) = -\th_2(z|i)$, $\th_3(z -1|i) = \th_3(z|i)$, and $\th_4(z -1|i) = \th_4(z|i)$ from \Eq{eq:thetadef} to rewrite \Eq{eq:siegeljac}, we find
\bea
\vartheta \lp \left. \frac{(k_1-\pi,k_2-\pi)}{2\pi} \right| \Phi \rp^2 &=  \theta_3(0|i)^2 \lp \theta_3(\frac{k_1}{\pi}|i) \theta_3(\frac{k_2}{\pi}|i)  - \frac{1}{\sqrt{2}} \Big( \theta_2(\frac{k_1}{\pi}|i) \theta_4(\frac{k_2}{\pi}|i) + \theta_4(\frac{k_1}{\pi}|i) \theta_2(\frac{k_2}{\pi}|i) \Big) \rp \ .
\eea
Recalling that $\sqrt{2} \th_4(0|i)^2 = \th_3(0|i)^2$, the claim in \Eq{eq:thetasquaredverify} is equivalent to
\bea
\label{eq:NTS8th}
\theta_3(2z|i) \theta_3(2w|i)  - \frac{1}{\sqrt{2}} \Big( \theta_2(2z|i) \theta_4(2w|i) + \theta_4(2z|i) \theta_2(2w|i) \Big) &=  \frac{\sqrt{2}}{\th_4(0|i)^6} \lp \th_3^2(z|i)\th_4^2(w|i) - \th_2^2(z|i)\th_3^2(w|i) \rp^2
\eea
where $z = k_1/2\pi, w = k_2/2\pi$ for brevity. The lefthand side can be simplified with the duplication formula
\bea
\label{eq:dupid}
\th_i(2z|\tau) \th_4(0|\tau)^3 &= \th_i(z|\tau)^4 - \th_1(z|\tau)^4, \qquad i = 2,4 \\
2^{3/4} \th_3(2z|\tau) \th_4(0|\tau)^3 &= \th_3(z|\tau)^4 + \th_1(z|\tau)^4
\eea
found in \Ref{CARLSON201142}. The identities in \Eq{eq:dupid} allow \Eq{eq:NTS8th} to be written as
\bea
\label{eq:NTSthnoi}
 \big( \th_3^2(z)\th_4^2(w) - \th_2^2(z)\th_3^2(w) \big)^2  &=  \frac{1}{4}(\theta^4_3(z)+\th^4_1(z))  (\theta^4_3(w)+\th^4_1(w))\\
& \qquad - \frac{1}{2} \Big( (\theta^4_2(z)-\th^4_1(z) )(\theta^4_4(w)-\th^4_1(w))+ (\theta^4_4(z)-\th^4_1(z) )(\theta^4_2(w)-\th^4_1(w)) \Big) \\
\eea
where for brevity we dropped the $i$ argument, e.g. $\th_3(z) \equiv \th_3(z|i)$. The next step is to use the sum-of-square identities at $\tau = i$  (see \url{https://dlmf.nist.gov/20.7}):
\bea
\th_1^2(z) &= \sqrt{2} \th_4^2(z) - \th_3^2(z) \\
\th_1^2(z) &=  \th_3^2(z)  - \sqrt{2} \th_2^2(z) \\
\sqrt{2} \th_1^2(z) &=  \th_4^2(z)  - \th_2^2(z) \\
\eea
which yield the identities
\bea
\label{eq:th1th3to24}
\th_1^2(z) &= \frac{1}{\sqrt{2}} ( \th_4^2(z)  - \th_2^2(z) ) \\
\th_3^2(z)  &= \frac{1}{\sqrt{2}}( \th_4^2(z) +\th_2^2(z) ) \ .
\eea
Upon plugging in \Eq{eq:th1th3to24} into \Eq{eq:NTSthnoi} so everything is written in terms of $ \th_2^2(z), \th_2^2(z),\th_2^2(w), \th_4^2(w)$, it is just a matter of algebra to gather like terms and verify that \Eq{eq:NTSthnoi} is true. Thus we have proven
\bea
\vartheta \lp \left. \frac{(k_1-\pi,k_2-\pi)}{2\pi} \right| \Phi \rp = \sqrt{2} \left| \th_1\lp \left. \frac{k}{2\pi} \right| i \rp \th_1 \left. \lp \frac{i \bar{k}}{2\pi} \right|  i \rp \right| \exp \lp - \frac{k \bar{k}}{4\pi} \rp \ ,
\eea
where $k = k_1 + i k_2$. It is also useful to have a shifted form of the identity:
\bea
\label{eq:siegelshifted1}
\vartheta \lp \left. \frac{(k_1,k_2)}{2\pi} \right| \Phi \rp = \sqrt{2} \left| \th_1\lp \left. \frac{k}{2\pi} + \frac{1+i}{2} \right| i \rp \th_1 \left. \lp \frac{i \bar{k}}{2\pi} + i \frac{1- i}{2}\right|  i \rp \right| \exp \lp - \frac{k \bar{k}}{4\pi} - \frac{1}{2}(k_1 + k_2) - \frac{\pi}{2}\rp \ .
\eea
To simplify the shifted theta functions, we recall from \Eq{eq:thetadef} that $\th_1(z | i) = -i e^{i \pi z - \pi/4} \th_3(z + \frac{1}{2} + \frac{i}{2} |i)$ which can be rewritten
\bea
\label{eq:th1th3relation}
\th_1(z + \frac{1+i}{2} | i) &= -i e^{i \pi z + i \pi/2 - \pi/2 - \pi/4} \th_3(z + 1 + i |i) = e^{\pi/4 - i \pi z} \th_3(z|i)
\eea
where in the second equality we used the quasi-periodicity of the theta functions. Plugging \Eq{eq:th1th3relation} into \Eq{eq:siegelshifted1}, we find
\bea
\label{eq:logNpos}
\boxed{
\vartheta \lp \left. \frac{(k_1,k_2)}{2\pi} \right| \Phi \rp = \sqrt{2} \left| \th_3\lp \left. \frac{k}{2\pi} \right| i \rp \th_3 \left. \lp \frac{i \bar{k}}{2\pi} \right|  i \rp \right| \exp \lp - \frac{k \bar{k}}{4\pi} \rp
}
\eea
where we used $|e^{- i \frac{k}{2} - i \frac{i\bar{k}}{2}} | = e^{k_2/2+k_1/2}$. As a byproduct, \Eq{eq:logNpos} shows trivially that $\vartheta \lp \left. \frac{(k_1,k_2)}{2\pi} \right| \Phi \rp \geq 0$, and also that the only zeros of $\vartheta \lp \left. \frac{(k_1,k_2)}{2\pi} \right| \Phi \rp$ occur when $\th_3(k/2\pi) \th_3(i\bar{k}/2\pi) = 0$. Up to multiplies of $2\pi$, the only zero is $k_1 = k_2 = \pi$ because $\th_3(1/2 + i/2|i) = 0$.

Let us now show that $\log \vartheta \lp \left. \frac{(k_1,k_2)}{2\pi} \right| \Phi \rp$ is a Green's function on the torus. This is a well-known result \cite{Polchinski:1998rq,Mamode2014FundamentalSO,2006math......8358L} using the Jacobi theta form (the righthand side of \Eq{eq:GFstari}), but we give a self-contained argument here.

First we recall that the Laplacian on the torus $\del^2 = \del_1^2 + \del_2^2$, where $\del_i = \frac{\del}{\del k_i}$ for brevity, can be rewritten as
\bea
\del^2 = 4 \del \bar{\del}, \qquad  2\del \equiv \del_1 - i \del_2, \ 2\bar \del \equiv \del_1 + i \del_2
\eea
where the holomorphic derivatives satisfy $\del k = 1, \bar \del k = 0$. Using \Eq{eq:GFstari}, we compute directly
\bea
\del^2 \log \vartheta \lp \left. \frac{(k_1-\pi,k_2-\pi)}{2\pi} \right| \Phi \rp &=  \del^2 \log \left[ \sqrt{2} \left| \th_1\lp \left. \frac{k}{2\pi} \right| i \rp \th_1 \left. \lp \frac{i \bar{k}}{2\pi} \right|  i \rp \right| \exp \lp - \frac{k \bar{k}}{4\pi} \rp  \right ] \\
&=  4 \del \bar{\del}  \left[\log \left| \th_1(k /2\pi | i )\right| + \log \left| \th_1(i \bar{k} /2\pi | i )\right|  - \frac{k \bar{k}}{4\pi}  \right] \\
&= 4 \del \bar{\del} \frac{k \bar{k}}{4\pi}  = -\frac{1}{\pi}, \qquad \text{ if } k_1, k_2 \neq 0
\eea
where in the last line we dropped the $\log \th_1$ terms when $k_1,k_2 \neq 0$ because $\del \bar{\del} f(k) = \del \bar{\del} f(\bar{k}) = 0$ if $f$ is differentiable. At $k_1,k_2 = 0$ however, $\th_1(k|i) = 0$ so $\log \th_1(k|i)$ is singular and will yield a delta function contribution. Near $k_1, k_2 =0$, we Taylor expand $\th_1(z|i) = z \th_1'(0|i) + \dots$ (see \url{https://dlmf.nist.gov/20.4}) to compute
\bea
\lim_{k,\bar{k} \to 0} \del^2 \log \vartheta \lp \left. \frac{(k_1-\pi,k_2-\pi)}{2\pi} \right| \Phi \rp &=  4 \del \bar{\del} \log \big( k \bar{k} \frac{\th'_1(0)^2}{(2\pi)^2} \big) \\
&=  \del^2 \log (k_1^2 + k_2^2) \\
&= 4\pi \delta(\mbf{k}) \\
\eea
using the 2D Green's function formula $\del^2 \lp \frac{1}{2\pi} \log |\mbf{k}| \rp = \delta(\mbf{k})$. Thus we have derived the formula
\bea
\label{eq:greensfunction}
\frac{1}{2} \del^2 \log \vartheta \lp \left. \frac{(k_1-\pi,k_2-\pi)}{2\pi} \right| \Phi \rp &= 2\pi \delta(\mbf{k}) - \frac{1}{2\pi} \\
\eea
which will play a crucial role in studying the Wilson loop of a Landau level state in \App{app:berry}.

\subsection{Completeness Relation}
\label{app:completeness}

In this section, we study the completeness of the magnetic translation group eigenstates $\ket{\mbf{k},n}$ defined in \Eq{eq:statedefAppa}. So far, \App{app:siegelnorm} has shown that the states
\bea
\ket{\mbf{k},n} = \frac{1}{\sqrt{N(\mbf{k})}} \sum_{\mbf{R}} e^{- i \mbf{k} \cdot \mbf{R}} T_{\mbf{a}_1}^{\mbf{R} \cdot \mbf{b}_1} T_{\mbf{a}_2}^{\mbf{R} \cdot \mbf{b}_2} \ket{n} \\
\eea
are orthonormal, i.e. $\braket{\mbf{k}',n'|\mbf{k},n} = (2\pi)^2 \delta(\mbf{k}-\mbf{k}') \delta_{n'n}$, and well-defined except at $\mbf{k}^* = \pi \mbf{b}_1 + \pi \mbf{b}_2$ where $\ket{\mbf{k}^*,n} = 0$ because of the Chern number obstruction to periodic states defined everywhere on the BZ. Note that the position of $\mbf{k}^*$ is arbitrary, and can be shifted by changing the overall phase of the magnetic translation operators (see \Eq{eq:shiftp}). In general, a complete basis satisfies
\bea
\label{eq:compeltenes1}
1 &= \sum_{\ell=0}^\infty \int \frac{d^2k}{(2\pi)^2} \ket{\mbf{k},\ell} \bra{\mbf{k},\ell} \qquad \text{(if complete)}
\eea
where the righthand side in interpreted as a projector onto all states of the Hilbert space. Although the states at $\mbf{k}^*$ vanish identically and do not appear in the integral, $\mbf{k}^*$ is a single point and thus is a set of measure zero. Thus the missing states can be neglected when \Eq{eq:compeltenes1} acts on a suitable test function which is a wave packet formed from a smooth superposition of $\ket{\mbf{k},n}$ states with finite weight on each $\ket{\mbf{k},m}$. (Note that this excludes momentum eigenstates, where an obvious pathological counterexample to \Eq{eq:compeltenes1} is a state of momentum $\mbf{k}^*$ which is projected out by the righthand side of \Eq{eq:compeltenes1}.) Quantitatively, we consider test functions in the form
\bea
\label{eq:rhostates}
\ket{\rho} = \sum_{n=0}^\infty \int \frac{d^2k}{(2\pi)^2} \rho_n(\mbf{k}) \ket{\mbf{k}, n} \\
\eea
for a suitable smooth function $\rho_n(\mbf{k})$, which excludes the case of $\rho_n(\mbf{k}) \sim \delta(\mbf{k} - \mbf{k}^*)$. Note that $\rho_n(\mbf{k})$ can always be defined locally at some $\mbf{k}$ by shifting the location of $\mbf{k}^*$ if necessary. The rigorous conditions $\rho_n(\mbf{k})$ should satisfy will not concern us here. Because $\rho_n(\mbf{k}^*)$ is necessarily finite and the states $\ket{\mbf{k}, n}$ are only undefined at $\mbf{k}^*$, we expect the completeness relation \Eq{eq:compeltenes1} holds up to a set of measure zero. We will now show that \Eq{eq:compeltenes1} is true when acting on test functions given by the continuum Landau levels $\propto a^{\dag n} b^{\dag m} \ket{0}$ which are of the type in \Eq{eq:rhostates} because they are delocalized in momentum space.

We will prove \Eq{eq:compeltenes1} using the known completeness relation \cite{Tao_1986,2013qher.book.....E} for the continuum Landau level operators:
\bea
\label{eq:mncomlee}
\sum_{m,n} \ket{m,n} \bra{m,n} = 1, \qquad \ket{m,n} = \frac{b^{\dag m}}{\sqrt{m!}} \frac{a^{\dag n}}{\sqrt{n!}} \ket{0}  \\
\eea
where $a \ket{0} = b \ket{0} = 0$. To study the completeness of the $\ket{\mbf{k}, n}$ basis, we will compute an expression for
\bea
\label{eq:Dnmnm}
\mathcal{I}_{m'm,n'n} = \bra{m',n'} \lp \sum_{\ell=0}^\infty \int \frac{d^2k}{(2\pi)^2} \ket{\mbf{k},\ell} \bra{\mbf{k},\ell} \rp \ket{m,n} \ .
\eea
Using \Eq{eq:mncomlee}, the $\ket{\mbf{k},n}$ basis is complete and \Eq{eq:compeltenes1} holds iff $\mathcal{I}_{m'm,n'n} = \delta_{mm'} \delta_{nn'}$. The rest of this section is devoted to the calculation of $\mathcal{I}_{m'm,n'n}$.

We begin by simplifying the $a$ oscillators in \Eq{eq:Dnmnm}. Because the $a$ and $b$ oscillators commute, we easily have
\bea
\label{eq:residNTS}
\bra{m',n'} \lp \sum_\ell \int \frac{d^2k}{(2\pi)^2} \ket{\mbf{k},\ell} \bra{\mbf{k},\ell} \rp \ket{m,n} = \delta_{nn'} \int \frac{d^2k}{(2\pi)^2} \bra{n} \frac{b^{m'}}{\sqrt{m'!}}  \ket{\mbf{k},n} \bra{\mbf{k},n}  \frac{b^{\dag m}}{\sqrt{m!}} \ket{n},\quad \text{ ($n$ unsummed)} \ .
\eea
The remaining overlaps can be evaluated using BCH. We expand the basis states with \Eq{eq:knexplicit} to find
\bea
\braket{n|\frac{b^{m'}}{\sqrt{m'!}} |\mbf{k},n} &= \frac{1}{\sqrt{\mathcal{N}(\mbf{k})}} \sum_{\mbf{R}} e^{- i \mbf{k} \cdot \mbf{R}+ i \pi R_1 R_2} \braket{0|\frac{b^{m'}}{\sqrt{m'!}} e^{i \mbf{Q} \cdot \mbf{R} }|0} \\
&= \frac{1}{\sqrt{\mathcal{N}(\mbf{k})}}\sum_{\mbf{R}} e^{- i \mbf{k} \cdot \mbf{R}+ i \pi R_1 R_2 - \frac{\phi}{4} R \bar{R}} \braket{0|\frac{b^{m'}}{\sqrt{m'!}} e^{i \sqrt{\phi/2} \bar{R} b^\dag}|0} \\
\eea
where in the second equality we used the normal ordering identity in \Eq{eqTranslationExpectation}. We now use the coherent state identity $b e^{x b^\dag}\ket{0} = x e^{x b^\dag}\ket{0}$ to find
\bea
\label{eq:gdagpowerkn}
\braket{n|\frac{b^{m'}}{\sqrt{m'!}} |\mbf{k},n}  &= \frac{1}{\sqrt{\mathcal{N}(\mbf{k})}} \sum_{\mbf{R}} e^{- i \mbf{k} \cdot \mbf{R}+ i \pi R_1 R_2 - \frac{\phi}{4} R \bar{R}} \frac{ (i \sqrt{\phi/2} \bar{R})^{m'}}{\sqrt{m'!}} \\
&=\frac{1}{\sqrt{\mathcal{N}(\mbf{k})}} \sum_{\mbf{R}}  \frac{ (-\sqrt{2\phi}\del)^{m'}}{\sqrt{m'!}} e^{- i \mbf{k} \cdot \mbf{R}+ i \pi R_1 R_2 - \frac{\phi}{4} R \bar{R}} \\
&= \frac{1}{\sqrt{\mathcal{N}(\mbf{k})}} \frac{ (-\sqrt{2\phi}\del)^{m'}}{\sqrt{m'!}} \mathcal{N}(\mbf{k})
\eea
where we introduced the holomorphic variables $k = k_1 + i k_2, \bar{k} = k_1 - i k_2$ and the holomorphic derivative $\del$ satisfying
\bea
\label{eq:holoderivids}
\bar{R} e^{-i \mbf{k} \cdot \mbf{R}} = \bar{R} e^{-i \frac{k \bar{R} + \bar{k}R}{2}} = 2i \del e^{-i \frac{k \bar{R} + \bar{k}R}{2}}, \qquad 2\del \equiv \frac{\del}{\del k_1} - i \frac{\del}{\del k_2} \equiv \del_1 - i \del_2
\eea
which obeys $\del k = 1, \del \bar{k} = 0$. The anti-holomorphic derivative $\bar{\del} = \del^*$ satisfies $\bar{\del} \bar k = 1, \bar{\del} k = 0$. Taking complex conjugates in \Eq{eq:gdagpowerkn}, we find
\bea
\braket{\mbf{k},n|\frac{b^{\dag m}}{\sqrt{m!}} |n}  &= \frac{1}{\sqrt{\mathcal{N}(\mbf{k})}} \sum_{\mbf{R}} e^{i \mbf{k} \cdot \mbf{R} - i \pi R_1 R_2 - \frac{\phi}{4} R \bar{R}} \frac{ (-i \sqrt{\phi/2} R)^{m}}{\sqrt{m!}} \\
&= \frac{1}{\sqrt{\mathcal{N}(\mbf{k})}} \sum_{\mbf{R}} e^{-i \mbf{k} \cdot \mbf{R} + i \pi R_1 R_2 - \frac{\phi}{4} R \bar{R}} \frac{ (i \sqrt{\phi/2} R)^{m}}{\sqrt{m!}} \\
&=\frac{1}{\sqrt{\mathcal{N}(\mbf{k})}} \sum_{\mbf{R}}  \frac{ (-\sqrt{2\phi}\bar \del)^{m}}{\sqrt{m!}} e^{- i \mbf{k} \cdot \mbf{R}+ i \pi R_1 R_2 - \frac{\phi}{4} R \bar{R}} \\
&= \frac{1}{\sqrt{\mathcal{N}(\mbf{k})}} \frac{ (-\sqrt{2\phi}\bar \del)^{m}}{\sqrt{m!}} \mathcal{N}(\mbf{k})
\eea

Returning to \Eq{eq:residNTS} with \Eq{eq:gdagpowerkn} and its Hermitian conjugate, we arrive at the expression
\bea
\label{eq:completenessint}
\mathcal{I}_{m'm,n'n} &=\delta_{nn'}  \int \frac{d^2k}{(2\pi)^2} \frac{1}{\mathcal{N}(\mbf{k})}\lp  \frac{(-\sqrt{2\phi}\del)^{m'}}{\sqrt{m'!}} \mathcal{N}(\mbf{k}) \rp \lp   \frac{ (-\sqrt{2\phi}\bar{\del})^{m}}{\sqrt{m!}} \mathcal{N}(\mbf{k})  \rp \\
&=\delta_{nn'} \frac{(2\phi)^{(m+m')/2}}{\sqrt{m!m'!}} (-1)^{m'+m} \int \frac{d^2k}{(2\pi)^2} \mathcal{N}^{-1}  (\del^{m'} \mathcal{N}) (\bar{\del}^m \mathcal{N}) \\
&=\delta_{nn'} \frac{(2\phi)^{(m+m')/2}}{\sqrt{m!m'!}} (-1)^{m'+m} \int \frac{d^2k}{(2\pi)^2} \mathcal{N} ( \mathcal{N}^{-1} \del^{m'} \mathcal{N} ) ( \mathcal{N}^{-1} \bar{\del}^m \mathcal{N} )
\eea
suppressing the $\mbf{k}$-dependence for brevity. In the last line, we suggestively added a factor of $\mathcal{N}^{-1} \mathcal{N}$ which we make use of shortly to introduce a covariant derivative structure.  \Eq{eq:completenessint} has reduced the computation of $\mathcal{I}_{m'm,n'n}$ to a single integral. Although it appears nontrivial, we can solve it exactly.

We now show that the integral can be evaluated at all $m,m'$ with an oscillator algebra of momentum space covariant derivatives to be introduced shortly. The only other result required is the integral at $m=m'=0$:
\bea
\label{eq:Nintnormal}
\mathcal{I}_{00,00} = \int \frac{d^2k}{(2\pi)^2} \mathcal{N}(\mbf{k}) &= \sum_{\mbf{n} \in \mathds{Z}^2} \int \frac{d^2k}{(2\pi)^2} e^{- i (k_1 n_1 + k_2 n_2) + \pi i \mbf{n} \cdot \Phi \cdot \mbf{n}} = \sum_{\mbf{n}\in\mathds{Z}^2} \delta_{\mbf{n},0} e^{\pi i \mbf{n} \cdot \Phi \cdot \mbf{n}} = 1 \ .
\eea
To perform the integral in \Eq{eq:completenessint} at general $m,m'$, we use the form of $\mathcal{N}(\mbf{k})$ in \Eq{eq:logNpos} which reads
\bea
\mathcal{N}(\mbf{k}) = \sqrt{2} \left| \th_3\lp \left. k/2\pi\right| i \rp \th_3 \left. \lp i \bar{k}/2\pi \right|  i \rp \right| \exp \lp - \frac{k \bar{k}}{4\pi} \rp \equiv \sqrt{2} f(k) \bar{f}(\bar{k}) e^{- k \bar{k}/4\pi}
\eea
where for brevity we defined $f(k) = | \th_3\lp \left. k/2\pi\right| i \rp |$ and $\bar f(\bar k) = | \th_3\lp \left. i \bar{k}/2\pi\right| i \rp |$. It should be noted that $f(k)$ and $\bar{f}(\bar{k})$ are not complex conjugates of each other. We will sometimes suppress the arguments of $f$ and $\bar f$ for brevity.

We can now simplify the expressions for the derivatives because $\mathcal{N}$ factors into holomorphic and anti-holomorphic parts, along with the exponential. First we note that the holomorphic derivative operators act in a simple way on $\mathcal{N}(\mbf{k})$:
\bea
\del \, (e^{- k \bar{k}/4\pi} f \bar{f} )&= e^{- k \bar{k}/8\pi} (\del - \bar{k}/8\pi)  (e^{- k \bar{k}/8\pi} f \bar{f} )=  e^{- k \bar{k}/8\pi} \nabla (e^{- k \bar{k}/8\pi} f \bar{f} ), \qquad \nabla = \del - \bar{k}/8\pi \\
\bar \del \, ( e^{- k \bar{k}/4\pi} f \bar{f}) &= e^{- k \bar{k}/8\pi} (\bar \del - k/8\pi) ( e^{- k \bar{k}/8\pi} f \bar{f} ) =  e^{- k \bar{k}/8\pi} \bar \nabla (e^{- k \bar{k}/8\pi} f \bar{f}), \qquad \bar \nabla = \bar \del - k/8\pi \\
\eea
where we defined the covariant derivatives $\nabla$ and $\bar{\nabla}$ which commute:
\bea
\null [\bar{\nabla}, \nabla] =  [ \bar{\del}, - \bar{k}/8\pi]  +  [- k/8\pi, \del] = -1/8\pi  + 1/8\pi = 0 \ . \\
\eea
We also note that $\nabla \bar f(\bar k) = (\del - \bar{k}/8\pi ) \bar f(\bar k) =  \bar f(\bar k) (\del - \bar{k}/8\pi ) =  \bar f(\bar k) \nabla$ and similarly $[ \bar \nabla, f(k) ] = 0$, so
\bea
 \mathcal{N}^{-1} \bar{\del}^m \mathcal{N} &= \frac{1}{f \bar{f}}  e^{k \bar{k}/4\pi- k \bar{k}/8\pi} \bar{\nabla}^m ( e^{- k \bar{k}/8\pi} f \bar{f} ) = \frac{1}{\bar{f}}  e^{k \bar{k}/8\pi} \bar{\nabla}^m  (e^{- k \bar{k}/8\pi} \bar{f} )\\
\mathcal{N}^{-1} \del^{m'} \mathcal{N} &= \frac{1}{f \bar{f}} e^{k \bar{k}/4\pi- k \bar{k}/8\pi} \nabla^{m'}  e^{- k \bar{k}/8\pi} f \bar{f}  = \frac{1}{f}  e^{k \bar{k}/8\pi} \nabla^{m'}  ( e^{- k \bar{k}/8\pi} f ) \ . \\
\eea
Thus the integrand of \Eq{eq:completenessint} can be written
\bea
\label{eq:oscilaltorform}
 \mathcal{N} ( \mathcal{N}^{-1} \del^{m'} \mathcal{N} ) ( \mathcal{N}^{-1} \bar{\del}^m \mathcal{N} )  &=  \sqrt{2} e^{- k \bar{k}/4\pi} f \bar{f} (\frac{1}{\bar{f}}  e^{k \bar{k}/8\pi} \bar{\nabla}^m  e^{- k \bar{k}/8\pi} \bar{f} ) (\frac{1}{f}  e^{k \bar{k}/8\pi} \nabla^{m'}  e^{- k \bar{k}/8\pi} f) \\
 &=  \sqrt{2} (\bar{\nabla}^m  e^{- k \bar{k}/8\pi} \bar{f} ) (\nabla^{m'}  e^{- k \bar{k}/8\pi} f) \ . \\
\eea
The simplicity of \Eq{eq:oscilaltorform} justifies our definition of the covariant derivatives $\nabla$ and $\bar \nabla$. To further develop \Eq{eq:oscilaltorform}, we will need an integration by parts identity. Observe that for test functions $u(\mbf{k}), v(\mbf{k})$
\bea
\int \frac{d^2k}{(2\pi)^2} u(\mbf{k}) \nabla v(\mbf{k}) &= \int \frac{d^2k}{(2\pi)^2}  (u(\mbf{k}) \del v(\mbf{k}) -  u(\mbf{k})v(\mbf{k}) \bar{k}/8\pi) \\
&= \int \frac{d^2k}{(2\pi)^2}  (u(\mbf{k}) \frac{\del_1 - i \del_2}{2}  v(\mbf{k}) -  u(\mbf{k})v(\mbf{k}) \bar{k}/8\pi) \\
&= \int \frac{d^2k}{(2\pi)^2}   (-v(\mbf{k}) \frac{\del_1 - i \del_2}{2}  u(\mbf{k}) -  v(\mbf{k})u(\mbf{k}) \bar{k}/8\pi) +  \int \frac{d^2k}{(2\pi)^2} (\del_1 - i \del_2) (u(\mbf{k})v(\mbf{k})) \\
&= - \int \frac{d^2k}{(2\pi)^2} (v(\mbf{k})\del u(\mbf{k}) + v(\mbf{k})u(\mbf{k}) \bar{k}/8\pi) \\
&= - \int \frac{d^2k}{(2\pi)^2} v(\mbf{k}) (\del + \bar{k}/8\pi) u(\mbf{k})  \\
\eea
where we have discarded the total derivative term because the integral is over the BZ which has no boundary. Hence we are led to define $\bar \nabla^\dag \equiv -\del - \bar{k}/8\pi$ which satisfies
\bea
\int \frac{d^2k}{(2\pi)^2} u \nabla v &= \int \frac{d^2k}{(2\pi)^2} v \bar \nabla^\dag u, \qquad \bar \nabla^\dag \equiv -(\del + \bar{k}/8\pi)
\eea
suppressing the $\mbf{k}$ dependence for brevity. An identical calculation shows that
\bea
\int \frac{d^2k}{(2\pi)^2} u \bar{\nabla} v &= \int \frac{d^2k}{(2\pi)^2} v \nabla^\dag u, \qquad \nabla^\dag \equiv -(\bar{\del} + k/8\pi) \ .
\eea
The full algebra of the covariant derivatives is
\bea
\null [\nabla, \bar{\nabla}] &=  [\nabla, \bar \nabla^\dag] =   [\nabla^\dag, \bar \nabla] =  [\nabla^\dag, \bar \nabla^\dag]  = 0 \\
\null [\nabla, \nabla^\dag] &=  [ \del - \bar k/8\pi, -(\bar \del + k /8\pi)] = -1/8\pi - 1/8\pi = -\frac{1}{4\pi} \\
\null [\bar{\nabla}, \bar \nabla^\dag] &=  [ \bar \del -  k/8\pi, -(\del + \bar k /8\pi)] = -1/8\pi - 1/8\pi = - \frac{1}{4\pi} \\
\eea
which form two decoupled oscillator algebra analogous to the $a$ and $b$ operators in real space. The last identity we need is
\bea
\label{eq:nablavacuum}
-(\nabla^\dag e^{- k \bar{k}/8\pi} f(k)) &= \bar{\del} (e^{- k \bar{k}/8\pi} f(k))+ e^{- k \bar{k}/8\pi} f(k) k/8\pi \\
&= f(k) \bar{\del} e^{- k \bar{k}/8\pi} + e^{- k \bar{k}/8\pi} f(k) k/8\pi \\
&= - f(k) e^{- k \bar{k}/8\pi} k /8\pi+ e^{- k \bar{k}/8\pi} f(k) k/8\pi \\
&= 0 \ . \\
\eea
In the analogy to the $a$ and $b$ real space algebra, we should think of $e^{- k \bar{k}/8\pi} f(k)$ as the vacuum of the $\nabla^\dag$ operator, and the states $\nabla^{m} (e^{- k \bar{k}/8\pi} f(k))$ as the (unnormalized) $m$th excited state. In an identical manner, we check that $\bar \nabla^\dag (e^{- k \bar{k}/8\pi} \bar{f}(\bar{k})) = 0$.

Integrating by parts in \Eq{eq:oscilaltorform}, we find
\bea
 \int \frac{d^2k}{(2\pi)^2} \mathcal{N} ( \mathcal{N}^{-1} \del^{m'} \mathcal{N} ) ( \mathcal{N}^{-1} \bar{\del}^m \mathcal{N} ) &=  \int \frac{d^2k}{(2\pi)^2}   \sqrt{2} (\bar{\nabla}^m  e^{- k \bar{k}/8\pi} \bar{f} ) (\nabla^{m'}  e^{- k \bar{k}/8\pi} f) \\
  &= \int \frac{d^2k}{(2\pi)^2}   \sqrt{2} (e^{- k \bar{k}/8\pi} \bar{f} ) \nabla^{\dag m} \nabla^{m'}  (e^{- k \bar{k}/8\pi} f) \\
\eea
which is analogous to the correlator $\braket{0| a^m a^{\dag m'}|0}$. In particular, the expression $   e^{- k \bar{k}/8\pi} \bar{f}(\bar{k}) \nabla^{\dag m} \nabla^{m'}  (e^{- k \bar{k}/8\pi} f(k) )$ can be evaluated with Wick's theorem as a standard textbook computation. Wick's theorem states that $\nabla^{\dag m} \nabla^{m'}$ is equal to the normal-ordered sum of all possible contractions, where in our case the normal ordering is defined by moving all $\nabla^{\dag}$ operators to the right, and a contraction replaces $\nabla^{\dag} \nabla$ by $\nabla^{\dag} \nabla - \nabla \nabla^{\dag} = [\nabla^{\dag},\nabla] = 1/4\pi$.  Then we just need the fact that any normal ordered string of operators $\nabla \dots \nabla^\dag$ obeys
\bea
 \int \frac{d^2k}{(2\pi)^2}   \sqrt{2}  e^{- k \bar{k}/8\pi} \bar{f} \nabla \dots \nabla^\dag (e^{- k \bar{k}/8\pi} f) = 0
\eea
because $\dots \nabla^\dag e^{- k \bar{k}/8\pi} f = 0$ on the righthand side and $\int d^2k \, e^{- k \bar{k}/8\pi} \bar{f}  \nabla \dots = 0$ on the lefthand side after an integration by parts because $\bar \nabla^\dag (e^{- k \bar{k}/8\pi} \bar{f} ) = 0$ (see \Eq{eq:nablavacuum}). Thus we find that the only nonzero term from Wick's theorem arises when $m = m'$ and all $\nabla$ and $\nabla^\dag$ operators can be fully contracted. Counting the $m!$ possible ways of totally contracting the $m$ $\nabla$ and $m$ $\nabla^\dag$ operators, we find
\bea
\label{eq:wickcompute}
 \int \frac{d^2k}{(2\pi)^2}   \sqrt{2}  e^{- k \bar{k}/8\pi} \bar{f}  \nabla^{\dag m} \nabla^{m'}  (e^{- k \bar{k}/8\pi} f) &= \delta_{mm'} m! \lp \frac{1}{4\pi} \rp^m \int \frac{d^2k}{(2\pi)^2}   \sqrt{2}  e^{- k \bar{k}/8\pi} \bar{f} (e^{- k \bar{k}/8\pi} f)  \\
 &= \delta_{mm'} m! \lp \frac{1}{4\pi} \rp^m \int \frac{d^2k}{(2\pi)^2} \sqrt{2}  \bar{f}(\bar k) f(k) e^{- k \bar{k}/4\pi} \\
  &= \delta_{mm'} m! \lp \frac{1}{4\pi} \rp^m \int \frac{d^2k}{(2\pi)^2}\mathcal{N}(\mbf{k}) \\
&= \delta_{mm'} m! \lp \frac{1}{4\pi} \rp^m \\
\eea
where in the last line we used \Eq{eq:Nintnormal}. We arrive at our final result by plugging \Eq{eq:wickcompute} into \Eq{eq:completenessint} to find
\bea
\mathcal{I}_{m'm,n'n}  &=\delta_{nn'} \frac{(2\phi)^{(m+m')/2}}{\sqrt{m!m'!}} (-1)^{m+m'}\delta_{mm'} m! \lp \frac{1}{4\pi} \rp^m =\delta_{nn'}  \delta_{mm'}
\eea
using $\phi = 2\pi$. Hence we have shown that
\bea
\label{eq:compeltenes2}
1 &= \sum_{\ell=0}^\infty \int \frac{d^2k}{(2\pi)^2} \ket{\mbf{k},\ell} \bra{\mbf{k},\ell}
\eea
is true when acting on the complete $\ket{n,m}$ basis. In position space, the states $\ket{n,m}$ are localized on the scale of the magnetic length \cite{yoshioka2013quantum}, and hence we expect them to be suitably well behaved such that \Eq{eq:compeltenes2} does not encounter pathological cases. This is confirmed by our direct calculation.

\subsection{Scattering Amplitudes}
\label{app:siegelcomplete}

To compute the effective Hamiltonian, we need expressions for the matrix elements. Our calculations are similar to those of Landau level overlaps in the symmetric gauge, where  $\psi_{l,n}(z) \propto (b^\dagger)^{l-n} (a^\dagger)^n e^{-\pi \frac{|z|^2}{2}}$, $l$ being the angular momentum quantum number.  Our states are different in that they are built from the momentum operators $T_\RR$ instead of $b^\dagger$ directly, and are manifestly gauge-invariant.

The choice of Landau level states (which are localized on the scale of the magnetic length $1/\sqrt{eB}$) in the definition of magnetic translation irreps (\Eq{eq:statedefAppa}) makes the kinetic term of the Hamiltonian very simple (see \App{app:square_lattice} for an example), so we focus on the potential term $U(\mbf{r})$. The potential term will create scattering between different Landau levels. Recall that $U(\mbf{r})$ is periodic so can be expanded as a Fourier series. Hence we need to compute the general scattering amplitude
\bea
\braket{\mbf{k},m| e^{- 2\pi i \mbf{G} \cdot \mbf{r} } | \mbf{k},n} \ .
\eea
with $\mbf{G} = G_1 \mbf{b}_1 + G_2 \mbf{b}_2, ~G_1,G_2 \in \mathds{Z}$. It is possible to do this exactly because $\mbf{G} \cdot \mbf{r}$ can be expressed simply in terms of  $\pmb{\pi}$ and $\mbf{Q}$ using $-(eB)^{-1}\eps_{\mu \nu}( \pi_\nu - Q_\nu) = -\eps_{\mu \nu}\eps_{\nu \rho} x_\rho = x_\mu$. We now define
\bea
\mbf{a}_i \cdot \pmb{\pi}  = \sqrt{\frac{\phi}{2}}\big( z_i a^\dag  + \bar{z}_i a\big), \quad z_i = \Omega^{-1/2} (\hat{x} + i \hat{y}) \cdot \mbf{a}_i, \bar{z}_i = \Omega^{-1/2} (\hat{x} - i \hat{y}) \cdot \mbf{a}_i
\eea
in terms of which we can write the Fourier harmonics
\bea
\label{eq:rintermsab}
\mbf{b}_1 \cdot \mbf{r} = \mbf{a}_2 \cdot \frac{\mbf{Q}- \pmb{\pi}}{eB \Omega} = \frac{1}{\sqrt{2\phi}} \lp i (b - b^\dag) - (z_2 a^\dag + \bar{z}_2 a) \rp, \qquad \mbf{b}_2 \cdot \mbf{r} = - \mbf{a}_1 \cdot \frac{\mbf{Q}- \pmb{\pi}}{eB \Omega}  = \frac{1}{\sqrt{2\phi}} \lp - (b + b^\dag) + z_1 a^\dag + \bar{z}_1 a \rp \ . \\
\eea
We made use of the relations $\mbf{b}_1 \times \mbf{v} = \frac{\mbf{a}_2}{\Omega} \cdot \mbf{v}$ and $\mbf{b}_2 \times \mbf{v} = - \frac{\mbf{a}_1}{\Omega} \cdot \mbf{v}$ for any $\mbf{v}$. This is easily verified by testing $\mbf{v} = \mbf{b}_1, \mbf{b}_2$ and using linearity of dot and cross products. The required matrix elements are in the form
\bea
\label{eq:matel}
\bra{m}e^{-i 2\pi \mbf{G} \cdot \mbf{r}} e^{i \mbf{R} \cdot \mbf{Q}}  \ket{n}
\eea
which can be evaluated exactly using the oscillator algebra. We now derive the normal-ordered form of \Eq{eq:matel}:
\bea
\label{eq:rgaugeinv}
e^{-i 2\pi \mbf{G} \cdot \mbf{r}} &= \exp \lp  -\frac{2\pi i G_1}{\sqrt{2\phi}} \lp i (b - b^\dag) - (z_2 a^\dag + \bar{z}_2 a) \rp -\frac{2\pi i G_2 }{\sqrt{2\phi}} \lp - (b + b^\dag) + z_1 a^\dag + \bar{z}_1 a \rp \rp  \\
&= \exp \lp -\frac{2\pi i}{\sqrt{2\phi}}i (Gb - \bar{G} b^\dag)  \rp  \exp \lp -\frac{2\pi i}{\sqrt{2\phi}} ( ( G_2 z_1 - G_1z_2) a^\dag +(G_2 \bar{z}_1 - G_1 \bar{z}_2) a )  \rp  , \qquad G = G_1 + i G_2\\
&= \exp \lp -\frac{2\pi i}{\sqrt{2\phi}}i (Gb - \bar{G} b^\dag)  \rp  \exp \lp \frac{i}{\sqrt{2\phi}} ( \bar{\gamma} a^\dag +\gamma a )  \rp
\eea
where in the last line we defined $\gamma = 2\pi \eps_{ij} G_i \bar{z}_j $ and we note $\gamma$ is dimensionless.  The exponentials separate in the last line as $[a,b] = [a, b^\dagger] = 0.$ Returning to \Eq{eq:rgaugeinv}, we use the BCH formula to find the normal-ordered form
\bea
\label{eq:normalorder}
e^{-i 2\pi \mbf{G} \cdot \mbf{r}} &= e^{- \frac{(2\pi)^2}{4\phi} \bar{G} G}e^{\frac{2\pi i}{\sqrt{2 \phi}} i \bar{G} b^\dag} e^{- \frac{2\pi i}{\sqrt{2 \phi}} i G b} e^{-\frac{\bar{\gamma}\gamma}{4\phi} }e^{\frac{i}{\sqrt{2 \phi}} \bar{\gamma} a^\dag} e^{\frac{i}{\sqrt{2 \phi}} \gamma a} \ . \\
\eea
It is now a simple matter to use BCH identities to reorder the $b$ oscillators to the vacuum state in \Eq{eq:matel}. To avoid clutter, we keep $\exp \lp \frac{i}{\sqrt{2\phi}} ( \bar{\gamma} a^\dag +\gamma a )  \rp$ in the following expression and will normal order it at a later stage. Using \Eq{eqTranslationExpectation}, as well as $[a^\dagger , b^\dagger] = [a, b^\dagger] = 0, b\ket{n} = \bra{m} b^\dagger = 0$, we compute
\bea
\label{eq:Hmatfactor}
\bra{m}e^{-i 2\pi \mbf{G} \cdot \mbf{r}} e^{i \mbf{R} \cdot \mbf{Q}}  \ket{n} &= e^{- \frac{\phi}{4} R\bar{R} }  \bra{m}  e^{-\frac{2\pi i}{\sqrt{2 \phi}} i \lp G b - \bar{G} b^\dag  \rp }  e^{i \sqrt{\phi/2} \bar{R} b^\dag} e^{\frac{i}{\sqrt{2\phi}} ( \bar{\gamma} a^\dag +\gamma a )} \ket{n} \\
&= e^{- \frac{\phi}{4} R\bar{R}}  e^{- \frac{(2\pi)^2}{4\phi} \bar{G} G} \bra{m}e^{-\frac{2\pi i}{\sqrt{2 \phi}} i G b}  e^{i \sqrt{\phi/2} \bar{R} b^\dag} e^{\frac{i}{\sqrt{2\phi}} ( \bar{\gamma} a^\dag +\gamma a )} \ket{n} \\
&= e^{- \frac{\phi}{4} R\bar{R}}  e^{- \frac{(2\pi)^2}{4\phi} \bar{G} G} e^{i\pi G  \bar{R} }  \braket{m| e^{\frac{i}{\sqrt{2\phi}} ( \bar{\gamma} a^\dag +\gamma a )} |n} \  . \\
\eea
We have factored out all of the $b$ operators into exponential factors depending on $\mbf{R}$ and $\mbf{G}$. This leaves only the $a$ operators in the correlator. We use \Eq{eq:normalorder} to normal order the $e^{\frac{i}{\sqrt{2\phi}} ( \bar{\gamma} a^\dag +\gamma a )}$ term and find
\bea
\bra{m}e^{-i 2\pi \mbf{G} \cdot \mbf{r}} e^{i \mbf{R} \cdot \mbf{Q}}  \ket{n} &= e^{- \frac{\phi}{4} R\bar{R}}  e^{- \frac{(2\pi)^2}{4\phi} \bar{G} G} e^{i\pi G  \bar{R} } e^{- \frac{\bar{\gamma} \gamma}{4\phi}}  \bra{m} e^{\frac{i}{\sqrt{2 \phi}}  \bar{\gamma} a^\dag} e^{\frac{i}{\sqrt{2 \phi}} \gamma a} \ket{n} \ . \\
\eea
We now use the Fock space identities $a \ket{n} = \sqrt{n} \ket{n-1}$ to prove the formula
\bea
 e^{x a} \ket{n} &= \sum_{k=0}^\infty \frac{1}{k!}x^k a^k\ket{n}  = \sum_{k=0}^n \frac{1}{k!} x^k \frac{\sqrt{n!}}{\sqrt{(n-k)!}} \ket{n-k} = \frac{1}{\sqrt{n!}} \sum_{k=0}^n x^k \frac{n!}{k! (n-k)!} a^{\dag (n-k)} \ket{0} \ \\
 &= \frac{1}{\sqrt{n!}} \sum_{k=0}^n x^{n-k} \frac{n!}{k! (n-k)!} a^{\dag k} \ket{0} \ . \\
\eea
Using the binomial formula, we find the expression
\bea
\bra{m}e^{\frac{i}{\sqrt{2 \phi}} \bar{\gamma} a^\dag} e^{\frac{i}{\sqrt{2 \phi}}  \gamma a} \ket{n} &=  \frac{1}{\sqrt{n! m!}} \sum_{l=0}^m \sum_{k=0}^n \lp \frac{i \bar{\gamma}}{\sqrt{2 \phi}}  \rp^{m-l} \lp \frac{i \gamma}{\sqrt{2 \phi}}  \rp^{n-k} \binom{n}{k}\binom{m}{l} \bra{0}a^{l} a^{\dag k} \ket{0} \\
&= \frac{1}{\sqrt{n! m!}} \sum_{l=0}^m \sum_{k=0}^n \lp \frac{i \bar{\gamma}}{\sqrt{2 \phi}} \rp^{m-l} \lp \frac{i \gamma}{\sqrt{2 \phi}} \rp^{n-k} \binom{n}{k}\binom{m}{l} \delta_{l,k} k! \braket{0|0}\\
&= \frac{1}{\sqrt{n! m!}} \sum_{k=0}^{\min(n,m)} \lp \frac{i \bar{\gamma}}{\sqrt{2 \phi}}  \rp^{m-k} \lp \frac{i \gamma}{\sqrt{2 \phi}}  \rp^{n-k} \frac{n! m!}{k!(n-k)!(m-k)!}  \braket{0|0} \ . \\
\eea
This function may be exactly evaluated using the Laguerre polynomial definition \cite{Abramowitz} \url{https://secure.math.ubc.ca/~cbm/aands/page_775.htm}:
\begin{align}
L^{(\alpha)}_n(x) = \sum_{i=0}^n (-1)^i { \dfrac{(n+\alpha)!}{(n-i)!(\alpha+i)!}} \frac{x^i}{i!},
\label{}
\end{align}
which allows us to write
\bea
\sum_{k=0}^{\min(n,m)}  x^{m-k} y^{n-k} \frac{n! m!}{k!(n-k)!(m-k)!}
&= \begin{cases}
m! y^{n-m} L_m^{|n-m|}(-xy), & n \geq m \\
n! x^{m-n} L_n^{|m-n|}(-xy), & m > n \\
\end{cases} \\
\eea
where $L_n^\al(x)$ are the associated Laguerre polynomials. We finally arrive at
\bea
\label{eq:aoscillators}
\bra{m} e^{\frac{i}{\sqrt{2 \phi}} \lp \gamma a + \bar{\gamma} a^\dag \rp } \ket{n} &=  e^{- \frac{ \bar{\gamma} \gamma}{4\phi}}  \begin{cases}
\sqrt{\frac{m!}{n!}} \lp \frac{i \gamma}{\sqrt{2 \phi}} \rp^{n-m} L_m^{|n-m|}\lp \frac{\bar{\gamma} \gamma}{2\phi} \rp, & n \geq m \\
\sqrt{\frac{n!}{m!}} \lp \frac{i\bar{\gamma}}{\sqrt{2 \phi}} \rp^{m-n} L_n^{|m-n|}\lp \frac{ \bar{\gamma}\gamma}{2\phi} \rp, & m > n \\
\end{cases} \\
\eea
We now have computed closed form expressions for all of the oscillator states.

Finally, we will give a formula for the matrix elements of the full momentum eigenstates. For brevity, we denote $T_{\mbf{R}} =T_{\mbf{a}_1}^{\mbf{R} \cdot \mbf{b}_1} T_{\mbf{a}_2}^{\mbf{R} \cdot \mbf{b}_2}$. The expression
\bea
\label{eq:scatteringsum}
\braket{\mbf{k},m| e^{-2\pi i \mbf{G} \cdot \mbf{r} } | \mbf{k},n} &= \frac{1}{\mathcal{N}(\mbf{k})} \sum_{\mbf{R},\mbf{R}'} e^{-i \mbf{k} \cdot (\mbf{R}-\mbf{R}')} \bra{m}T^\dag_{\mbf{R}'} e^{-2\pi i \mbf{G} \cdot \mbf{r}} T_{\mbf{R}}\ket{n} \\
\eea
can be simplified using the fact that $T^\dag_{\mbf{R}'}$ commutes with $e^{-2\pi i \mbf{G} \cdot \mbf{r}}$ because  $e^{-2\pi i \mbf{G} \cdot \mbf{r}}$ is periodic in $\mbf{a}_i$. We then use BCH (as in \Eq{eq:rdotq}) to simplify the $T_{\mbf{R}}$ operators in the first line and \Eq{eq:Hmatfactor} and the BCH formula in the second to find:
\bea
\label{eq:TGexp}
\bra{m}e^{-2\pi i \mbf{G} \cdot \mbf{r}} T^\dag_{\mbf{R}'} T_{\mbf{R}}\ket{n} &= e^{i \frac{\phi}{2} R_1 R_2 -i \frac{\phi}{2} R'_1 R'_2 } \bra{m}e^{-2\pi i \mbf{G} \cdot \mbf{r}} e^{-i \mbf{R}' \cdot \mbf{Q}} e^{i \mbf{R} \cdot \mbf{Q}} \ket{n}  \\
&= e^{i \frac{\phi}{2} R_1 R_2 - i \frac{\phi}{2} R'_1 R'_2 + i \frac{\phi}{2} (R_1R_2'-R_2R_1')}  e^{- \frac{\phi}{4} (R-R')(\bar{R}-\bar{R}')}  e^{- \frac{(2\pi)^2}{4\phi} \bar{G} G} e^{i\pi G  (\bar{R}-\bar{R}') } \bra{m} e^{\frac{i}{\sqrt{2 \phi}} \lp \gamma a + \bar{\gamma} a^\dag \rp } \ket{n} \ .
\eea
Recall that $R = R_1 + iR_2$. Collecting terms, we find
\bea
&\braket{\mbf{k},m| e^{-2\pi i \mbf{G} \cdot \mbf{r} } | \mbf{k},n} \\
&= \frac{ \bra{m} e^{\frac{i}{\sqrt{2 \phi}} \lp \gamma a + \bar{\gamma} a^\dag \rp } \ket{n} }{\mathcal{N}(\mbf{k})} e^{- \frac{(2\pi)^2}{4\phi} \bar{G} G}  \sum_{\mbf{R},\mbf{R}'} e^{-i \mbf{k} \cdot (\mbf{R}-\mbf{R}')}  e^{i \frac{\phi}{2} R_1 R_2 +i \frac{\phi}{2} R'_1 R'_2 -i \frac{\phi}{2} (R_1R_2'-R_2R_1')}  e^{- \frac{\phi}{4} (R-R')(\bar{R}-\bar{R}') +i\pi G  (\bar{R}-\bar{R}') }
\eea
We need only perform the sum of the $\mbf{R},\mbf{R}'$ coordinates. We perform the same change of variables as in the normalization calculation (\Eq{eq:cofv}) to  $\mbf{s} = (\mbf{R}+ \mbf{R}')/2$ and $\mbf{d} = \mbf{R}-\mbf{R}'$. The calculation is identical to \Eq{eq:cofv} but with the extra $i \pi G \bar{d}$ term in the exponent which is linear and so does not enter the quadratic form. Explicitly, we have shown in Eq.~\ref{eq:siegel_sum_simplified} that
\bea
&\sum_{\mbf{R},\mbf{R}'} e^{-i \mbf{k} \cdot \mbf{R} +i \mbf{k}' \cdot \mbf{R}' +   i \frac{\phi}{2} (\mbf{R} \cdot \mbf{b}_1)(\mbf{R} \cdot \mbf{b}_2)- i \frac{\phi}{2} (\mbf{R}' \cdot \mbf{b}_1)(\mbf{R}' \cdot \mbf{b}_2)  - \frac{\phi}{4} \bar{R}R- \frac{\phi}{4} \bar{R}'R'+ \frac{\phi}{2} R' \bar{R} } \\
&= (2\pi)^2 \delta(\mbf{k}-\mbf{k}') \vartheta \lp \left. \frac{(k_1,k_2)}{2\pi} \right|  \Phi \rp \ . \\
\eea
To evaluate the sum
\begin{align}
\sum_{\mbf{R},\mbf{R}'} e^{-i \mbf{k} \cdot (\mbf{R}-\mbf{R}')}  e^{i \frac{\phi}{2} R_1 R_2 -i \frac{\phi}{2} R'_1 R'_2 -i \frac{\phi}{2} (R_1R_2'-R_2R_1')}  e^{- \frac{\phi}{4} (R-R')(\bar{R}-\bar{R}') +i\pi G  (\bar{R}-\bar{R}')},
\label{}
\end{align}
we note that we can simply absorb the extra term with $G$ into the momentum:
\bea
&\sum_{\mbf{R},\mbf{R}'} e^{-i \mbf{k} \cdot (\mbf{R}-\mbf{R}')}  e^{i \frac{\phi}{2} R_1 R_2 -i \frac{\phi}{2} R'_1 R'_2 -i \frac{\phi}{2} (R_1R_2'-R_2R_1')}  e^{- \frac{\phi}{4} (R-R')(\bar{R}-\bar{R}') +i\pi G  (\bar{R}-\bar{R}')} \\
&= \sum_{\mbf{R},\mbf{R}'} e^{-i \mbf{k} \cdot (\mbf{R}-\mbf{R}')}  e^{i \frac{\phi}{2} R_1 R_2 -i \frac{\phi}{2} R'_1 R'_2 - \frac{\phi}{4} R\bar{R} - \frac{\phi}{4} R'\bar{R}' + \frac{\phi}{2}R'\bar{R}} e^{i\pi (G, -iG) \cdot (\RR-\RR')} \\
&=\sum_{\mbf{R},\mbf{R}'} e^{-i (\mbf{k}-(\pi G, -i\pi G)) \cdot (\mbf{R}-\mbf{R}')}  e^{i \frac{\phi}{2} R_1 R_2 -i \frac{\phi}{2} R'_1 R'_2 - \frac{\phi}{4} R\bar{R} - \frac{\phi}{4} R'\bar{R}' + \frac{\phi}{2}R'\bar{R}} \\
&=  \vartheta \lp \left. \frac{(k_1 - \pi G,k_2 + i \pi G)}{2\pi} \right|  \Phi \rp (2\pi)^2 \delta(0) \ .
\label{eq:siegelwithG}
\eea
Returning the Landau level factor in \Eq{eq:aoscillators} and using \Eq{eq:siegelwithG} gives the closed-form expression
\bea
\label{eq:Hmat1}
 \braket{\mbf{k},m| e^{-2\pi i \mbf{G} \cdot \mbf{r} } | \mbf{k},n} &=(2\pi)^2 \delta(0) \frac{ \vartheta \lp \frac{(k_1 - \pi G,k_2 + i \pi G)}{2\pi} \left| \Phi \right. \rp}{ \vartheta \lp \frac{(k_1,k_2)}{2\pi} \left| \Phi \right. \rp}  e^{- \frac{(2\pi)^2}{4\phi} \bar{G} G}  e^{ - \frac{\bar{\gamma} \gamma}{4\phi} }  \begin{cases}
\sqrt{\frac{m!}{n!}} \lp \frac{ i \gamma}{\sqrt{2 \phi}} \rp^{n-m} L_m^{|n-m|}\lp \frac{\bar{\gamma} \gamma}{2\phi}\rp, & n \geq m \\
\sqrt{\frac{n!}{m!}} \lp \frac{ i \bar{\gamma}}{\sqrt{2 \phi}} \rp^{m-n} L_n^{|m-n|}\lp \frac{\bar{\gamma} \gamma}{2\phi}  \rp, & m > n \\
\end{cases}  \ .
\eea
Recall that  $\gamma = 2\pi \eps_{ij} G_i \bar{z}_j$, $G = (\mbf{a}_1 + i \mbf{a}_2) \cdot \mbf{G}, \bar{z}_i = (\hat{x} - i \hat{y}) \cdot \mbf{a}_i /\sqrt{\Omega}$. The most important feature is the factorization of the momentum $\mbf{G}$ dependence and the Landau level $n,m$ dependence. This expression can actually be simplified considerably using the quasi-periodicity of the Siegel theta functions.  Let $\mbf{n} = (G_2, -G_1) \in \mathds{Z}^2$ such that
\bea
(\frac{k_1}{2\pi} - G/2, \frac{k_2}{2\pi} + i G/2) + \Phi \mbf{n} &= (\frac{k_1}{2\pi} , \frac{k_2}{2\pi}- G_2) \ \quad \Phi \mbf{n} = \frac{1}{2}(G,-i\bar{G}) .
\eea
Using the quasi-periodicity property \Eq{eq:thetaquasiper}, we can transform the $G$ dependence in the amplitude:
\bea
\vartheta \lp \left. \frac{(k_1,k_2)}{2\pi} \right|  \Phi \rp &= \vartheta \lp \left. \frac{(k_1 ,k_2 - 2\pi G_2)}{2\pi} \right|  \Phi \rp = \vartheta \lp \left. \frac{(k_1 - \pi G,k_2 + i \pi G)}{2\pi} + \Phi \mbf{n} \right|  \Phi \rp  \\
&= e^{- i \pi \mbf{n}^T \Phi \mbf{n} - i (k_1 - \pi G,k_2 + i \pi G) \cdot \mbf{n}} \vartheta \lp \left. \frac{(k_1 - \pi G,k_2 + i \pi G)}{2\pi}  \right|  \Phi \rp \ .  \\
\eea
Remarkably, we find that the ratio of theta functions appearing in the amplitude can be written simply at $\phi = 2\pi$ as
\bea
\label{eq:thetaratio}
\frac{ \vartheta \lp \frac{(k_1 - \pi G,k_2 + i \pi G)}{2\pi} \left| \Phi  \right. \rp}{ \vartheta \lp \frac{(k_1,k_2)}{2\pi} \left| \Phi \right. \rp} &= \exp \lp i \pi \mbf{n}^T \Phi \mbf{n} + i (k_1 - \pi G,k_2 + i \pi G) \cdot \mbf{n} \rp \\
 &=  \exp \lp \frac{\pi}{2} G \bar{G}+ i \pi G_1 G_2 + i (k_1 G_2 - k_2 G_1) \rp \ . \\
\eea
The most important feature of this calculation is the resulting exponential function in the second line of \Eq{eq:thetaratio}. Hence we have proven that the zero at $(\pi,\pi)$ in the denominator of \Eq{eq:thetaratio} has canceled (it is a removable singularity), so the amplitude is well-defined everywhere in the BZ despite the wavefunction being defined on patches. Thus the final expression for the scattering amplitude is
\bea
\label{eq:Hmat}
\boxed{
 \braket{\mbf{k},m| e^{-2\pi i \mbf{G} \cdot \mbf{r} } | \mbf{k},n} = (2\pi)^2 \delta(0) \exp \Big( - i \pi G_1 G_2 - i (G_1 k_2 - G_2 k_1)\Big) \mathcal{H}^{2\pi \mbf{G}}_{mn}
 }
\eea
and we defined the general Landau level scattering matrix in terms of $\gamma_q = \eps_{ij} q_i \bar{z}_j$ and $q_i = \mbf{a}_i \cdot \mbf{q}$
\bea
\label{eq:Hcaldef}
\boxed{
\mathcal{H}^{\mbf{q}}_{mn} =  \bra{m} \exp \lp i \frac{\gamma_q a + \bar{\gamma}_q a^\dag }{\sqrt{2 \phi}} \rp \ket{n}  =  e^{ - \frac{ \bar{\gamma}_q \gamma_q}{4\phi}}  \begin{cases}
\sqrt{\frac{m!}{n!}} \lp \frac{i \gamma_q}{\sqrt{2 \phi}} \rp^{n-m} L_m^{|n-m|}\lp \frac{\bar{\gamma}_q\gamma_q}{2\phi}  \rp, & n \geq m \\
\sqrt{\frac{n!}{m!}} \lp \frac{i \bar{\gamma}_q}{\sqrt{2 \phi}} \rp^{m-n} L_n^{|m-n|}\lp \frac{\bar{\gamma}_q\gamma_q}{2\phi}  \rp, & m > n \\
\end{cases}
} \ .
\eea
 It is useful to think of $\mathcal{H}^{\mbf{q}}$ as a matrix on the Landau levels indices. It is a unitary matrix: $\mathcal{H}^{\mbf{q} \, \dag} \mathcal{H}^{\mbf{q}} = \mathbb{1}$  and $\mathcal{H}^{\mbf{q} \, \dag} = \mathcal{H}^{-\mbf{q}}$. The latter is easily proved from \Eq{eq:Hcaldef} by realizing $\frac{\gamma_q a + \bar{\gamma}_q a^\dag }{\sqrt{2 \phi}}$ is a Hermitian operator which is odd in $\mbf{q}$. To prove unitarity, we write
\bea
\null [\mathcal{H}^{\mbf{q} \, \dag} \mathcal{H}^{\mbf{q}}]_{mn} &= \sum_{r=0}^\infty \bra{m} \exp \lp -i \frac{\gamma_q a + \bar{\gamma}_q a^\dag }{\sqrt{2 \phi}} \rp \ket{r} \bra{r} \exp \lp i \frac{\gamma_q a + \bar{\gamma}_q a^\dag }{\sqrt{2 \phi}} \rp \ket{n}\\
&= \bra{m} \exp \lp -i \frac{\gamma_q a + \bar{\gamma}_q a^\dag }{\sqrt{2 \phi}} \rp \left[ \sum_{r=0}^\infty \ket{r} \bra{r} \right] \exp \lp i \frac{\gamma_q a + \bar{\gamma}_q a^\dag }{\sqrt{2 \phi}} \rp \ket{n} \\
&= \bra{m} \exp \lp -i \frac{\gamma_q a + \bar{\gamma}_q a^\dag }{\sqrt{2 \phi}} \rp \left[ \sum_{r=0}^\infty \sum_{s=0}^\infty  \frac{{b^\dag}^s}{\sqrt{s!}} \ket{r}\bra{r}  \frac{b^s}{\sqrt{s!}} \right] \exp \lp i \frac{\gamma_q a + \bar{\gamma}_q a^\dag }{\sqrt{2 \phi}} \rp \ket{n} \\
&= \bra{m} \exp \lp -i \frac{\gamma_q a + \bar{\gamma}_q a^\dag }{\sqrt{2 \phi}} \rp \exp \lp i \frac{\gamma_q a + \bar{\gamma}_q a^\dag }{\sqrt{2 \phi}} \rp \ket{n} \\
&= \delta_{mn}
\label{eq:Hcalunitproof}
\eea
where we inserted the $s$ sum of the $b$ and $b^\dag$ operators, which commute with the $a$ operators, because $b^s \ket{n} = 0$ for all $s\neq 0$, and $b^s = 1$ for $s=0$. Then we used that the $a^\dag,b^\dag$ operators form a complete set:
\begin{align}
\ket{r, s} = \frac{(a^\dagger)^r (b^\dagger)^s}{\sqrt{r!s!}}\ket{0}, ~\sum_{r,s} \ket{r, s} \bra{r, s} = 1.
\label{}
\end{align}
The matrix $\mathcal{H}^\mbf{q}$ will reappear throughout the paper. We will sometimes use the alternative representation (see \Eq{eq:Z} of the Main Text) where $\gamma_q = \eps_{ij} q_i \bar{z}_j$ is expanded to yield
\bea
\label{eq:HcaldefZ}
\boxed{
\mathcal{H}^{\mbf{q}}_{mn} =  \bra{m} \exp \lp i \eps_{ij} q_i Z_j \rp \ket{n} = [e^{i \eps_{ij} q_i \tilde{Z}_j}]_{mn},
\qquad  Z_j= \frac{\bar{z}_j a + z_j a^\dag }{\sqrt{2 \phi}}, \quad [\tilde{Z}_j]_{mn} = \braket{m|Z_j|n}
}
\eea
where $Z_j$ is an operator and $\tilde{Z}_j$ is a matrix on the Landau level indices.  To calculate the Wilson loop expressions in \App{app:berry}, we will need the commutation relations of $\tilde{Z}$ matrices, which are (recall $z_i = \frac{1}{\sqrt{\Omega}} (\hat{x} + i\hat{y}) \cdot \aaa_i$)
\bea
\label{eq:Zprop}
\null \big[ [\tilde{Z}_i,\tilde{Z}_j] \big]_{mn} &= \frac{1}{2\phi}\braket{m| [z_i a^\dag +\bar{z}_i a,z_j a^\dag +\bar{z}_j a] |n} \\
&= \frac{1}{2\phi}  (\bar{z}_i z_j - \bar{z}_j z_i) \delta_{mn} \\
&= \frac{1}{2\phi}  \delta_{mn} \Omega^{-1} (\hat{x} - i \hat{y})^T (\mbf{a}_i \mbf{a}_j^T - \mbf{a}_j \mbf{a}_i^T) (\hat{x} + i \hat{y}) \\
&= \frac{1}{2\phi}  \delta_{mn} \eps_{ij} (\hat{x} - i \hat{y}) \times (\hat{x} + i \hat{y}) \\
&= \frac{i}{\phi}  \delta_{mn} \eps_{ij}
\eea
which we can write in matrix notation as $[\tilde{Z}_i,\tilde{Z}_j] = \mathbb{1} \frac{i}{\phi} \eps_{ij} $ where $\mathbb{1}$ acts on the Landau levels and we use $\mbf{v}^T \mbf{u} = \mbf{v} \cdot \mbf{u}$ as well as the cross product identity
\bea
\eps_{ij} [\mbf{a}_i \mbf{a}_j^T - \mbf{a}_j \mbf{a}_i^T]_{\mu\nu} = \Omega \eps_{\mu \nu}
\eea
which follows from the Levi-Civita identities for matrix determinants.

With minor modifications, we can compute a more general correlator: $ \braket{\mbf{k}',m| e^{-i \mbf{q} \cdot \mbf{r} } | \mbf{k},n} $ where $\mbf{q}$ is an arbitrary momentum in $\mathbb{R}^2$ which may connect different momenta $\mbf{k},\mbf{k}'\in BZ$.  This calculation will be essential for the many-body form factors in \App{app:gcgii}.

We follow the same steps as in  \Eq{eq:scatteringsum}. We first expand out the basis states:
\bea
\braket{\mbf{k}',m| e^{-i \mbf{q} \cdot \mbf{r} } | \mbf{k},n} &= \frac{1}{\sqrt{\mathcal{N}(\mbf{k})\mathcal{N}(\mbf{k}')}} \sum_{\mbf{R},\mbf{R}'} e^{-i \mbf{k} \cdot \mbf{R} + i \mbf{k}' \cdot \mbf{R}'} \bra{m}T^\dag_{\mbf{R}'} e^{-i \mbf{q} \cdot \mbf{r}} T_{\mbf{R}}\ket{n} \\
&= \frac{1}{\sqrt{\mathcal{N}(\mbf{k})\mathcal{N}(\mbf{k}')}} \sum_{\mbf{R},\mbf{R}'} e^{-i \mbf{k} \cdot \mbf{R} + i \mbf{k}' \cdot \mbf{R}'} \bra{m} e^{[-i \RR' \cdot \QQ, -i\qq \cdot \rr]} e^{-i \mbf{q} \cdot \mbf{r}} T^\dag_{\mbf{R}'} T_{\mbf{R}}\ket{n} \\
&= \frac{1}{\sqrt{\mathcal{N}(\mbf{k})\mathcal{N}(\mbf{k}')}} \sum_{\mbf{R},\mbf{R}'} e^{-i \mbf{k} \cdot \mbf{R} + i (\mbf{k}' + \mbf{q}) \cdot \mbf{R}'} \bra{m} e^{-i \mbf{q} \cdot \mbf{r}} T^\dag_{\mbf{R}'} T_{\mbf{R}}\ket{n}\ . \\
\eea
The correlator is expanded following the same steps in \Eq{eq:TGexp}, replacing $\mathbf{G} \rightarrow \frac{\qq}{2\pi}$:
\bea
\bra{m}e^{-i \mbf{q} \cdot \mbf{r}} T^\dag_{\mbf{R}'} T_{\mbf{R}}\ket{n} &= e^{i \frac{\phi}{2} R_1 R_2 - i \frac{\phi}{2} R'_1 R'_2 } \bra{m}e^{-i \mbf{q} \cdot \mbf{r}} e^{-i \mbf{R}' \cdot \mbf{Q}} e^{i \mbf{R} \cdot \mbf{Q}} \ket{n}  \\
&= e^{i \frac{\phi}{2} R_1 R_2 - i \frac{\phi}{2} R'_1 R'_2 -i \frac{\phi}{2} (R_1R_2'-R_2R_1')}  e^{- \frac{\phi}{4} (R-R')(\bar{R}-\bar{R}')}  e^{i \frac{q}{2}  (\bar{R}-\bar{R}') }  e^{- \frac{\bar{q}q}{4\phi} } \bra{m} e^{i \frac{ \gamma_q a + \bar{\gamma}_q a^\dag}{\sqrt{2 \phi}}} \ket{n} \ . \\
\eea
The final two factors do not depend on $\mbf{R}$ and factor out of the sum. The sum itself reads
\bea
\label{eq:sumq}
&\sum_{\mbf{R},\mbf{R}'} e^{-i \mbf{k} \cdot \mbf{R} + i (\mbf{k}' + \mbf{q}) \cdot \mbf{R}'} \bra{m} e^{-i \mbf{q} \cdot \mbf{r}} T^\dag_{\mbf{R}'} T_{\mbf{R}}\ket{n} \nonumber \\
&= \sum_{\mbf{R},\mbf{R}'} e^{-i \mbf{k} \cdot \mbf{R} + i (\mbf{k}' + \mbf{q}) \cdot \mbf{R}'}  e^{i \frac{\phi}{2} R_1 R_2 -i \frac{\phi}{2} R'_1 R'_2 -i \frac{\phi}{2} (R_1R_2'-R_2R_1')}  e^{- \frac{\phi}{4} (R-R')(\bar{R}-\bar{R}')}  e^{i \frac{q}{2}  (\bar{R}-\bar{R}') } e^{- \frac{\bar{q}q}{4\phi} } \bra{m} e^{i \frac{ \gamma_q a + \bar{\gamma}_q a^\dag}{\sqrt{2 \phi}}} \ket{n}
\eea
which is in theta function form. As in \Eq{eq:TGexp}, the additional $q$ dependence can be absorbed into the momentum dependence:
\begin{align}
&\sum_{\mbf{R},\mbf{R}'} e^{-i \mbf{k} \cdot \mbf{R} +i(\mbf{k}' +\qq)\cdot \mbf{R}'}  e^{i \frac{\phi}{2} R_1 R_2 -i \frac{\phi}{2} R'_1 R'_2 -i \frac{\phi}{2} (R_1R_2'-R_2R_1')}  e^{- \frac{\phi}{4} (R-R')(\bar{R}-\bar{R}') +i\frac{q}{2} (\bar{R}-\bar{R}')} \\
&= \sum_{\mbf{R},\mbf{R}'} e^{-i \mbf{k} \cdot \mbf{R} +i(\mbf{k}' +\qq)\cdot \mbf{R}'}  e^{i \frac{\phi}{2} R_1 R_2 -i \frac{\phi}{2} R'_1 R'_2 - \frac{\phi}{4} R\bar{R} - \frac{\phi}{4} R'\bar{R}' + \frac{\phi}{2}R'\bar{R}} e^{i (\frac{q}{2}, -i\frac{q}{2}) \cdot (\RR-\RR')} \\
&=\sum_{\mbf{R},\mbf{R}'} e^{-i (\mbf{k} - (\frac{q}{2}, -i\frac{q}{2})) \cdot \mbf{R} +i(\mbf{k}' +\qq  - (\frac{q}{2}, -i\frac{q}{2}))\cdot \mbf{R}'}  e^{i \frac{\phi}{2} R_1 R_2 -i \frac{\phi}{2} R'_1 R'_2 - \frac{\phi}{4} R\bar{R} - \frac{\phi}{4} R'\bar{R}' + \frac{\phi}{2}R'\bar{R}} \\
&=  \vartheta \lp \left. \frac{(k_1 - q/2,k_2 + iq/2)}{2\pi} \right|  \Phi \rp (2\pi)^2 \delta(\kk - \kk' -\qq)
\label{}
\end{align}
When $\kk = \kk'$ and $\qq = 2\pi \mbf{G}$, this expression must reduce to Eq.~\eqref{eq:siegelwithG}, which indeed it does.  Note that the delta function is defined modulo a reciprocal lattice vector $2\pi \mbf{G}$. We arrive at the formula
\bea
\label{eq:qformfactorderiv}
\braket{\mbf{k}',m| e^{-i \mbf{q} \cdot \mbf{r}} | \mbf{k},n} &= (2\pi)^2 \delta(\mbf{k}-\mbf{k'} - \mbf{q})\frac{e^{- \frac{ \bar{q} q}{4\phi}}  \vartheta \lp \left. \frac{(k_1 - q/2 ,k_2+ i q/2)}{2\pi} \right| \Phi \rp  }{\sqrt{\vartheta \lp \left. \frac{(k_1 ,k_2)}{2\pi} \right| \Phi \rp \vartheta \lp \left. \frac{(k_1 - q_1 ,k_2 - q_2)}{2\pi} \right| \Phi \rp}}  \bra{m} e^{i \frac{\gamma_q a + \bar{\gamma}_q a^\dag }{\sqrt{2 \phi}}} \ket{n} \\
&\equiv (2\pi)^2 \delta(\mbf{k}-\mbf{k'} - \mbf{q}) e^{i \xi_\mbf{q}(\mbf{k})} \mathcal{H}^{\mbf{q}}_{mn} \ .
\eea
Note that the operator $e^{-i \mbf{q} \cdot \mbf{r}}$ is unitary acting on position-space wavefunctions. Its representation on the $\ket{\mbf{k}, m}$ basis is also unitary, as must be the case, which is proved by showing $e^{i \xi_\mbf{q}(\mbf{k})}$ is a pure phase (see \Eq{eq:Hcalunitproof}) and that $\mathcal{H}^{\mbf{q}}_{mn} $ is a unitary matrix on the Landau level indices (see \Eq{eq:xiphasevariousproofs}).

\subsection{Berry Connection and the Chern Number of a Lattice Landau level}
\label{app:berry}

Our basis of magnetic translation eigenstates is built from continuum Landau levels. These states are known to carry a Chern number, and it will be important to see how this arises in our formalism. To do this carefully, we need to compute the continuum Berry connection
\bea
(2\pi)^2 \delta(\mbf{k}-\mbf{k}') \mathcal{A}^{mn}(\mbf{k}) &= \braket{\mbf{k}',m| \mbf{r}| \mbf{k},n} \ . \\
\eea
We will calculate the expression directly using \Eq{eq:rgaugeinv}. Using the identity
\begin{align}
  x_\mu = (eB)^{-1}\eps_{\mu \nu}(Q_\nu - \pi_\nu),
\label{}
\end{align}
the position operator $\mbf{r}$ can be expressed in terms of the oscillator operators via \Eq{eq:rintermsab} which reads
\bea
\mbf{q} \cdot \mbf{r} =  \frac{1}{\sqrt{2\phi}} \Big( i (qb - \bar{q} b^\dag) - (\bar{\gamma}_q a^\dag +\gamma_q a) \Big), \qquad q = (\mbf{a}_1 + i \mbf{a}_2) \cdot \mbf{q}, \quad \gamma_q  =  \eps_{ij} q_i \bar{z}_j
\eea
where we have introduced the \emph{arbitrary} vector $\mbf{q}$ as a technical device to aid the calculation. First we will show that the $a,a^\dag$ terms give a simple term:
\bea
\label{eq:acontribberry}
\braket{\mbf{k}',m| \frac{\bar{\gamma}_q a^\dag +\gamma_q a}{\sqrt{2\phi}}  | \mbf{k},n} = (2\pi)^2 \delta(\mbf{k}-\mbf{k}') \eps_{ij} q_i \braket{m|Z_j|n} =  (2\pi)^2 \delta(\mbf{k}-\mbf{k}') \eps_{ij} q_i [\tilde{Z}_j]_{mn},
\eea
using \Eqs{eq:Hcaldef}{eq:HcaldefZ}. Note that, interestingly, $\exp \lp i \eps_{ij} q_i [\tilde{Z}_j]_{mn}  \rp = {\cal H}^\qq_{mn}$.  Note that \Eq{eq:acontribberry} gives a contribution to $\mathcal{A}(\mbf{k})$ which is independent of $\mbf{k}$ and hence will not contribute to the Berry curvature or any winding when we consider only a single Landau level. However, it will be important for multi-band effects as we consider afterwards. Physically, the $1/\sqrt{\phi}$ dependence of this term signals a contribution due to the magnetic length. It is a non-Abelian term because localizing an electron below the magnetic length requires many Landau levels, implying that many other different bands $n \neq m$ contribute to the Berry curvature of a band $m$.

Moving onwards, we consider the $b,b^\dag$ oscillators in $\mbf{q} \cdot \mbf{r}$. Using \Eq{eq:knexplicit}, we need to compute
\bea
\bra{\mbf{k}', m} i(qb - \bar{q} b^\dag) \ket{\mbf{k}, n} &= \frac{1}{\sqrt{\mathcal{N}(\mbf{k})\mathcal{N}(\mbf{k}')}} \sum_{\mbf{R},\mbf{R}'} e^{-i \mbf{k} \cdot \mbf{R}+i \mbf{k}' \cdot \mbf{R}' +  i \frac{\phi}{2} R_1 R_2- i \frac{\phi}{2} R_1' R_2'} \bra{m}  e^{-i\mbf{R}' \cdot \mbf{Q} } i(qb - \bar{q} b^\dag)e^{i\mbf{R} \cdot \mbf{Q} }\ket{n}  \ . \\
\eea
Using the coherent state identities $b e^{x b^\dag}\ket{0} = x e^{x b^\dag}\ket{0}$, as well as BCH, this correlator can be evaluated as
\bea
\bra{m}  e^{-i\mbf{R}' \cdot \mbf{Q} } i(qb - \bar{q} b^\dag)e^{i\mbf{R} \cdot \mbf{Q} }\ket{n} &= \delta_{mn} e^{- \phi/4 \bar{R} R- \phi/4 \bar{R}' R'} \braket{0 |
e^{-i \sqrt{\phi/2} \bar{R}' b^\dagger} e^{-i \sqrt{\phi/2} R' b}  i(qb - \bar{q} b^\dag) e^{i \sqrt{\phi/2} \bar{R} b^\dag} e^{i \sqrt{\phi/2} R b} |0} \\
&= \delta_{mn} e^{- \phi/4 \bar{R} R- \phi/4 \bar{R}' R'} \braket{0 |
e^{-i \sqrt{\phi/2} R' b}  i(qb - \bar{q} b^\dag) e^{i \sqrt{\phi/2} \bar{R} b^\dag} |0} \\
&= -\sqrt{\phi/2}  \delta_{mn} e^{- \phi/4 \bar{R} R- \phi/4 \bar{R}' R'} \Big(\bar{q} R' + q \bar{R}  \Big) \braket{0|
e^{-i \sqrt{\phi/2} R' b}   e^{i \sqrt{\phi/2} \bar{R} b^\dag} |0} \\
&= -\sqrt{\phi/2}  \delta_{mn} e^{- \phi/4 \bar{R} R- \phi/4 \bar{R}' R'} \Big(\bar{q} R' + q \bar{R}  \Big) \braket{0|
  e^{\frac{\phi}{2} R' \bar{R} [b,b^\dagger]} e^{i \sqrt{\phi/2} \bar{R} b^\dag} e^{-i \sqrt{\phi/2} R' b}|0} \\
&= -\sqrt{\phi/2}  \delta_{mn} e^{- \phi/4 \bar{R} R- \phi/4 \bar{R}' R' + \frac{\phi}{2} R'  \bar{R}} \Big(\bar{q} R' + q \bar{R}  \Big).\\
\eea

We are left with an infinite sum:
\bea
\label{eq:sumderout}
\bra{\mbf{k}', m} \frac{i}{\sqrt{\phi/2}}(qb - \bar{q} b^\dag) \ket{\mbf{k}, n} &= -\frac{\delta_{mn}}{\sqrt{\mathcal{N}(\mbf{k})\mathcal{N}(\mbf{k}')}} \sum_{\mbf{R},\mbf{R}'}  (\bar{q} R' + q \bar{R} )  e^{-i \mbf{k} \cdot \mbf{R}+i \mbf{k}' \cdot \mbf{R}' +  i \frac{\phi}{2} R_1 R_2 - i \frac{\phi}{2} R_1' R_2'- \frac{\phi}{4} \bar{R} R- \frac{\phi}{4} \bar{R}' R' + \frac{\phi}{2}R' \bar{R}} \\
&= -\frac{\delta_{mn} }{\sqrt{\mathcal{N}(\mbf{k})\mathcal{N}(\mbf{k}')}} \sum_{\mbf{R},\mbf{R}'}  (-2 \bar{q} i \bar{\del}' + 2 i q \del )  e^{-i \mbf{k} \cdot \mbf{R}+i \mbf{k}' \cdot \mbf{R}' +  i \frac{\phi}{2} R_1 R_2 - i \frac{\phi}{2} R_1' R_2'- \frac{\phi}{4} \bar{R} R- \frac{\phi}{4} \bar{R}' R' + \frac{\phi}{2} R' \bar{R}} \ . \\
\eea
Here we have used the holomorphic variables $\bar{R}, R$, and momentum derivatives $\partial, \bar{\partial}$ in \Eq{eq:holoderivids} which we reproduce for convenience below:
\bea
\bar{R} e^{-i \mbf{k} \cdot \mbf{R}} = \bar{R} e^{-i \frac{k \bar{R} + \bar{k}R}{2}} = 2i \del e^{-i \frac{k \bar{R} + \bar{k}R}{2}}, \qquad 2\del \equiv \frac{\del}{\del k_1} - i \frac{\del}{\del k_2} \equiv \del_1 - i \del_2
\eea
and similarly for the anti-holomorphic derivative $\bar{\del} = \del^*$ which satisfy $\del k = \bar{\del} \bar{k} = 1$. The primed derivatives act on the $k'$ and $\bar{k}'$. Taking the derivatives out of the sum (which is easily justified because it converges absolutely) leaves \Eq{eq:sumderout} in terms of exactly the theta function expression in \Eq{eq:TpT}:
\bea
 \sum_{\mbf{R},\mbf{R}'}  e^{-i \mbf{k} \cdot \mbf{R}+i \mbf{k}' \cdot \mbf{R}' +  i \frac{\phi}{2} R_1 R_2+  i \frac{\phi}{2} R_1' R_2'- \frac{\phi}{4} \bar{R} R- \frac{\phi}{4} \bar{R}' R' + \frac{\phi}{2} R' \bar{R}} &= (2\pi)^2 \delta(\mbf{k}-\mbf{k}') \vartheta \lp \left. \frac{(k_1+k_1',k_2+k_2')/2}{2\pi} \right| \Phi \rp \ .
\eea
It is convenient to keep the symmetrized argument $(\mbf{k}+\mbf{k}')/2$ as appears in \Eq{eq:TpTsym} because we need to take derivatives of the Dirac delta function. Explicitly, we must calculate
\bea
\bra{\mbf{k}', m} \frac{i}{\sqrt{2\phi}}(qb - \bar{q} b^\dag) \ket{\mbf{k}, n}  &= \frac{\delta_{mn} (2\pi)^2 }{\sqrt{\mathcal{N}(\mbf{k})\mathcal{N}(\mbf{k}')}} i(\bar{q} \bar{\del}' - q \del ) \left[ \delta(\mbf{k}-\mbf{k}')  \mathcal{N}(\mbf{k}'/2+\mbf{k}/2) \right], \qquad   \mathcal{N}(\mbf{k})  = \vartheta \lp \left. \frac{(k_1,k_2)}{2\pi} \right| \Phi \rp \ . \\
\eea  Note the $\frac{i}{\sqrt{2\phi}}$ prefactor in this expression compared to Eq.~\eqref{eq:sumderout}.

Expanding the derivatives yields
\bea
 i(\bar{q} \bar{\del}' - q \del ) \left[ \delta(\mbf{k}-\mbf{k}')  \mathcal{N}\lp\frac{\mbf{k}'+\mbf{k}}{2}\rp \right] &= \mathcal{N}\lp\frac{\mbf{k}'+\mbf{k}}{2}\rp i(\bar{q} \bar{\del}' - q \del ) \delta(\mbf{k}-\mbf{k}')  + \delta(\mbf{k}-\mbf{k}')    i(\bar{q} \bar{\del}' - q \del ) \left[ \mathcal{N}\lp\frac{\mbf{k}'+\mbf{k}}{2}\rp \right] \\
 &= \mathcal{N}\lp\frac{\mbf{k}'+\mbf{k}}{2}\rp i(-\bar{q} \bar{\del} - q \del ) \delta(\mbf{k}-\mbf{k}')  + \frac{1}{2} \delta(\mbf{k}-\mbf{k}')  i \left.(\bar{q} \bar{\del}_z - q \del_z) \mathcal{N}(\mbf{z})\right|_{\mbf{z} = \frac{\mbf{k}'+\mbf{k}}{2}} \\
 &= \mathcal{N}(\mbf{k})  i(-\bar{q} \bar{\del} - q \del ) \delta(\mbf{k}-\mbf{k}')  + \frac{1}{2} \delta(\mbf{k}-\mbf{k}')  i(\bar{q} \bar{\del} - q \del ) \mathcal{N}(\mbf{k}) \\
\eea
where we have used the chain rule to rewrite the $\del'$ operators in terms of $\del$ and then set $\mbf{k} = \mbf{k}'$ after performing the derivatives. The manipulation in the second line uses the chain rule to write, e.g. $\del_k [ f(\frac{k+k' }{2}) ]= f'(\frac{k+k' }{2}) \del_k \frac{k+k'}{2} = \frac{1}{2} f'(\frac{k+k' }{2}) $ where as usual $f'(x) = \del_x f(x)$.

Finally, we have the following identities
\bea
\bar{q} \bar{\del} + q \del  = q_i \del_i, \quad i(\bar{q} \bar{\del} - q \del)  = -\eps_{ij} q_i \del_j
\eea
which give the expression
\bea
\label{eq:bcontribberry}
\bra{\mbf{k}', m} \frac{i}{ \sqrt{2\phi} }(qb - \bar{q} b^\dag) \ket{\mbf{k}, n}  &=  \delta_{mn} (2\pi)^2 q_i \Big(- i  \del_i \delta(\mbf{k}-\mbf{k}') - \frac{1}{2} \delta(\mbf{k}-\mbf{k}') \eps_{ij} \del_j \log \mathcal{N}(\mbf{k}) \Big) \ .
\eea
The two terms in this expression have different physical consequences. The first term appears in the Berry connection at zero flux:
\bea
 \int d^2r \ e^{-i (\mbf{k}'-\mbf{G}') \cdot \mbf{r}} \mbf{r} \, e^{i (\mbf{k}-\mbf{G}) \cdot \mbf{r}} &= (-i \del_{\mbf{k}})   \int d^2r \ e^{-i (\mbf{k}'-\mbf{G}') \cdot \mbf{r}} \mbf{r} \, e^{i (\mbf{k}-\mbf{G}) \cdot \mbf{r}}  = -i (2\pi)^2 \delta_{\mbf{G},\mbf{G}'} \del_{\mbf{k}} \delta(\mbf{k} - \mbf{k}')
\eea
where the functions $e^{i (\mbf{k}-\mbf{G}) \cdot \mbf{r}}$ are a basis of the Bloch states at zero flux. When there are nontrivial Bloch functions, the $-i\del_{\mbf{k}} $ is responsible for the conventional $U^\dag (i \del_\mbf{k}) U$ contribution to the Berry curvature where $U(\mbf{k})$ are the matrix eigenvectors.  To see this, recall that the Bloch eigenstates in the $M$th band are defined
\begin{align}
\ket{\kk, M}_{\phi=0} = \sum_m \ket{\kk, \mbf{G}} U^M_\mbf{G}(\kk),
\label{}
\end{align}
so the Berry connection in this basis yields
\bea
\braket{\kk', M | \rr | \kk, N}_{\phi=0}  &= \sum_{\mbf{G}\mbf{G}'} [U^\dagger  (\kk') ]^M_{\mbf{G}'}{\braket{\kk', \mbf{G}' | \rr | \kk, \mbf{G}} U^N_{\mbf{G}}(\kk)} \\
&= \sum_{\mbf{G} \mbf{G}'} [U^\dagger  (\kk') ]^M_{\mbf{G}'}  [-i(2\pi)^2 \delta_{\mbf{G}' \mbf{G}} \partial_\kk \delta(\kk - \kk')] U^N_{\mbf{G}}(\kk) \\
&= (2\pi)^2 \delta(\kk - \kk') \sum_{\mbf{G} \mbf{G}'} [U^\dagger  (\kk') ]^M_{\mbf{G}'} [\delta_{\mbf{G} \mbf{G}'}  i\partial_\kk U^N_\mbf{G}(\kk)] \\
&= (2\pi)^2 \delta(\kk - \kk') [U^\dagger(\kk) (i\partial_\kk) U(\kk)]^{MN} \ .
\label{eq:kmUsum}
\eea
where we used the delta function identity $f(\mbf{k}) \del_{\mbf{k}} \delta(\mbf{k} - \mbf{k}') = - \delta(\mbf{k} - \mbf{k}')\del_{\mbf{k}} f(\mbf{k})$. If the basis includes orbital degrees of freedom carrying Landau levels, such as sublattice or layer as in TBG, then \Eq{eq:kmUsum} should be summed over the orbital indices as well.

\begin{figure}
\includegraphics[width=0.36\textwidth, trim = 0 .3cm 0 .3cm, clip]{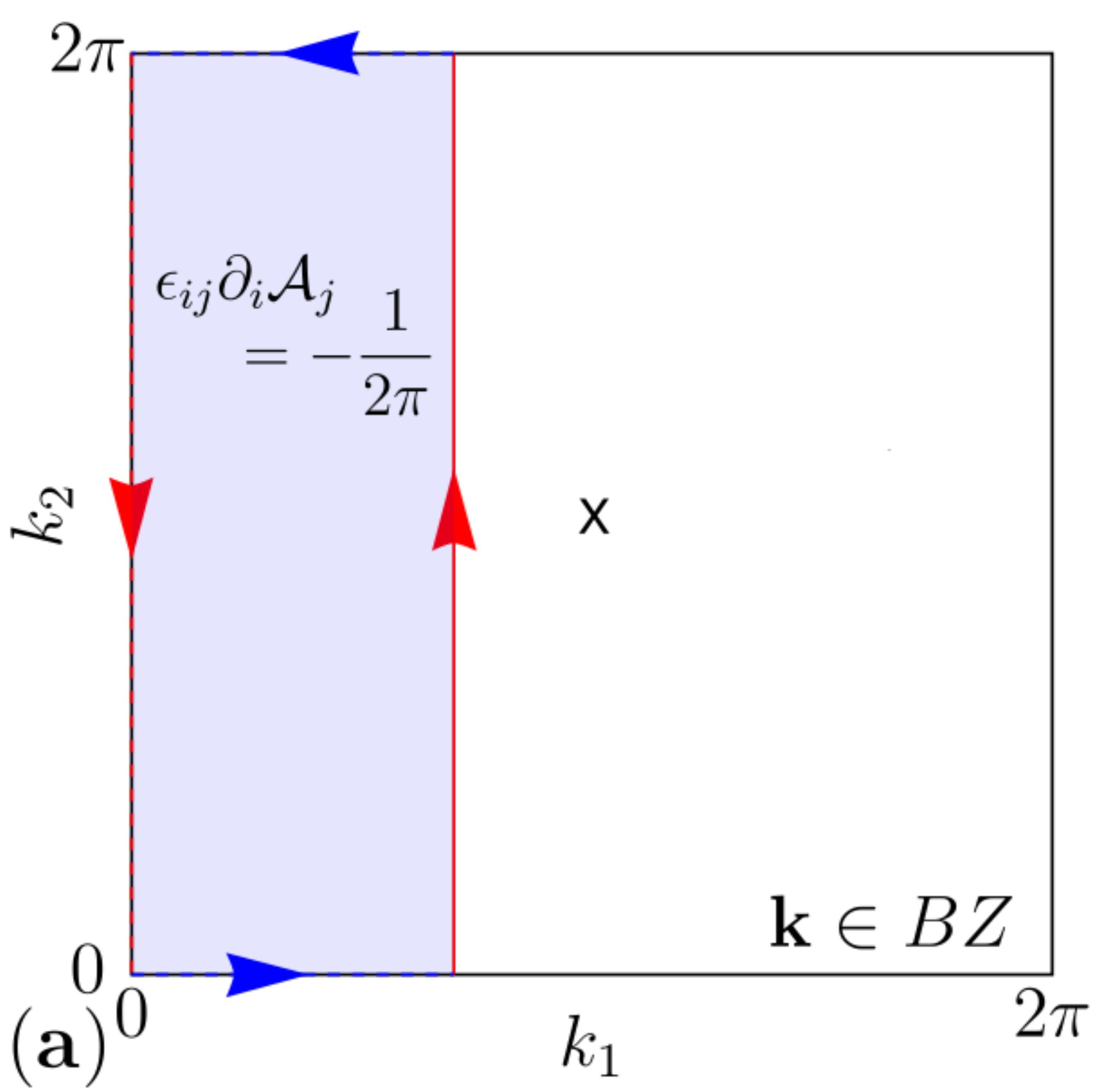} \qquad
\includegraphics[width=0.35\textwidth]{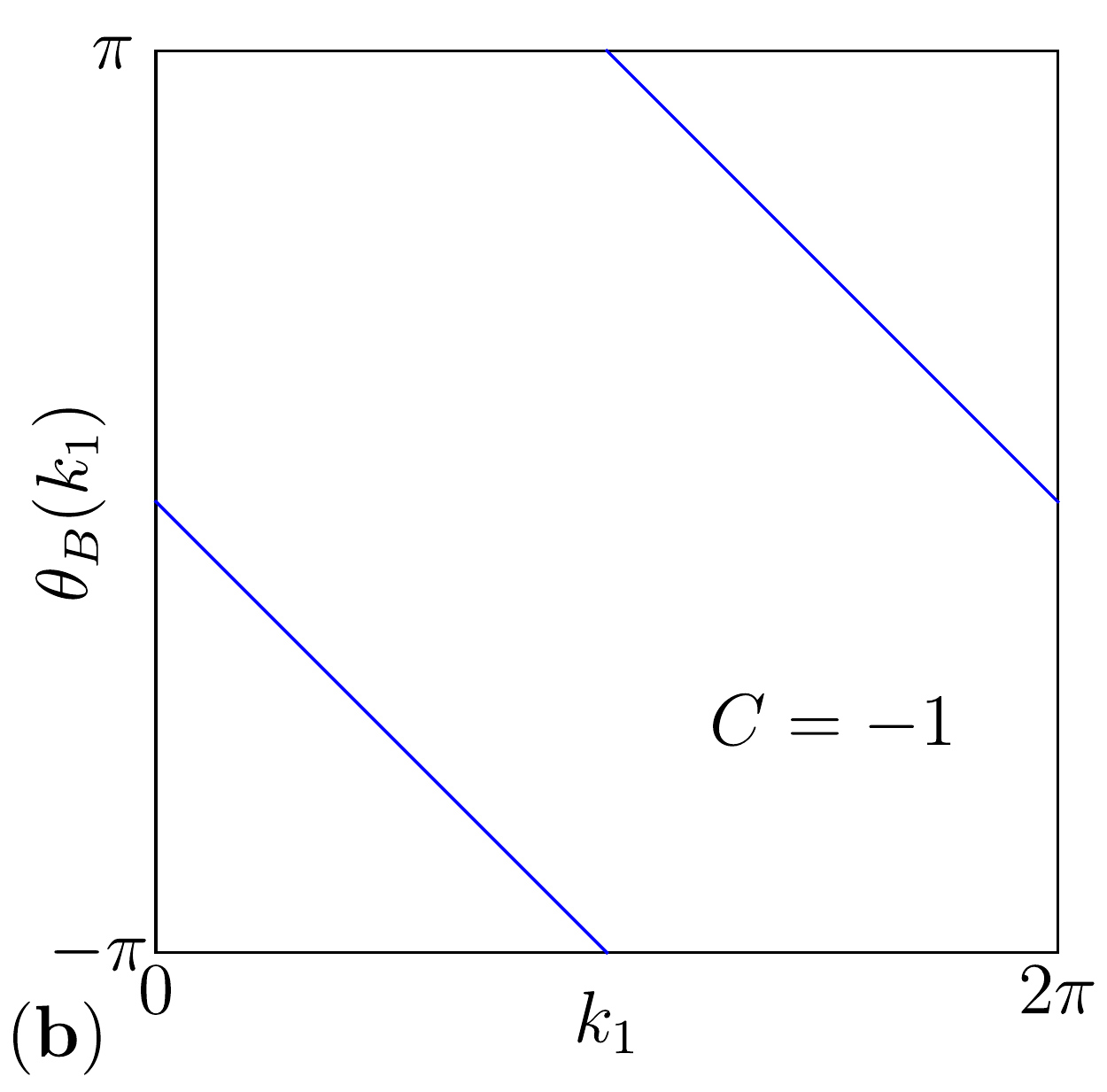}
\caption{Landau Level Wilson Loops. $(a)$ The Berry phase can be calculated using Stokes's theorem in \Eq{eq:WLcalcLandau}. The red arrows represent Wilson loops integrated along $k_2$, and the blue arrows represent two additional paths (which sum to zero because $k_2 = 0$ and $k_2 = 2\pi$ are identified) which form a closed loop. There is an additional delta function marked by the X at $(k_1, k_2) = (\pi,\pi)$ which adds $2\pi$ to the Berry phase as $k_1$ crosses $\pi$. $(b)$ We plot the Berry phase obtained from the Abelian Wilson loop in \Eq{eq:WLcalcLandau}, from which we see that the winding gives $C=-1$.
}
\label{fig:WLLandau}
\end{figure}

The $\eps_{ij} \del_j \log \mathcal{N}(\mbf{k})$ term in \Eq{eq:bcontribberry} has no zero-flux analogue. It arises entirely from the normalization factor $\mathcal{N}(\mbf{k})$. This term is responsible for the nonzero Chern number of the basis states as we now show. The Abelian Berry connection of the $m$th Landau level is independent of $m$ and is given by
\bea
\mathcal{A}_i^{mm} = \mbf{b}_i \cdot \mathcal{A}^{mm} &= - \frac{1}{2} \eps_{ij} \del_j \log \mathcal{N}(\mbf{k}) \ . \\
\eea
Using \Eq{eq:greensfunction}, we have the following fact about the Abelian Berry curvature:
\bea
\label{eq:BC2}
\eps_{ij} \del_i \mathcal{A}_j^{mm} =  - \frac{1}{2} \eps_{jk} \eps_{ij} \del_i \del_k \log \mathcal{N}(\mbf{k}) =  \frac{1}{2} \del^2 \log \mathcal{N}(\mbf{k}) = -\frac{1}{2\pi} + 2\pi \delta(\mbf{k} - \pi \mbf{b}_1 - \pi \mbf{b}_2) \\
\eea
which shows that there are two contributions to the Berry curvature. The first term $-\frac{1}{2\pi}$ is the perfectly flat Berry curvature expected for a Landau level \cite{2021arXiv210507491W}. Integrating  $-\frac{1}{2\pi}$ over the BZ of area $(2\pi)^2$ gives $-2\pi$ corresponding to a Chern number $C=-1$. The second term in \Eq{eq:BC2} is a singular delta function contribution to the Berry curvature which is appears at the singular point $\mbf{k}^* = \pi \mbf{b}_1 + \pi \mbf{b}_2$ where $\mathcal{N}(\mbf{k}^*) =0$ and $\ket{\mbf{k}^*,n}$ is undefined. Note that $\mbf{k}^*$ can be chosen arbitrarily via a gauge transformation (see \Eq{eq:shiftp}) and thus the delta function cannot be a physical contribution to the Berry curvature. We show this in two ways. First we observe that $\eps_{ij} \del_i \mathcal{A}_j^{mm}  =  \frac{1}{2} \del^2 \log \mathcal{N}(\mbf{k})$ is a total divergence so \emph{if} the wavefunction could be continuously defined, then the Berry curvature would integrate to zero over the BZ. However, a Chern number forbids a globally well-defined gauge, which in our case is manifested by the undefined states where $\mathcal{N}(\mbf{k}^*)=0$. To compute the Chern number, we define the wavefunction locally in patches. Let patch 1 be $BZ$ with a small neighborhood near $\mbf{k}^*$ removed in the gauge where $\mathcal{N}(\mbf{k}^*) = 0$, and let patch 2 be the small neighborhood around $\mbf{k}^*$ in a gauge where  $\mathcal{N}(\mbf{p}^*) = 0$ for some $\mbf{p}^*$ outside the neighborhood. The Berry curvature is non-singular in both patches and integrates to give $C=-1$. A more physical way to understand this comes from computing the Wilson loop. Because the Berry connection appears in the exponential, the $2\pi$ delta function is unobservable as we now show. The Abelian Wilson loop of a single Landau level band integrated along the $k_2$ direction is \cite{Alexandradinata:2012sp}
\bea
W(k_1) = e^{i \th_B(k_1)} = \exp \lp i \oint_{(k_1,0)}^{(k_1,2\pi)} dk_2 \mathcal{A}^{mm}_2(\mbf{k}) \rp
\eea
where $\th_B(k_1)$ is the Berry phase whose winding determines the Chern number. We will calculate an exact expression for $W(k_1)$. To start, it is easy to show that $\th_B(0) = 0$ for the Wilson loop integrated along the $k_2$ direction because $dk_2 \mathcal{A}^{mm}_2(0,k_2) = \frac{1}{2}dk_2 \del_1 \log \mathcal{N}(0,k_2) = 0$, which follows from the fact that $\mathcal{N}(k_1,k_2)  = \mathcal{N}(-k_1,k_2)$ so $\del_1 \mathcal{N}|_{k_1= 0} = 0$. This is easily observed from \Eq{eq:siegeljac} which shows $\mathcal{N}(k_1,k_2)$ is even in both its arguments. We can now calculate directly
\bea
\label{eq:WLcalcLandau}
\th_B(k_1) = \th_B(k_1) - \th_B(0) &= \lp \oint_{(k_1,0)}^{(k_1,2\pi)} - \oint_{(0)}^{(0,2\pi)}  \rp dk_i \mathcal{A}_i^{mm}(\mbf{k}) \mod 2\pi \\
&=  \int_{(0,k_1) \times (0,2\pi)} d^2k \, \eps_{ij} \del_i \mathcal{A}_j^{mm}(\mbf{k}) \mod 2\pi   \\
&= -\frac{2\pi k_1}{2\pi}  + 2\pi \Theta(k_1-\pi) \mod 2\pi \\
&= \begin{cases}
-k_1, & 0\leq k_1 < \pi \\
2\pi -k_1, & \pi < k_1 \leq 2\pi \\
\end{cases}
\eea
where the second equality follows from adding paths in opposite direction separated by $2\pi$ which cancel, and then applying Stokes' theorem (see \Fig{fig:WLLandau}(a)). Finally $\Theta(x)$ is the Heavyside step function: $\del_x \Theta(x) = \delta(x)$. The winding of $\th_B(k_1)$ is plotted in \Fig{fig:WLLandau}(b). We see that the $2\pi$ discontinuity from the delta function in \Eq{eq:BC2} is unobservable because $\th_B(k_1)$ is only defined mod $2\pi$. 

Having discussed the Abelian terms from the Landau level basis states, we now consider the full non-Abelian Wilson loop with contributions from the band eigenvectors. We gather the terms from \Eq{eq:acontribberry} and \Eq{eq:bcontribberry}. Incorporating the prefactors, we find
\bea
\label{eq:wilsonterms}
\braket{\mbf{k}',m| \mbf{b}_i \cdot \mbf{r}| \mbf{k},n} &=(2\pi)^2  \delta_{mn}  (- i  \del_i) \delta(\mbf{k}-\mbf{k}') \\
&\qquad - (2\pi)^2 \delta(\mbf{k}-\mbf{k}') \eps_{ij}  \Big( \frac{1}{2}   \delta_{mn}  \del_j \log \mathcal{N}(\mbf{k}) +[\tilde{Z}_j]_{mn} \Big) \ .
\eea
The explicit matrix elements are
\bea
\label{eqLZexptrace}
\null [\tilde{Z}_j]_{mn}= \braket{m|\frac{z_j a^\dag +\bar{z}_j a}{\sqrt{2\phi}} |n} =  \frac{1}{\sqrt{2\phi}} (z_j \sqrt{m} \delta_{m-1,n} + \bar{z}_j \sqrt{n} \delta_{m,n-1} ) \ .
\eea

Finally, we obtain an explicit expression for the non-Abelian Berry connection $ \mathcal{A}^{MN}(\mbf{k})$ in the occupied bands indexed by $M,N$ defined by
\bea
(2\pi)^2 \delta(\mbf{k}-\mbf{k}') \mathcal{A}^{MN}(\mbf{k}) &= \sum_{mn} [U^\dag]^{M}_{m} \braket{\mbf{k}',m| \mbf{r}| \mbf{k},n} U_{n}^N(\mbf{k}) \\
\eea
 where $U(\mbf{k})$ is the $N_{LL} \times N_{occ}$ matrix of eigenvectors, $N_{occ}$ is the number of occupied bands, and $N_{LL}$ is the dimension of the matrix Hamiltonian after the Landau levels are truncated above some cutoff. Plugging in \Eq{eq:bcontribberry}, we find the final expression
\bea
\label{eq:Acalform}
\boxed{
\mathcal{A}_i^{MN}(\mbf{k}) = \bbb_i \cdot \mathcal{A}^{MN}(\mbf{k}) = [U^\dag (i \del_i - \eps_{ij} \tilde{Z}_j) U]^{MN} - \frac{\delta^{MN}}{2} \eps_{ij}   \del_j \log \vartheta \lp \left. \frac{(k_1,k_2)}{2\pi} \right| \Phi \rp
}
\eea
where we performed an identical manipulation to \Eq{eq:kmUsum} to act the $\del_i$ derivative on $U(\mbf{k})$.

To compute Wilson loops numerically, we need to be able to write the Wilson loop as a discretized product. To do so, we need to deal with the $\tilde{Z}_j$ term using the commutation relations in \Eq{eq:Zprop}. We need the following identity
\bea
\label{eq:derexp}
e^{- i \eps_{jl} k_j \tilde{Z}_l} \del_i e^{i \eps_{jl} k_j \tilde{Z}_l} &= i \eps_{il} \tilde{Z}_l - \frac{1}{2}  [i \eps_{jl} k_j \tilde{Z}_l, \del_i (i \eps_{j'l'} k_{j'} \tilde{Z}_{l'})] + \text{higher commutators} \\
&= i \eps_{il} \tilde{Z}_l + \frac{1}{2}  \eps_{jl} \eps_{il'} k_j [\tilde{Z}_l,  \tilde{Z}_{l'}] \\
&= i \eps_{ij} \tilde{Z}_j + \frac{i}{2\phi}  \eps_{ll'} \eps_{jl} \eps_{il'} k_j \\
&= i \eps_{ij} \tilde{Z}_j - \frac{i}{2\phi} \eps_{ij} k_j \\
\eea
where the first line is the formula for the derivative of the exponential map (see Wikipedia for instance \url{https://en.wikipedia.org/wiki/Derivative_of_the_exponential_map}) and the higher commutator terms vanish because $[\tilde{Z}_i,\tilde{Z}_j] \propto \mathbb{1}$ is central. The additional $- \frac{i}{2\phi} \eps_{ij} k_j$ term has nonzero constant curl and so cannot be absorbed into the exponential. However, $- \frac{i}{2\phi} \eps_{ij} k_j$ is proportional to the identity so it can be factored out. Using \Eq{eq:derexp}, we have
\bea
U^\dag(\mbf{k})  e^{- i \eps_{jl} k_j \tilde{Z}_l} i\del_i \lp e^{i \eps_{jl} k_j \tilde{Z}_l} U(\mbf{k})  \rp &= U^\dag(\mbf{k})  \lp e^{- i \eps_{jl} k_j \tilde{Z}_l} i\del_i  e^{i \eps_{jl} k_j \tilde{Z}_l} \rp  U(\mbf{k})  + U^\dag(\mbf{k}) i\del_i U(\mbf{k}) \\
&= U^\dag(\mbf{k}) i\del_i U(\mbf{k})  + U^\dag(\mbf{k})  \lp- \eps_{ij} \tilde{Z}_j + \frac{1}{2\phi} \eps_{ij} k_j \rp  U(\mbf{k})  \\
&= U^\dag(\mbf{k}) \lp i\del_i- \eps_{ij} \tilde{Z}_j  \rp U(\mbf{k})  + \frac{1}{2\phi} \eps_{ij} k_j  U^\dag(\mbf{k}) U(\mbf{k})  \\
\eea
and because $U^\dag(\mbf{k}) U(\mbf{k}) = \mathbb{1}$ due to the orthonormality of the eigenvectors, we arrive at
\bea
\label{eq:Acalanom}
\mathcal{A}_i^{MN}(\mbf{k}) &=  [\tilde{U}^\dag i \del_i \tilde{U}]^{MN}  - \frac{\delta^{MN}}{2\phi} \eps_{ij} k_j - \frac{\delta^{MN}}{2} \eps_{ij}   \del_j \log \vartheta \lp \left. \frac{(k_1,k_2)}{2\pi} \right| \Phi \rp , \qquad \tilde{U}(\mbf{k}) \equiv e^{i \eps_{ij} k_i {\tilde{Z}}_j} U(\mbf{k}) = \mathcal{H}^{\mbf{k}} U(\mbf{k})
\eea
and we used $e^{i \eps_{ij} k_i \tilde{Z}_j} = \mathcal{H}^\mbf{k}$ which is a unitary matrix. As such, note that $ \tilde{U}^\dag(\mbf{k})  \tilde{U}(\mbf{k}) =  U^\dag(\mbf{k}) {\mathcal{H}^{\mbf{k}}}^\dag  \mathcal{H}^{\mbf{k}} U(\mbf{k}) = U^\dag  (\mbf{k})U(\mbf{k}) = \mathbb{1}$. The exponential factor $e^{i\epsilon_{ij}k_i \tilde{Z}_j}$ has nontrivial commutation relations and cannot be absorbed into the definition of $U(\kk)$ via a gauge transformation;  we will keep this term throughout our calculations. The first term in \Eq{eq:Acalanom} is in the canonical Berry connection form. The second term is the extraneous Abelian term (which we will show is canceled at the end of the calculation), and the third term of \Eq{eq:Acalanom} is the winding of the Chern basis. As with the standard zero-flux Wilson loop, we want to express the Wilson loop as a product of projectors. In constructing the projector product, we will find that a term appears to exactly cancel the $ - \frac{\delta^{MN}}{2\phi} \eps_{ij} k_j $ extraneous term. The end result will be a simple product formula in terms of $U(\mbf{k})$. We define the path-ordered non-Abelian Wilson loop along the path $\mathcal{C}$ to be \cite{Alexandradinata:2012sp}
\bea
\label{eq:nonabwilstart}
\null [W_{\mathcal{C}}]^{MN} &= \left[ \exp \lp i \oint_{\mathcal{C}} d\mbf{k} \cdot \mathcal{A}(\mbf{k})  \rp \right]^{MN} \\
\eea
We now plug in \Eq{eq:Acalanom} and factor out the Abelian contributions:
\bea
\label{eq:Winitialfull}
W_{\mathcal{C}} &= e^{-i \oint_{\mathcal{C}} d\mbf{k} \times \frac{1}{2}\pmb{\nabla} \log \vartheta \lp \left. \frac{(k_1,k_2)}{2\pi} \right| \Phi \rp } \left[ e^{-\frac{i}{2\phi} \int_{\mathcal{C}} dk_i \eps_{ij} k_j} \exp \lp i \oint_{\mathcal{C}} dk_i \, \tilde{U}^\dag i \del_i \tilde{U} \rp \right] \ .
\eea
Note that we have grouped the extraneous Abelian term with the $\tilde{U}$ term inside the brackets. Our goal is now to produce a discretized expression for the bracketed term. To do so, we break the path $\mathcal{C}$ into $L$ increments labeled by $d\mbf{k}_n, n = 1, \dots L$ where $d\mbf{k}_n = \mbf{k}_{n} - \mbf{k}_{n-1}$. Because $\mathcal{C}$ is a closed path, $\mbf{k}_{0} \equiv \mbf{k}_L$. Taking $L \to \infty$, we find
\bea
\exp \lp i \oint_{\mathcal{C}} dk_i \, \tilde{U}^\dag i \del_i \tilde{U} \rp &= \exp \lp i \sum_{n=1}^L  \tilde{U}^\dag(\mbf{k}_n) i (\tilde{U}(\mbf{k}_{n})-\tilde{U}(\mbf{k}_{n-1}) )\rp \\
&= \exp \lp i \sum_{n=1}^L  i (\mathbb{1} - \tilde{U}^\dag(\mbf{k}_n) \tilde{U}(\mbf{k}_{n-1})) \rp \\
\eea
and now using BCH (neglecting terms of $O(d\mbf{k}^2)$ which vanish as $L \to \infty$), we have
\bea
\exp \lp i \oint_{\mathcal{C}} dk_i \, \tilde{U}^\dag i \del_i \tilde{U} \rp &\to \prod_{n}^{L \leftarrow 1} \exp \lp  (-\mathbb{1} + \tilde{U}^\dag(\mbf{k}_n) \tilde{U}(\mbf{k}_{n-1})) \rp \\
&= \prod_{n}^{L \leftarrow 1} \lp \mathbb{1} + (-\mathbb{1} + \tilde{U}^\dag(\mbf{k}_n) \tilde{U}(\mbf{k}_{n-1})) + \dots \rp \\
&= \prod_{n}^{L \leftarrow 1}  \tilde{U}^\dag(\mbf{k}_n) \tilde{U}(\mbf{k}_{n-1}) \\
\eea
where in the second to last line, we expanded $e^M = 1 + M + \dots$ where $M = \tilde{U}^\dag(\mbf{k}_n) \tilde{U}(\mbf{k}_{n-1}) - \mathbb{1}$ is a matrix of order $ O(d\mbf{k})$. Terms of higher order vanish in the $L \to \infty$ limit. So far, this argument is identical to the standard Wilson loop.  Regrouping terms in the product gives the standard Wilson loop form
\bea
\label{eq:Utildewilsloop}
\exp \lp i \oint_{\mathcal{C}} dk_i \, \tilde{U}^\dag i \del_i \tilde{U} \rp &= \tilde{U}^\dag(\mbf{k}_{L})  \lp \prod_{n}^{L-1 \leftarrow 1}   \tilde{U}(\mbf{k}_n) \tilde{U}^\dag(\mbf{k}_{n}) \rp \tilde{U}(\mbf{k}_{0})  \ .
\eea
The factors in parentheses are a product of $\tilde{U}$ projectors. By plugging in $\tilde{U}(\mbf{k}) = \mathcal{H}^{\mbf{k}} U(\mbf{k})$, we will arrive at an expression written only in terms of the matrix eigenvectors $U(\mbf{k})$:
\bea
\label{eq:prodUH}
\prod_{n}^{L-1 \leftarrow 1}   \tilde{U}(\mbf{k}_n) \tilde{U}^\dag(\mbf{k}_{n})  &=   \mathcal{H}^{\mbf{k}_{L-1}} P(\mbf{k}_{L-1}) \mathcal{H}^{-\mbf{k}_{L-1}}  \dots  \mathcal{H}^{\mbf{k}_2} P(\mbf{k}_2)  \mathcal{H}^{-\mbf{k}_2} \mathcal{H}^{\mbf{k}_1} P(\mbf{k}_1)  \mathcal{H}^{-\mbf{k}_1}, \quad P(\mbf{k}) \equiv U(\mbf{k}) U^\dag(\mbf{k}) \ .
\eea
To simplify the Wilson loop further, we focus on factors in the form $\mathcal{H}^{-\mbf{k}_{n+1}} \mathcal{H}^{\mbf{k}_n}$. To do so, recall that $\mathcal{H}^\mbf{k} = e^{i \eps_{ij} k_i \tilde{Z}_j} $ and use the Baker-Campbell-Hausdorff identity $e^{X+Y} = e^{- \frac{1}{2}[X,Y]} e^X e^Y$ for $[X,Y]$ a $c$-number to show
\bea
\label{eq:HkHkn1}
\mathcal{H}^{-\mbf{k}_{n+1}} \mathcal{H}^{\mbf{k}_n} = \mathcal{H}^{- d\mbf{k}_{n+1} -\mbf{k}_n} \mathcal{H}^{\mbf{k}_n} &= \exp \lp - \frac{1}{2} [-i \eps_{ij} dk^{n+1}_{i} \tilde{Z}_j,-i \eps_{i'j'} k^{n}_{i'} \tilde{Z}_{j'}]\rp  \mathcal{H}^{ - d\mbf{k}_{n+1} } \mathcal{H}^{-\mbf{k}_n} \mathcal{H}^{\mbf{k}_n}
\eea
where we raised the $n$ indices on $d\mbf{k}$ and $\mbf{k}$ to avoid confusion with the vector indices $i,j$. The commutator is direct to evaluate with \Eq{eq:Zprop}:
\bea
\null  [-i \eps_{ij} dk^{n+1}_{i} \tilde{Z}_j,-i \eps_{i'j'} k^{n}_{i'} \tilde{Z}_{j'}] &= - \eps_{ij} \eps_{i'j'} dk^{n+1}_{i}   k^{n}_{i'}[\tilde{Z}_j,  \tilde{Z}_{j'}] \\
&= - \eps_{ij} \eps_{i'j'} dk^{n+1}_{i}   k^{n}_{i'}  \frac{i}{\phi} \eps_{jj'} \\
&= - \frac{i}{\phi} \eps_{ij} dk^{n+1}_{i}   k^{n}_{j}   \\
&= - \frac{i}{\phi} d\mbf{k}_{n+1} \times \mbf{k}_{n}  \ . \\
\eea
Returning to \Eq{eq:HkHkn1}, we now have the simple relation
\bea
\mathcal{H}^{-\mbf{k}_{n+1}} \mathcal{H}^{\mbf{k}_n} &= e^{\frac{i}{2\phi} d\mbf{k}_{n+1} \times \mbf{k}_{n}} \mathcal{H}^{-d\mbf{k}_{n+1}} \\
\eea
using $\mathcal{H}^{-\mbf{k}_n} \mathcal{H}^{\mbf{k}_n} = \mathbb{1}$ by unitarity. With this result, \Eq{eq:prodUH} reads
\bea
\prod_{n}^{L-1 \leftarrow 1}   \tilde{U}(\mbf{k}_n) \tilde{U}^\dag(\mbf{k}_{n})  &= e^{\frac{i}{2\phi} \sum_{n=1}^{L-2} d \mbf{k}_{n+1} \times \mbf{k}_n} \mathcal{H}^{\mbf{k}_{L-1}} P(\mbf{k}_{L-1}) \dots  \mathcal{H}^{-d\mbf{k}_3} P(\mbf{k}_2) \mathcal{H}^{-d\mbf{k}_2} P(\mbf{k}_1)  \mathcal{H}^{-\mbf{k}_1} \\
\tilde{U}^\dag(\mbf{k}_{L}) \lp \prod_{n}^{L-1 \leftarrow 1}   \tilde{U}(\mbf{k}_n) \tilde{U}^\dag(\mbf{k}_{n}) \rp \tilde{U}(\mbf{k}_{n})  &= e^{\frac{i}{2\phi} \sum_{n=0}^{L-1} d \mbf{k}_{n+1} \times \mbf{k}_n} U^\dag(\mbf{k}_L) \mathcal{H}^{-d\mbf{k}_{L}} P(\mbf{k}_{L-1}) \dots  \mathcal{H}^{-d\mbf{k}_3} P(\mbf{k}_2) \mathcal{H}^{-d\mbf{k}_2} P(\mbf{k}_1)  \mathcal{H}^{-d\mbf{k}_1} U(\mbf{k}_0) \\
\eea
where we grouped the Abelian  $- \frac{i}{\phi} d\mbf{k}_{n+1} \times \mbf{k}_{n} $ terms into an overall prefactor. Now we have a simple expression for the $\tilde{U}$ Wilson loop in \Eq{eq:Utildewilsloop} which reads
\bea
\exp \lp i \oint_{\mathcal{C}} dk_i \, \tilde{U}^\dag i \del_i \tilde{U} \rp &= e^{\frac{i}{2\phi} \sum_{n=0}^{L-1} d \mbf{k}_{n+1} \times \mbf{k}_n}  U^\dag(\mbf{k}_{L})  \mathcal{H}^{-d\mbf{k}_{L}} \lp \prod_{n}^{L-1 \leftarrow 1}  P(\mbf{k}_{n})  \mathcal{H}^{-d\mbf{k}_n} \rp U(\mbf{k}_{0})  \ .
\eea
Because we have taken $L \to \infty$, the Abelian overall phase becomes
\bea
\lim_{L\to \infty} \frac{i}{2\phi} \sum_{n=0}^{L-1} d \mbf{k}_{n+1} \times \mbf{k}_n  = \frac{i}{2\phi} \oint_{\mathcal{C}} dk_i \eps_{ij} k_j \ .
\eea
This prefactor exactly cancels the extraneous Abelian term in \Eq{eq:Winitialfull}, which now reads
\bea
\label{eq:Wilsonloopfinal}
\boxed{
W_{\mathcal{C}} = e^{-i \oint_{\mathcal{C}} d\mbf{k} \times \frac{1}{2}\pmb{\nabla} \log \vartheta ( \frac{(k_1,k_2)}{2\pi} | \Phi ) }  U^\dag(\mbf{k}_{L})  \mathcal{H}^{-d\mbf{k}_{L}} P(\mbf{k}_{L-1})  \dots  P(\mbf{k}_{2})  \mathcal{H}^{-d\mbf{k}_2} P(\mbf{k}_{1})  \mathcal{H}^{-d\mbf{k}_1} U(\mbf{k}_{0}) }
\eea
showing that the Wilson loop at $\phi = 2\pi$ factors into an Abelian winding factor that gives each Landau level a nonzero Chern number, and a non-Abelian product of projectors with insertions of $\mathcal{H}^{-d\mbf{k}}$. To understand the effect of the insertions, we consider the Berry connection (\Eq{eq:Acalform}) when \emph{all} bands are occupied, so $U(\mbf{k}) = \mathbb{1}$. Because $U(\mbf{k})$ is independent of $\mbf{k}$ in this case, the $\mbf{k}$ derivative term disappears and
\bea
\label{eq:Akconst}
\mathcal{A}_i^{MN}(\mbf{k}) &= [(- \eps_{ij} \tilde{Z}_j)]^{MN} - \frac{\delta^{MN}}{2} \eps_{ij}   \del_j \log \vartheta \lp \left. \frac{(k_1,k_2)}{2\pi} \right| \Phi \rp \ .
\eea
In a crystal at zero flux, the Berry connection of fully occupied bands is identically zero, but we see from the above expression that two contributions survive at $\phi = 2\pi$. Using \Eq{eqLZexptrace} to show that $\tilde{Z}_j$ is traceless, we see that only the Abelian term (the second term in \Eq{eq:Akconst}) contributes to the Berry curvature and therefore the winding. However, the non-Abelian term is nontrivial. We compute the Wilson loop along a straight contour $\mathcal{C} = (k_1, 0) \to (k_1, 2\pi)$, finding:
\bea
\label{eq:wilocc1}
W_{\mathcal{C}} =  \exp \lp  i \oint_{\mathcal{C}} dk_i \mathcal{A}_i(\mbf{k}) \rp  &= e^{i \oint_{\mathcal{C}} dk_i (- \eps_{ij} \tilde{Z}_j)} e^{-i \oint_{\mathcal{C}} d\mbf{k} \times \frac{1}{2}\pmb{\nabla} \log \vartheta \lp \left. \frac{(k_1,k_2)}{2\pi} \right| \Phi \rp }  \\
&= e^{-i  (2\pi \mbf{b}_2)_i \eps_{ij} \tilde{Z}_j} e^{-i \oint_{\mathcal{C}} d\mbf{k} \times \frac{1}{2}\pmb{\nabla} \log \vartheta \lp \left. \frac{(k_1,k_2)}{2\pi} \right| \Phi \rp }  \\
&= \mathcal{H}^{- 2\pi \mbf{b}_2} e^{-i \oint_{\mathcal{C}} d\mbf{k} \times \frac{1}{2}\pmb{\nabla} \log \vartheta \lp \left. \frac{(k_1,k_2)}{2\pi} \right| \Phi \rp }  \\
\eea
showing that $\mathcal{H}^{-2\pi \mbf{b}_2}$ can be interpreted as the non-Abelian factor of the Wilson loop in the fully occupied limit (note that it does not depend on $k_1$ so does not contribute to the winding). Importantly, this result agrees with \Eq{eq:Wilsonloopfinal} which in the $U(\mbf{k}) = \mathbb{1}$ limit reads
\bea
\label{eq:wilocc2}
W_{\mathcal{C}} &= e^{-i \oint_{\mathcal{C}} d\mbf{k} \times \frac{1}{2}\pmb{\nabla} \log \vartheta \lp \left. \frac{(k_1,k_2)}{2\pi} \right| \Phi \rp }   \mathcal{H}^{-d\mbf{k}_{L}} \lp \prod_{n}^{L-1 \leftarrow 1}   \mathcal{H}^{-d\mbf{k}_n} \rp \\
&= e^{-i \oint_{\mathcal{C}} d\mbf{k} \times \frac{1}{2}\pmb{\nabla} \log \vartheta \lp \left. \frac{(k_1,k_2)}{2\pi} \right| \Phi \rp } \prod_{n}^{L \leftarrow 1}  \mathcal{H}^{-\frac{2\pi}{L} \mbf{b}_2} \\
&= e^{-i \oint_{\mathcal{C}} d\mbf{k} \times \frac{1}{2}\pmb{\nabla} \log \vartheta \lp \left. \frac{(k_1,k_2)}{2\pi} \right| \Phi \rp }  \mathcal{H}^{-2\pi \mbf{b}_2} \\
\eea
using \Eq{eq:HcaldefZ} in the last line. The physical picture of \Eq{eq:wilocc1} and \Eq{eq:wilocc2} is that, unlike the atomic limits of zero-flux crystals, the fully occupied Landau level state has a nontrivial Wilson loop where every band winds identically due to the $\eps_{ij} \del_j \log \vartheta$ term. The $k_1$-independent $\mathcal{H}^{-2\pi \mbf{b}_2}$ term can be diagonalized to obtain the Wannier centers which are the Wilson loop eigenvalues \cite{Alexandradinata:2012sp,2012RvMP...84.1419M}.

\section{Square Lattice Calculations}
\label{app:square_lattice}

This section is brief, pedagogically oriented review of how Bloch's theorem is used to produce a matrix Hamiltonian on a plane wave basis at zero flux (\App{eq:zerofluxsquare}) and how our magnetic Bloch theorem is used to produce a matrix Hamiltonian on a Landau level basis at $2\pi$ flux (\App{app:squarelattice2pifluxHam}). We use a simple $p^2/(2m)$ kinetic Hamiltonian in a cosine potential on the square lattice.

\subsection{Zero Flux}
\label{eq:zerofluxsquare}

The zero-flux Hamiltonian is chosen to be
\begin{align}
H^{\phi=0}(\rr) = -\frac{1}{2}{\pmb{\nabla}}^2 + \frac{w}{2}(e^{-2\pi i\bbb_1 \cdot \rr} + e^{-2\pi i\bbb_2 \cdot \rr} + H.c.)
\end{align}
If $w = 0$, the solution to the Hamiltonian is simple -- the eigenstates are plane waves of the form
\begin{align}
\tilde{\psi}_{\tilde \kk}(\rr) = e^{i{\tilde \kk} \cdot \rr},
\label{eq:original_basis_square}
\end{align} where ${\tilde \kk}$ runs over the entire plane $\mathbb{R}^2$.  When the lattice potential is added, the continuous translation symmetry is broken down to a discrete symmetry indexed by $\kk$, where $\kk$ is defined in first Brillouin zone $\kk = k_1 \bbb_1 + k_2 \bbb_2, k_{1,2} \in (-\pi, \pi]$
. States at $\mbf{k} + 2\pi \mbf{G}$ where $\GG = 2\pi G_1 \bbb_1 + 2\pi G_2 \bbb_2, G_{1,2} \in \mathbb{Z}$ can be scattered to states at $\mbf{k}$ by the periodic potential. The states
\begin{align}
\psi_{\kk, \GG} (\rr) = \tilde{\psi}_{\kk - \GG} (\rr).
\label{}
\end{align}
form a basis of the Hilbert space on which we can find a representation of the Hamiltonian, which will necessarily be diagonal in $\mbf{k}$. This selection rule arises because $\psi_{\kk, \GG} (\rr)$ is an eigenstate of the translation operator: $\psi_{\kk, \GG} (\rr+\mbf{a}_i)  = e^{i (\mbf{k} + 2\pi \mbf{G}) \cdot \mbf{a}_i}\psi_{\kk, \GG} (\rr) = e^{i \mbf{k} \cdot \mbf{a}_i}\psi_{\kk, \GG} (\rr)$ using the fact that $\mbf{G} \cdot \mbf{a}_i \in \mathds{Z}$. For each crystal momentum $\kk$ in the first BZ, the plane wave index $\GG$ labels different states with the same crystal momentum.

We calculate the matrix elements of the Hamiltonian in the $\ket{\kk, \GG}$ basis, where $\langle \rr | \kk, \GG \rangle = \psi_{\kk, \GG} (\rr)$. Working in the $\kk, \GG$ basis, the Hamiltonian reads
\bea
\bra{\kk, \GG} H^{\phi = 0} \ket{\kk', \GG'}  &= \int d^2r d^2r' \, \braket{\kk, \GG | \rr} \bra{\rr} H^{\phi = 0} \ket{\rr'} \braket{\rr' | \kk', \GG'}\\
&= \int d^2r d^2r' \, \psi^*_{\kk, \GG}(\rr) H^{\phi = 0} (\rr) \delta(\rr - \rr') \psi_{\kk', \GG'}(\rr') \\
&= \int d^2r \, e^{-i(\kk - \GG) \cdot \rr} H^{\phi = 0} (\rr) e^{i(\kk' - \GG') \cdot \rr} \\
&= \int d^2r \, e^{-i(\kk - \GG) \cdot \rr} \left[\frac{1}{2}(\kk' - \GG')^2 + \frac{w}{2}(e^{-2\pi i\bbb_1 \cdot \rr} + e^{-2\pi i\bbb_2 \cdot \rr} + H.c.) \right] e^{i(\kk' - \GG') \cdot \rr} \\
\eea
The kinetic term has no position dependence, so
\bea
 \int d^2r \, e^{-i(\kk - \GG) \cdot \rr} \left[\frac{1}{2}(\kk' - \GG')^2 \right]  e^{i(\kk' - \GG') \cdot \rr} = (2\pi)^2 \delta(\mbf{k}-\mbf{k}')\left[\frac{1}{2}(\kk' - \GG')^2  \delta_{\mbf{G},\mbf{G}'}  \right]
\eea
which is diagonal in $\mbf{k}$ and $\mbf{G}$. The potential term does have position dependence and will not be diagonal in $\mbf{G}$:
\bea
&\frac{w}{2} \int d^2r \, e^{-i(\kk - \GG) \cdot \rr} \left[e^{-2\pi i\bbb_1 \cdot \rr} + e^{-2\pi i\bbb_2 \cdot \rr} + H.c. \right] e^{i(\kk' - \GG') \cdot \rr} \\
&\qquad\qquad\qquad\qquad =   \frac{w}{2} \int d^2r \, \left[e^{-i(\kk - \GG - \kk' + \GG' - 2\pi \bbb_1)\cdot \rr}  + e^{-i(\kk - \GG - \kk' + \GG'  - 2\pi \bbb_2)\cdot \rr} + H.c. \right]  \\
&\qquad\qquad\qquad\qquad = (2\pi)^2 \delta(\mbf{k}-\mbf{k}') \frac{w}{2} \left[ \delta_{\mbf{G}+2\pi \mbf{b}_1,\mbf{G}'} +\delta_{\mbf{G}'+2\pi \mbf{b}_1,\mbf{G}}  + \delta_{\mbf{G}+2\pi \mbf{b}_2,\mbf{G}}  + \delta_{\mbf{G}'+2\pi \mbf{b}_2,\mbf{G}}  \right] \ .
\eea
This term is still diagonal in $\mbf{k}$ as ensures by the translation symmetry. We derive the Bloch Hamiltonian
\bea
\bra{\kk, \GG} H^{\phi = 0} \ket{\kk', \GG'} &= (2\pi) \delta(\mbf{k}-\mbf{k}') H^{\phi=0}_{\mbf{G},\mbf{G}'}(\mbf{k}), \\
 H^{\phi=0}_{\mbf{G},\mbf{G}'}(\mbf{k}) &= \frac{1}{2}(\kk' - \GG')^2  \delta_{\mbf{G},\mbf{G}'}  + \frac{w}{2} (\delta_{\mbf{G}+2\pi \mbf{b}_1,\mbf{G}'} +\delta_{\mbf{G}'+2\pi \mbf{b}_1,\mbf{G}}  + \delta_{\mbf{G}+2\pi \mbf{b}_2,\mbf{G}}  + \delta_{\mbf{G}'+2\pi \mbf{b}_2,\mbf{G}} ) \ .
\eea
The Bloch Hamiltonian can be thought of as an infinite matrix on the $\mbf{G}$ basis as a function of $\mbf{k}$. By imposing a cutoff on the $\mbf{G}$ plane wave states, this matrix can be diagonalized and the spectrum obtained.

\subsection{$2\pi$-flux}
\label{app:squarelattice2pifluxHam}
The Hamiltonian at $2\pi$ flux is created via canonical substitution, sending $-i\pmb{\nabla} \rightarrow \pmb{\pi}$:
\begin{align}
H^{\phi = 2\pi}(\rr) = \frac{1}{2}{\pmb{\pi}}^2 + \frac{w}{2}(e^{-2\pi i\bbb_1 \cdot \rr} + e^{-2\pi i\bbb_2 \cdot \rr} + H.c.).
\label{}
\end{align}
At $2\pi$ flux, the magnetic translation operators commute with $H^{\phi = 2\pi}(\rr)$ and allow us to represent $H^{\phi = 2\pi}(\rr)$ as a matrix acting on the $\ket{\mbf{k},n}$ eigenstates of the magnetic translation operators. We see that the Landau levels play the role of the plane waves at zero flux. The substantial difference is that, instead of simple Bloch states, the magnetic translation eigenstates are given in \Eq{eq:statedefAppa}.

\begin{figure}
\includegraphics[width=0.5\textwidth]{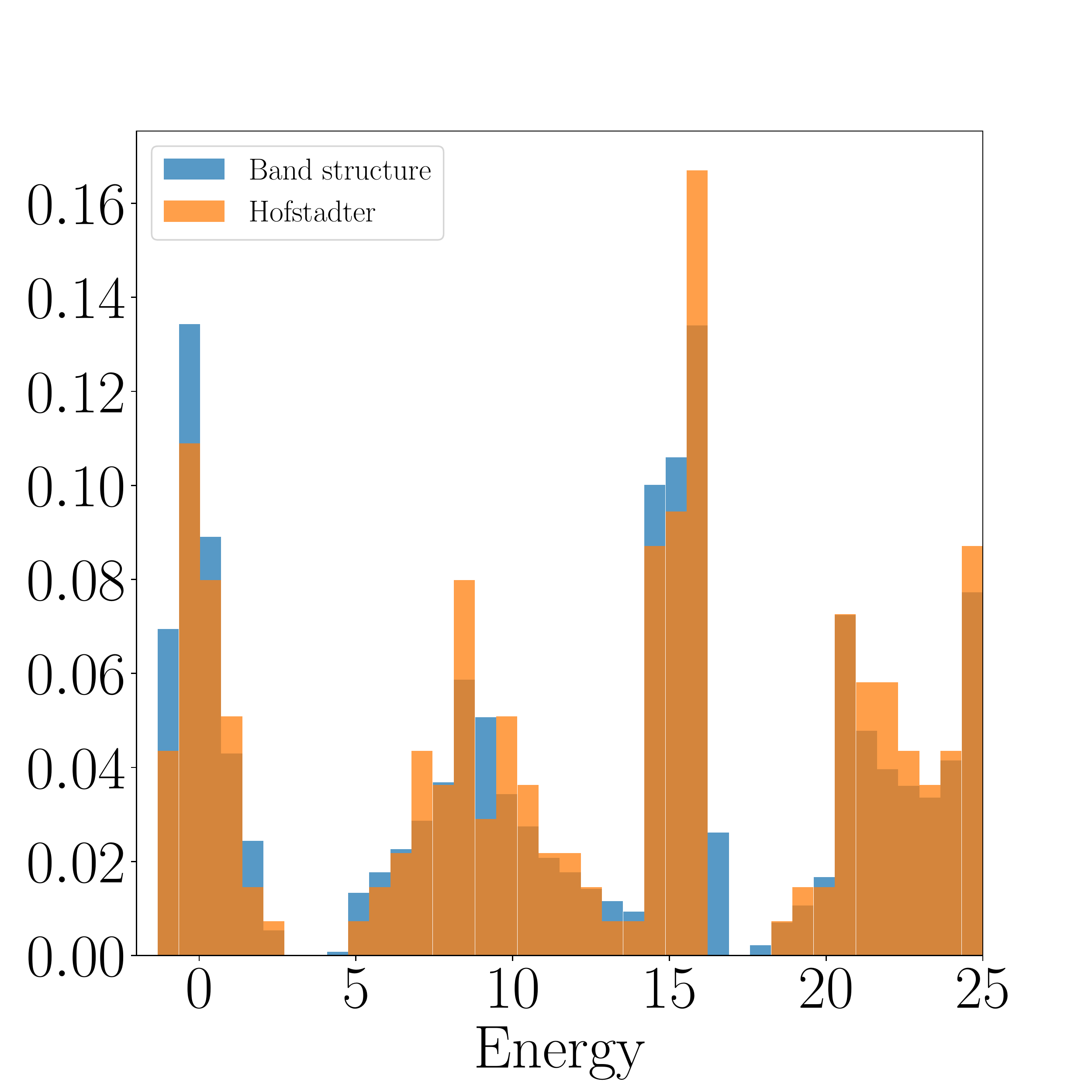}
\caption{Density of states computed from the band structure (blue) versus the open momentum space method (orange) developed in Ref.~\onlinecite{2021PhRvB.103L1405L}. The small discrepancies  can be improved by sampling more points in both methods.
}
\label{fig:square_comparison}
\end{figure}

The matrix elements in the magnetic translation basis read
\bea
\bra{\kk,m} H^{\phi = 2\pi}(\rr) \ket{\kk',n} &= \bra{\kk,m} \phi(a^\dagger a + \frac{1}{2}) + \frac{w}{2}(e^{-2\pi i\bbb_1 \cdot \rr} + e^{-2\pi i\bbb_2 \cdot \rr} + H.c.) \ket{\kk',n}  \\
&= (2\pi)^2 \delta(\kk - \kk')\left[ \phi(m +\frac{1}{2}) \delta_{mn} + \frac{w}{2} (e^{- i k_2}  \mathcal{H}^{2\pi \mbf{b}_1}_{mn} + e^{i k_1} \mathcal{H}^{2\pi \mbf{b}_2}_{mn} + H.c.) \right] \\
\label{}
\eea
where we used $\frac{1}{2}{\pmb{\pi}}^2 = \phi(a^\dag a + \frac{1}{2})$ which acts diagonal on the $m,n$ Landau level indices because $a^\dag a \ket{n} = n\ket{n}$. The matrix elements of the potential term were computed in \App{app:siegelcomplete}, and an expression for $\mathcal{H}^{2\pi \mbf{G}}$ may be found in \Eq{eq:Hcaldef}. Thus the magnetic Bloch Hamiltonian $H^{\phi=2\pi}_{mn}(\kk)$ is defined
\bea
\bra{\kk,m} H^{\phi = 2\pi}(\rr) \ket{\kk',n}  &= (2\pi)\delta(\mbf{k}-\mbf{k}') H^{\phi=2\pi}_{mn}(\kk)  \ .
\eea
In analogy to the zero flux case, $H^{\phi=2\pi}_{mn}(\kk)$ is a matrix on the $mn$ Landau level indices as a function of $\mbf{k}$, and can be diagonalized by imposing a Landau level cutoff.  See Fig.~\ref{fig:square_comparison} for a comparison for the density of states obtained by our band structures versus the open momentum space method in Ref.~\onlinecite{2021PhRvB.103L1405L}.

\section{Coulomb Interaction in a Gauge-Invariant Formalism}
\label{app:gcgii}

In this Appendix, we show how to write the Coulomb interaction
\bea
H_{int} &= \frac{1}{2} \int d^2r d^2r' \, n(\mbf{r}) V(\mbf{r}-\mbf{r}') n(\mbf{r}')
 = \frac{1}{2} \int \frac{d^2q}{(2\pi)^2} \,  V(\mbf{q}) \rho_{-\mbf{q}} \rho_{\mbf{q}},
 \eea
 in the terms of the magnetic translation group eigenstates in a gauge-invariant manner. We keep $V(\mbf{q})$ general but require the Fourier transform is even, $V(\mbf{q}) = V(-\mbf{q})$, and is non-negative real, $V(\mbf{q}) \geq 0$, to preserve the positive semi-definite structure. We require a formula for the Fourier components of the density
\bea
\rho_{\mbf{q}}=  \int d^2r \, e^{-i \mbf{q} \cdot \mbf{r}} n(\mbf{r})
\eea
written in terms of the magnetic translation group eigenstates. The single-particle magnetic translation group eigenstates are
\bea
\psi_{\mbf{k},n} =   \frac{1}{\sqrt{\mathcal{N}(\mbf{k})}} \sum_{\mbf{R}} e^{-i \mbf{k}\cdot \mbf{R}} T_{\mbf{a}_1}^{\mbf{R}\cdot \mbf{b}_1}T_{\mbf{a}_2}^{\mbf{R}\cdot \mbf{b}_2} \frac{a^{\dag n}}{\sqrt{n!}} \psi_{0}, \qquad a  \psi_{0} = b  \psi_{0} = 0  \ .
\eea
The single-particle eigenstates arises from diagonalizing the single-particle Hamiltonian on this basis, yielding matrix eigenvectors $U^N_n(\mbf{k})$ where $N$ is the band index and $n$ is the Landau level orbital index. The continuum electron operators obey $\{c^\dag(\mbf{r}), c(\mbf{r}') \} = \delta(\mbf{r}-\mbf{r}')$, and the creation operators for the magnetic translation group basis states are denoted $\gamma^\dag_{\mbf{k},N} = \sum_n U^N_n(\mbf{k}) \psi^\dag_{\mbf{k},n}$.  The magnetic translation group eigenstates obey
\bea
\braket{\mbf{r}|\psi^\dag_{\mbf{k},n} |0} = \braket{0| c(\mbf{r}) \psi^\dag_{\mbf{k},n} |0}  = \psi_{\mbf{k},n}(\mbf{r}) \ .
\eea
Our goal now is to express the density operator $n(\mbf{r}) = c^\dag(\mbf{r})c(\mbf{r})$ in terms of the single-particle eigenstates.  If there are any internal indices of $c$, like spin or valley in TBG, they are implicitly summed over:  $n(\mbf{r}) = \sum_\alpha c_\alpha^\dag(\mbf{r})c_\alpha(\mbf{r})$. Because $\rho(\mbf{q})$ is bilinear in fermion operators, we only need to calculate its single-particle matrix elements. Inserting resolutions of the identity and using $\braket{\mbf{r}'|n(\mbf{r})|\mbf{r}''} = \delta(\mbf{r}-\mbf{r}'')$, we find
\bea
\braket{0 |\psi_{\mbf{k}',m}  \rho_{\mbf{q}} \psi^\dag_{\mbf{k},n} | 0} &= \int d^2r \, e^{-i\qq \cdot \rr} \braket{0 |\psi_{\mbf{k}',m}  c^\dagger(\rr) c(\rr) \psi^\dag_{\mbf{k},n} | 0}  \\
 &= \int d^2r \, e^{-i\qq \cdot \rr}  \braket{0 |\psi_{\mbf{k}',m}  c^\dagger(\rr) |0 } \braket{ 0 | c(\rr) \psi^\dag_{\mbf{k},n} | 0} \\
 &= \int d^2r \, e^{-i\qq \cdot \rr}  \psi^*_{\mbf{k}',m} (\rr ) \psi_{\mbf{k},n} (\rr ) \\
&= \braket{\mbf{k}',m |e^{-i \mbf{q} \cdot \mbf{r}} |\mbf{k},n} \ .
\eea
where in the second line we have inserted a resolution of identity, but because of fermion conservation only the vacuum survives. Hence we obtain an explicit expression for
\bea
\rho_{\mbf{q}} &= \sum_{mn} \int \frac{d^2kd^2k'}{(2\pi)^4} \braket{\mbf{k}',m| e^{-i \mbf{q} \cdot \mbf{r} } | \mbf{k},n} \psi^\dag_{\mbf{k}',m}  \psi_{\mbf{k},n} \\
\eea
where the matrix element $\braket{\mbf{k}',m| e^{-i \mbf{q} \cdot \mbf{r} } | \mbf{k},n} $ was computed in \Eq{eq:qformfactorderiv}. This is essentially the density operator $\rho_\qq$ in the Landau level basis. Thus we have determined the coefficients in the expansion of $\rho_\mbf{q}$:
\bea
\rho_{\mbf{q}} &= \sum_{mn} \int \frac{d^2kd^2k'}{(2\pi)^4} \braket{\mbf{k}',m| e^{-i \mbf{q} \cdot \mbf{r} } | \mbf{k},n} \psi^\dag_{\mbf{k}',m}  \psi_{\mbf{k},n} =\sum_{mn} \int \frac{d^2k}{(2\pi)^2} e^{i \xi_\mbf{q}(\mbf{k})} \psi^\dag_{\mbf{k} - \mbf{q},m} \mathcal{H}^{\mbf{q}}_{mn} \psi_{\mbf{k},n} \ . \\
\eea
For convenience, we recall that
\bea
e^{i \xi_\mbf{q}(\mbf{k})} &= \frac{e^{- \frac{ \bar{q} q}{4\phi}}  \vartheta \lp \left. \frac{(k_1 - q/2 ,k_2+ i q/2)}{2\pi} \right| \Phi \rp  }{\sqrt{\vartheta \lp \left. \frac{(k_1 ,k_2)}{2\pi} \right| \Phi \rp \vartheta \lp \left. \frac{(k_1 - q_1 ,k_2 - q_2)}{2\pi} \right| \Phi \rp}} \\
\mathcal{H}^{\mbf{q}}_{mn} &=  \bra{m} \exp \lp i \frac{\gamma_q a + \bar{\gamma}_q a^\dag }{\sqrt{2 \phi}} \rp \ket{n} =  e^{ - \frac{ \bar{\gamma}_q \gamma_q}{4\phi}}  \begin{cases}
\sqrt{\frac{m!}{n!}} \lp \frac{i \gamma_q}{\sqrt{2 \phi}} \rp^{n-m} L_m^{|n-m|}\lp \frac{\bar{\gamma}_q\gamma_q}{2\phi}  \rp, & n > m \\
\sqrt{\frac{n!}{m!}} \lp \frac{i \bar{\gamma}_q}{\sqrt{2 \phi}} \rp^{m-n} L_n^{|m-n|}\lp \frac{\bar{\gamma}_q\gamma_q}{2\phi}  \rp, & m > n \\
\end{cases}
\eea
which are both unitary. It is simple now to write $\rho_{\mbf{q}}$ in terms of the single-particle eigenstate basis by recalling that in the $N$th band
\bea
\gamma^\dag_{\mbf{k},N} &= \sum_m U^N_m(\mbf{k}) \psi^\dag_{\mbf{k},m} \ . \\
\eea
The creation operator $\gamma_{\kk, N}^\dagger$ has no relation to the momentum factor $\gamma_q = \epsilon_{ij} q_i \bar{z}_j$.  If our Landau levels carried additional indices $\alpha$, the unitary $U$ would also be a matrix in $\alpha, \beta$.

Orthogonality of the eigenvectors at a given $\mbf{k}$ gives $\sum_N U^{N *}_n(\mbf{k}) \gamma^\dag_{\mbf{k},N}  = \psi^\dag_{\mbf{k},n}$ and hence
\bea
\rho_{\mbf{q}} &=  \int \frac{d^2k}{(2\pi)^2} e^{i \xi_\mbf{q}(\mbf{k})} \sum_{MN} \gamma^\dag_{\mbf{k}- \mbf{q},M}  [U^\dag(\mbf{k}-\mbf{q}) \mathcal{H}^{\mbf{q}} U(\mbf{k})]_{MN} \gamma_{\mbf{k},N} \ . \\
\eea
To project the Hamiltonian into a set of low energy bands, one merely restricts the sum over $MN$. Our final step is to define the form factor
\bea
\label{eq:formfactorM}
M_{MN}(\mbf{k},\mbf{q}) \equiv e^{i \xi_\mbf{q}(\mbf{k})} [U^\dag(\mbf{k}-\mbf{q}) \mathcal{H}^{\mbf{q}} U(\mbf{k})]_{MN} \ .
\eea
A few comments about $M(\mbf{k},\mbf{q})$ are in order. For all $\mbf{q}$, the eigenvalues of $M(\mbf{k},\mbf{q})$ have magnitudes less than or equal to $1$ because $e^{i \xi_\mbf{q}(\mbf{k})} \mathcal{H}^{\mbf{q}} $ is unitary and $U(\mbf{k})$ is composed from normalized eigenvectors. At generic $\mbf{q}$, $M(\mbf{k},\mbf{q})$ does not have a gauge-invariant spectrum because $U(\mbf{k})$ and $U(\mbf{k}-\mbf{q})$ are eigenvector matrices and each come with arbitrary phases. At $\mbf{q} = 2\pi \mbf{G}$, $U(\mbf{k})$ and $U(\mbf{k}+2\pi\mbf{G})$ are identical because $\mbf{k}$ is periodic on the BZ. This is because the states $\psi_{\mbf{k},n}(\mbf{r})$ are explicitly periodic in $\mbf{k}$, so Hamiltonian is explicitly translation-invariant. Thus $M(\mbf{k},2\pi\mbf{G})$ has a gauge-invariant spectrum. This eigenvector gauge-invariance is important because later expressions will depend on the spectrum of $M(\mbf{k},2\pi\mbf{G})$. Using $\mathcal{H}^{-\mbf{q}}  = \mathcal{H}^{\mbf{q} \, \dag} $, we find
\bea
\label{eq:Mherm}
M(\mbf{k},\mbf{q})^\dag = e^{-i \xi_\mbf{q}(\mbf{k})} U^\dag(\mbf{k}) \mathcal{H}^{-\mbf{q}}  U(\mbf{k}-\mbf{q}) = e^{-i \xi_\mbf{q}(\mbf{k}) - i \xi_{-\mbf{q}}(\mbf{k}-\mbf{q})} M(\mbf{k}- \mbf{q},-\mbf{q}) = M(\mbf{k}- \mbf{q},-\mbf{q})
\eea
where we used the theta function identity $e^{-i \xi_\mbf{q}(\mbf{k}) - i \xi_{-\mbf{q}}(\mbf{k}-\mbf{q})} = 1$. This is proved by observing
\bea
\label{eq:xiidentity}
e^{i \xi_{\mbf{q}}(\mbf{k}) + i \xi_{-\mbf{q}}(\mbf{k}-\mbf{q}) } &= \frac{e^{- \frac{ \bar{q} q}{4\phi}}  \vartheta \lp \left. \frac{(k_1 - q/2 ,k_2+ i q/2)}{2\pi} \right| \Phi \rp  }{\sqrt{\vartheta \lp \left. \frac{(k_1 ,k_2)}{2\pi} \right| \Phi \rp \vartheta \lp \left. \frac{(k_1 - q_1 ,k_2 - q_2)}{2\pi} \right| \Phi \rp}} \frac{e^{- \frac{ \bar{q} q}{4\phi}}  \vartheta \lp \left. \frac{(k_1 -q_1 + q/2 ,k_2 -q_2 - i q/2)}{2\pi} \right| \Phi \rp  }{\sqrt{\vartheta \lp \left. \frac{(k_1 - q_1 ,k_2 - q_2)}{2\pi} \right| \Phi \rp \vartheta \lp \left. \frac{(k_1 ,k_2)}{2\pi} \right| \Phi \rp }}  \\
&= e^{- \frac{ \bar{q} q}{2\phi}}  \frac{\vartheta \lp \left. \frac{(k_1 - q/2 ,k_2+ i q/2)}{2\pi} \right| \Phi \rp   \vartheta \lp \left. \frac{(k_1 - \bar{q}/2 ,k_2 - i \bar{q}/2)}{2\pi} \right| \Phi \rp }{\vartheta \lp \left. \frac{(k_1 ,k_2)}{2\pi} \right| \Phi \rp \vartheta \lp \left. \frac{(k_1 - q_1 ,k_2 - q_2)}{2\pi} \right| \Phi \rp}  \\
&= e^{- \frac{ \bar{q} q}{2\phi}}  \frac{\left |\vartheta \lp \left. \frac{(k_1 - q/2 ,k_2+ i q/2)}{2\pi} \right| \Phi \rp  \right|^2}{\vartheta \lp \left. \frac{(k_1 ,k_2)}{2\pi} \right| \Phi \rp \vartheta \lp \left. \frac{(k_1 - q_1 ,k_2 - q_2)}{2\pi} \right| \Phi \rp} >0 \\
\eea
where we have used \Eq{eq:realtheta} which states $\theta(\mbf{z}^*|\Phi)$  =  $\theta(\mbf{z}|\Phi)^*$ and that $\theta(\mbf{z}|\Phi)$ is real and positive for $\mbf{z} \in \mathbb{R}^2$. Because $|e^{i \xi_{\mbf{q}}(\mbf{k}) + i \xi_{-\mbf{q}}(\mbf{k}-\mbf{q}) }| = 1$ and  $e^{i \xi_{\mbf{q}}(\mbf{k}) + i \xi_{-\mbf{q}}(\mbf{k}-\mbf{q}) }$ is real and positive, we find $e^{i \xi_{\mbf{q}}(\mbf{k}) + i \xi_{-\mbf{q}}(\mbf{k}-\mbf{q}) } = 1$. \Eq{eq:Mherm} is the Hermiticity of form factor, as described in Ref.~\onlinecite{2020arXiv200912376B} in zero flux where it is also true that $M^\dagger(\kk,\qq) = M(\kk-\qq,-\qq)$.

Finally, we remark that $M(\mbf{k} + 2\pi \mbf{G},\mbf{q}) = M(\mbf{k},\mbf{q})$ because $U(\mbf{k} + 2\pi \mbf{G}) = U(\mbf{k})$ and $e^{i \xi_{\mbf{q}}(\mbf{k} + 2\pi \mbf{G})} = e^{i \xi_{\mbf{q}}(\mbf{k})}$ as follows from the periodicity of the Siegel theta functions in $\mbf{k}$. However, $ M(\mbf{k},\mbf{q})$ is not periodic in $\mbf{q}$, which is expected because $\mbf{q} \in \mathbb{R}^2$ is a continuum momentum.
As a function of $\mbf{q}$ at fixed $mn$, $\mathcal{H}^\mbf{q}_{mn}$ is the product of a decaying exponential factor in $\gamma_\mbf{q} \bar \gamma_\mbf{q} \sim |\mbf{q}|^2$ and a polynomial factor in $\gamma_\mbf{q}$. Thus for large $\mbf{q}$, this term will decay exponentially but with subleading power law growth.

We end this section with the explicit form of the Coulomb interaction Hamiltonian. The density operator is written in terms of the form factor via
\bea
\rho_{\mbf{q}} &=  \int_{BZ} \frac{d^2k}{(2\pi)^2} \sum_{MN} M_{MN}(\mbf{k},\mbf{q}) \gamma^\dag_{\mbf{k}- \mbf{q},M}  \gamma_{\mbf{k},N}
\eea
which leads to the expression
\bea
H_{int} = \frac{1}{2} \int \frac{d^2q}{(2\pi)^2} \,  V(\mbf{q}) \rho_{-\mbf{q}} \rho_{\mbf{q}} &=  \frac{1}{2} \int \frac{d^2q}{(2\pi)^2} \,   O_{-\mbf{q}} O_{\mbf{q}}, \qquad O_\mbf{q} = \sqrt{V(\mbf{q})} \int_{BZ} \frac{d^2k}{(2\pi)^2} \sum_{MN} M_{MN}(\mbf{k},\mbf{q}) \gamma^\dag_{\mbf{k}- \mbf{q},M}  \gamma_{\mbf{k},N}
\eea
which is positive semi-definite because $O_{-\mbf{q}} = O_{\mbf{q}}^\dag$ as follows from \Eq{eq:Mherm}. For numerical calculations, it is convenient to discretize the momentum integrals into a sum over $L^2$ terms according to
\bea
 \int_{BZ} \frac{d^2k}{(2\pi)^2} f(\mbf{k}) \to \frac{1}{\Omega_{tot}} \sum_{\mbf{k} \in BZ} f(\mbf{k}), \qquad  \int_{\mathbb{R}^2} \frac{d^2q}{(2\pi)^2} f(\mbf{q}) \to \frac{1}{\Omega_{tot}} \sum_{\mbf{G}} \sum_{\mbf{k} \in BZ} f(\mbf{k} +2\pi \mbf{G})
 \label{eq:discretization_sums}
\eea
where $\Omega_{tot} = L^2 \Omega$ is the total area of an $L\mbf{a}_1 \times L \mbf{a}_2$ sample, and there are $L^2$ terms in the BZ sums, and the sum $\mbf{G}$ is over the reciprocal lattice vectors $\mbf{G} = \mathds{Z} \mbf{b}_1 + \mathds{Z} \mbf{b}_2$. Notice that in Eq.~\eqref{eq:discretization_sums}, the $\kk$ sum goes over the BZ. The $\mbf{q}$ integral is over all of $\mathbb{R}^2$. It is useful to write $\mbf{q} = \mbf{k} + 2\pi \mbf{G}$ where $\mbf{k}$ is integrated over the BZ and $\mbf{G}$ is summed over the reciprocal lattice. \Eq{eq:discretization_sums} follows by approximating the $BZ$ integral as a sum.

\section{Bistritzer-MacDonald Hamiltonian at $2\pi$ Flux}
\label{app:BMHam}

In this Appendix, we discuss the Bistritzer-MacDonald (BM) model of TBG which is the central physical motivation for this work. In \App{app:spBMham}, we lay out our conventions for the moir\'e unit cell, the Bistritzer-MacDonald (BM) Hamiltonian, and the explicit form of the single-particle Hamiltonian at $2\pi$ flux. We then discuss the symmetries of the full two-valley system which are relevant for the many-body calculations in \App{eq:BMsymm}. A discussion of the strong coupling expansion used to treat the Coulomb interaction is given in \App{app:coulham}. \App{eq:TBGBMgs} contains a derivation of the exact eigenstates at even integer fillings. 

\subsection{Moir\'e Lattice Conventions and Single-particle Hamiltonian}
\label{app:spBMham}

Let us set our conventions for the geometry of the moir\'e twist unit cell. First, the graphene unit cell has a lattice vector of length $a_{g} = .246$nm and an area $\Omega_{g}  = a_{g}^2 \frac{\sqrt{3}}{2}$. The gaphene K point is $\mbf{K}_{g} = \frac{2\pi}{a_g}(0, 2/3)$. The moir\'e vectors are defined by the difference in momentum space of the rotated layers' $K$ points:
\bea
2\pi \mbf{q}_1 = (R_{\theta/2}-R_{-\theta/2})\mbf{K}_{g}, \quad \mbf{q}_j = C_3 \mbf{q}_{j-1}, \quad 2\pi |\mbf{q}_j| =  k_\th = 2 \sin \frac{\th}{2} \, |\mbf{K}_{g}| =  \frac{8\pi  \sin \frac{\th}{2} }{3a_{g}} \ . \\
\eea
Note that in this convention, we keep a factor of $2\pi$ explicit. The moir\'e reciprocal lattice vectors are defined
\bea
\mbf{b}_j = \mbf{q}_j - \mbf{q}_3, \quad \mbf{b}_1 \times \mbf{b}_2 &= \frac{(2 \sin \frac{\th}{2})^2}{\Omega_{g}}, \qquad \mbf{k} = k_1 \mbf{b}_1 + k_2 \mbf{b}_2, \ k_i \in (-\pi,\pi) \ . \\
\eea
Note that the moir\'e BZ has area $(2\pi)^2 \mbf{b}_1 \times \mbf{b}_2$ which is smaller than the graphene BZ by a factor of $(2 \sin \frac{\th}{2})^2 \sim \th^2$. This is because the lattice constant of the moir\'e unit cell is larger by a factor of $\th$. The moire lattice is defined by $\mbf{a}_i \cdot \mbf{b}_j = \delta_{ij}$ which yields in cartesian coordinates
\bea
\mbf{a}_1 = \frac{a_{g}}{2 \sin \frac{\th}{2}}  \lp -\frac{\sqrt{3}}{2}, - \frac{1}{2} \rp \quad \mbf{a}_2 = \frac{a_{g}}{2 \sin \frac{\th}{2}} \lp \frac{\sqrt{3}}{2},  -\frac{1}{2} \rp, \qquad \Omega = \mbf{a}_1 \times \mbf{a}_2 = \frac{\Omega_{g}}{(2 \sin \frac{\th}{2})^2} \ . \\
\eea
In the range of the magic angle at $\theta = 1.05^\circ$, the moir\'e unit cell is 3000 times larger than the graphene unit cell. The magnetic translation group is commutative when $\phi = \frac{e B \Omega }{\hbar} = 2\pi$ where we temporarily restore $\hbar = h/(2\pi)$. In physical units, the corresponding magnetic field is in the range
\bea
B &= \frac{h }{e\Omega} = \frac{h }{e\Omega_{g}} \theta^2 \in (25,32)\text{T for } \theta \in (1.03^\circ, 1.15^\circ) \ .
\eea
The BM Hamiltonian at $\phi = 2\pi$ can be written (with $\sigma_i$ and $T_j$ acting on the sublattice indices)
\bea
\label{eq:Hmoirepot}
H^{\phi}_{BM}(\mbf{r}) &= \bpm
v_F \pmb{\pi} \cdot \pmb{\sigma} &  T(\mbf{r})^\dag \\
T(\mbf{r})  & v_F \pmb{\pi} \cdot \pmb{\sigma} \\
\epm, \qquad T(\mbf{r}) = \sum_{j=1}^3 T_j e^{i 2\pi \mbf{q}_{j} \cdot \mbf{r}}, \quad T_{j+1} = \bpm w_0 & w_1 e^{-\frac{2\pi i}{3} j }\\ w_1 e^{\frac{2\pi i}{3} j } & w_0 \\ \epm \ .
\eea
In this gauge, Hamiltonian is not in Bloch form because the off-diagonals are do not have the lattice periodicity. This is because of the offset of the layer BZs. To remedy this, we perform a unitary transformation with the momentum shift matrix $\text{diag}(e^{i \pi \mbf{q}_1 \cdot \mbf{r}}, e^{-i \pi \mbf{q}_1 \cdot \mbf{r}})$ which puts the Hamiltonian in Bloch form
\bea
\label{eq:HqshiftK}
H^{\phi}(\mbf{r}) &= \bpm
v_F \pmb{\pi} \cdot \pmb{\sigma} - \pi v_F \mbf{q}_1 \cdot \pmb{\sigma}&  T_1 +T_2 e^{-i 2\pi \mbf{b}_1 \cdot \mbf{r}} + T_3 e^{-i 2\pi \mbf{b}_2 \cdot \mbf{r}}\\
T_1 +T_2 e^{i 2\pi \mbf{b}_1 \cdot \mbf{r}} + T_3 e^{i 2\pi \mbf{b}_2 \cdot \mbf{r}} & v_F \pmb{\pi} \cdot \pmb{\sigma} + \pi v_F \mbf{q}_1 \cdot \pmb{\sigma} \\
\epm, \qquad \mbf{b}_j = \mbf{q}_{j+1} - \mbf{q}_1 \ .
\eea
We now need to compute the overlaps of this matrix with the magnetic Bloch irrep states. A suitable basis is $\ket{\mbf{k}, l,\al, n} \equiv  \ket{l} \otimes \ket{\al} \otimes \ket{\mbf{k},n} $ where $l = 0,1$ is the layer index and $\al = 0,1$ is the sublattice index. The Hamiltonian in this basis reads
\bea
\label{eq:magblochham}
H^{\phi = 2\pi}(\mbf{k}) = \bpm  v_F k_\th (\sqrt{\frac{\phi}{2\pi}}  h(\pmb{\pi}) - \frac{1}{2}\sigma_2) &  T_1 + T_2 e^{-i k_2} \mathcal{H}^{2\pi\mbf{b}_1} + T_3  e^{i k_1} \mathcal{H}^{2\pi\mbf{b}_2} \\
T_1 + T_2 e^{i k_2} \mathcal{H}^{-2\pi\mbf{b}_1} + T_3  e^{-i k_1} \mathcal{H}^{-2\pi\mbf{b}_2}  & v_F k_\th (\sqrt{\frac{\phi}{2\pi}}  h(\pmb{\pi})  + \frac{1}{2} \sigma_2)
\epm, \quad h(\pmb{\pi}) =   \lp\frac{3\sqrt{3}}{2\pi} \rp^{1/2}  \bpm
0 & a^\dag \\
a& 0 \\
\epm
\eea
where $a,a^\dag, \mathcal{H}^{2\pi \mbf{b}_1}$ are all matrices in the Landau level basis. We compute the band structure of this matrix by truncating the Landau levels from $n=0,\dots, n_{LL}$. Because of the truncation, the highest Landau level will be annihilated by the kinetic term, and introduces a two spurious states at each $\mbf{k}$ (one per layer). These bands are easily removed.

\subsection{Symmetries}
\label{eq:BMsymm}

In this section, we discuss the single-particle BM Hamiltonians in the $K$ and $K'$ valleys and demonstrate the $U(1)$ valley symmetry, $C_{2z}$, and $P$ symmetries which survive at all flux and are essential for the Coulomb Hamiltonian in \App{app:coulham}. We also discuss the chiral symmetry $C$.

The BM Hamiltonian at zero flux reads
\bea
H_K^{\phi = 0} = \bpm
-iv_F \pmb{\nabla} \cdot \pmb{\sigma}  & T(\rr)^\dagger \\
T(\rr) & -iv_F  \pmb{\nabla} \cdot \pmb{\sigma}
\epm,
\label{eq:BMsymlay}
\eea
where the $T(\mbf{r})$ moir\'e potentials are given in \Eq{eq:Hmoirepot}. In particular, $T_1 = w_0 \sigma_0 + w_1 \sigma_1$ and $T_j = e^{\frac{2\pi i (j-1)}{3} \sigma_3} T_1 e^{-\frac{2\pi i (j-1)}{3} \sigma_3}$. The matrices $\sigma_i$ are Pauli matrices acting on the graphene sublattice index. The layer index is often in matrix notation, as in \Eq{eq:BMsymlay}, and we will use $\tau_i$ Pauli matrices to denote the layer index as well. For many-body calculations, we must also study the Hamiltonian in the $K'$ valley, which is related to the $K$ valley Hamiltonian at $\phi =0$ by spinless time reversal $T$ which acts as complex conjugation:
\bea
H_{K'}^{\phi = 0} = \bpm
iv_F \pmb{\nabla} \cdot \pmb{\sigma}^*  & T^*(\rr)^\dagger \\
T^*(\rr) & iv_F  \pmb{\nabla} \cdot \pmb{\sigma}^*
\epm \ .
\label{eq:BMKprime}
\eea
The two Hamiltonians in \Eq{eq:BMsymlay} and \Eq{eq:BMKprime} can be written as
\bea
H^{\phi = 0}_K &= -iv_F \tau_0 (\sigma_1 \del_x + \sigma_2 \del_y) + \sum_{j=1}^3 \tau_1 \cos(2\pi \qq_j \cdot \rr) T_j + \tau_2 \sin(2\pi \qq_j \cdot \rr) T_j \\
H^{\phi = 0}_{K'} &= iv_F \tau_0 (\sigma_1 \del_x - \sigma_2 \del_y) + \sum_{j=1}^3 \tau_1 \cos(2\pi \qq_j \cdot \rr) T_j^* - \tau_2 \sin(2\pi \qq_j \cdot \rr) T_j^*.
\label{eq:Hkkp}
\eea
We will use $\mu_i$ as the Pauli matrices acting on the graphene valley degree of freedom while $\tau_i$ are the Pauli matrices acting on the layer. In this notation, \Eq{eq:Hkkp} can be written as one equation:
\bea
\label{eq:HtwoV}
H^{\phi = 0}(\mbf{r}) &= -iv_F \tau_0 (\mu_3 \sigma_1 \del_x + \mu_0 \sigma_2 \del_y) + \sum_{j=1}^3 (\mu_0 \tau_1  \cos 2\pi \qq_j \cdot \rr + \mu_3 \tau_2  \sin 2\pi \qq_j \cdot \rr )  e^{\frac{2\pi i}{3} (j-1) \mu_3 \sigma_3} T_1 e^{-\frac{2\pi i}{3} (j-1) \mu_3 \sigma_3}  \\
\eea
which is the full two-valley Hamiltonian. It is direct to check that time reversal is a symmetry of $H^{\phi = 0}$ (which it must be because $H^{\phi = 0}$ is simply the $K$ valley with its time-reversed partner) by using the form $\mathcal{T} = \mu_1 K, \mathcal{T}^2 = +1$ on the two-valley model:
\bea
\label{eq:TRrealspace}
\mathcal{T} H^{\phi = 0}(\mbf{r})  \mathcal{T}^{-1} &= \mu_1 \lp iv_F \tau_0 (\mu_3 \sigma_1 \del_x -\mu_0 \sigma_2 \del_y) + \sum_{j=1}^3 (\mu_0 \tau_1  \cos 2\pi \qq_j \cdot \rr - \mu_3 \tau_2  \sin 2\pi \qq_j \cdot \rr )  e^{-\frac{2\pi i}{3} (j-1) \mu_3 \sigma_3} T_1 e^{\frac{2\pi i}{3} (j-1) \mu_3 \sigma_3} \rp \mu_1 \\
&=  iv_F \tau_0 (-\mu_3 \sigma_1 \del_x -\mu_0 \sigma_2 \del_y) + \sum_{j=1}^3 (\mu_0 \tau_1  \cos 2\pi \qq_j \cdot \rr + \mu_3 \tau_2  \sin 2\pi \qq_j \cdot \rr )  e^{(-1)^2\frac{2\pi i}{3} (j-1) \mu_3 \sigma_3} T_1 e^{- \frac{2\pi i}{3} (j-1) \mu_3 \sigma_3} \\
&= H^{\phi = 0}(\mbf{r}) \ .
\eea
It is now a simple matter to obtain the real space Hamiltonian at nonzero flux with the canonical substitution $-i\pmb{\nabla} \to \pmb{\pi}$ yielding
\bea
\label{eq:HtwoVflux}
H^{\phi}(\mbf{r}) &= v_F \tau_0 (\mu_3 \sigma_1 \pi_x + \mu_0 \sigma_2 \pi_y) + \sum_{j=1}^3 (\mu_0 \tau_1  \cos 2\pi \qq_j \cdot \rr + \mu_3 \tau_2  \sin 2\pi \qq_j \cdot \rr )  e^{\frac{2\pi i}{3} (j-1) \mu_3 \sigma_3} T_1 e^{-\frac{2\pi i}{3} (j-1) \mu_3 \sigma_3}  \\
\eea
Eq. \ref{eq:HtwoVflux} is the central result of this section, making it a simple matter to determine the symmetries of the model at all flux, and then transform them into a momentum or Landau level basis. It is direct to check that
\bea
\mathcal{T} H^{\phi}(\mbf{r})  \mathcal{T}^{-1} = H^{-\phi}(\mbf{r})
\eea
because $\mathcal{T} (-i \pmb{\nabla} + e \mbf{A}(\mbf{r}) ) \mathcal{T}^{-1} =  i \pmb{\nabla} + e \mbf{A}(\mbf{r})$, so $-i\pmb{\nabla}$ and $\mbf{A}$ do not transform the same way unless the magnetic field is flipped. This is expected because a magnetic field is known to break time reversal.

The first and most important symmetry is $U(1)$ valley \footnote{More carefully, TBG possesses a $U(2) \times U(2)$ symmetry: spin and number conservation within each valley.} which is a continuous symmetry generated by $\mu_3$ (note that $\mu_1$ and $\mu_2$ do not appear in  \Eq{eq:HtwoVflux} because the valleys are decoupled) which remains at all flux. There are three discrete symmetries of interest to us: inversion $C_{2z}$, unitary particle-hole $P$, and chiral symmetry $C$ when $w_0 = 0$. All three survive in nonzero flux. The actions of these symmetries on the \emph{full two-valley} Hamiltonian are
\bea
C_{2z} H^{\phi}(\mbf{r}) C_{2z}^\dag &= H^{\phi}(\mbf{r}) , \qquad C_{2z}^2 = +1 \\
P H^{\phi}(\mbf{r}) P^\dag &= -H^{\phi}(\mbf{r}) , \qquad P^2 = -1 \\
C H^{\phi}(\mbf{r}) C&= -H^{\phi}(\mbf{r}) , \qquad C^2 = +1, \qquad \text{when $w_0 = 0$} \ . \\
\eea
We now derive the forms of these symmetries in real space for all flux, momentum space at zero flux, and on the Landau level basis of the magnetic Bloch Hamiltonian at $2\pi$ flux.

\subsubsection{Real Space, all flux}
\label{app:realspaceallflux}

We begin by proving the real-space form of the symmetries which can be derived from first principles in the continuum model. The results of this section alone are sufficient to check the physical symmetries of the BM model in flux. In \App{eq:momspacesym}, we check that the same symmetries can also be written in momentum space, matching the expressions derived in \Ref{Song_2019}.

To start, we analyze the $C_{2z}$ which acts as a $\pi$ rotation around the center of the graphene honeycomb unit cell. Hence, it takes $\mbf{r} \to - \mbf{r}$ (switching the sublattice index) and hence inverts the momentum (switching the graphene valley). $C_{2z}$ performs a rotation around the vector normal to the plane, and thus it does \emph{not} act on the layer index. From this geometry, we arrive at
\bea
\label{eq:defC2zealspace}
C_{2z} &= \sigma_1 \mu_1 e^{i\pi L_z}
\eea
where $L_z$  is the angular momentum operator given by
\begin{align}
L_z = \frac{\pmb{\pi}^2 - \mbf{Q}^2}{2 eB} = -i \epsilon_{\mu \nu} x_\mu \partial_\nu
\label{}
\end{align}
where in the second equality we used the symmetric gauge. By working in the symmetric gauge, one can see explicitly that the $L_z$ operator is the same as at zero flux. However, we prefer to use the gauge-invariant $L_z$ operator for the sake of generality. In this case it is easy to show in any gauge that the gauge-invariant inversion operator $e^{i \pi L_z}$ reverses the canonical momentum
\bea
e^{i \pi L_z} \pmb{\pi} e^{-i \pi L_z} = - \pmb{\pi} \ .
\eea
We check directly that $D[C_{2z}]$ is a symmetry by computing
\bea
C_{2z} H^{\phi}(\mbf{r}) C_{2z}^\dag &= \sigma_1 \mu_1 \lp -v_F \tau_0 (\mu_3 \sigma_1 \pi_x + \mu_0 \sigma_2 \pi_y) + \sum_{j=1}^3 (\mu_0 \tau_1  \cos 2\pi \qq_j \cdot \rr - \mu_3 \tau_2  \sin 2\pi \qq_j \cdot \rr )  e^{\frac{2\pi i}{3} (j-1) \mu_3 \sigma_3} T_1 e^{-\frac{2\pi i}{3} (j-1) \mu_3 \sigma_3} \rp \sigma_1 \mu_1 \\
&=  -v_F \tau_0 (-\mu_3 \sigma_1 \del_x - \mu_0 \sigma_2 \del_y) + \sum_{j=1}^3 (\mu_0 \tau_1  \cos 2\pi \qq_j \cdot \rr + \mu_3 \tau_2  \sin 2\pi \qq_j \cdot \rr )  e^{(-1)^2 \frac{2\pi i}{3} (j-1) \mu_3 \sigma_3} T_1 e^{-(-1)^2\frac{2\pi i}{3} (j-1) \mu_3 \sigma_3}  \\
&= H^{\phi}(\mbf{r})
\eea
which is the correct behavior for the two-valley Hamiltonian. We now discuss the unitary particle-hole symmetry $P$ which acts as inversion in real space and also interchanges the layers, so it takes the form
\bea
\label{eq:Pfirstapp}
P &= i \tau_2 \mu_3 e^{i\pi L_z}  \ .
\eea
The factor of $i$ ensures $P$ is real and commutes with $\mathcal{T}$. We check the action of $P$ directly:
\bea
P H^{\phi}(\mbf{r}) P^\dag &= \tau_2 \mu_3\lp -v_F \tau_0 (\mu_3 \sigma_1 \pi_x + \mu_0 \sigma_2 \pi_y) + \sum_{j=1}^3 (\mu_0 \tau_1  \cos 2\pi \qq_j \cdot \rr - \mu_3 \tau_2  \sin 2\pi \qq_j \cdot \rr )  e^{\frac{2\pi i}{3} (j-1) \mu_3 \sigma_3} T_1 e^{-\frac{2\pi i}{3} (j-1) \mu_3 \sigma_3} \rp \tau_2 \mu_3 \\
&= -v_F \tau_0 (\mu_3 \sigma_1 \pi_x + \mu_0 \sigma_2 \pi_y) + \sum_{j=1}^3 (-\mu_0 \tau_1  \cos 2\pi \qq_j \cdot \rr - \mu_3 \tau_2  \sin 2\pi \qq_j \cdot \rr )  e^{\frac{2\pi i}{3} (j-1) \mu_3 \sigma_3} T_1 e^{-\frac{2\pi i}{3} (j-1) \mu_3 \sigma_3}  \\
&= - H^{\phi}(\mbf{r}) \ .
\eea
An identical calculation shows that $P' = i \tau_2 \mu_0 e^{i\pi L_z}$ (and thus any diagonal $\mu$ matrix) also anti-commutes with $H^\phi(\mbf{r})$, which shows that $P$ is not only an anti-commuting symmetry of the two-valley model, but also of each valley individually. This is due to the $U(1)$ valley symmetry. Lastly, it is trivial to show that the chiral symmetry
\bea
\label{eq:chiralsymreal}
C = \tau_0 \sigma_3 \mu_0
\eea
anti-commutes with $H^\phi(\mbf{r})$ when $w_0 = 0$ (the first chiral limit) because
\bea
\sigma_3 e^{\frac{2\pi i}{3} (j-1) \mu_3 \sigma_3} T_1 e^{-\frac{2\pi i}{3} (j-1) \mu_3 \sigma_3} \sigma_3 &= w_1 e^{\frac{2\pi i}{3} (j-1) \mu_3 \sigma_3} \sigma_3 \sigma_1 \sigma_3 e^{-\frac{2\pi i}{3} (j-1) \mu_3 \sigma_3}  \\
&= -w_1 e^{\frac{2\pi i}{3} (j-1) \mu_3 \sigma_3} \sigma_1 e^{-\frac{2\pi i}{3} (j-1) \mu_3 \sigma_3}  \\
\eea
and $\sigma_3$ trivially anti-commutes with the kinetic term because it only contains $\sigma_1$ and $\sigma_2$ matrices. Note that $\tau_0 \sigma_3 \mu_3$ also anti-commutes with $H^\phi(\mbf{r})$, so $C$ is an anti-commuting symmetry of both valleys individually.

Our calculations establish that $C_{2z}$, $P$, and the chiral symmetry $C$ at $w_0 = 0$ are unbroken by flux. Because $\mathcal{T}$ reverses the flux, $C_{2z}\mathcal{T}$ is not a symmetry at $\phi \neq 0$, which has a significant effect on the topology. Finally, it is direct to check that $\{P, C_{2z}\} = 0$ arising from the anti-commutation of the valley factors $\mu_1$ and $\mu_3$ and hence $(C_{2z} P)^2 = - P^2 C_{2z}^2 = +1$. The most important observation is that $C_{2z} P = \mu_2 \tau_2 \sigma_1$ which is local in real space: the $e^{i \pi L_z}$ operators have canceled.

\subsubsection{Zero flux, Momentum space}
\label{eq:momspacesym}

Although it is not required for this work, we now connect the real space symmetry operators we have derived in \App{app:realspaceallflux} to the momentum space symmetries familiar in zero flux.  We will follow heavily Ref.~\onlinecite{2020arXiv200912376B}.

First, let us ignore interlayer coupling in the BM model. The two sheets of graphene have four Dirac cones (two for layer and two for spin.)  If the standard sheet of graphene has Dirac cones at $\tilde{\kk} = \KK_g, -\KK_g$, then in twisted bilayer graphene they occur at $\KK_+ = M_{\th/2} \KK_g, -\KK_+ = -M_{\th/2} \KK_g,$ for the top layer,  and $\KK_- = M_{-\th/2} \KK_g, -\KK_- = -M_{-\th/2} \KK_g$ for the bottom. Here $M_{\pm\th/2}$ is a rotation matrix. Adding interlayer coupling reduces the graphene translation symmetry to the (approximate) moir\'e translation symmetry;  momenta separated by reciprocal lattice vectors $\QQ = 2\pi m \bbb_1 + 2\pi n \bbb_2$ become identified. In the folded moir\'e BZ, the four Dirac cones are located at the moir\'e $K$ points $\KK_M = \pi \qq_1, -\KK_M = -\pi \qq_1$ which each host $2$ Dirac cones (one per valley).  Define the two ${\cal Q}$ sublattices to be
\begin{align}
{\cal Q}_+ &= \{\pi \qq_1 + 2\pi m \bbb_1 + 2\pi n \bbb_2| m,n \in \mathbb{Z} \} \\
{\cal Q}_- &= \{-\pi \qq_1 + 2\pi m \bbb_1 + 2\pi n \bbb_2| m,n \in \mathbb{Z} \}.
\label{}
\end{align}
The two Dirac points $\KK_+, -\KK_- \in {\cal Q}_+$, while $\KK_-, -\KK_+ \in {\cal Q}_-$.  Thus, if we define
\begin{align}
\zeta_\QQ = \begin{cases}
+1, ~~\QQ \in {\cal Q}_+ \\
-1, ~~\QQ \in {\cal Q}_-,
\end{cases}
\label{}
\end{align}
then $\zeta_\QQ = \eta l$, with $\eta$ as the valley and $l$ the layer. We are now in a position to define the Fourier-transformed states.
\begin{align}
\label{eq:cdagmoire}
c^\dagger_{\kk, \QQ, \eta, \alpha, s} = \frac{1}{\sqrt{N}} \sum_{\tilde{\RR} \text{ in layer }l} e^{i(\kk + \eta \KK_l - \QQ) \cdot (\tilde{\RR} + \pmb{t}_\alpha)} c_{\tilde{\RR}, l, \alpha, s}^\dagger, \qquad \text{with $\mbf{Q}, \eta,$ and $l$ obeying } \ l = \zeta_\QQ \eta,
\end{align}
where $\pmb{t}_\alpha $ is the graphene sublattice vector, $\al$ is the sublattice index, $l = \pm$ is the layer, $\tilde{\mbf{R}}$ is the graphene Bravais lattice, and $s$ is the spin. The decoupled valleys are indexed by $\eta = \pm$.

We now study the matrix form of the symmetry operators.  Define the unitary component of the symmetry $g$ as $D(g)$, where
\begin{align}
g c_{\kk, \QQ, \eta, \alpha, s}^\dagger g^{-1} = c_{g\kk, \QQ', \eta', \alpha', s'}^\dagger [D(g)]_{\QQ, \eta, \alpha, s, \QQ', \eta', \alpha', s'} \ .
\label{}
\end{align}
Applying $T$ to \Eq{eq:cdagmoire} gives
\begin{align}
&{\cal T} c^\dagger_{\kk, \QQ, \eta, \alpha, s} {\cal T}^{-1} = \frac{1}{\sqrt{N}} \sum_{\tilde{\RR} \text{ in layer }l} e^{-i(\kk + \eta \KK_l - \QQ) \cdot (\tilde{\RR} + \pmb{t}_\alpha)} {\cal T} c_{\tilde{\RR}, l, \alpha, s}^\dagger {\cal T}^{-1} \\
&= c^\dagger_{-\kk, -\QQ, -\eta, \alpha, s}
\label{}
\end{align}
so $D[{\cal T}]= \delta_{\QQ, -\QQ'} \mu_1$ and $\mathcal{T}$ takes $\kk \rightarrow -\kk$. Because both $\zeta_\QQ$ and $\eta$ change, the layer remains the same. This matches the real-space picture because $\mathcal{T}$ is local.

A similar calculation holds for $C_{2z}$ and $C$.  For $C$, which only acts as $\pm 1$ on the sublattice, define $\xi_\alpha = +1$ for sublattice A and $-1$ for sublattice B.  Then
\begin{align}
  &{C} c^\dagger_{\kk, \QQ, \eta, \alpha, s} {C}^{-1} = \frac{1}{\sqrt{N}} \sum_{\tilde{\RR} \text{ in layer }l} e^{-i(\kk + \eta \KK_l - \QQ) \cdot (\tilde{\RR} + \pmb{t}_\alpha)} {C} c_{\tilde{\RR}, l, \alpha, s}^\dagger {C}^{-1} \\
  &= \frac{1}{\sqrt{N}} \sum_{\tilde{\RR} \text{ in layer }l} e^{-i(\kk + \eta \KK_l - \QQ) \cdot (\tilde{\RR} + \pmb{t}_\alpha)} c_{\tilde{\RR}, l, \alpha, s}^\dagger \xi_\alpha \\
\label{}
\end{align}
so $D[{C}] = \sigma_3$ and $C$ takes $\kk \rightarrow \kk$. For $C_{2z}$, we use $\mbf{t}_\alpha = -\mbf{t}_\beta$ when $\alpha \neq \beta$:
\begin{align}
  &{C_{2z}} c^\dagger_{\kk, \QQ, \eta, \alpha, s} {C_{2z}}^{-1} = \frac{1}{\sqrt{N}} \sum_{\tilde{\RR} \text{ in layer }l} e^{-i(\kk + \eta \KK_l - \QQ) \cdot (\tilde{\RR} + \pmb{t}_\alpha)} {C_{2z}} c_{\tilde{\RR}, l, \alpha, s}^\dagger {C_{2z}}^{-1} \\
  &= \frac{1}{\sqrt{N}} \sum_{\tilde{\RR} \text{ in layer }l} e^{-i(\kk + \eta \KK_l - \QQ) \cdot (\tilde{\RR} + \pmb{t}_\alpha)} c_{\tilde{-\RR}, l, \beta, s}^\dagger [\sigma_1]_{\beta\alpha} \\
  &= \frac{1}{\sqrt{N}} \sum_{\tilde{\RR} \text{ in layer }l} e^{-i(\kk + \eta \KK_l - \QQ) \cdot (-\tilde{\RR} - \pmb{t}_\beta)} c_{\tilde{\RR}, l, \beta, s}^\dagger [\sigma_1]_{\beta\alpha} \\
  &= c^\dagger_{-\kk, -\QQ, -\eta, \beta, s} [\sigma_1]_{\beta\alpha} \\
\label{}
\end{align}
so $D[{C_{2z}}] = \delta_{\QQ,-\QQ'} \mu_1 \sigma_1$ and takes $\kk \rightarrow -\kk$. It is important to note that, like $\mathcal{T}$, $C_{2z}$ switches the sign of valley and $\mbf{Q}$, the layer remains the same. This matches the real-space picture because $C_{2z}$ is a rotation in the plane, and does not reverse the layer.

The particle-hole symmetry $P$ differs from $\mathcal{T}$ and $C_{2z}$ because $P$ (which takes $\mbf{R} \to - \mbf{R}$ and $l\to -l$ as in \Eq{eq:Pfirstapp}) is not a true symmetry of moir\'e superlattice \cite{Song_2019}, and only emerges when $\th \to 0$ in the continuum BM model. As such, $P$ does not have a well-defined action on the $c_{\tilde{\RR}, l, \alpha, s}$ because taking $\mbf{R} \to - \mbf{R}$ and $l \to - l$ is not an exact symmetry of the moir\'e lattice. We will simply give the momentum-space form of $P$ in a single valley from \Ref{Song_2019}, which is
\begin{align}
D[P] = \zeta_\QQ \delta_{\QQ, -\QQ'} \mu_3,~\kk \rightarrow -\kk \ .
\label{}
\end{align}
The crucial observation is that $D[P]$ acts on the layer index $l$ (determined by $\mbf{Q} \in \eta \mathcal{Q}_l$) as $i \tau_2$ because $\zeta_\QQ$ takes opposite values for opposite $l$ and taking $\mbf{Q} \to - \mbf{Q}$ reverses the layer. This exactly matches the real space symmetry derived in \Eq{eq:Pfirstapp}.


\subsubsection{$2\pi$-flux, Magnetic Bloch Hamiltonian}
\label{eq:C2zPLL}

We now consider the magnetic Bloch Hamiltonian at $2\pi$ flux, which is written as a matrix with layer, sublattice, and Landau level indices in a given valley. We derive the forms of the particle-hole $P$ and inversion symmetries $C_{2z}$ whose product $C_{2z} P$ is a local symmetry of the two-valley Bistrizter-MacDonald (BM) Hamiltonian. Recall that at $\phi = 2\pi$, the matrix Hamiltonian in the $K$ graphene valley is
\bea
\label{eq:hamrealqlabe}
H^{\phi =2\pi}_{K}(\mbf{r}) &= \bpm
v_F \pmb{\pi} \cdot \pmb{\sigma} - \pi v_F \mbf{q}_1\cdot \pmb{\sigma} &  T_1 + T_2 e^{-2\pi i \mbf{b}_1 \cdot \mbf{r}} + T_3 e^{-2\pi i \mbf{b}_2 \cdot \mbf{r}} \\
T_1 + T_2 e^{2\pi i \mbf{b}_1 \cdot \mbf{r}} + T_3 e^{2\pi i \mbf{b}_2 \cdot \mbf{r}} & v_F \pmb{\pi} \cdot \pmb{\sigma} + \pi v_F \mbf{q}_1\cdot \pmb{\sigma} \\
\epm \ .
\eea
which in the Landau level basis reads
\bea
\label{eq:magblochham}
H_K^{\phi = 2\pi}(\mbf{k}) = \bpm  v_F k_\th (\sqrt{\frac{\phi}{2\pi}}  h(\pmb{\pi}) - \frac{1}{2}\sigma_2) &  T_1 + T_2 e^{-i k_2} \mathcal{H}^{2\pi\mbf{b}_1} + T_3  e^{i k_1} \mathcal{H}^{2\pi\mbf{b}_2} \\
T_1 + T_2 e^{i k_2} \mathcal{H}^{-2\pi\mbf{b}_1} + T_3  e^{-i k_1} \mathcal{H}^{-2\pi\mbf{b}_2}  & v_F k_\th (\sqrt{\frac{\phi}{2\pi}}  h(\pmb{\pi})  + \frac{1}{2} \sigma_2)
\epm, \quad h(\pmb{\pi}) =   \lp\frac{3\sqrt{3}}{2\pi} \rp^{1/2}  \bpm
0 & a^\dag \\
a& 0 \\
\epm \ .
\eea
We will also need an explicit expression for the magnetic Bloch Hamiltonian in the $K'$ valley which we obtained from \Eq{eq:Hkkp}, which in Bloch form reads
\bea
\label{eq:magblochhamprime}
H^{\phi =2\pi}_{K'}(\mbf{r}) &= \bpm
-v_F \pmb{\pi} \cdot \pmb{\sigma}^* + \pi v_F \mbf{q}_1\cdot \pmb{\sigma} &  T_1 + T_2^* e^{2\pi i \mbf{b}_1 \cdot \mbf{r}} + T_3^* e^{2\pi i \mbf{b}_2 \cdot \mbf{r}} \\
T_1 + T_2' e^{- 2\pi i \mbf{b}_1 \cdot \mbf{r}} + T_3' e^{- 2\pi i \mbf{b}_2 \cdot \mbf{r}} & -v_F \pmb{\pi} \cdot \pmb{\sigma}^* - \pi v_F \mbf{q}_1\cdot \pmb{\sigma} \\
\epm \ .
\eea
As explained in \Eq{eq:HqshiftK}, we have performed a momentum shift between the two layers to account for the $-2\pi\mbf{q}_1$ displacement between the Dirac cones in the $K'$ valley (compared to a $+2\pi\mbf{q}_1$ displacement in the $K$ valley).

Now following the same steps as in \Eq{eq:magblochham}, we obtain the magnetic Bloch Hamiltonian:
\bea
\label{eq:HmagKpblochLL}
H_{K'}^{\phi = 2\pi}(\mbf{k}) = \bpm  v_F k_\th (\sqrt{\frac{\phi}{2\pi}}  h'(\pmb{\pi}) + \frac{1}{2}\sigma_2) &  T_1 + T_2^* e^{i k_2} \mathcal{H}^{-2\pi\mbf{b}_1} + T_3^* e^{-i k_1} \mathcal{H}^{-2\pi\mbf{b}_2} \\
T_1 + T_2^* e^{-i k_2} \mathcal{H}^{2\pi\mbf{b}_1} + T_3^* e^{i k_1} \mathcal{H}^{2\pi\mbf{b}_2}  & v_F k_\th (\sqrt{\frac{\phi}{2\pi}}  h'(\pmb{\pi}) - \frac{1}{2} \sigma_2)
\epm, \quad h'(\pmb{\pi}) =  -\lp\frac{3\sqrt{3}}{2\pi} \rp^{1/2}  \bpm
0 & a\\
a^\dag& 0 \\
\epm \ .
\eea
 In particular, the kinetic term is obtained via
\bea
h'(\pmb{\pi}) &= - \pi_x \sigma_x + \pi_y \sigma_y \\
&= - (\pi_x \sigma_x - \pi_y \sigma_y) \\
&=  - \lp\frac{3\sqrt{3}}{2\pi} \rp^{1/2}  \bpm
0 & a\\
a^\dag & 0 \\
\epm \\
&= - \sigma_1  h(\pmb{\pi}) \sigma_1 \\
\eea
and the potential term is merely complex conjugated (it does not have explicit flux dependence). The identities $T_2^* = T_3, T_3^* = T_2$ and $T_1^* = T_1$ will be used throughout the section. We now derive the action of the continuum symmetry operators in \Eq{app:realspaceallflux} on the magnetic translation group irreps. The only nontrivial action is that of the rotation operator $e^{i \pi L_z}$ which obeys
\bea
e^{i \pi L_z} \pmb{\pi} e^{-i \pi L_z} &= - \pmb{\pi}, \\
e^{i \pi L_z} \mbf{Q} e^{-i \pi L_z} &= - \mbf{Q},
\eea
and thus $\{e^{i \pi L_z}, a^\dag\} = 0$ because $a^\dag= (\pi_x + i \pi_y)/\sqrt{2eB}$. Acting on \Eq{eq:knexplicit}, we derive
\bea
\label{eq:LLang}
e^{i \pi L_z} \ket{\mbf{k}, n} &= \frac{1}{\sqrt{\mathcal{N}(\mbf{k})}} \sum_{\mbf{R}} e^{-i \mbf{k} \cdot \mbf{R} +  i \frac{\phi}{2} (\mbf{R} \cdot \mbf{b}_1)(\mbf{R} \cdot \mbf{b}_2)} e^{i \pi L_z} e^{i\mbf{R} \cdot \mbf{Q}} e^{-i \pi L_z} e^{i \pi L_z}\ket{n} \\
&= \frac{1}{\sqrt{\mathcal{N}(\mbf{k})}} \sum_{\mbf{R}} e^{-i \mbf{k} \cdot \mbf{R} +  i \frac{\phi}{2} (\mbf{R} \cdot \mbf{b}_1)(\mbf{R} \cdot \mbf{b}_2)}  e^{i\mbf{R} \cdot (-\mbf{Q})} e^{i \pi L_z}\frac{a^{\dag n}}{\sqrt{n!}} \ket{0} \\
&= \frac{1}{\sqrt{\mathcal{N}(\mbf{k})}} \sum_{\mbf{R}} e^{-i \mbf{k} \cdot \mbf{R} +  i \frac{\phi}{2} (\mbf{R} \cdot \mbf{b}_1)(\mbf{R} \cdot \mbf{b}_2)}  e^{i\mbf{R} \cdot (-\mbf{Q})} (-1)^n \ket{n} \\
&= (-1)^n \frac{1}{\sqrt{\mathcal{N}(\mbf{k})}} \sum_{\mbf{R}} e^{-i \mbf{k} \cdot (-\mbf{R}) +  i \frac{\phi}{2} (-\mbf{R} \cdot \mbf{b}_1)(-\mbf{R} \cdot \mbf{b}_2)}  e^{i\mbf{R} \cdot \mbf{Q}}  \ket{n} \\
&= (-1)^n \ket{-\mbf{k}, n}
\eea
which is expected because the Landau level state $\ket{n}$ has angular momentum $n$. With this result, we can determine the action of the $C_{2z}$ and $P$ symmetries on the $\ket{\mbf{k},n}$ basis. It will be useful to note $(-1)^{a^\dag a} \ket{-\mbf{k}, n}  = (-1)^n \ket{-\mbf{k}, n}$.

First we prove that particle-hole symmetry in the $K$ valley is implemented by the operator
\bea
\label{eq:PHsym}
P_K= \bpm
0 & 1 \\
-1 & 0 \\
\epm (-1)^{a^\dag a} = i \tau_2 \sigma_0 (-1)^{a^\dag a}, \qquad [(-1)^{{a}^\dag a}]_{mn} = (-1)^m \delta_{mn} \\
\eea
which is determined from \Eq{eq:Pfirstapp} in real space using \Eq{eq:LLang} to derive the Landau level parity operator $(-1)^{a^\dagger a}$, which obeys $\{ (-1)^{{a}^\dag a}, a \} = \{ (-1)^{{a}^\dag a}, a^\dag \} = 0$ because $[a^\dag a, a] = -1$. We abuse notation and refer to $(-1)^{{a}^\dag a}$ as the matrix representation of the Landau level operators on the Landau level basis. Thus $(-1)^{a^\dag  a}$ anti-commutes with $h(\pmb{\pi})$ which is linear in $a,a^\dag$. We also need  $(-1)^{{a}^\dag a} \mathcal{H}^\mbf{q} (-1)^{{a}^\dag a} = \mathcal{H}^{-\mbf{q}}$ as is apparent from \Eq{eq:Hcaldef}. Using these results, we have
\bea
\label{eq:PHmatrixcheck}
P_K H_K^{\phi=2\pi}(\mbf{k}) P_K^\dag &= \bpm
0 & 1 \\
-1 & 0 \\
\epm   \bpm  v_F k_\th (-\sqrt{\frac{\phi}{2\pi}}  h(\pmb{\pi}) - \frac{1}{2}\sigma_2) &  T_1 + T_2 e^{-i k_2} \mathcal{H}^{-2\pi\mbf{b}_1} + T_3  e^{i k_1} \mathcal{H}^{-2\pi\mbf{b}_2} \\
T_1 + T_2 e^{i k_2} \mathcal{H}^{2\pi\mbf{b}_1} + T_3  e^{-i k_1} \mathcal{H}^{2\pi\mbf{b}_2}  & v_F k_\th (-\sqrt{\frac{\phi}{2\pi}}  h(\pmb{\pi})  + \frac{1}{2} \sigma_2)
\epm \bpm
0 & -1 \\
1 & 0 \\
\epm \\
&=  \bpm  v_F k_\th (-\sqrt{\frac{\phi}{2\pi}}  h(\pmb{\pi})  + \frac{1}{2} \sigma_2) & -(T_1 + T_2 e^{i k_2} \mathcal{H}^{2\pi\mbf{b}_1} + T_3  e^{-i k_1} \mathcal{H}^{2\pi\mbf{b}_2} )  \\
-( T_1 + T_2 e^{-i k_2} \mathcal{H}^{-2\pi\mbf{b}_1} + T_3  e^{i k_1} \mathcal{H}^{-2\pi\mbf{b}_2}) & v_F k_\th (-\sqrt{\frac{\phi}{2\pi}}  h(\pmb{\pi}) - \frac{1}{2}\sigma_2) \\
\epm \\
&= - \bpm  v_F k_\th (\sqrt{\frac{\phi}{2\pi}}  h(\pmb{\pi})  - \frac{1}{2} \sigma_2) & T_1 + T_2 e^{i k_2} \mathcal{H}^{2\pi\mbf{b}_1} + T_3  e^{-i k_1} \mathcal{H}^{2\pi\mbf{b}_2}  \\
 T_1 + T_2 e^{-i k_2} \mathcal{H}^{-2\pi\mbf{b}_1} + T_3  e^{i k_1} \mathcal{H}^{-2\pi\mbf{b}_2} & v_F k_\th (\sqrt{\frac{\phi}{2\pi}}  h(\pmb{\pi}) + \frac{1}{2}\sigma_2) \\
\epm \\
&= -H_K^{\phi=2\pi}(-\mbf{k}) \ .
\eea
We now study the $K'$ valley where particle-hole is implemented by the operator
\bea
\label{eq:PHsymK'}
P_{K'} = - \bpm
0 & 1 \\
-1 & 0 \\
\epm (-1)^{a^\dag a} = - i \tau_2 \sigma_0 (-1)^{a^\dag a}
\eea
which differs from \Eq{eq:PHsym} only by an overall minus. We are free to choose the sign of $P$ independently in each valley because of the $U(1)$ valley symmetry. We check \Eq{eq:PHsym} directly following the same steps in \Eq{eq:PHmatrixcheck}:
\bea
P_{K'} H_{K'}^{\phi=2\pi}(\mbf{k}) P_{K'}^\dag &= \bpm
0 & 1 \\
-1 & 0 \\
\epm   \bpm  v_F k_\th ( - \sqrt{\frac{\phi}{2\pi}} h'(\pmb{\pi}) + \frac{1}{2}\sigma_2) &  T_1 + T_2^* e^{i k_2} \mathcal{H}^{+2\pi\mbf{b}_1} + T_3^* e^{-i k_1} \mathcal{H}^{+2\pi\mbf{b}_2} \\
T_1 + T_2^* e^{-i k_2} \mathcal{H}^{- 2\pi\mbf{b}_1} + T_3^* e^{i k_1} \mathcal{H}^{- 2\pi\mbf{b}_2}  & v_F k_\th (-\sqrt{\frac{\phi}{2\pi}}  h'(\pmb{\pi}) - \frac{1}{2} \sigma_2)
\epm \bpm
0 & -1 \\
1 & 0 \\
\epm \\
&= -  \bpm  v_F k_\th (\sqrt{\frac{\phi}{2\pi}}  h'(\pmb{\pi}) + \frac{1}{2}\sigma_2) &  T_1 + T_2^* e^{-i k_2} \mathcal{H}^{-2\pi\mbf{b}_1} + T_3^* e^{i k_1} \mathcal{H}^{-2\pi\mbf{b}_2} \\
T_1 + T_2^* e^{i k_2} \mathcal{H}^{2\pi\mbf{b}_1} + T_3^* e^{-i k_1} \mathcal{H}^{2\pi\mbf{b}_2}  & v_F k_\th (\sqrt{\frac{\phi}{2\pi}}  h'(\pmb{\pi}) - \frac{1}{2} \sigma_2)
\epm \\
&= -H_{K'}^{\phi=2\pi}(-\mbf{k}) \ .
\eea
Thus we have checked the particle-hole operator in both valleys, yielding the final expression
\bea
\label{eq:Pvalley2}
P = \mu_3 \bpm
0 & 1 \\
-1 & 0 \\
\epm (-1)^{a^\dag a} = i \mu_3 \tau_2 \sigma_0 (-1)^{a^\dag a}
\eea
where $\mu_3$ is a Pauli matrix on the valley index. The fact that we can choose the $\pm$ sign freely in different valleys is a feature of the $U(1)$ valley quantum number. The same is true in real space as in \Eq{eq:Pfirstapp}.

We now study $C_{2z}$ which is not a symmetry of the one-valley BM model because it reverses the valley. Using \Eq{eq:LLang}, the expression for $C_{2z}$ on the Landau level basis is
\bea
\label{eq:C2zsym}
C_{2z} = \mu_1 \tau_0 \sigma_1 (-1)^{a^\dag a} \\
\eea
which acts trivially on the layer indices. We now compute the action of $C_{2z}$ on $H_K^{\phi=2\pi}$ directly:
\bea
\label{eq:HKpCC2zcom}
C_{2z} H^{\phi=2\pi}_K(\mbf{k}) C_{2z}^\dag &= \sigma_1 \bpm  v_F k_\th (-\sqrt{\frac{\phi}{2\pi}}  h(\pmb{\pi}) - \frac{1}{2}\sigma_2) &  T_1 + T_2 e^{-i k_2} \mathcal{H}^{-2\pi\mbf{b}_1} + T_3  e^{i k_1} \mathcal{H}^{-2\pi\mbf{b}_2} \\
T_1 + T_2 e^{i k_2} \mathcal{H}^{2\pi\mbf{b}_1} + T_3  e^{-i k_1} \mathcal{H}^{2\pi\mbf{b}_2}  & v_F k_\th (-\sqrt{\frac{\phi}{2\pi}}  h(\pmb{\pi})  + \frac{1}{2} \sigma_2)
\epm \sigma_1  \\
&=  \bpm  v_F k_\th (-\sqrt{\frac{\phi}{2\pi}}  \sigma_1 h(\pmb{\pi}) \sigma_1 + \frac{1}{2}\sigma_2) &  T_1 + T_3 e^{-i k_2} \mathcal{H}^{-2\pi\mbf{b}_1} + T_2  e^{i k_1} \mathcal{H}^{-2\pi\mbf{b}_2} \\
T_1 + T_3 e^{i k_2} \mathcal{H}^{2\pi\mbf{b}_1} + T_2  e^{-i k_1} \mathcal{H}^{2\pi\mbf{b}_2}  & v_F k_\th (-\sqrt{\frac{\phi}{2\pi}}  \sigma_1 h(\pmb{\pi}) \sigma_1 - \frac{1}{2} \sigma_2)
\epm  \\
\eea
where we used that $\sigma_1 T_2 \sigma_1 = T_3, \sigma_1 T_3 \sigma_1 = T_2$. Comparing \Eq{eq:HKpCC2zcom} with \Eq{eq:HmagKpblochLL} and using $h'(\pmb{\pi}) = - \sigma_1  h(\pmb{\pi}) \sigma_1$, we find that $C_{2z} H_{K}(\mbf{k}) C_{2z}^\dag = H_{K'}(-\mbf{k})$ and thus
\bea
C_{2z} &= \mu_1 \tau_0 \sigma_1 (-1)^{a^\dag a} \\
\eea
which anti-commutes with $P$ due to the valley matrices.

Lastly, we study the chiral symmetry on the $\ket{\mbf{k},n}$ basis. Chiral symmetry is very simple because it acts trivially on $\mbf{r}$ (see \Eq{eq:chiralsymreal}). It is direct to see that when $w_0 = 0$, the chiral symmetry
\bea
C_K = C_{K'} = \tau_0 \sigma_3 \mathbb{1}
\eea
(where $\mathbb{1}$ acts on the Landau levels) obeys
\bea
\{C_K, H^{\phi=2\pi}_K(\mbf{k}) \} = \{C_{K'}, H^{\phi=2\pi}_{K'}(\mbf{k}) \} = 0 ,
\eea
because \Eqs{eq:magblochham}{eq:magblochhamprime} have only $\sigma_1$ and $\sigma_2$ matrices when $w_0 = 0$. Thus we can choose
\bea
C = \mu_0 \tau_0 \sigma_3 \mathbb{1}
\eea
to be the chiral symmetry on the two-valley model. Again, we emphasize that any diagonal $\mu$ matrix is allowed, and the choice of $\mu_0$ carries no physical significance. This is because of the $U(1)$ valley symmetry.

The most important symmetry for the many-body properties of TBG is $C_{2z} P$ which acts as $ (\mu_1 \tau_0 \sigma_1 (-1)^{a^\dag a}) (i \mu_3 \tau_2 \sigma_0 (-1)^{a^\dag a}) = \mu_2 \tau_2 \sigma_1 \mathbb{1}$ on the magnetic translation group eigenbasis (see \Eq{eq:C2zsym} and \Eq{eq:Pvalley2}) and takes $\mbf{k} \to \mbf{k}$.

\subsection{Coulomb Interaction at  $\phi = 2\pi$}
\label{eq:appcoulomb}
\label{app:coulham}

We now study interaction effects within the flat bands. We will neglect the Zeeman splitting which is $\leq 2$meV. Although this is comparable to the flat band kinetic energy, the gap to the passive bands is $\geq 40$meV which justifies a strong coupling expansion where the single-particle flat bands are taken to be exactly flat at zero energy. In this case, there is an $SU(2)$ spin symmetry, and and an obvious $U(1)$ charge symmetry. Together, the  $U(1)$ charge symmetry, $U(1)$ valley symmetry, the $SU(2)$ spin symmetry, and the product $ C_{2z} P$ symmetry are all preserved at nonzero flux. These symmetries are essential in the many-body physics of TBG as we now discuss.

First, we set up the many-body notation. Explicitly, the Hilbert space is spanned by the operators $\gamma^\dag_{\mbf{k}, M, s, \eta}$ where $M= \pm1$ refers to the two flat bands, $s$ to the spin, and $\eta$ to the valley. The wavefunction of the $M$th band is
\bea
\label{eq:basisconstrc}
\braket{\mbf{r}, s,\eta| \gamma^\dag_{\mbf{k}, M, s, \eta}|0} = \sum_{l \al n} U_{l \al n}^{M, \eta}(\mbf{k}) \psi_{\mbf{k},l,\al,n}, \quad \psi_{\mbf{k},l,\al,n} =   \frac{1}{\sqrt{\mathcal{N}(\mbf{k})}} \sum_{\mbf{R}} e^{-i \mbf{k}\cdot \mbf{R}} T_{\mbf{R}} \frac{a^{\dag n}}{\sqrt{n!}} \psi_{0, l,\al} (\mbf{r}) \\
\eea
where $ U_{l \al n}^{M,\eta}(\mbf{k})$ are the matrix eigenvectors of the single-particle Hamiltonian in the $\eta$ valley and $\psi_{0,l,\al}$ is the zeroth Landau level state in the $l = \pm1$ layer and $\al = A,B$ sublattice. Note that the single-particle Hamiltonian is explicitly periodic: $H^{\phi=2\pi}(\mbf{k}+2\pi \mbf{G}) = H^{\phi=2\pi}(\mbf{k})$ and the states $\psi_{\mbf{k}, l, \al,n}$ are also explicitly periodic, so $\gamma^\dag_{\mbf{k}+2\pi \mbf{G}, M, s, \eta} = \gamma^\dag_{\mbf{k}, M, s, \eta}$. The two valleys are related by $C_{2z} P = \mu_2 \tau_2 \sigma_1 \mathbb{1}$ (see \App{eq:C2zPLL}), where $\mathbb{1}$ acts on the Landau level indices and obeys
\bea
- H^{\phi=2\pi}_{-\eta}(\mbf{k}) = (i \tau_2 \sigma_1 \mathbb{1})^\dag H^{\phi=2\pi}_{\eta}(\mbf{k}) (i \tau_2 \sigma_1 \mathbb{1}) \ .
\eea
which relates the Hamiltonians between the two valleys. Note that $C_{2z} P$ acts trivially on the Landau level indices.

We will now use $C_{2z} P$ to relate the eigenvectors of the two valleys. We focus on the two flat bands which we index by $M = \pm$. We denote by $U^\eta(\mbf{k})$ the matrix of occupied eigenvectors where each column $U_N^\eta(\mbf{k})$ is one of the flat band eigenvectors in the $\eta$ valley. Because $C_{2z} P$ anti-commutes with the single-particle Hamiltonian at each $\mbf{k}$, $C_{2z} P$ switches the energies of the flat bands and we have
\bea
\label{eq:etagauge}
i \tau_2 \sigma_1 \mathbb{1} U_N^{\eta=K}(\mbf{k}) &= \sum_M U^{\eta=-K}_M(\mbf{k}) [\nu_1]_{MN}
\eea
where $\nu_1$ is a Pauli matrix which exchanges the two bands. \Ref{2020arXiv200912376B} used $\xi_i$ to represent these Pauli matrices, but we reserve $\xi$ for $\xi_\mbf{q}(\mbf{k})$ the phase fact.  One can think of \Eq{eq:etagauge} as a $C_{2z} P$ gauge-fixing procedure for the eigenstates equivalent to the sewing matrices of \Ref{2020arXiv200912376B}. We point out that the sewing matrices were vital to constructing the Chern basis at zero flux in \Refs{2020arXiv200912376B}{2020arXiv200913530L} which is protected by $C_{2z}\mathcal{T}$. In the gauge fixing of \Ref{2020arXiv200912376B}, it was convenient to set the sewing matrix of $C_{2z}\mathcal{T}$ to be the identity. However at $2\pi$ flux, we will not need to make use of the sewing matrices because $C_{2z} \mathcal{T}$ is broken by the flux and there is no Chern basis.

Note that the Hamiltonian is independent of the spin, so the eigenstates at different spin are identical. Hence we adapt the form factor from \Eq{eq:formfactorM} to TBG in the $\eta$ valley via
\bea
\label{eq:formfactorMTBG}
\bar{M}^\eta_{MN}(\mbf{k},\mbf{q}) \equiv e^{i \xi_\mbf{q}(\mbf{k})} [U_\eta^\dag(\mbf{k}-\mbf{q}) (\tau_0  \sigma_0  \mathcal{H}^{\mbf{q}}) U_\eta(\mbf{k})]_{MN} \ .
\eea
As explained in \Eq{eq:formFactor}, $\bar{M}^\eta_{MN}(\mbf{k},\mbf{q})$ is not gauge-invariant because the eigenvectors in the columns of $U_\eta(\mbf{k})$ and $U_\eta(\mbf{k}-\mbf{q})$ are only defined up to an overall phase (or arbitrary unitary transformations at band touchings), meaning $U_\eta(\mbf{k}) \to U_\eta(\mbf{k}) V$ and $U_\eta(\mbf{k}-\mbf{q}) \to U_\eta(\mbf{k}-\mbf{q}) W$ where $V$ and $W$ are arbitrary $2\times 2$ unitary matrices at each $\mbf{k}$. Hence $\bar{M}^\eta_{MN}(\mbf{k},\mbf{q})$ is only defined up to the eigenvector gauge transformations $\bar{M}^\eta(\mbf{k},\mbf{q}) \to W^\dag \bar{M}^\eta(\mbf{k},\mbf{q}) V$. Under this transformation, only the singular values of $\bar{M}^\eta(\mbf{k},\mbf{q})$ are gauge-invariant. The singular values of $M$ are the eigenvalues of $M^\dag M$, which are invariant because $M^\dag M \to V^\dag M^\dag W W^\dag M V = V^\dag M^\dag M V$ has the same spectrum as $M^\dag M$. There is notable simplification at $\mbf{q} = 2\pi \mbf{G}$ where
\bea
\label{eq:MGinv}
\bar{M}^\eta_{MN}(\mbf{k},2\pi \mbf{G}) \equiv e^{i \xi_{2\pi \mbf{G}}(\mbf{k})} [U_\eta^\dag(\mbf{k}) (\tau_0  \sigma_0  \mathcal{H}^{\mbf{q}}) U_\eta(\mbf{k})]_{MN}
\eea
using the fact that $U_\eta(\mbf{k}) = U_\eta(\mbf{k}+2\pi \mbf{G})$ because the magnetic Bloch Hamiltonian $H^{\phi=2\pi}(\mbf{k})$ is explicitly $2\pi \mbf{G}$ periodic. In this case $\bar{M}^\eta_{MN}(\mbf{k},2\pi \mbf{G})$ is defined up to the eigenvector gauge transformation $\bar{M}^\eta(\mbf{k},\mbf{q}) \to V^\dag \bar{M}^\eta(\mbf{k},\mbf{q}) V$, and hence its \emph{eigenvalues} are gauge-invariant.

For brevity, we will not write the identity factors $\tau_0  \sigma_0$ going forward. Because the valleys are related by $C_{2z} P$ symmetry, \Eq{eq:etagauge} shows (writing $\eta$ as a subscript for convenience)
\bea
\bar{M}^{-\eta}(\mbf{k},\mbf{q}) &= e^{i \xi_\mbf{q}(\mbf{k})} U_{-\eta}^\dag(\mbf{k}-\mbf{q}) \mathcal{H}^{\mbf{q}} U_{-\eta}(\mbf{k})  \\
 &= \nu_1 e^{i \xi_\mbf{q}(\mbf{k})} U_{\eta}^\dag(\mbf{k}-\mbf{q}) \tau_2 \sigma_1\mathcal{H}^{\mbf{q}} \tau_2 \sigma_1 U_\eta(\mbf{k}) \nu_1 \\
&= \nu_1 e^{i \xi_\mbf{q}(\mbf{k})} U_{\eta}^\dag(\mbf{k}-\mbf{q}) \mathcal{H}^{\mbf{q}} U_\eta(\mbf{k}) \nu_1 \\
&= \nu_1 \bar{M}^{\eta}(\mbf{k},\mbf{q}) \nu_1 \\
\eea
so we see that $\bar{M}^{-\eta}(\mbf{k},\mbf{q})$ is related to $\bar{M}^{\eta}(\mbf{k},\mbf{q})$ by a unitary transformation.

The symmetries of $H_{int}$ are essential to understanding its groundstates. The $U(1)$ charge conversation and $SU(2)$ spin rotation in each valley give a $U(2) \times U(2)$ symmetry group that commutes with $O_\mbf{q}$. The two copies of $U(2)$ in separate valleys form $U(4)$ when the $C_{2z} P$ symmetry (which is not broken by magnetic field) is added to the symmetry group because $C_{2z} P$ interchanges the valleys. We refer the reader to  \Ref{2020arXiv200912376B} for a comprehensive treatment of the $U(4)$ algebra and irreps. Note that $C_{2z} P$ takes $\mbf{k}$ to $(-1)^2 \mbf{k}$ and so it commutes with Fourier modes of the density operator at all momenta.  (The anti-unitary symmetries $C_{2z} \mathcal{T}$ and $\mathcal{T}$ which broken in flux are not part of the $U(4)$ algebra.) The $U(4)$ symmetry of $H_{int}$ is a symmetry of the full Hamiltonian as well if we set the single particle Hamiltonian $H_0$ to zero ---otherwise $C_{2z} P$ anti-commutes with $H_0$. In general, incorporating kinetic energy will split the $U(4)$ irreps into $U(2) \times U(2)$ irreps which an energy difference on the order of the bandwidth $\sim 1$meV.

We now define the Coulomb interaction in terms of the charge density $\bar{n}(\mbf{r})$ where the bar indicates that the density is measured with respect to charge neutrality. This is equivalent to choosing the chemical potential of the system at half-filling. Discretizing $\mbf{k}$ for numerical convenience, the density modes are
\bea
\label{eq:rhoCNP}
 \bar{\rho}_{\mbf{q}}&=   \int d^2r \, e^{-i \mbf{q} \cdot \mbf{r}} \bar{n}(\mbf{r}) =  \sum_{\mbf{k}\in BZ} \sum_{MN, \eta,s} \bar{M}^\eta_{MN}(\mbf{k},\mbf{q}) \lp \gamma^\dag_{\mbf{k}- \mbf{q},M,\eta,s}  \gamma_{\mbf{k},N, \eta,s} - \frac{1}{2} \delta_{MN} \delta_{\mbf{q},0} \rp \ . \\
\eea
Note that $\mbf{k}$ is summed over the moir\'e BZ $k_i \in (-\pi,\pi)$ and $\mbf{q}$ is an arbitrary momentum in $\mathbb{R}^2$. The $\frac{1}{2} \delta_{MN} \delta_{\mbf{q},0} $ term shifts the eigenvalues of the density operator from $\{0,1\}$ to $\{-\frac{1}{2},\frac{1}{2}\}$.
From  \Ref{2020arXiv200912376B}, the Hamiltonian is
\bea
H_{int} &= \frac{1}{2} \int d^2r d^2r' \, \bar{n}(\mbf{r}) V(\mbf{r}-\mbf{r}') \bar{n}(\mbf{r}')  
 = \frac{1}{2} \int \frac{d^2q}{(2\pi)^2} \,  V(\mbf{q}) \bar{\rho}_{-\mbf{q}} \bar{\rho}_{\mbf{q}}, \quad \ V(\mbf{q}) = \pi \xi^2 U_\xi \frac{\tanh \xi |\mbf{q}|/ 2}{\xi |\mbf{q}|/ 2} \\
\eea
where the parameters of the screened Coulomb interaction are $\xi = 10$nm, $U_{\xi} = e^2/(\eps \xi) = 24$meV where $\eps$ is the dielectric constant. Because $\mbf{K}_{g} \gg 1/\xi$, intervalley scattering is strongly suppressed which justifies our decoupling of the valleys.

To understand the phase of TBG at fillings $\nu \in (-4,4)$ where $H_{int}$ describes the leading order electronic behavior, one must solve a strongly interacting problem. Our strategy to do so is to project $H_{int}$ into the flat bands. This is straightforwardly done by keeping only the terms with $M= \pm1$ in \Eq{eq:rhoCNP}.

Finally, we give the expression for the full Coulomb Hamiltonian:
\bea
H_{int} = \frac{1}{2 \Omega_{tot}} \sum_{\mbf{q}} V(\mbf{q}) \bar\rho_{-\mbf{q}} \bar\rho_{\mbf{q}} &=  \frac{1}{2  \Omega_{tot}} \sum_{\mbf{G}} \sum_{\mbf{q} \in BZ} O_{-\mbf{q},-\mbf{G}} O_{\mbf{q}, \mbf{G}}, \\
 O_{\mbf{q}, \mbf{G}} &= \sqrt{V(\mbf{q}+2\pi \mbf{G})} \sum_{\mbf{k}\in BZ} \sum_{\eta, s} \sum_{MN} \bar{M}^\eta_{MN}(\mbf{k},\mbf{q}+2\pi \mbf{G}) (\gamma^\dag_{\mbf{k}- \mbf{q},M, \eta,s}  \gamma_{\mbf{k},N, \eta,s}  - \frac{1}{2} \delta_{MN} \delta_{\mbf{q},0} )\ .
\eea
It is now a simple matter to project $H_{int}$ into the flat bands by restricting the sum to $M,N = \pm1$, the two approximately zero energy flat bands. To good approximation, the single particle Hamiltonian vanishes when projected to the flat bands because the bandwidth is $<2$meV in comparison to the $26$meV scale of the screened Coulomb interaction. Thus, the low energy Hamiltonian consists entirely of the projected $H_{int}$ operator.

\subsection{Exact Insulator Groundstates}
\label{eq:TBGBMgs}

To derive eigenstates of the interacting Hamiltonian as in \Ref{2020arXiv200913530L}, we rewrite $H_{int}$ by introducing a parameter $\la_{\mbf{G}}$:
\bea
\label{eq:inteig}
H_{int} &= \frac{1}{2  \Omega_{tot}} \sum_{\mbf{G}} \sum_{\mbf{q} \in BZ} O_{-\mbf{q},-\mbf{G}} O_{\mbf{q}, \mbf{G}} \\
&= \frac{1}{2  \Omega_{tot}} \sum_{\mbf{G}} \lp \la_{-\mbf{G}} O_{0,\mbf{G}} + \la_{\mbf{G}} O_{0,-\mbf{G}} - \la_{-\mbf{G}} \la_{\mbf{G}} + \sum_{\mbf{q}\in BZ} (O_{-\mbf{q},-\mbf{G}} -  \la_{-\mbf{G}} \delta_{\mbf{q},0}) (O_{\mbf{q}, \mbf{G}} -  \la_{\mbf{G}} \delta_{\mbf{q},0}) \rp
\eea
for any $\la_{\mbf{G}}$ which satisfies the Hermiticity condition $\la_{\mbf{G}} = \la^*_{-\mbf{G}}$. The purpose of introducing $\la_{\mbf{G}}$ is to make use of the flat metric condition \cite{2020arXiv200911301B} which is the approximation
\bea
\label{eq:FMCapp}
\text{flat metric condition: } \quad \bar{M}^\eta_{MN}(\mbf{k},2\pi \mbf{G}) = m_{\mbf{G}} \delta_{MN},
\eea
in other words that $\bar{M}^\eta(\mbf{k},2\pi \mbf{G})$ is proportional to the $2\times2$ identity at each $\mbf{G}$. (Importantly, \Eq{eq:MGinv} shows that the flat metric condition is ``gauge-invariant" under unitary rescalings of the eigenvectors.) If the flat metric condition is satisfied, we will be able to analytically construct groundstates of $H_{int}$ as in \Ref{2020arXiv200913530L}. To explain this, let us study $O_{0, \mbf{G}}$ which acts diagonally on $\mbf{k}$:
\bea
O_{0, \mbf{G}} =  \sqrt{V(2\pi \mbf{G})} \sum_{\mbf{k}\in BZ} \sum_{\eta, s} \sum_{MN} \bar{M}^\eta_{MN}(\mbf{k},2\pi \mbf{G}) (\gamma^\dag_{\mbf{k},M, \eta,s}  \gamma_{\mbf{k},N, \eta,s}  - \frac{1}{2} \delta_{MN}) \ .
\eea
Because $O_{-\mbf{q}, -\mbf{G}} = O_{\mbf{q},\mbf{G}}^\dag$, the term $\la_{-\mbf{G}} O_{0,\mbf{G}} + \la_{\mbf{G}} O_{0,-\mbf{G}}$ in \Eq{eq:inteig} is Hermitian and acts diagonally on $\mbf{k}$. If the flat metric condition in \Eq{eq:FMCapp} is satisfied, then
\bea
\label{eq:O0mu}
O_{0, \mbf{G}} &= \sqrt{V(2\pi \mbf{G})} \sum_{\mbf{k}\in BZ} \sum_{\eta, s} \sum_{MN} m_{\mbf{G}} \delta_{MN} (\gamma^\dag_{\mbf{k},M, \eta,s}  \gamma_{\mbf{k},N, \eta,s}  - \frac{1}{2} \delta_{MN} ) \\
&= \sqrt{V(2\pi \mbf{G})} \sum_{\mbf{k}\in BZ} \sum_{\eta, s} m_{\mbf{G}}(\sum_M \gamma^\dag_{\mbf{k},M, \eta,s}  \gamma_{\mbf{k},M, \eta,s}  - 1 ) \\
&= \sqrt{V(2\pi \mbf{G})} \sum_{\mbf{k}\in BZ} \sum_{\eta, s} m_{\mbf{G}}(\sum_M \gamma^\dag_{\mbf{k},M, \eta,s}  \gamma_{\mbf{k},M, \eta,s}  - 1 ) \\
\eea
which is proportional to the total particle number $N = \sum_{\mbf{k} \in BZ} \sum_{\eta,s, M} \gamma^\dag_{\mbf{k},M, \eta,s}  \gamma_{\mbf{k},M, \eta,s}$. This will be important for computing the chemical potential in \Eq{eq:mucompute}. In this case, the only nontrivial part of $H_{int}$ is the final $\mbf{q}$ sum in \Eq{eq:inteig}, which is also positive semi-definite. Because it is positive semi-definite, any state $\ket{\Psi}$ satisfying
\bea
\label{eq:Oqeig0}
(O_{\mbf{q}, \mbf{G}} - \la_{\mbf{G}} \delta_{\mbf{q},0}) \ket{\Psi} &= 0  \\
\eea
for some $\la_{\mbf{G}}$ is a groundstate. Note that states $\ket{\Psi}$ satisfying \Eq{eq:Oqeig0} are still eigenstates of $O_{\mbf{q}, \mbf{G}}$ and hence $H_{int}$ without the flat metric condition, but we cannot prove they are groundstates because $ \la_{-\mbf{G}} O_{0,\mbf{G}} + \la_{\mbf{G}} O_{0,-\mbf{G}}$ is not in general proportional to the identity.\footnote{Note that the flat metric condition in \Eq{eq:FMCapp} can be weakened to the requirement that $\sum_\mbf{G} (\la_{-\mbf{G}} O_{0,\mbf{G}} + \la_{\mbf{G}} O_{0,-\mbf{G}} )\propto N$ up to constant terms. } As shown in \Fig{fig:FMC}, that the flat metric condition holds to excellent accuracy for all $\mbf{G}$ except the first shell $\mbf{G} = \pm \mbf{b}_1, \pm \mbf{b}_2, \pm (\mbf{b}_1-\mbf{b}_2)$. On these momenta, the flat metric condition is only weakly violated. Hence the flat metric condition is a reliable approximation, and we can justify that the exact eigenstates in \Eq{eq:Oqeig0} are in fact groundstates at fixed filling.

We now construct states satisfying \Eq{eq:Oqeig0} at integer even density $\nu = -4, -2,0,2, 4$ around the charge neutrality point. Our trial state with $\nu + 4$ occupied flat bands takes the form
\bea
\label{eq:exacteigstate}
\ket{\Psi_{\nu}} &= \prod_{\mbf{k}}  \prod_{j}^{(\nu+4)/2}  \gamma^\dag_{\mbf{k},+, s_j, \eta_j}  \gamma^\dag_{\mbf{k},-, s_j, \eta_j}  \ket{0}
\eea
where the spin and valley indices are arbitrary and is implicit in the state. The operator $ O_{\mbf{q},\mbf{G}}$ has a simple action on $\ket{\Psi_{\nu}}$ because
\bea
\gamma^\dag_{\mbf{k}- \mbf{q},M, \eta,s}  \gamma_{\mbf{k},N, \eta,s} \ket{\Psi_\nu} = 0 \text{ if } \mbf{q} \neq 0
\eea
because both $M = \pm1$ bands are fully occupied or unoccupied at every $\mbf{k}$. We also use the fact
\bea
\sum_{MN} \bar{M}^\eta_{MN}(\mbf{k},2\pi\mbf{G}) \gamma^\dag_{\mbf{k},M, \eta,s}  \gamma_{\mbf{k},N, \eta,s} \gamma^\dag_{\mbf{k}, +, \eta, s}\gamma^\dag_{\mbf{k}, -, \eta, s}\ket{0} &= \sum_{MN} \bar{M}^\eta_{MN}(\mbf{k},2\pi\mbf{G}) \gamma^\dag_{\mbf{k},M, \eta,s} (\delta_{N,+} \gamma^\dag_{\mbf{k}, -, \eta,s} - \delta_{N,-}\gamma^\dag_{\mbf{k}, +, \eta, s}  )\ket{0} \\
&= \sum_{M}  (\bar{M}^\eta_{M+}(\mbf{k},2\pi\mbf{G}) \gamma^\dag_{\mbf{k},M, \eta,s}  \gamma^\dag_{\mbf{k}, -, \eta,s} - \bar{M}^\eta_{M-}(\mbf{k},2\pi\mbf{G}) \gamma^\dag_{\mbf{k},M, \eta,s} \gamma^\dag_{\mbf{k}, +, \eta, s}  )\ket{0} \\
&= ( \bar{M}^\eta_{++}(\mbf{k},2\pi\mbf{G}) \gamma^\dag_{\mbf{k},+, \eta,s}  \gamma^\dag_{\mbf{k}, -, \eta,s} - \bar{M}^\eta_{--}(\mbf{k},2\pi\mbf{G}) \gamma^\dag_{\mbf{k},-, \eta,s} \gamma^\dag_{\mbf{k}, +, \eta, s}  )\ket{0} \\
&= \Tr [\bar{M}^\eta(\mbf{k},2\pi\mbf{G})] \gamma^\dag_{\mbf{k},+, \eta,s}  \gamma^\dag_{\mbf{k}, -, \eta,s}\ket{0} \\
\eea
which we use to calculate (with $\mbf{q}\in BZ$ and recalling that $\ket{\Psi_\nu}$ contains flavors $s_j$ and $\eta_j$)
\bea
\label{eq:Metadisc}
 O_{\mbf{q},\mbf{G}} \ket{\Psi_\nu}  &= \delta_{\mbf{q},0} \sqrt{V(2\pi\mbf{G})} \sum_{\mbf{k}\in BZ} \sum_{\eta, s} \sum_{MN} \bar{M}^\eta_{MN}(\mbf{k},2\pi\mbf{G}) (\gamma^\dag_{\mbf{k},M, \eta,s}  \gamma_{\mbf{k},N, \eta,s}  - \frac{1}{2} \delta_{MN}) \ket{\Psi_\nu} \\
 &= \delta_{\mbf{q},0} \sqrt{V(2\pi\mbf{G})} \sum_{\mbf{k}\in BZ} \sum_{\eta, s}  (\sum_j \delta_{s,s_j} \delta_{\eta,\eta_j} \Tr  [\bar{M}^\eta(\mbf{k},2\pi\mbf{G})]  - \frac{1}{2}  \Tr  [\bar{M}^\eta(\mbf{k},2\pi\mbf{G})]) \ket{\Psi_\nu} \\
 &= \delta_{\mbf{q},0} \sqrt{V(2\pi\mbf{G})} \sum_{\mbf{k}\in BZ} (\frac{\nu+4}{2} \Tr  [\bar{M}^\eta(\mbf{k},2\pi\mbf{G})]  - \frac{4}{2} \Tr  [\bar{M}^\eta(\mbf{k},2\pi\mbf{G})] ) \ket{\Psi_\nu} \\
 &= \nu \delta_{\mbf{q},0}  \sqrt{V(2\pi\mbf{G})} \sum_{\mbf{k}\in BZ} \frac{1}{2} \Tr \bar{M}^\eta(\mbf{k},2\pi\mbf{G}) \ket{\Psi_\nu}  \\
 \eea
 where we used that $ \Tr \bar{M}^{-\eta}(\mbf{k},2\pi\mbf{G})  =  \Tr \nu_1 \bar{M}^\eta(\mbf{k},2\pi\mbf{G}) \nu_1 =  \Tr \bar{M}^\eta(\mbf{k},2\pi\mbf{G})$. We use the abbreviation
 \bea
 \Tr \bar{M}(\mbf{k},2\pi\mbf{G}) =  \Tr \bar{M}^{-\eta}(\mbf{k},2\pi\mbf{G}) =  \Tr \bar{M}^{\eta}(\mbf{k},2\pi\mbf{G})
 \eea
 to emphasize that the trace is independent of the valley. Consulting \Eq{eq:Oqeig0}, we find that $\ket{\Psi_\nu}$ is an exact eigenstate provided we choose \cite{2020arXiv200913530L}
\bea
\label{eq:mabdadef}
\la_{\mbf{G}} =  \nu \sqrt{V(2\pi\mbf{G})} \sum_{\mbf{k}\in BZ}  \frac{1}{2} \Tr \bar M(\mbf{k},2\pi\mbf{G}) \ .
\eea
Using \Eq{eq:Mherm}, we find $\la_{\mbf{G}} = \la_{-\mbf{G}}^*$. Returning to \Eq{eq:inteig} and acting on the state $\ket{\Psi_\nu}$, we find
\bea
H_{int} \ket{\Psi_\nu} &=  \frac{1}{2  \Omega_{tot}} \sum_{\mbf{G}} \la_{-\mbf{G}} O_{0,\mbf{G}} \ket{\Psi_\nu} + \la_{\mbf{G}} O_{0,-\mbf{G}}\ket{\Psi_\nu} - \la_{-\mbf{G}} \la_{\mbf{G}} \ket{\Psi_\nu} \\
&=  \frac{1}{2 \Omega_{tot}} \lp \sum_{\mbf{G}} \la_{-\mbf{G}} \la_{\mbf{G}}  + \la_{\mbf{G}}\la_{-\mbf{G}}  -  \la_{-\mbf{G}} \la_{\mbf{G}} \rp \ket{\Psi_\nu} \\
&= \lp  \frac{1}{2 \Omega_{tot}} \sum_{\mbf{G}}  \la_{-\mbf{G}} \la_{\mbf{G}} \rp \ket{\Psi_\nu} \\
\eea
which gives the exact energy of the state (not including the chemical potential):
\bea
\frac{E_{int}}{\Omega_{tot}} =  \frac{\nu^2}{2} \sum_{\mbf{G}} V(2\pi\mbf{G}) \left| \frac{1}{ \Omega_{tot}}\sum_{\mbf{k}\in BZ}  \frac{1}{2} \Tr \bar M(\mbf{k},2\pi\mbf{G}) \right|^2 \ .
\eea
The dominant contribution is from the $\mbf{G} = 0$ term where $ \Tr \bar M^\eta(\mbf{k},0) = \Tr U_\eta^\dag(\mbf{k})U_\eta(\mbf{k})  = 2$. Numerically calculating the sums at $\mbf{G} \neq 0$, we find that only the first shell of $\mbf{G}$ consisting of $\pm \mbf{b}_1,\pm \mbf{b}_2, \pm (\mbf{b}_1-\mbf{b}_2)$ contributes significantly to the sum due to the exponential fall-off of $M(\mbf{k},\mbf{q})$ with $\mbf{q}$. In general, we cannot guarantee that $\ket{\Psi_\nu}$ is the groundstate without the flat metric condition. This assumption can be tested using exact diagonalization studies for small systems \cite{2020arXiv201000588X}, which we leave for future work. However, $E_{int}=0$ at $\nu = 0$ so it must be a groundstate because $H_{int}$ is positive semi-definite.

If the FMC holds such that $M(\mbf{k},2\pi\mbf{G}) = m_\mbf{G} \mathbb{1}$, we can obtain an exact expression for the chemical potential $\mu$ at even fillings. Using \Eq{eq:O0mu}, we compute
\bea
\label{eq:mucompute}
 \frac{1}{2  \Omega_{tot}} \sum_{\mbf{G}} \lp \la_{-\mbf{G}} O_{0,\mbf{G}} + \la_{\mbf{G}} O_{0,-\mbf{G}} \rp &= \frac{1}{2\Omega_{tot}} \lp \sum_\mbf{G} \sqrt{V(2\pi \mbf{G})}  \la_{-\mbf{G}}m_\mbf{G} \rp N + h.c. + \text{const}  = \mu N + \text{const}, \\
 \eea
 where the chemical potential $\mu$ is given by
 \bea
 \label{eq:muwithFMC}
 \mu &= \frac{1}{\Omega_{tot}} \sum_\mbf{G} \sqrt{V(2\pi \mbf{G})}  (\la_{-\mbf{G}}m_\mbf{G} +\la_{\mbf{G}}m_{-\mbf{G} })/2 \\
 &= \nu \sum_\mbf{G} \Omega^{-1} V(2\pi \mbf{G})  (m_{-\mbf{G}} m_\mbf{G} +m_{\mbf{G}}m_{-\mbf{G} })/2 \\
  &= \nu \sum_\mbf{G} \Omega^{-1} V(2\pi \mbf{G})  |m_{\mbf{G}}|^2 \\ 
\eea
and we used \Eq{eq:mabdadef} and the fact that $\frac{1}{2}\Tr M(\mbf{k},2\pi \mbf{G}) = m_\mbf{G}$. Note that $\mu = 0$ at $\nu=0$ with or without the FMC because $\la_\mbf{G} \propto \nu$. Because $H_{int} - \mu N$ is positive definite (see \Eq{eq:inteig}), we can guarantee $\ket{\Psi_\nu}$ are many-body groundstates at chemical potential $\mu$ in the FMC. If we do not make the FMC approximation, then we cannot determine an exact expression for the chemical potential. The approximation we make is to compute an average value of $m_\mbf{G}$ over the BZ:
\bea
 \mu \approx \nu \sum_\mbf{G} \Omega^{-1} V(2\pi \mbf{G}) \left| \frac{1}{N_{M}} \sum_{\mbf{k} \in BZ} \frac{1}{2} \Tr M(\mbf{k},2\pi \mbf{G}) \right|^2 
\eea
which reduces to \Eq{eq:muwithFMC} if the FMC holds.

\setcitestyle{numbers,square}

\end{document}